\documentclass[a4paper,11pt]{article}
\pdfoutput=1
\usepackage{jheppub}
\makeatletter
\gdef\@fpheader{}
\makeatother
\usepackage[T1]{fontenc}

\usepackage{amsmath,amssymb,mathtools,bm}
\usepackage{booktabs}
\usepackage{microtype}
\usepackage{enumitem}
\usepackage{xcolor}
\usepackage{cancel}
\usepackage{tikz}
\usepackage{float}
\usepackage{needspace}
\usepackage[nameinlink,capitalise]{cleveref}
\crefname{section}{Sec.}{Secs.}
\Crefname{section}{Sec.}{Secs.}
\crefname{subsection}{Sec.}{Secs.}
\Crefname{subsection}{Sec.}{Secs.}
\crefname{subsubsection}{Sec.}{Secs.}
\Crefname{subsubsection}{Sec.}{Secs.}
\crefname{equation}{Eq.}{Eqs.}
\Crefname{equation}{Eq.}{Eqs.}
\crefrangeformat{equation}{Eqs.~(#3#1#4--#5#2#6)}
\Crefrangeformat{equation}{Eqs.~(#3#1#4--#5#2#6)}
\numberwithin{equation}{section}

\newcommand{\M}{\mathcal M}

\newcommand{\Acal}{\mathcal A}
\newcommand{\Dcal}{\mathcal D}

\newcommand{\Ocal}{\mathcal O}
\newcommand{\Rcal}{\mathcal R}

\newcommand{\Disc}{\operatorname{Disc}}

\newcommand{\Tr}{\operatorname{Tr}}
\newcommand{\dd}{\mathrm d}
\newcommand{\ii}{\mathrm i}
\newcommand{\e}{\mathrm e}

\newcommand{\sumint}{\mathop{\vcenter{\hbox{\ooalign{%
  \hfil$\displaystyle\sum$\hfil\cr
  \hfil$\displaystyle\int$\hfil\cr}}}}\displaylimits}

\title{\boldmath Infrared-Finite Positivity Bounds in 4D Gravity}

\author{Ziyu Dong}
\affiliation{The Abdus Salam International Centre for Theoretical Physics,\\
Strada Costiera 11, 34151 Trieste, Italy}
\emailAdd{zdong@ictp.it}

\abstract{
We formulate an S-matrix bootstrap in an exactly flat half-space, where
boundary reflection provides geometric infrared control for four-dimensional
gravity. Under the channel positivity conditions, we obtain an
infrared-finite positivity bound on local four-scalar Wilson
coefficients that includes a finite gravitational correction of order $G_N$
and is derived from the full UV absorptive spectrum without specifying a UV
theory, imposing a spectral ansatz, requiring weak coupling or expanding the
spectrum perturbatively. At fixed dimensionless external state parameters,
the coefficient of the leading gravitational correction is independent of
the EFT scale and requires neither an arbitrary soft resolution
scale nor a model-dependent UV scale.

Gaussian wave-packet scattering localizes interactions within the flat region
and directly probes
bulk Wilson coefficients while exponentially suppressing sensitivity to the
boundary. Long-range gravitons remain sensitive to the boundary and
give a finite contribution. The smearing functional passes numerical
positivity tests with general half-space absorptive kernels in both crossed
channels. An explicit string example shows how boundary reflection suppresses
the collective contribution of string resonances to the absorptive part of
the amplitude, connecting infrared control of low-energy scattering to the
UV spectrum of quantum gravity in the same dispersion relation.
The framework opens a route to infrared-finite constraints on effective
interactions across particle physics and gravity.
}


\begin{document}
\maketitle
\raggedbottom

\section{Introduction}
\label{sec:introduction}

Effective field theory (EFT) describes heavy physics through local interactions among the particles
that remain accessible at low energies. Nonredundant operator bases systematically organize these
local interactions through Hilbert-series and on-shell methods
\cite{Henning:2015daa,Henning:2019enq,Dong:2021vxo,Dong:2022mcv,Harlander:2023ozs,Dong:2026swi}. A
central question is which of these interactions can arise from a consistent microscopic theory.
Scattering provides a way to answer this question without specifying that theory in detail.
Unitarity, analyticity and crossing relate low-energy Wilson coefficients to physical states at
higher energies, constraining both the strength of an interaction and the scale at which new physics
must appear \cite{Adams:2006sv,Arkani-Hamed:2020blm,Caron-Huot:2021rmr}. With dynamical gravity,
however, a local interaction is measured in a scattering process that also samples arbitrarily long
distances. In four dimensions this long-range contribution can make a dispersive EFT bound depend on
an infrared (IR) regulator. The question addressed here is how to retain the information about local
interactions while defining an infrared-finite scattering problem on both sides of the dispersion
relation.

This question already matters for the lowest nontrivial three- and four-point couplings.
Higher-derivative graviton interactions constrain departures from general relativity (GR) and the
onset of additional states \cite{Camanho:2014apa,Caron-Huot:2022ugt}. Photon interactions connect
light-by-light scattering to gravitational corrections and charged-black-hole extremality
\cite{Henriksson:2022oeu}. Related dispersive methods constrain Standard Model effective
interactions, pion amplitudes at large $N_c$ and anomaly-induced couplings, including the
chiral-gravitational anomaly \cite{Remmen:2019cyz,Albert:2023jtd,Dong:2024omo,Dong:2025dpy}. For the
lowest EFT coefficients, the gravitational contribution cannot in general be avoided by taking more
energy derivatives without also losing the quantity one intended to constrain. A treatment of the
infrared problem is therefore part of the physical content of such bounds.

The basic dispersive argument starts from probability conservation. Writing $S=1+\ii T$, unitarity
implies, for a normalized incoming state $|\Psi\rangle$,
\begin{equation}
  2\,\operatorname{Im}\langle\Psi|T|\Psi\rangle
  =\sumint_X\bigl|\langle X|T|\Psi\rangle\bigr|^2\geq0.
  \label{eq:intro-optical-theorem}
\end{equation}
The sum includes the phase space of every allowed intermediate state. Analyticity expresses a
low-energy contour integral in terms of discontinuities on physical cuts, and crossing relates those
cuts to different scattering channels. Provided the amplitude has sufficiently controlled growth at
large complex energy, the outer contour can be discarded. Suitable EFT coefficients are then related
to positive moments of the physical spectrum. In this way a low-energy measurement constrains the
possible ultraviolet (UV) completion, while the positivity of high-energy probabilities restricts
the allowed low-energy interactions.

Forward elastic scattering gives the simplest implementation because its absorptive part has the
sign in \eqref{eq:intro-optical-theorem}. Partial-wave unitarity supplies more detailed information,
resolving the intermediate states by angular momentum. The angular functions are Legendre
polynomials for scalars and Wigner matrices for spinning particles \cite{Jacob:1959at}.
Momentum-transfer derivatives extend forward bounds \cite{deRham:2017avq}, while full crossing
symmetry relates different low-energy expansions and yields additional constraints on their spectral
moments \cite{Tolley:2020gtv,Arkani-Hamed:2020blm,Sinha:2020win}. Smearing over spacelike transfer
provides positive combinations that retain finite-angle information
\cite{Caron-Huot:2021rmr,Caron-Huot:2022ugt}. These methods can constrain a region of
Wilson-coefficient space rather than the sign of a single forward coefficient
\cite{EliasMiro:2022xaa,Bonnefoy:2025uzf}. When light-particle loops contribute, their cuts must
also be included or subtracted consistently \cite{Bellazzini:2021oaj,Chang:2025cxc}.

Gravity obstructs this argument in a particularly direct way. Tree-level graviton exchange produces
a forward pole proportional to $G_Ns^2/(-t)$ and a logarithmic eikonal phase in the two-dimensional
transverse plane. In the usual four-dimensional flat-space construction, transfer smearing leaves an
infrared logarithm in the gravitational contribution to the bound
\cite{Caron-Huot:2021rmr,Caron-Huot:2022ugt}. Beyond tree level, soft virtual gravitons also make
exclusive amplitudes between free Fock states regulator dependent at nonzero transfer
\cite{Weinberg:1965nx,Akhoury:2011kq,Naculich:2011ry}.

The local EFT Lagrangian remains meaningful, but its polynomial contact terms are only part of the
amplitude. For example, the first few four-scalar contact terms are
\begin{equation}
  \M_{\rm contact}(s,t)
  =\frac{g_2}{\Lambda_{\rm EFT}^4}(s^2+t^2+u^2)
  +\frac{g_3}{\Lambda_{\rm EFT}^6}stu+\cdots,
  \qquad s+t+u=0.
  \label{eq:intro-local-contact}
\end{equation}
Massless exchange and light-particle thresholds add nonanalytic terms that can extend to zero energy
\cite{Donoghue:1994dn}. A bound on a local coefficient must therefore specify how the complete
scattering quantity, including these long-distance contributions, is defined.

The same issue has a classical manifestation in gravitational time delay. A probe continues to
accumulate a delay along its long-range trajectory, and comparison with free propagation has
logarithmic endpoint dependence. An actual radar measurement instead compares specified emitters and
receivers at finite separation, as in the Shapiro proposal and the Cassini experiment
\cite{Shapiro:1964uw,Bertotti:2003rm}. Quantum measurements likewise resolve only part of the
radiation produced in scattering. Inclusive probabilities sum over unresolved quanta, whereas
dressed-state constructions incorporate long-range fields into the asymptotic states. These examples
show why the choice of external states and the measurement resolution are physical inputs to a
scattering problem. A propagation distance, a detector's energy resolution and a dressing profile
encode different information, so translating any of them into a dispersive prescription requires an
explicit construction.

Faddeev--Kulish (FK) dressing provides an important approach to infrared-finite scattering
\cite{Kulish:1970ut,Ware:2013zja}. It accompanies hard particles with coherent soft clouds and
modifies the long-range asymptotic phase. Such a cloud contains components with arbitrarily many
soft quanta, and its mean particle number can diverge while its total soft energy remains finite.
The profile specifying the unresolved sector is part of the state. In particular, an energy or
spatial-momentum resolution is defined relative to a frame, as in treatments of four-dimensional
soft radiation and its logarithmic corrections \cite{Sahoo:2018lxl}. For a rotationally invariant
dressing profile fixed in a timelike frame, the Lorentz symmetry $SO(1,3)$ is reduced to the spatial
rotation group $SO(3)$ of that frame, while translations in all four spacetime directions are
retained. Simultaneously transforming the frame and all momenta still gives a covariant description,
but a boost of the hard configuration with the frame held fixed changes the scattering data.
Consequently, at fixed external masses and with conserved hard momenta, a generic four-scalar
dressed amplitude depends on five independent kinematic variables rather than only $s$ and $t$.

Using such an object in a bootstrap requires crossing the complete dressed external states and
controlling its analytic continuation. Infrared cancellation by itself does not determine the
large-complex-energy behavior needed in a dispersion relation. In particular, the assumptions
entering Regge bounds must be checked for the scattering object under consideration
\cite{Haring:2022cyf}, and a leading soft approximation need not remain uniform as $s$ becomes large
at fixed $t$. Analyticity and positivity also require separate analysis in systems with
spontaneously broken Lorentz invariance \cite{Creminelli:2022onn,Creminelli:2023kze}, where the
preferred frame has a different origin from a prescribed dressing profile. These questions motivate
distinguishing the existence of infrared-finite scattering from its use in a positive dispersive sum
rule. We discuss the flat-space obstruction with this distinction in mind, before turning to a
scattering construction in which infrared control comes from spatial geometry.

Anti-de Sitter (AdS) space offers a successful geometric approach. Boundary correlators encode bulk
interactions \cite{Witten:1998qj,Heemskerk:2009pn,Penedones:2010ue}, and positive conformal field
theory (CFT) dispersion relations give sharp constraints on local bulk couplings, including
infrared-finite bounds in $\mathrm{AdS}_4$ \cite{Caron-Huot:2021enk}. Curvature supplies an infrared
scale, while a sufficiently large radius separates the local interaction from curvature corrections.
Sending the radius to infinity also removes this geometric infrared scale, so flat-space infrared
sensitivity returns in the scattering limit. The relation to flat-space scattering is studied
through suitable limits of the AdS states and correlators.

A general curvature scale does not automatically provide the same infrared control. For example, the
Euclidean problem associated with a massless minimally coupled scalar on de Sitter (dS) space
retains a normalizable constant zero mode \cite{Allen:1987tz,Folacci:1992xc}. A zero mode is a
zero-eigenvalue mode of the relevant propagation operator allowed by its boundary conditions,
including those at infinity on a noncompact geometry. For gravitational shock waves, the transverse
operator whose inverse defines the Green function contains a curvature term. On the spherical
wavefront of $\mathrm{dS}_4$, its zero modes are the $\ell=1$ harmonics rather than the constant
mode, and defining its inverse requires a compatible source and a prescription for these zero modes.
On the hyperbolic wavefront of $\mathrm{AdS}_4$, the corresponding operator instead has a positive
normalizable spectrum with the usual decaying boundary condition \cite{Podolsky:2001vu}. Thus,
placing scattering in $\mathrm{dS}_4$ with curvature radius equal to the Hubble radius does not by
itself justify using that radius as a phenomenological infrared cutoff. For a general time-dependent
Friedmann--Lema\^itre--Robertson--Walker (FLRW) geometry or another evolving background, whether a
cosmological scale provides an effective infrared cutoff must be determined from the propagation
problem, the states and the observable under study.

We pursue a complementary possibility, in which an exterior geometry controls long-distance
propagation while the interaction region remains exactly flat and admits asymptotic scattering
states. The bulk Wilson coefficients can then be probed through localized wavefunction overlaps
without requiring a curvature expansion of the local interaction. Flat-space approaches using
specified Regge assumptions and finite-energy sum rules also yield infrared-finite bounds on
low-energy EFT Wilson coefficients in the presence of gravity \cite{Noumi:2022wwf}.
Resolution-dependent amplitudes have also been proposed as a route to infrared-safe positivity
bounds in four-dimensional theories with long-range interactions, with the radiation resolution
exponentially smaller than the hard scales \cite{Bellazzini:2025bay}. Here the infrared prescription
is built into the scattering states and propagators. The same geometric problem must supply the
low-energy EFT matrix elements, the physical absorptive data and the high-energy behavior used to
relate them.

To identify a useful geometry, we first examine eikonal propagation, where long-range exchange is
controlled by a transverse Green function. Product spacetimes $\mathbb M_{1,1}\times\Sigma_2$, with
$\mathbb M_{1,1}$ denoting two-dimensional Minkowski spacetime, isolate the relevant transverse
spectral question. When $\Sigma_2$ is the compact sphere $S^2$, its Laplacian has a discrete
spectrum that includes a constant zero mode. In contrast, on the hyperbolic plane $H_2$ the
normalizable spectrum has a positive lower edge, with no normalizable zero mode. Removing the
spherical zero mode by a uniform compensating source changes the source background and produces a
sign-changing Green function. This extra source modification is unnecessary on $H_2$, where the
decaying Green function follows from the geometry itself. The spherical examples below show that a
discrete spectrum from compactness, even together with real-axis unitarity of the
zero-mode-subtracted amplitude, is insufficient to justify the required dispersion relation. To keep
the interaction region exactly flat while retaining a hyperbolic exterior, one can instead join a
flat core to an asymptotically hyperbolic geometry. Its scattering modes, however, depend on the
details of the joining region \cite{Mazzeo:1987resolvent,Guillarmou:2005resolvent}. These
comparisons lead us to seek a simpler geometry with explicit modes that retains as much of the
four-dimensional Poincar\'e symmetry as possible.

The resulting construction uses the flat half-space $y>0$ with a Dirichlet boundary at $y=0$. The
boundary preserves $ISO(1,2)$ along the wall, and scalar modes are sine waves formed from a plane
wave and its reflected image. The image also makes the long-range cancellation explicit. In the
two-dimensional transverse problem, with $X=(x,y)$ and $\bar X'=(x',-y')$, the Green function is
\begin{equation}
 G_h(X,X')=\frac{1}{2\pi}\log\frac{|X-\bar X'|}{|X-X'|}.
 \label{eq:intro-image-green}
\end{equation}
The infrared divergence cancels between the direct and image terms. For scattering localized at
finite distance from the wall, the remaining long-range field has a decaying power-law tail and
gives finite exchange. Moving the wave packets far from the wall recovers flat-space infrared
behavior while keeping the spacetime background and its bulk S-matrix operator fixed.

The boundary has a geometric realization through a flat interior joined to an AdS exterior. In the
limit described in the body of the paper, the boundary-side AdS region induces the Dirichlet
condition. Inserting a flat strip between the boundary-side and horizon-side AdS regions gives the
Dirichlet--Neumann (DN) completion in the same limit. This distinction matters beyond the localized
tree-level calculation. The half-space has no spectral gap, and a freely spreading packet can
eventually explore distances arbitrarily far from the wall. The resulting access to the soft sector
matters when gravitational loops are included. The finite DN strip supplies a gap that controls this
residual soft sector. Both constructions retain an exactly flat interaction region in which the
local EFT interactions are evaluated. In the explicit bound we calculate the leading low-energy
gravitational contribution, appropriate when $G_N\Lambda_{\rm EFT}^2\ll1$. No corresponding
weak-coupling expansion is imposed on the ultraviolet spectrum, which may contain contributions at
arbitrary loop order and coupling strength.

The wall removes normal-momentum conservation and changes the kinematics of a four-point amplitude to
\begin{equation}
 \M_h(s_\parallel,t_\parallel;k_1,k_2,k_3,k_4),
 \qquad m_{i\parallel}^2=m_i^2+k_i^2.
 \label{eq:intro-six-invariants}
\end{equation}
The six invariants comprise two parallel Mandelstam variables and four real normal-mode momenta. At
fixed $k_i$, the parallel problem has the effective masses $m_{i\parallel}$, which organize analytic
continuation and crossing while the normal labels remain real. Reflecting boundary conditions
conserve probability, and local propagation in the bulk provides the setting for the scattering
assumptions discussed below. We also develop the construction for spinning states, giving explicit
wave-packet and polarization conventions for massive and massless integer-spin particles
\cite{Jacob:1959at}. Reflection acts on polarization indices together with momentum, preserving the
corresponding transversality conditions. This supplies the state construction needed to extend the
example of scalar scattering to polarized channels. Lorentz-covariant scalar and spinor wave packets
are discussed in \cite{Oda:2021tiv,Oda:2023qek}.

The next step is to identify positive smearing functionals for these states. The smearing functional
pairs the transition operator $T$ with a smearing kernel $K$ of correlated two-particle
wavefunctions, including the momentum integrals. It need not describe four independently chosen
packets. A positive semidefinite (PSD) kernel is a nonnegatively weighted sum of projectors onto
two-particle states. Each term pairs the same incoming and outgoing state, so its absorptive pairing
is positive by the optical theorem. Starting from flat-space forward or partial-wave kernels, we
construct a linear map $V$ that localizes the common normal position and a reflection map $\Pi_h$
that imposes the boundary condition. Positivity is preserved by
\begin{equation}
 K\succeq0
 \quad\Longrightarrow\quad
 K_h=\Pi_h V K V^\dagger\Pi_h^\dagger\succeq0.
 \label{eq:intro-positive-map}
\end{equation}
This gives a systematic construction of positive kernels for a physical channel. It also identifies
the momentum-space phase-space measure for smearing the $t$-channel momentum transfer, whose
wall-parallel and wall-normal components must now be integrated separately.

We then use this structure to define a common smearing ansatz for scalar scattering in the
dispersive variables and evaluate its responses in the crossed $s$ and $u$ physical channels. The
numerical tests cover momentum-conserving resonance blocks and general half-space absorptive blocks
with nonconserving normal profiles. They vary the profile shape, spectral energy and angular
momentum. The sampled responses are positive in both channels. Complementary analytic checks use a
small width of the wave packets in normal momentum, whose conjugate position-space width is
inversely proportional to that momentum width. They establish positivity at leading order in this
small-momentum-width limit and positive asymptotic behavior at large mass and large angular
momentum. Together these results support the half-space positivity bootstrap and its application to
an explicit EFT bound.

For four identical massless scalars, we evaluate the local contact terms and the full tensor
graviton exchange with the same normalized functional. For the correlated wave packets specified by
this functional, the gravitational contribution is infrared finite and equals $G_N$ times a finite
number. The leading contact term then gives a bound on the dimensionless coefficient $g_2$ with a
finite gravitational allowance of order $G_N\Lambda_{\rm EFT}^2$, as made explicit in
\eqref{eq:halfspace-computed-bound}. The momentum-space smearing measure and wave-packet profiles
have a particularly transparent interpretation in the eikonal limit. Their parameters set the
position-space beam widths parallel and normal to the wall and the beam's distance from it. The
functional itself also applies at finite scattering angles, where it describes correlated
wavefunctions without requiring a narrow-beam picture. Boundary sensitivity of local vertices is
controlled by the exponentially suppressed tails of the wavefunction overlap.

Infrared-finite bounds also raise a question about ultraviolet physics. In both AdS correlator
bootstrap and the half-space S-matrix bootstrap, what does infrared control of the low-energy
amplitude imply for the ultraviolet completion of gravity? A dispersion relation reconstructs the
long-range graviton contribution from an integral over physical intermediate states
\cite{Haring:2024wyz}. Its infrared singularity is therefore also encoded in the integrated
high-energy spectrum. A consistent geometric prescription must also act on the UV-completed theory
in the same scattering or correlator problem. In eikonal $2\to2$ scattering, a large center-of-mass
energy $\sqrt{s}$ does not confine sensitivity to the UV completion to small impact parameters $b$.
The absorptive part of the amplitude can contain collective contributions extending over large
transverse distances. It is useful here to distinguish the elastic intermediate states that
unitarize the leading gravitational eikonal from microscopic resonances of a UV completion. Their
contributions realize different spectral mechanisms for reconstructing the long-range interaction
\cite{Haring:2024wyz}.

We examine this resonance mechanism explicitly using a tree-level closed-superstring amplitude. In
its Regge regime, summing the string resonance tower produces a transverse diffusion profile whose
mean-square width grows logarithmically with energy
\cite{Amati:1987uf,Brower:2006ea,DiVecchia:2023frv}. This extended profile is a collective property
of the tower, while each individual fixed-spin resonance can be approximated by a point-particle
exchange. Integrating the tower's spectral contribution reproduces the infrared singularity of the
low-energy graviton pole. The half-space boundary acts on this collective propagation as well. When
the diffusion length becomes comparable to the distance of the external wave packets from the wall,
direct and reflected contributions cancel in the region where the interaction is probed. We track
this suppression in both the real and absorptive Regge kernels. It gives a physical interpretation
of infrared-finite gravitational sum rules in which the same geometry controls both sides of the
dispersion relation.

Section~\ref{sec:flat-obstruction} explains the flat-space infrared obstruction.
Section~\ref{sec:transverse-modes} compares candidate geometries through eikonal propagation and
their implications for dispersion relations. Section~\ref{sec:halfspace-scattering} develops the
half-space and DN scattering states, including their kinematics and polarization transport.
Section~\ref{sec:halfspace-positive-functionals} constructs positive state densities from correlated
wave packets, tests the common smearing functional and computes the EFT bound.
Section~\ref{sec:extended-uv-states} studies the ultraviolet Regge mechanism, and
Section~\ref{sec:conclusion} discusses physical implications and future applications. The appendices
cover conventions, beam spreading and logarithmic estimates, construction using relative-state
envelopes, and AdS Regge diffusion.

\section{Flat-space infrared obstructions}
\label{sec:flat-obstruction}

Can flat-space scattering provide an infrared-finite four-point amplitude
that is still described by ordinary $2\to2$ scattering between
finite-particle Fock states and can therefore be used in an amplitude
bootstrap?  The required smearing functional must satisfy two conditions.
First, it must preserve the PSD absorptive part furnished by
the partial-wave expansion of ordinary $2\to2$ Fock-state scattering.
Second, the resulting amplitude must possess the analyticity, crossing
symmetry and unitarity required for a dispersion-relation bootstrap.  The
flat-space constructions considered below have not realized both conditions
in a single infrared-finite object.

Once momentum conservation is imposed, the test wavefunction in a
four-point smearing functional has correlated momenta on its four external
legs. It is not a product of four independent one-particle wave packets.  A
nonfactorizing momentum-space wavefunction for two initial legs or two final
legs can describe entanglement.  Between an initial and a final leg, the
corresponding relation is a correlation between the bra and ket.  We refer to
the full four-leg profile in either description as a correlated wave packet.

For ordinary finite-particle amplitudes, this program is well developed.
Positive smearing functionals for nonforward amplitudes can be built from the
absorptive partial-wave expansion
\cite{Caron-Huot:2021rmr,Beadle:2025cdx}.  Subtracting calculable low-energy
contributions gives improved dispersion relations
\cite{deRham:2017avq,Bellazzini:2021oaj}, while full crossing symmetry
supplies further null constraints and crossing-symmetric functionals
\cite{Tolley:2020gtv,Caron-Huot:2021rmr}.  Related four-point
bootstrap methods have been applied to pion and large-$N_c$ quantum chromodynamics (QCD) amplitudes,
to Standard Model effective field theory (SMEFT) scattering and to gauge--gravitational amplitudes
\cite{Albert:2023jtd,Remmen:2019cyz,Dong:2024omo}.

These constructions start from an ordinary finite-particle $2\to2$ $S$
matrix.  In flat space with massless gauge fields or gravity, however, an
infrared-finite transition is instead defined between dressed states that
contain sectors with arbitrarily many unresolved soft quanta.  It is not a
$2\to2$ Fock-state amplitude, so the usual partial-wave positivity of
finite-particle scattering cannot be applied to it directly.  The rest of
this section makes this obstruction explicit.

\subsection{Dressed states and finite-resolution scattering}

Consider exclusive scattering of two hard, massless scalar particles into
two hard, massless scalar particles.  The incoming and outgoing states are
normalizable two-particle Fock states of finite norm.  Virtual photons or
gravitons exchanged pairwise among the four external lines generate
infrared-divergent loop contributions.  We collect our metric, all-incoming
momentum convention and Mandelstam conventions in
appendix~\ref{app:conventions}.

For a pair of external lines $i,j$, write the corresponding virtual factor as
\begin{equation}
 W_{ij}
 =
 \exp\!\left[
 i\int_{\lambda_{\rm IR}}^{\Lambda_{\rm loop}}
 \frac{\dd^4\ell}{(2\pi)^4}\,
 \frac{N_{ij}(p_i,p_j,\ell)}
 {(\ell^2+i\epsilon)[(p_i-\ell)^2+i\epsilon]
 [(p_j+\ell)^2+i\epsilon]}
 \right].
 \label{eq:soft-factor-schematic}
\end{equation}
The endpoints restrict the spatial magnitude of $\ell$ in the detector rest
frame.  The lower endpoint regulates the infrared divergence and will tend to
zero after dressing.  The upper matching scale $\Lambda_{\rm loop}$ keeps the
recoil too small to change the hard external momenta by order one.
The parameter $\epsilon$ specifies the Feynman propagator prescription.  In
Feynman gauge for the photon and de Donder gauge for the graviton, the numerators for scalar
external particles are
\begin{align}
 N_{ij}^{\gamma}(p_i,p_j,\ell)
 &=-q_iq_j(2p_i-\ell)\cdot(2p_j+\ell)
 \xrightarrow{\rm S.L.}-4q_iq_j\,p_i\cdot p_j ,
 \nonumber\\
 N_{ij}^{h}(p_i,p_j,\ell)
 &=\kappa^2\!\left[
 (p_i\cdot p_j)^2
 +(p_i\cdot p_j)(p_i-p_j)\cdot\ell
 -\frac12(p_i\cdot p_j)\ell^2\right]
 \xrightarrow{\rm S.L.}\kappa^2(p_i\cdot p_j)^2 .
 \label{eq:scalar-pair-numerators}
\end{align}
Here S.L. denotes the soft limit and $\kappa^2=32\pi G_N$.
Before taking this
limit, the complete loop graph must be evaluated with its recoil-dependent
propagators and vertices. It does not in general factorize or exponentiate.
Only when the recoil is soft can one hold the hard subdiagram fixed, pull it
outside the loop and exponentiate the leading contribution. This ordering
underlies leading soft factorization and exponentiation
\cite{Sahoo:2018lxl,White:2011yy,Naculich:2011ry}.

The eikonal denominators have the standard ray representation
\begin{equation}
 \lim_{\epsilon\to0^+}\frac{1}{p\cdot\ell\pm i\epsilon}
 =
 \mp i\lim_{\epsilon\to0^+}
 \int_0^\infty\dd\tau\,
 e^{\pm i\tau(p\cdot\ell\pm i\epsilon)} .
 \label{eq:ray-denominator-identities}
\end{equation}
Using this identity after
$(p_i-\ell)^2\simeq-2p_i\cdot\ell$ and
$(p_j+\ell)^2\simeq2p_j\cdot\ell$, the two linear denominators inherit
opposite $i\epsilon$ prescriptions, $p_i\cdot\ell-i\epsilon$ and
$p_j\cdot\ell+i\epsilon$, after a harmless rescaling of $\epsilon$.
The leading gravitational contribution then becomes
\begin{equation}
 \log W_{ij}
 =
 -8\pi G_N(p_i\cdot p_j)^2
 \int_0^\infty\dd\tau_i\int_0^\infty\dd\tau_j\,
 D_{F,\lambda_{\rm IR}}^{\Lambda_{\rm loop}}
 (\tau_i p_i-\tau_j p_j) .
 \label{eq:virtual-loop-rays}
\end{equation}
Here
$D_{F,\lambda_{\rm IR}}^{\Lambda_{\rm loop}}(x)
=\int_{\lambda_{\rm IR}}^{\Lambda_{\rm loop}}\dd^4\ell\,
 i e^{-i\ell\cdot x}/[(2\pi)^4(\ell^2+i\epsilon)]$
is the regulated massless Feynman propagator.  Eq.~\eqref{eq:virtual-loop-rays}
describes the interaction between two straight external worldlines.  It is
the position-space form of the triangle only at leading soft order, where the
recoil-dependent graph factorizes.

The failure of these rays to become free at large distance is now explicit.
When the momentum regulator is removed,
$D_F(x)=1/[4\pi^2(-x^2+i\epsilon)]$.  For massless external momenta,
$p_i^2=p_j^2=0$, its argument is
$(\tau_i p_i-\tau_jp_j)^2=-2\tau_i\tau_j p_i\cdot p_j$, so the
long-distance part contains
\begin{equation}
 \int_{\tau_0}^{R/|n\cdot p_i|}\frac{\dd\tau_i}{\tau_i}
 \int_{\tau_0}^{R/|n\cdot p_j|}\frac{\dd\tau_j}{\tau_j}
 =
 \log\!\frac{R}{\tau_0|n\cdot p_i|}
 \log\!\frac{R}{\tau_0|n\cdot p_j|}
 \xrightarrow{R\to\infty}\infty .
 \label{eq:worldline-log}
\end{equation}
Individual null pairs contain a double logarithm.  In the sum over all six
pairs the leading double logarithm cancels by momentum conservation, but the
remaining single logarithm is nonzero for generic wide-angle scattering
\cite{Weinberg:1965nx,Naculich:2011ry}.  Consequently the complete virtual
factor $W=\prod_{i<j}W_{ij}$ suppresses the exclusive Fock-state amplitude
$\mathcal M_{\rm ex}$ to zero as $\lambda_{\rm IR}\to0$.

A spatially finite detector cuts off the same accumulation in a different
way.  Once a hard particle reaches the detector and is absorbed, its
worldline does not continue as the same freely propagating external ray.
Replacing the upper limit by $R/|n\cdot p_i|$ gives an effective
$\lambda_{\rm IR}\sim R^{-1}$ away from collinear directions.  This identifies
the relevant infrared scale but is not an exact equality between a sharp
momentum cutoff and a finite segment.  A complete stationary detector model
must also include the endpoint current or stress tensor.  The momentum-space
Faddeev--Kulish (FK) construction remains invariant under spatial translations, whereas a
detector at a fixed position does not.  A stationary rotationally symmetric
detector instead preserves time translations and $SO(3)$ about its center.

To cancel the infrared divergence of the virtual factor, one uses
dressed asymptotic states. The incoming and outgoing modifiers
$D_{\rm in}$ and $D_{\rm out}$ are specified by a fixed profile
$f(n,|\bm k|/\omega)$, where $|\bm k|$ is the spatial momentum magnitude
in the detector rest frame. Each modifier contains a coherent FK
radiation factor and a phase factor. The dressed matrix element is
\begin{equation}
 \mathcal M_{\rm dr}^{\lambda_{\rm IR},\omega}(p_i;n,f)
 ={}_{\rm Fock}\langle p_1,p_2|
 D_{\rm in}^{\dagger}S_D^{\lambda_{\rm IR}}D_{\rm out}
 |-p_3,-p_4\rangle_{\rm Fock}.
 \label{eq:fk-matrix-element}
\end{equation}
The bra and ket displayed in \eqref{eq:fk-matrix-element} are two-particle
Fock states. The central operator $S_D^{\lambda_{\rm IR}}$ is the regulated
Dyson $S$ matrix. Its transition operator $T_D$ is defined by
$S_D=1+\ii T_D$, and the subscript $D$ denotes Dyson. The modifiers
$D_{\rm in}$ and $D_{\rm out}$ attach the dressing to these Fock states.

Normal ordering the full matrix element, including contractions between the
radiation factors and $S_D$, cancels the real exponential suppression in $W$,
while the FK phase
factor cancels its long-range imaginary phase
\cite{Kulish:1970ut,Ware:2013zja,Choi:2017ylo,Choi:2019rlz}.  The infrared
cutoff can then be taken to zero, defining the finite dressed amplitude
$\mathcal M_{\rm dr}^{\omega}=\lim_{\lambda_{\rm IR}\to0}
\mathcal M_{\rm dr}^{\lambda_{\rm IR},\omega}$.  For the sharp profile at
generic wide angle, normal ordering gives
\begin{equation}
 \mathcal M_{\rm ex}^{\lambda_{\rm IR}}
 =\exp\!\left[\sum_{i<j}\theta_{ij}(\lambda_{\rm IR},\omega)\right]
 \mathcal M_{\rm dr}^{\omega}(p_i;n,f)
 \left[1+\mathcal{O}\!\left(
 \frac{\omega}{\sqrt{\min\{|s|,|t|,|u|\}}}
 \right)\right].
 \label{eq:exclusive-dressed-matching}
\end{equation}
This factorization is controlled only when the momenta in the dressed clouds
lie in the common soft domain of all six external-line pairs.
In cross-section language, an inclusive result
is obtained by squaring the hard--soft transition amplitudes, summing over the
unresolved occupation numbers and polarizations, and integrating over their
momenta.  The bootstrap is formulated at amplitude level, so neither the soft
momenta carried by the dressing profile nor the soft-particle numbers created
by its creation operators can be traced out.
Ordinary Lehmann--Symanzik--Zimmermann (LSZ) reduction assumes free Fock evolution on the external rays.
Eqs.~\eqref{eq:virtual-loop-rays} and \eqref{eq:worldline-log} show why
this assumption fails.  The long-range interaction continues to accumulate
at arbitrarily large separation.  The modifiers in
\eqref{eq:fk-matrix-element} supply the required long-range asymptotic
dynamics.  The resulting state contains arbitrarily many unresolved quanta.
It is not a finite-particle $2\to2$ Fock state.

At fixed dressing profile $f$ and resolution scale $\omega$, the dressed
amplitude is a function of five kinematic variables formed from the
Mandelstam invariants and the dressing-frame vector $n$,
\begin{equation}
 \mathcal M_{\rm dr}^{\omega,f}
 \bigl(s,t;n_s,n_t,n_u\bigr),
 \qquad
 n_s=n\cdot(p_1+p_2),\quad
 n_t=n\cdot(p_1+p_3),\quad
 n_u=n\cdot(p_1+p_4).
 \label{eq:five-invariants}
\end{equation}
The dependence on $n$ specifies the detector frame and the attached soft
cloud.  The result is not a Lorentz-invariant function of $(s,t)$ alone.
In general, the clouds can exchange momentum of order $\omega$ with the
hard lines, so the four hard momenta need not sum to zero by themselves.
Eq.~\eqref{eq:five-invariants} uses external states for which the
all-incoming hard momenta and cloud momenta separately sum to
zero.

At wide angle, \eqref{eq:exclusive-dressed-matching} requires
$\omega^2\ll\min\{|s|,|t|,|u|\}$.  This estimate is not uniform in the
forward or eikonal limit.  In the eikonal limit, ordinary soft factorization
requires the virtual loop momentum $\ell$ to satisfy
\begin{equation}
 \sqrt{s}\,|\ell^+|\ll |t|,
 \qquad
 \sqrt{s}\,|\ell^-|\ll |t|,
 \qquad
 \boldsymbol\ell_\perp^2\ll |t|.
 \label{eq:longitudinal-soft-validity}
\end{equation}
At leading eikonal power the virtual factor reduces to a phase from
off-shell graviton exchange in the Glauber region.  The poles of the nearly
on-shell hard lines select
$|\ell^\pm|\sim\boldsymbol\ell_\perp^2/\sqrt{s}$ and
$\ell^2\simeq-\boldsymbol\ell_\perp^2$.  For the full ladder,
$\boldsymbol\ell_\perp^2\sim|t|$ gives $|\ell^\pm|\sim|t|/\sqrt{s}$.
For the imaginary part of the virtual-loop exponent canceled by the
FK phase factor, the longitudinal conditions follow from
\begin{equation}
 \boldsymbol\ell_\perp^2\lesssim\omega^2\ll|t|.
 \label{eq:transverse-resolution-validity}
\end{equation}
This phase captures the IR-divergent part of conservative first post-Minkowskian (1PM) dynamics.
Dissipation through radiated energy--momentum first appears at 3PM, so
the 1PM eikonal has no radiative damping, consistently with the vanishing
real part of the virtual-loop exponent at leading eikonal
power~\cite{DiVecchia:2021bdo}.

At either wide or small angle, leaving the appropriate common soft domain
allows order-one or larger corrections to
\eqref{eq:exclusive-dressed-matching}.  The full FK dressed amplitude still
exists, but the stripped amplitude $\mathcal M_{\rm ex}/W$ is no longer a
controlled approximation to it.  After dressing cancels virtual modes
below $\omega$, the remaining integral associated with
\eqref{eq:soft-factor-schematic} starts at $\omega$.  If this lower scale
reaches or exceeds the upper soft matching scale $\Lambda_{\rm loop}$,
there is no common soft interval left.  Extending $W$ beyond it cannot be
justified by the leading-soft approximation.

Unitarity must therefore act on complete dressed asymptotic states, which
belong to the von Neumann space of soft-cloud representations rather than
the original Hilbert-space representation of finite-particle Fock states.
Their soft occupation is divergent but their cloud energy remains
finite~\cite{Kulish:1970ut,Ware:2013zja}. For electrically and magnetically
charged particles, the Lorentz transformation of FK-dressed states has also
been related to the pairwise little group \cite{Csaki:2022tvb}. In gravity, this change of
infrared representation is described by the soft Bondi--Metzner--Sachs (BMS) supertranslation
data~\cite{Choi:2017ylo,Donnay:2026urd}.

To extend a positive two-particle smearing functional to these states,
keep a finite IR regulator and use the Dyson transition operator $T_D$
defined above, with all dressing attached to the states.
Choose normalized two-particle Fock packets $|\Phi_a\rangle$ and a
positive density kernel
\begin{equation}
 K_{2\to2}=\int\dd a\,\rho(a)|\Phi_a\rangle\langle\Phi_a|.
 \label{eq:hard-packet-density}
\end{equation}
Here $\rho(a)\geq0$ are its spectral weights and
$\int\dd a\,\rho(a)=1$, with discrete sums included when appropriate.
Initial labels are written on the left. The initial dressed bra is
$ {}_{\rm dr}\langle\Phi_a|=\langle\Phi_a|\mathcal D_{\rm in}^{\dagger}$.
On the matched forward slice the outgoing dressed ket is its adjoint,
$|\Phi_a\rangle_{\rm dr}=\mathcal D_{\rm out}|\Phi_a\rangle
=({}_{\rm dr}\langle\Phi_a|)^\dagger$.
For the corresponding dressed kernel $K_{\rm dr}$, the optical theorem gives
\begin{equation}
 \begin{aligned}
 2\operatorname{Tr}\bigl(K_{\rm dr}\operatorname{Im}T_D\bigr)
 &=2\int\dd a\,\rho(a)\operatorname{Im}
 {}_{\rm dr}\langle\Phi_a|T_D|\Phi_a\rangle_{\rm dr}\\
 &=\int\dd a\,\rho(a)\,
 \sum_X|{}_{\rm dr}\langle\Phi_a|T_D|X\rangle|^2\geq0 .
 \end{aligned}
 \label{eq:packet-unitarity-flat}
\end{equation}
The operator imaginary part is
$\operatorname{Im}T_D=(T_D-T_D^\dagger)/(2\ii)$.
The sum on the right includes all final soft-particle numbers
and polarizations and integrals over their momenta, together with all
allowed hard final states.
Correlated two-particle wave packets and their positive mixtures therefore
give PSD smearing functionals.  Attaching the clouds to
the ket and its matching bra is precisely the congruence
\begin{equation}
 \begin{aligned}
 K_{2\to2}\succeq0\quad\Longrightarrow\quad
 K_{\rm dr}
 &=\int\dd a\,\rho(a)|\Phi_a\rangle_{\rm dr}
 {}_{\rm dr}\langle\Phi_a|\\
 &=\mathcal D_{\rm in}K_{2\to2}
 \mathcal D_{\rm in}^{\dagger}\succeq0 .
 \end{aligned}
 \label{eq:dressed-positive-congruence}
\end{equation}
The modifiers contain the profile $f$ and reference vector $n$.
The initial bra dressing may depend only on the incoming hard state and the fixed
dressing-profile data, and the outgoing ket dressing only on its own hard state and those
data.  Neither may depend on the independently chosen state on the other
side.  The same linear state-preparation rule must give the ket and its adjoint
bra, as imposed above.  These requirements make
\eqref{eq:dressed-positive-congruence} a congruence and ensure the
positivity in \eqref{eq:packet-unitarity-flat}.

The $2\to2$ smearing kernel generalizes plane-wave forward elastic
scattering at $t=0$ to forward scattering of a correlated wave packet.
A nonzero analytic PSD packet kernel has nonzero
diagonal weight.  Local analyticity then ensures support arbitrarily close
to zero momentum transfer.
In four dimensions, positivity at all impact parameters requires the
smearing profile $\phi(t)$ in the $\dd t$ convention to
satisfy $\lim_{t\to0^-}\phi(t)>0$~\cite{Caron-Huot:2021rmr}.
This is a requirement on the functional itself.  Its action on the
gravitational pole consequently contains $\int\dd t/t$.
The dressed kernel in \eqref{eq:dressed-positive-congruence} must therefore
sample $|t|\lesssim\omega^2$, where the factorization in
\eqref{eq:exclusive-dressed-matching} is not controlled.  Its
PSD smearing functional must act on the full dressed
amplitude, rather than the stripped amplitude
$\mathcal M_{\rm ex}/W$.
Consequently, dressed amplitudes do not directly inherit the ordinary
$2\to2$ Fock-state partial-wave positivity.

\subsection{Zero-mode obstruction}
\label{sec:constant-mode-no-go}

Let $\psi_1,\psi_2\to\psi_3,\psi_4$ denote the four external wavefunctions.
At fixed $s=(p_1+p_2)^2$ for massless scalar hard particles, the
$t$-smearing functional assigns partial-wave weights that define an
$SO(3)$-invariant kernel on the relative orientations of the two-particle
states.  This kernel is PSD when the weights are nonnegative.
Write $\Omega$ and $\Omega'$ for the incoming and outgoing relative
directions, with $\cos\theta=\Omega'\cdot\Omega$ and
$t=-s(1-\cos\theta)/2$.
\begin{equation}
 \begin{aligned}
 a_J(s)&=\int_{-t_{\max}}^{0}\dd t\,\phi(t)
       P_J\!\left(1+\frac{2t}{s}\right)\geq0,\\
 K_{2\to2}(\Omega,\Omega')
 &=\sum_J\frac{2J+1}{4\pi}a_J(s)d^J_{00}(\theta)\\
 &=\sum_J a_J(s)\sum_{m=-J}^{J}
   Y_{Jm}(\Omega)^*Y_{Jm}(\Omega') .
 \end{aligned}
 \label{eq:angular-positive-kernel}
\end{equation}
Here $d^J_{00}(\theta)=P_J(\cos\theta)$ is the Wigner function when both the initial and final
helicity differences vanish,
and the last equality uses the spherical-harmonic addition theorem.
Thus smearing the absorptive amplitude uses $a_J(s)$ to weight the
nonnegative partial-wave absorptive coefficients, or equivalently to pair
the initial and final spherical waves with positive weights.
In the large-$J$ eikonal limit, $b\sim2J/\sqrt{s}$, angular-momentum
diagonality becomes impact-parameter diagonality.  Positivity then takes
the form of a PSD density kernel in transverse impact-parameter
space~\cite{Caron-Huot:2021rmr,DiVecchia:2023frv}.

Within this eikonal description, choose matching packets
$\psi_3=\psi_1$ and $\psi_4=\psi_2$.  Their pair densities are
\begin{equation}
 \rho_{13}(x)=\psi_1(x)\psi_3(x)^*\geq0,
 \qquad
 \rho_{24}(x)=\psi_2(x)\psi_4(x)^*\geq0 .
 \label{eq:pair-densities}
\end{equation}
The inequalities implement the positive diagonal pairing above.
For normalizable probability densities we choose
\begin{equation}
 \int\dd^2x\,\rho_{13}(x)=1>0,
 \qquad
 \int\dd^2x\,\rho_{24}(x)=1>0 .
 \label{eq:positive-charges}
\end{equation}
We use the phase-shift convention $S=e^{2\ii\delta}$~\cite{DiVecchia:2023frv}.
At leading eikonal order, tree-level massless exchange gives the smeared
phase
\begin{equation}
 \delta_G(s)
 =\frac12\int\dd^2x\,\dd^2x'\,
 \rho_{13}(x)G(x,x';s)\rho_{24}(x').
 \label{eq:green-bilinear}
\end{equation}
For massless graviton exchange, the two packet pairs couple through
conserved rank-two stress-tensor currents.  At center-of-mass energy
$\sqrt{s}$, their leading eikonal contraction gives the position-space
form of the gravitational tree-level pole,
$G(x,x';s)=4\pi G_Ns\,G(x,x')$.  The wavefunction-overlap densities in
\eqref{eq:pair-densities} carry the packet profiles after this tensor
contraction.\footnote{Photon exchange between charged particles uses
conserved vector currents in place of the stress-tensor currents.
The same zero-mode ambiguity comes from the massless propagator
pole, while the numerator fixes the charge or tensor
coefficient~\cite{DiVecchia:2023frv}.}
Here $G(x,x')$ is the transverse inverse Laplacian used below.
On the translationally invariant plane, $G(x,x')=G(x-x')$.
For GR in the eikonal limit, the four packets have the
$S$-matrix element
\begin{equation}
 S(s)=\int\dd^2x\,\dd^2x'\,
 \rho_{13}(x)\e^{\ii G(x,x';s)}\rho_{24}(x')
 \simeq\e^{2\ii\delta_G(s)} .
 \label{eq:packet-eikonal-phase}
\end{equation}
Let $b$ be the distance between the centers of the pair densities and
$\Delta b$ their characteristic position-space width.  When $\Delta b\ll b$, the 1PM phase varies by
order $G_Ns\,\Delta b/b$ across the
packets.  For $G_Ns\,\Delta b/b\ll1$, impact-parameter diagonality
therefore allows the density convolution to be approximated by the
exponential of twice the phase in \eqref{eq:green-bilinear}.

Suppose the transverse geometry admits a constant zero mode, allowing the
Green function to shift as $G(x,x')\mapsto G(x,x')+C$.
At fixed $s$, the gravitational phase then changes by
\begin{equation}
 \delta_{G+C}(s)-\delta_G(s)
 =2\pi G_Ns\,C
   \left(\int\dd^2x\,\rho_{13}(x)\right)
   \left(\int\dd^2x'\,\rho_{24}(x')\right)
 =2\pi G_Ns\,C .
 \label{eq:constant-shift}
\end{equation}
For positive normalized densities, this shift is independent of the
packet shape.  Reshaping the wave packet, i.e.\ changing the corresponding
smearing functional within this positive class, cannot remove it.
In a formal low-energy expansion of the exclusive $S$-matrix element at
$G_Ns\ll1$, \eqref{eq:constant-shift} shifts the coefficient linear in
$s$ by $4\pi\ii G_NC$.  The same zero-mode ambiguity appears
in the corresponding momentum-space amplitude, obstructing an unambiguous
definition of the Wilson coefficients extracted from its low-energy
EFT expansion.

Requiring either density to have zero integral would remove the constant
contribution, but a nonzero density must then take both positive and
negative values across the transverse space.  It no longer represents
the positive probability density required here.
The same condition can be seen in momentum space.  On the translationally
invariant plane, the convolution in \eqref{eq:green-bilinear} becomes a
product of Fourier transforms.  With
$\widetilde\rho(q)=\int\dd^2x\,\e^{-\ii q\cdot x}\rho(x)$ and the same
IR regularization understood, the zero-transfer value of the rotationally
invariant smearing profile $\phi$ in \eqref{eq:angular-positive-kernel}
is proportional to the same numerator,
\begin{equation}
 \begin{aligned}
 \delta_G(s)
 &=2\pi G_Ns\int\frac{\dd^2q}{(2\pi)^2}\,
   \frac{\widetilde\rho_{13}(-q)\widetilde\rho_{24}(q)}{q^2},\\
 \phi(0)&\propto\widetilde\rho_{13}(0)\widetilde\rho_{24}(0)
 =\left(\int\dd^2x\,\rho_{13}(x)\right)
   \left(\int\dd^2x'\,\rho_{24}(x')\right)=1 .
 \end{aligned}
 \label{eq:constant-mode-momentum}
\end{equation}
Thus eliminating the constant contribution requires the numerator to
vanish at $q=0$, contrary to the second line for positive normalized
densities.  Equivalently, it would require $\phi(0)=0$.

On $\mathbb R^2$ the massless Green function is
\begin{equation}
  G_{\mathbb R^2}(x,x')
  =-\frac{1}{2\pi}\log\frac{|x-x'|}{L_{\rm IR}},
  \label{eq:R2-green}
\end{equation}
where changing the regulator length $L_{\rm IR}\sim\lambda_{\rm IR}^{-1}$
is a constant shift.  Eq. \eqref{eq:constant-shift} therefore relates
the flat-space IR ambiguity directly to the constant zero mode sampled by
the normalized packet densities.  Smooth reshaping of the packets cannot
remove this obstruction.  Fourier transformation rewrites the
momentum-space $1/t$ pole as the position-space logarithm.
The correlated wave packets corresponding to the smearing functional
convolve this logarithm with their densities and retain its logarithmic
IR divergence.  Smearing changes its representation without removing
the divergence.

One possibility is to bootstrap the EFT in a spacetime without a
zero mode, such as AdS$_4$.  With the usual decaying boundary
condition, the negatively curved transverse geometry has a well-defined
resolvent, but curvature also modifies the local EFT.  Curvature insertions
in its derivative expansion produce corrections of order
$\mathcal O(R_{\rm AdS}^{-2}\Lambda_{\rm EFT}^{-2})$ and higher, where
$R_{\rm AdS}$ is the curvature radius.  These power corrections
contaminate the extraction of flat-space Wilson-coefficient data at
finite radius.  For an AdS$_4$ EFT that remains weakly coupled up to
its UV cutoff $\Lambda_{\rm EFT}$, the large-gap and large-central-charge
conditions, together with their associated curvature and gravitational-loop
corrections, are~\cite{Heemskerk:2009pn,Caron-Huot:2021enk}
\begin{equation}
 \begin{aligned}
 &\Delta_{\rm gap}\sim R_{\rm AdS}\Lambda_{\rm EFT}\gg1,
 &&\mathcal O\!\left(\frac{1}{R_{\rm AdS}^2\Lambda_{\rm EFT}^2}\right)
 =\mathcal O(\Delta_{\rm gap}^{-2}),\\
 &C_T\sim\frac{R_{\rm AdS}^2}{G_N}\gg\Delta_{\rm gap}^2,
 &&\mathcal O(G_N\Lambda_{\rm EFT}^2)
 =\mathcal O\!\left(\frac{\Delta_{\rm gap}^2}{C_T}\right).
 \end{aligned}
 \label{eq:ads-eft-control}
\end{equation}
Here $\Delta_{\rm gap}$ is the single-trace gap above the retained light
fields, including the higher-spin gap, and $C_T$ is the stress-tensor
two-point normalization of the CFT$_3$.  At fixed energies in AdS units,
the gravitational loop correction is of order $1/C_T$.

At finite AdS radius, the timelike boundary prevents the ordinary
asymptotically separated free in and out states of flat-space scattering.
Extracting a flat-space amplitude requires a state-preparation prescription
and a large-radius limit~\cite{Penedones:2010ue}.
For long-range gauge and gravitational interactions, matching that limit
to the appropriate dressed asymptotic states remains an active
problem~\cite{Duary:2022afn}.  When the curvature corrections
vanish in the flat-space limit, the transverse zero-mode ambiguity
reappears in the exclusive object considered here.
A regulator better suited to this construction would exclude the
zero mode by modifying only the far IR geometry, while retaining
the bulk EFT in an exactly flat interaction region.

The Hubble scale in the asymptotically de Sitter future of cosmology with a positive cosmological constant
and cold dark matter ($\Lambda$CDM)~\cite{Wald:1983ky} does not by itself
remove this zero-mode obstruction.
In the simple transverse model $\mathbb M_{1,1}\times S^2$, take the
sphere radius to be $H^{-1}$.  The eigenvalues of $-\Delta_{S^2}$ are
$\ell(\ell+1)H^2$, so the first nonzero eigenvalue is $2H^2$, while the
$\ell=0$ constant zero mode remains.  The Euclidean section of
dS$_4$ is $S^4$.  For a minimally coupled massless scalar, its Laplacian
has eigenvalues $\ell(\ell+3)H^2$, again including a constant zero mode.  This mode requires
separate treatment before
Euclidean continuation can be used to infer Lorentzian
correlators~\cite{Folacci:1992xc}.
The gap above zero does not make the massless inverse well defined
while the zero mode is retained, as the next section shows
explicitly.

\section{Eikonal positivity and geometric infrared control}
\label{sec:transverse-modes}

We seek a spacetime with no zero mode and an exactly flat
interaction region, and ask whether scattering between ordinary two-particle
Fock asymptotic states admits an IR-finite smearing functional with a PSD kernel.
The aim here is to establish feasibility, without optimizing the resulting
bounds.  For a smearing functional on $-t_{\max}<t<0$, we can further
tune its form to impose nonnegative partial-wave eigenvalues in the
respective $s$- and $u$-channel regions $s,u\sim t_{\max}$.
Here we focus on the respective limits $s,u\gg t_{\max}$, asking whether
positivity can be maintained without an IR divergence.

In this eikonal regime we use the product geometry
$\mathbb M_{1,1}\times\Sigma_2$.  The large-angular-momentum limit of
the partial-wave pairing in \eqref{eq:angular-positive-kernel} gives a
PSD density kernel in impact-parameter space.  The longitudinal modes
remain plane waves, while the transverse space $\Sigma_2$ may be finite
and need not have translation symmetry.  Its eigenmodes therefore need
not be plane waves, and its spectrum may include or exclude a zero mode.
These modes resolve the transverse Green function of the tree-level
$t$-channel exchange, which forms the two-particle-irreducible rung of
the leading eikonal ladder.  Separating the transverse sector thus
isolates the Green operator controlling long-range exchange.
This reduction is inspired by AdS/CFT and
Brower--Polchinski--Strassler--Tan (BPST) Regge analyses~\cite{Cornalba:2006xk,Brower:2006ea},
without assuming equivalence to scattering in AdS$_4$ or dS$_4$.

\subsection{Wave packets, transverse exchange, and local EFT}

Let $x_{\mathrm L}=(t,z)$ denote the longitudinal coordinates and
$b\in\Sigma_2$ a transverse point.  The area measure is
$\dd^2b\equiv\sqrt{g(b)}\,\dd b^1\dd b^2$ throughout this section,
where $g$ is the determinant of the transverse metric in local
coordinates $b^1,b^2$. Choose a complete
orthonormal eigenbasis of the transverse Laplacian satisfying
\begin{equation}
 \begin{gathered}
 -\Delta_{\Sigma_2}\phi_\alpha(b)
 =\lambda_\alpha\phi_\alpha(b),\\
 \int_{\Sigma_2}\dd^2b\,
 \phi_\alpha(b)^*\phi_\beta(b)=\delta_{\alpha\beta},
 \qquad
 \sum_\alpha\phi_\alpha(b)\phi_\alpha(b')^*
 =\delta_{\Sigma_2}(b,b').
 \end{gathered}
 \label{eq:transverse-eigenproblem}
\end{equation}
The second line gives orthogonality between modes and completeness in
the position basis.  A scalar one-particle mode is
\begin{equation}
 \Psi_{p_{\mathrm L},\alpha}(x_{\mathrm L},b)
 =\frac{1}{\sqrt{2E}}\,
  \e^{\ii p_{\mathrm L}\cdot x_{\mathrm L}}\phi_\alpha(b),
 \qquad
 E^2-p_z^2=m^2+\lambda_\alpha.
 \label{eq:product-mode}
\end{equation}
Thus $\lambda_\alpha$ enters longitudinal kinematics as an effective
mass shift.  In flat space the one-particle Fock basis consists of momentum
eigenstates.  On $\mathbb M_{1,1}\times\Sigma_2$ it is labeled by
$p_{\mathrm L}$ and $\alpha$ and diagonalizes the free kinetic operator.

\nopagebreak[4]

The eikonal simplification of a PSD smearing functional follows from
\eqref{eq:angular-positive-kernel}. At a given center-of-mass energy
$\sqrt{s}$, consider large angular momentum $J$ and small scattering
angle $\theta$, with $q=\sqrt{-t}$ and $b\sim2J/\sqrt{s}$.
Replacing the Legendre polynomial in the partial waves by
the Wigner function $d^J_{h_ih_f}(\theta)$, where $h_i$ and $h_f$ are the initial and final
helicity differences, gives
\begin{equation}
 \begin{aligned}
 &\lim_{J\to\infty}d^J_{h_ih_f}(bq/J)
   =J_{h_i-h_f}(bq),\\
 &a_J^{h_ih_f}(s)\simeq a^{h_i-h_f}(b)
 =2\int_0^{\sqrt{t_{\max}}}\dd q\,q\,\phi(-q^2)
 J_{h_i-h_f}(bq).
 \end{aligned}
 \label{eq:eikonal-common-weight}
\end{equation}
Here the leading eikonal approximation uses the small deflection angle
$\theta\sim2q/\sqrt{s}\sim bq/J$~\cite{Bellazzini:2022wzv}.
For a given smearing profile $\phi$, the resulting impact-parameter weight
depends only on the difference $h_i-h_f$ between the initial and final
helicities, and no longer on $s$.
For scalar external particles, $h_i=h_f=0$.

\begin{figure}[H]
 \centering
\begin{tikzpicture}[x=0.88cm,y=0.88cm,>=stealth,
  every node/.style={font=\normalsize}]
 \definecolor{beamblue}{RGB}{30,117,160}
 \definecolor{beamgold}{RGB}{189,111,37}
 \definecolor{cavity}{RGB}{159,181,193}
 \begin{scope}[rotate=12]
  \shade[top color=cavity!4,bottom color=cavity!16,middle color=white,
    opacity=.72]
    (-4.25,1.48)--(4.25,1.48)
    arc[start angle=90,end angle=-90,x radius=.60,y radius=1.48]
    --(-4.25,-1.48)
    arc[start angle=-90,end angle=-270,x radius=.60,y radius=1.48];
  \fill[cavity!7,opacity=.38] (-4.25,0) ellipse[x radius=.60,y radius=1.48];
  \draw[cavity!40,line width=.55pt,opacity=.60]
    (-4.25,1.48)--(4.25,1.48);
  \draw[cavity!45,line width=.55pt,opacity=.55]
    (-4.25,-1.48)--(4.25,-1.48);
  \draw[cavity!50,line width=.55pt,opacity=.50]
    (-4.25,0) ellipse[x radius=.60,y radius=1.48];
  \draw[cavity!50,line width=.55pt,opacity=.50]
    (4.25,0) ellipse[x radius=.60,y radius=1.48];
  \foreach \width/\alpha in {24/.018,20/.025,16/.035,12/.055,8/.085,4/.15}{
    \draw[beamblue,line width=\width pt,opacity=\alpha,line cap=round]
      (-4.00,.53)--(4.00,.53);
    \draw[beamgold,line width=\width pt,opacity=\alpha,line cap=round]
      (4.00,-.53)--(-4.00,-.53);
  }
  \draw[beamblue!65,line width=.65pt,opacity=.65] (-4.00,.53)--(4.00,.53);
  \draw[beamgold!65,line width=.65pt,opacity=.65] (4.00,-.53)--(-4.00,-.53);
  \draw[->,beamblue,line width=1.15pt] (-.85,.53)--(.85,.53)
    node[midway,above=7pt,text=beamblue] {$v_1$};
  \draw[->,beamgold,line width=1.15pt] (.85,-.53)--(-.85,-.53)
    node[midway,below=7pt,text=beamgold] {$v_2$};
  \node[beamblue,above=5pt] at (-3.62,.53) {$1$};
  \node[beamblue,above=5pt] at (3.62,.53) {$3$};
  \node[beamgold,below=5pt] at (3.62,-.53) {$2$};
  \node[beamgold,below=5pt] at (-3.62,-.53) {$4$};
  \node[text=black!55] at (-4.80,1.90) {$\Sigma_2$};
  \draw[black!25,line width=.45pt] (-4.65,1.63)--(-4.35,1.30);
  \draw[->,black!40,line width=.55pt] (2.5,-1.98)--(3.7,-1.98)
    node[right,text=black!60] {$z$};
 \end{scope}
\end{tikzpicture}
 \caption{Schematic spatial view of two counterpropagating beams.
 A disk represents the general transverse space $\Sigma_2$.
 Legs $1\to3$ and $2\to4$ follow the two separated beams at leading
 eikonal order. The arrows show physical velocities.
 The $s$--$u$ crossing exchanges labels $1$ and $3$ on the upper beam.}
 \label{fig:eikonal-cylinder}
\end{figure}
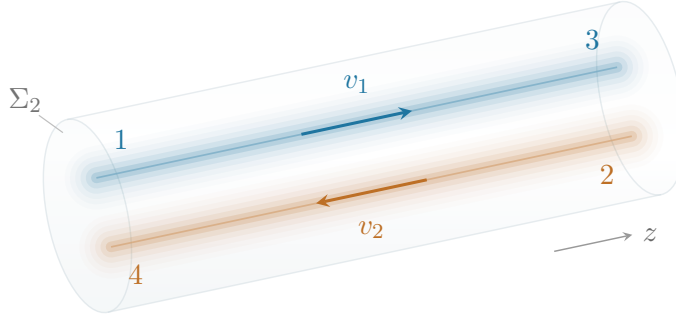

In the all-incoming convention, the initial and final helicity
differences are
\begin{equation}
 \begin{aligned}
 &h_{i,s}=h_1-h_2,\qquad h_{f,s}=h_4-h_3,\\
 &h_{i,u}=h_3-h_2,\qquad h_{f,u}=h_4-h_1,\\
 &h_{i,s}-h_{f,s}=h_{i,u}-h_{f,u}
 =h_1+h_3-h_2-h_4 .
 \end{aligned}
 \label{eq:eikonal-helicity-crossing}
\end{equation}
The physical $u$ channel therefore gives the same Bessel transform
with $b\sim2J/\sqrt{u}$ and the crossed external helicities.
For helicity-preserving elastic scattering, $h_3=-h_1$ and $h_4=-h_2$,
so the Bessel function is $J_0$ and the positivity condition is
$a_J(s)\geq0$, or $a(b)\geq0$ at leading order.
For transitions between different two-particle helicity channels,
the weights form a matrix at each $J$, which must be PSD.
The leading smearing weight in \eqref{eq:eikonal-common-weight} has
no remaining $s$ dependence. Its analytic continuation along a complex
$s$ contour with $|s|\gg t_{\max}$ consequently uses the same
$a^{h_i-h_f}(b)$ throughout. The common weight on the physical
$s$- and $u$-channel boundaries expresses this continuation after
the helicity relabelling in \eqref{eq:eikonal-helicity-crossing}.

Consider focused transverse packets whose beam widths are small compared
with the geometric length scales of $\Sigma_2$. In eikonal scattering,
the transferred transverse momentum is small compared with the hard
longitudinal momentum. At leading eikonal order, we use matching
transverse profiles $\psi_1\sim\psi_3$ and
$\psi_2\sim\psi_4$, with localized centers $b_1\sim b_3$ and
$b_2\sim b_4$. In the leading eikonal approximation, nearly on-shell
hard propagators keep successive local interaction vertices on the same
ray. The tree exchange is the irreducible rung coupling these two fixed
rays~\cite{Cornalba:2007zb}.
Transverse diffraction enters beyond leading order.

The transition operator $T$ is defined by $S=1+\ii T$. We use
$\mathcal T(s,b)$ for its impact-parameter response, so that
$S(s,b)=1+\ii\mathcal T(s,b)$. The Fourier transform of the
momentum-space amplitude $\M(s,t)$ includes the kinematic factor
$1/(2s)$~\cite{DiVecchia:2023frv}. Denoting the two beam positions
by $b_{13}$ and $b_{24}$ gives
\begin{equation}
 \begin{aligned}
 \mathcal T(s,b)
 &\equiv\frac{1}{2s}\int\frac{\dd^2q}{(2\pi)^2}\,
       \e^{\ii q\cdot b}\M(s,-q^2),\\
 \mathcal T(s,b)&\quad\longrightarrow\quad\mathcal T(s,b_{13},b_{24}).
 \end{aligned}
 \label{eq:eikonal-amplitude-two-points}
\end{equation}
A general $\Sigma_2$ need not have transverse translation symmetry,
so the transition response depends on $b_{13}$ and $b_{24}$ separately,
rather than only on their difference.

On $\Sigma_2$, Fourier transformation is replaced by expansion in the
transverse eigenbasis. For the gravitational massless pole used in
Sec.~\ref{sec:flat-obstruction}, the tree phase is proportional to
the two-point Green function $G$ of $L=-\Delta_{\Sigma_2}$.
We fix boundary conditions that make the transverse kinetic operator,
and hence the free Hamiltonian, self-adjoint.
For nonzero modes, projection onto \eqref{eq:transverse-eigenproblem}
gives the flat-space massless pole its spectral counterpart
$1/q^2\longmapsto G_{\alpha\beta}=\delta_{\alpha\beta}/\lambda_\alpha$.
The Green function is therefore diagonal in this basis.
For the elastic channel considered here, the phase at fixed
$b_{13},b_{24}$ is a c-number and the leading ladder sum exponentiates
pointwise~\cite{Cornalba:2007zb},
\begin{equation}
 \begin{aligned}
 \delta_{\rm tree}(s,b_{13},b_{24})
 &=2\pi G_Ns\,G(b_{13},b_{24})
 =2\pi G_Ns\sum_\alpha
       \frac{\phi_\alpha(b_{13})\phi_\alpha(b_{24})^*}{\lambda_\alpha},\\
 \ii\mathcal T(s,b_{13},b_{24})
 &=\e^{2\ii\delta_{\rm tree}(s,b_{13},b_{24})}-1\\
 &=\ii\sum_{\alpha,\beta}\mathcal T_{\alpha\beta}(s)
       \phi_\alpha(b_{13})\phi_\beta(b_{24})^*.
 \end{aligned}
 \label{eq:eikonal-amplitude-mode-transform}
\end{equation}
Thus $\mathcal T_{\rm tree}=2\delta_{\rm tree}$.
Here $\alpha$ resolves the $t$-channel exchanged transverse mode,
not the four external states. Depending on the geometry of $\Sigma_2$,
$\alpha$ may be discrete and summed over, or continuous and integrated over.
Position-space multiplication corresponds to a convolution weighted by
eigenfunction overlaps in mode space, just as pointwise exponentiation
of the flat-space eikonal phase corresponds to convolution in momentum space.

\nopagebreak[4]

For the matched elastic packets, the common longitudinal phases cancel
between bra and ket in the leading straight-ray approximation.
The remaining transverse profiles give the densities
\begin{equation}
 \rho_{13}(b)=\psi_1(b)\psi_3(b)^*=|\psi_1(b)|^2\geq0,
 \qquad
 \rho_{24}(b')=\psi_2(b')\psi_4(b')^*=|\psi_2(b')|^2\geq0.
 \label{eq:transverse-packet-densities}
\end{equation}
The densities obey
$\int\dd^2b\,\rho_{13}(b)=1$ and
$\int\dd^2b'\,\rho_{24}(b')=1$,
and describe the transverse distributions of the two beams over
$b=b_{13}$ and $b'=b_{24}$.
In flat space, an eikonal smearing in $t$ becomes an impact-parameter
smearing with the Fourier weight $\widetilde\phi(b)$.
In flat space with translation symmetry, these packets give
$\widetilde\phi(b)=\int\dd^2b'\,\rho_{13}(b+b')\rho_{24}(b')$.
On a general $\Sigma_2$, the momentum transform becomes a mode expansion
with discrete or continuous labels, while the position pairing remains
an integral over two independent points. The arrows below indicate the
correspondence between the smeared impact-parameter and mode responses.
The momentum-space smearing includes its normalization explicitly,
\begin{equation}
 \begin{aligned}
 &\int\dd^2b\,\widetilde\phi(b)\mathcal T(s,b)
 \quad\longmapsto\quad
 \int_{\Sigma_2}\dd^2b\,\dd^2b'\,
       \rho_{13}(b)\rho_{24}(b')\mathcal T(s,b,b'),\\
 &\frac{1}{8\pi s}\int\dd t\,\phi(t)\M(s,t)
 \quad\longmapsto\quad
       \sum_{\alpha,\beta}\rho_{13,\alpha}^*
       \mathcal T_{\alpha\beta}(s)\rho_{24,\beta}.
 \end{aligned}
 \label{eq:eikonal-spectral-smearing}
\end{equation}
Here $\rho_{13,\alpha}=\int\dd^2b\,\phi_\alpha(b)^*\rho_{13}(b)$,
and $\rho_{24,\beta}$ is defined in the same way.
The Fourier convention is
$\widetilde\phi(b)=\int\dd^2q\,\e^{\ii q\cdot b}\phi(-q^2)/(2\pi)^2$.
The coefficients $\mathcal T_{\alpha\beta}(s)$ come from the mode transform
after eikonal exponentiation in
\eqref{eq:eikonal-amplitude-mode-transform}.

The connection to partial-wave positivity can be seen directly from the
spherical-harmonic pairing in \eqref{eq:angular-positive-kernel}.
In the leading eikonal continuum limit, both $\Omega$ and $\Omega'$
approach the beam direction $\hat z$. Write their transverse momenta as
$\mathbf k=\sqrt{s}\,\Omega_{\mathrm T}/2$ and
$\mathbf k'=\sqrt{s}\,\Omega'_{\mathrm T}/2$, with magnitudes $k,k'$
and azimuths $\varphi,\varphi'$. The spherical harmonics contract to
Bessel functions, and the sum over $J$ becomes an integral over
$b\sim2J/\sqrt{s}$~\cite{Caron-Huot:2021rmr}.
\begin{equation}
 \begin{aligned}
 K_{2\to2}(\Omega,\Omega')
 &\simeq\frac{s}{2}\int_0^\infty\dd b\,b\,\widetilde\phi(b)
 \sum_{m\in\mathbb Z}J_m(bk)J_m(bk')
       \e^{\ii m(\varphi'-\varphi)}\\
 &=\frac{s}{4\pi}\int\dd^2\mathbf b\,
       \widetilde\phi(b)\e^{\ii\mathbf b\cdot(\mathbf k'-\mathbf k)}.
 \end{aligned}
 \label{eq:eikonal-density-contraction}
\end{equation}
The factor $s$ comes from converting the angular measure $\dd\Omega$
to the transverse momentum measure $\dd^2\mathbf k$.
The kernel is diagonal in relative impact parameter in the eikonal
limit, with nonnegative weight $\widetilde\phi(b)$.
The nonnegative partial-wave weights thus become a nonnegative
relative density $\widetilde\phi(b)$. For the matched packets above,
this density is obtained by integrating
$\rho_{13}(b+b')\rho_{24}(b')$ over the common beam position $b'$.

On a general $\Sigma_2$, the matched wavefunctions in
\eqref{eq:transverse-packet-densities} give the same construction directly.
Their squared magnitudes are the two nonnegative beam densities.
To evaluate the imaginary part of the smeared transition response in
\eqref{eq:eikonal-spectral-smearing}, take the normalized two-particle
packet $|\psi_{12}\rangle=|\psi_1,\psi_2\rangle$.
For paraxial beams whose transverse momentum spread is set by the
widths $\Delta b$ and $\Delta b'$ of $\rho_{13}$ and $\rho_{24}$,
$\Delta b,\Delta b'\gg|s|^{-1/2}$ makes longitudinal recoil subleading.
The matched longitudinal plane-wave phases then cancel pairwise at
leading order, leaving factors independent of $x_{\mathrm L}$ but still
dependent on $b,b'$. In position space, the optical theorem
at leading eikonal order gives
\begin{equation}
 \begin{aligned}
 &\int_{\Sigma_2}\dd^2b\,\dd^2b'\,
       \rho_{13}(b)\rho_{24}(b')\operatorname{Im}\mathcal T(s,b,b')
       =\operatorname{Im}\langle\psi_{12}|T|\psi_{12}\rangle\\
 &\qquad=\frac12\sum_X\left|
       \int_{\Sigma_2}\dd^2b\,\dd^2b'\,
       \psi_1(b)^*\psi_2(b')^*\langle b,b'|T|X\rangle
       \right|^2\geq0.
 \end{aligned}
 \label{eq:single-packet-optical-theorem}
\end{equation}
Here $X$ runs over a complete set of intermediate states. The sum includes
the allowed particle numbers, momenta with transverse momentum
resolved into the modes of \eqref{eq:transverse-eigenproblem}, and
polarizations, with sums or integrals as appropriate.
The factors $\rho_{13}$ and $\rho_{24}$ on the left are the matched
bra and ket profiles in \eqref{eq:transverse-packet-densities}.
Nonnegative weighted integrals over parameters of these wave packets
also preserve the PSD density matrix and the inequality.

Using the eigenmodes of $\Sigma_2$ as external states makes even tree
amplitudes and their eikonal ladder sums depend on nontrivial mode
overlaps. A four-point tree amplitude is built from a local quartic
vertex or two cubic vertices joined by an $s$-, $t$-, or $u$-channel
propagator. We instead choose localized external packets whose overlap
at each local vertex lies in a flat bulk region, where their
wavefunctions agree with the flat-space profiles. If the packet tails
decay exponentially outside that region, local vertex-overlap corrections
are exponentially suppressed at each fixed derivative order, provided
the differentiated profiles obey the corresponding tail bounds.
Gaussian packets are one example. Other localized choices include
$\mathrm{C}^{\infty}$ profiles with strictly compact momentum support and
stretched-exponential position tails.\footnote{Smooth compact profiles have faster-than-power
decay, and suitable bump profiles give stretched-exponential
tails~\cite{Johnson:2015bump}.}
The long-range propagator can nevertheless change at order one.
A geometry that makes massless exchange IR finite can therefore
regulate long-range gravity while preserving the local bulk EFT.

Consider the local three-point vertex producing the $t$-channel
exchange in Figure~\ref{fig:eikonal-cylinder}. For four-dimensional
wavefunctions depending on $x_{\mathrm L}$ and $b$, its tensor components are
\begin{equation}
 \M_{13X^{\mu\cdots}}
 =g\int\dd^2x_{\mathrm L}\dd^2b\,
 \psi_1^{h_1}\,
 \Dcal(-\ii\partial_{\mathrm L},\nabla_b)
 (\psi_3^\dagger)^{h_3}\,\psi_X^{\mu\cdots}.
 \label{eq:scalar-cubic-action}
\end{equation}
Here $\psi^h$ is a spinning wavefunction of helicity $h$.
We use Lorentz indices for $\psi_X^{\mu\cdots}$ to describe a spin-$J$
on-shell state or off-shell exchanged field, such as the graviton
field $h^{\mu\nu}$. The indices on $X$ label the open tensor components
of the amputated vertex. The differential vertex $\Dcal$ specifies the
local coupling and its index contractions. Its derivatives
$\partial_{\mathrm L}$ and $\nabla_b$ may act on any of the three
wavefunctions, on their longitudinal and transverse coordinates
respectively. In the all-incoming convention, with standard external-leg
normalization factored out, the three-point mode integrals give
\begin{subequations}
 \label{eq:cubic-mode-pair}
 \begin{align}
 \M_{1_\alpha^{h_1}\,3_\beta^{h_3}\,X_\gamma^{\mu\cdots}}
 &=(2\pi)^2\delta^{(2)}
 (p_{1\mathrm L}+p_{3\mathrm L}+p_{X\mathrm L})
 \,g_{\alpha\beta\gamma}^{h_1h_3,\mu\cdots},
 \label{eq:mode-cubic-amplitude}\\
 g_{\alpha\beta\gamma}^{h_1h_3,\mu\cdots}
 &=g\int_{\Sigma_2}\dd^2b\,
 \phi_\alpha^{h_1}(b)\,
 \Dcal(p_{\mathrm L},\nabla_b)
 (\phi_\beta^\dagger)^{h_3}(b)\,\phi_\gamma^{\mu\cdots}(b).
 \label{eq:mode-overlap-vertex}
 \end{align}
\end{subequations}
The functions $\phi_\alpha^h(b)$ and $\phi_\gamma^{\mu\cdots}(b)$
are transverse modes of the spinning field. The labels
$\alpha,\beta,\gamma$ specify the modes, while $h$ or the Lorentz indices
specify the helicity or polarization tensor. For an on-shell mode, its
eigenvalue fixes the longitudinal mass shell, as in \eqref{eq:product-mode}.
The eigenmode overlap in \eqref{eq:mode-overlap-vertex} replaces the
transverse momentum-conservation delta function in flat momentum space.

For a $t$-channel spin-$J$ tensor exchange of mass $M_X$, the tree-level
amplitude on $\Sigma_2$ is written in the transverse spectral basis
of the exchanged propagator as
\begin{equation}
 \M_t^{\rm tree}
 =\sum_\gamma
 \frac{g_{13X_\gamma}^{\mu\cdots}g_{X_\gamma24,\mu\cdots}}
 {(t_{\mathrm L}-\lambda_\gamma)-M_X^2+\ii\epsilon},
 \qquad t_{\mathrm L}=(p_{1\mathrm L}+p_{3\mathrm L})^2.
 \label{eq:spin-exchange-mode-sum}
\end{equation}
The two couplings join the $1$--$3$ and $2$--$4$ waves to the off-shell
field $X$. The lower-index coupling includes the tensor contraction and
sign of the exchange line. Lorentz indices are contracted using the metric
of $\mathbb M_{1,1}\times\Sigma_2$.
The absence of transverse translation symmetry means that the exchanged
transverse momentum is no longer fixed by legs $1$ and $3$ alone,
and the flat-space $t_{\mathrm T}$ is replaced by a sum over the transverse
exchange spectrum $\lambda_\gamma$.

The spinning eigenfunctions and their overlaps are generally complicated.
For bootstrap we instead choose localized packets with controlled tails,
matched as $\psi_1\sim\psi_3$ and $\psi_2\sim\psi_4$ at leading eikonal
order. Let $B$ be a characteristic length of $\Sigma_2$, such as its
curvature radius or the distance from the center of a flat region
to its boundary. For paraxial profiles whose transverse spectral scale
is set by their widths, we require
\begin{equation}
 \begin{aligned}
 &|s|^{-1}\ll\lambda_\alpha^{-1}
       \sim(\Delta b)^2,(\Delta b')^2<B^2,\\
 &p_z(\lambda_\alpha)=\sqrt{E^2-\lambda_\alpha}
 =E-\frac{\lambda_\alpha}{2E}
   +\mathcal O\!\left(\frac{\lambda_\alpha^2}{E^3}\right),\\
 &t_{\mathrm L}=-(\delta p_z)^2
       \sim-\frac{(\lambda_\beta-\lambda_\alpha)^2}{s}.
 \end{aligned}
 \label{eq:packet-width-longitudinal-recoil}
\end{equation}
The first line keeps the transverse transfer scale below the hard energy
and the packet widths below the geometric scale $B$.
The remaining lines give the massless eikonal expansion on
$\mathbb M_{1,1}\times\Sigma_2$ for an elastic $1$--$3$ pair at common
physical energy $E\sim\sqrt{s}/2$, with
$\delta p_z=p_z(\lambda_\beta)-p_z(\lambda_\alpha)$.
For characteristic exchange modes $\lambda_\gamma\sim\lambda_\alpha$,
$t_{\mathrm L}$ is suppressed relative to the transverse contribution by
$\lambda_\alpha/|s|$. The recoil in \eqref{eq:spin-exchange-mode-sum}
is therefore dominated by $\lambda_\gamma$, while the longitudinal
plane-wave dependence remains common at leading order.

The condition $\Delta b<B$ does not require a parametric hierarchy.
For example, a normalized Gaussian density proportional to
$\e^{-r^2/(\Delta b)^2}$ in the transverse plane has integrated tail
$\e^{-B^2/(\Delta b)^2}$ beyond radius $B$. At $\Delta b=B/10$ this
is $\e^{-100}$.
For the compact-momentum $\mathrm{C}^{\infty}$ profile, $q$ is the transverse
momentum magnitude and $\Delta b$ its width parameter. The profile
vanishes for $q\Delta b\geq1$ and otherwise has the form below.
Its Fourier transform gives the leading large-distance decay envelope
of the density $\rho=|\psi|^2$~\cite{Johnson:2015bump},
\begin{equation}
 \widetilde\psi(q)\propto
       \exp\!\left[-\frac{1}{1-q^2(\Delta b)^2}\right],\qquad
 \underset{r/\Delta b\to\infty}{\rho(r)}\propto
       \left(\frac{\Delta b}{r}\right)^{5/2}
       \e^{-2\sqrt{r/\Delta b}}.
 \label{eq:compact-bump-tail}
\end{equation}
Thus $\Delta b=B/100$ gives a density suppression factor of order
$10^{-14}$ at $r=B$.

If the third leg is also a localized external packet with its vertex
overlap in the flat region, \eqref{eq:mode-overlap-vertex} becomes the
flat-space packet overlap up to these tail corrections. If instead $X$
is an internal line, summing the exchanged modes reconstructs the two-point
Green function $G_X(t_{\mathrm L};b,b')$. The amplitude is its convolution
with the local $1$--$3$ and $2$--$4$ currents, including the derivatives
and spin contractions in $\Dcal$. The currents retain their matched
flat-space values up to tail corrections, while the Green function can
change at order one. The same local-overlap argument applies to quartic
vertices. Both Gaussian packets and compact-momentum bumps give the
exponential corrections displayed above.
Short-range exchange described by a massive Green function is also
insensitive to sufficiently distant IR geometry. The transverse Yukawa
tail scales as $\e^{-Mr}/\sqrt{Mr}$ for $Mr\gg1$.
For $M\sim\Lambda_{\rm EFT}$ and $B\sim10/\Lambda_{\rm EFT}$,
propagation to the modified region carries an exponential factor
$\e^{-10}\sim10^{-5}$.
Derivatives in local interactions and spin contractions change the
prefactor but not the exponential suppression.

\newpage
\subsection{From zero-mode obstructions to geometric screening}

A PSD packet weight does not by itself make massless exchange IR finite.
IR finiteness also depends on whether the transverse geometry admits a zero mode.
We first consider geometries with a zero mode, including compact connected
surfaces without boundary, such as $S^2$ and $T^2$, and compact disks or boxes
with Neumann boundary conditions. They have finite area $A$ and a gap above
the zero mode. The normalized constant eigenfunction is
$\phi_0=A^{-1/2}$, with $\lambda_0=0$ and
$\int_{\Sigma_2}\dd^2b\,|\phi_0|^2=1$.
These models isolate why a gap above zero does not remove the zero-mode
obstruction. The minimally coupled massless scalar on the Euclidean
section of dS$_4$, discussed in Sec.~\ref{sec:flat-obstruction},
illustrates this issue~\cite{Folacci:1992xc}.
Below, a $\mathbb M_{1,1}\times S^2$ toy model will also illustrate
why de Sitter-like positive curvature alone does not remove the IR divergence.
We will then compare these geometries with the complete hyperbolic plane
$H_2$ and Dirichlet disks or boxes. The Dirichlet boundary condition, or
the negative-curvature geometry with decay at infinity, excludes the
constant zero mode. Their massless Green functions are IR finite and
pointwise positive in impact-parameter space, as shown below.

The Green function is the position kernel of the inverse of $-\Delta$.
On a compact surface without boundary, or in a region with the Neumann
condition of vanishing normal derivative,
$-\Delta_bG(b,b')=\delta_{\Sigma_2}(b,b')$ has no solution. Integrating over
$b$ gives zero on the left by the divergence theorem and one on the right.
Removing the zero mode gives the transverse Green function
\begin{equation}
  G_{\cancel{0}}(b,b')
  =\sum_{\alpha\neq0}
  \frac{\phi_\alpha(b)\phi_\alpha^*(b')}{\lambda_\alpha}.
  \label{eq:pseudoinverse-spectral}
\end{equation}
The slash through zero denotes the omitted mode. We interpret this
subtraction as the transverse spectral counterpart of the infrared
cancellation between FK dressing and the virtual soft
factor. A dressing profile that resolves only the isolated zero mode
below the transverse gap leaves the higher nonzero modes unchanged.
The zero-mode-subtracted Green function obeys the equation and
zero-average convention
\begin{equation}
  -\Delta_bG_{\cancel{0}}(b,b')=\delta_{\Sigma_2}(b,b')-\frac1A,
  \qquad
  \int_{\Sigma_2}\dd^2b\,G_{\cancel{0}}(b,b')=0.
  \label{eq:pseudoinverse-equation}
\end{equation}
The matched nonnegative densities in
\eqref{eq:transverse-packet-densities} give the impact-parameter smearing
of \eqref{eq:eikonal-spectral-smearing}, with the common $s$- and
$u$-channel PSD weight described by
\eqref{eq:eikonal-common-weight}--\eqref{eq:eikonal-helicity-crossing}.
Within this class, two separated localized beams can be chosen such that
\begin{equation}
 \int_{\Sigma_2}\dd^2b\,\dd^2b'\,
 \rho_{13}(b)G_{\cancel{0}}(b,b')\rho_{24}(b')<0.
 \label{eq:compact-negative-pairing}
\end{equation}
Indeed, the kernel takes both signs. Its local behavior is
$-(2\pi)^{-1}\log[r(b,b')/\ell]$ plus a finite term, where $r(b,b')$
is geodesic distance and $\ell$ is a geometric IR length. It is positive
at sufficiently small separation, while the zero area average in
\eqref{eq:pseudoinverse-equation} forces a negative region elsewhere.
Two separated beams can be supported in that region. The gap makes the
zero-mode-subtracted inverse IR finite, but does not make its position
kernel everywhere nonnegative.

We now calculate examples of this Green function and the resulting
leading eikonal phase. For the gravitational exchange in
\eqref{eq:eikonal-amplitude-mode-transform}, this is the 1PM phase
$2\pi G_NsG_{\cancel{0}}(b,b')$ of the reduced transverse model.
A zero mode with a gap above it occurs both on closed surfaces, such
as $S^2$ and the flat $T^2$, and on finite domains with Neumann boundary
conditions. We use the round $S^2$ and a flat Neumann disk as explicit examples.

On a sphere of radius $R$, choose a pole and use geodesic polar
coordinates $r=R\theta$ and azimuth $\varphi$. The $SO(3)$ isometry
allows either source point to be chosen as the pole, so the Green
function depends only on the separation $r$,
\begin{equation}
 \begin{aligned}
 G_{S^2,\cancel{0}}(r)
 &=\frac1{4\pi}\sum_{l=1}^{\infty}
   \frac{2l+1}{l(l+1)}P_l\!\left(\cos\frac rR\right)\\
 &=\frac1{4\pi}\left[-2\log\sin\frac r{2R}-1\right],
 \qquad 0<r\leq\pi R.
 \end{aligned}
 \label{eq:S2-pseudoinverse-closed}
\end{equation}
The omitted $l=0$ term has exactly zero eigenvalue. Its IR divergence
comes from this discrete mode, not merely from a limiting sequence of
small nonzero transfers. After its removal,
\begin{equation}
 G_{S^2,\cancel{0}}(r)<0
 \quad\Longleftrightarrow\quad
 \frac rR>2\arcsin(\e^{-1/2})\simeq1.303.
 \label{eq:S2-negative-region}
\end{equation}
Take normalized $\rho_{13}$ and $\rho_{24}$ supported in polar caps
around $r/R=0$ and $\pi$. Packet radii satisfying
$|s|^{-1/2}\ll\Delta b,\Delta b'\lesssim0.919R$
allow eikonal sampling of the negative region. For narrower caps,
$\Delta b,\Delta b'\ll R$, the expansion around antipodal separation gives
\begin{equation}
 \int\dd^2b\,\dd^2b'\,
 \rho_{13}(b)G_{S^2,\cancel{0}}(b,b')\rho_{24}(b')
 =-\frac1{4\pi}
  +\Ocal\!\left(\frac{(\Delta b+\Delta b')^2}{R^2}\right)<0.
 \label{eq:S2-positive-packets-negative-limit}
\end{equation}
Thus the smeared tree phase is negative even for packets that remain
within the eikonal regime. For $G_N|s|\ll1$, this tree term controls the
low-energy expansion. At larger $G_Ns$, the position-space amplitude
in \eqref{eq:eikonal-amplitude-mode-transform} is exponentiated pointwise
before smearing. Define the average of a function of $b,b'$ as below.
For the antipodal caps and real $s$, the standard complex-exponential
remainder bound~\cite{Shevtsova:2014exponent} gives the simple estimate
\begin{equation}
 \begin{aligned}
 \langle f\rangle_\phi
 &\equiv\int\dd^2b\,\dd^2b'\,\rho_{13}(b)\rho_{24}(b')f(b,b'),\\
 \left|\langle\e^{2\ii\delta}\rangle_\phi
       -\e^{2\ii\langle\delta\rangle_\phi}\right|
 &\leq G_N|s|\frac{(\Delta b+\Delta b')^2}{R^2}.
 \end{aligned}
 \label{eq:packet-exponentiation-error}
\end{equation}
Here $\delta=2\pi G_NsG_{\cancel{0}}(b,b')$.
For $\Delta b\sim\Delta b'$, the condition
$G_N|s|(\Delta b)^2/R^2\ll1$ therefore permits the approximation
$\langle\e^{2\ii\delta}\rangle_\phi\simeq\e^{2\ii\langle\delta\rangle_\phi}$.

For a flat disk of radius $R$ with Neumann boundary conditions, the
zero-average Green function is~\cite{Bressloff:2024diffusion}
\begin{equation}
 \begin{aligned}
 G_{D^2_{\rm N},\cancel{0}}(b,b')
 &=-\frac1{2\pi}\log\!\left[
   \frac{|z-z'|}{R}\left|1-\frac{z\bar z'}{R^2}\right|\right]
   +\frac{|z|^2+|z'|^2}{4\pi R^2}-\frac3{8\pi},\\
 G_{D^2_{\rm N},\cancel{0}}(r,0)
 &=\frac1{2\pi}\left[-\log\frac rR+\frac{r^2}{2R^2}-\frac34\right].
 \end{aligned}
 \label{eq:Neumann-disk-green}
\end{equation}
Here $z=b^1+\ii b^2$ and $r=|z|$ in the second line. Place one beam
at the center. The negative region begins at $r=r_*$ below.
Place the other beam at $r=0.8R$, away from both the sign-change radius
and the boundary. For normalized packets symmetric about their
respective centers, with
$|s|^{-1/2}\ll\Delta b\sim\Delta b'\ll0.2R$, one obtains
\begin{equation}
 \begin{aligned}
 &G_{D^2_{\rm N},\cancel{0}}(r,0)<0,
   \qquad 0.549R\simeq r_*<r\leq R,\\
 &\Bigl\langle G_{D^2_{\rm N},\cancel{0}}\Bigr\rangle_\phi
 =-3.292\times10^{-2}
  +\Ocal\!\left(\frac{(\Delta b)^2}{R^2}\right).
 \end{aligned}
 \label{eq:Neumann-disk-negative-packets}
\end{equation}
Other geometries, including $T^2$ and a Neumann box, exhibit the same
zero-mode obstruction. We do not repeat their calculations here.
On any connected compact geometry of this class, the zero-average
Green function has a negative open region. Sufficiently focused positive
beam densities realize \eqref{eq:compact-negative-pairing} whenever the
eikonal scale separation is available. With small phase variation across
the packets, the ladder sum can be estimated using
$\e^{2\ii\langle\delta\rangle_\phi}$, as in \eqref{eq:packet-exponentiation-error}.

The dispersive consequence is particularly transparent in the leading
eikonal model of GR. Keep a finite inner contour
$\mathcal C_0$ of radius $s_0$, with
$R^{-2}\ll t_{\max}\ll s_0\ll G_N^{-1}$.
Localized Gaussian packets, or smooth compact packets as above, give
the same leading $s$- and $u$-channel weight after crossing, by
\eqref{eq:eikonal-common-weight}--\eqref{eq:eikonal-helicity-crossing}.
With the impact-parameter normalization of
\eqref{eq:eikonal-amplitude-mode-transform}, the smeared transition
response $\mathcal T_\phi$ has the Hermitian-analytic boundary values
below, understood as $\epsilon\to0^+$,
\begin{equation}
 \begin{aligned}
 \mathcal T_\phi(s\pm\ii\epsilon)
 &=\pm\ii\Bigl\langle
  1-\e^{\pm4\pi\ii G_NsG_{\cancel{0}}(b,b')}
  \Bigr\rangle_\phi,\\
 \frac{\Disc\mathcal T_\phi(s)}{2\ii}
 &=\Bigl\langle1-\cos\!\left[4\pi G_NsG_{\cancel{0}}(b,b')\right]
  \Bigr\rangle_\phi\geq0.
 \end{aligned}
 \label{eq:compact-eikonal-cut-positivity}
\end{equation}
The phase is linear in $s$. Crossing continues from the physical
$s$-channel at $s=m_s^2+\ii\epsilon$ through the upper half-plane to
$s=-m_u^2+\ii\epsilon+\Ocal(t_{\max})$, without crossing a branch cut.
The relative corrections $\Ocal(t_{\max}/|s|)$ are omitted at leading
eikonal order, giving the same positive physical $u$-channel discontinuity.

\begin{figure}[H]
 \centering
 \begin{tikzpicture}[x=1cm,y=1cm,>=stealth,font=\small]
 \definecolor{contourblue}{RGB}{34,94,153}
 \definecolor{cutorange}{RGB}{196,102,30}
 \draw[->,gray!70] (-4.2,0)--(4.35,0) node[right] {$\operatorname{Re}s$};
 \draw[->,gray!70] (0,-3.55)--(0,3.8) node[above] {$\operatorname{Im}s$};
 \draw[contourblue,thick] (4:3.1) arc[start angle=4,end angle=176,radius=3.1];
 \draw[contourblue,thick] (184:3.1) arc[start angle=184,end angle=356,radius=3.1];
 \draw[contourblue,thick] (6:1.3) arc[start angle=6,end angle=174,radius=1.3];
 \draw[contourblue,thick] (186:1.3) arc[start angle=186,end angle=354,radius=1.3];
 \draw[contourblue,thick,->] (35:3.1) arc[start angle=35,end angle=59,radius=3.1];
 \draw[contourblue,thick,->] (213:3.1) arc[start angle=213,end angle=237,radius=3.1];
 \draw[contourblue,thick,->] (41:1.3) arc[start angle=41,end angle=75,radius=1.3];
 \draw[contourblue,thick,->] (221:1.3) arc[start angle=221,end angle=255,radius=1.3];
 \draw[cutorange,line width=2.1pt] (1.3,0)--(3.9,0);
 \draw[cutorange,line width=2.1pt] (-1.3,0)--(-3.9,0);
 \draw[cutorange,line width=1.0pt] (.10,0)--(1.3,0);
 \draw[cutorange,line width=1.0pt] (-.10,0)--(-1.3,0);
 \draw[cutorange,->] (1.65,.14)--(2.55,.14);
 \draw[cutorange,->] (2.55,-.14)--(1.65,-.14);
 \draw[cutorange,->] (-2.55,.14)--(-1.65,.14);
 \draw[cutorange,->] (-1.65,-.14)--(-2.55,-.14);
 \node[contourblue,fill=white,inner sep=3pt] at (-2.1,2.4) {$\mathcal C_\infty$};
 \node[contourblue,fill=white,inner sep=2pt] at (-.85,1.02) {$\mathcal C_0$};
 \node[cutorange,fill=white,inner sep=2pt] at (2.15,.5) {$s$-channel};
 \node[cutorange,fill=white,inner sep=2pt] at (-2.15,.5) {$u$-channel};
 \node[below right,fill=white,inner sep=1.5pt] at (1.3,-.2) {$s_0$};
 \node[below left,fill=white,inner sep=1.5pt] at (-1.3,-.2) {$-s_0$};
 \node[fill=white,inner sep=3pt] at (0,.45) {$\displaystyle\oint_{\mathcal C_0}<0$};
 \node[fill=white,inner sep=2pt,font=\footnotesize] at (0,-2.05) {$t_{\max}\ll s_0\ll G_N^{-1}$};
\end{tikzpicture}
 \caption{The finite low-energy contour $\mathcal C_0$ and the large
 contour $\mathcal C_\infty$, both oriented counterclockwise.
 The physical cuts continue inside $\mathcal C_0$
 (thin orange). For $G_Ns_0\ll1$, their leading eikonal contribution
 comes from outside (thick orange). The negative inner arc and positive
 cut integrals require a nonzero negative large-contour contribution.}
 \label{fig:compact-dispersion-contours}
\end{figure}
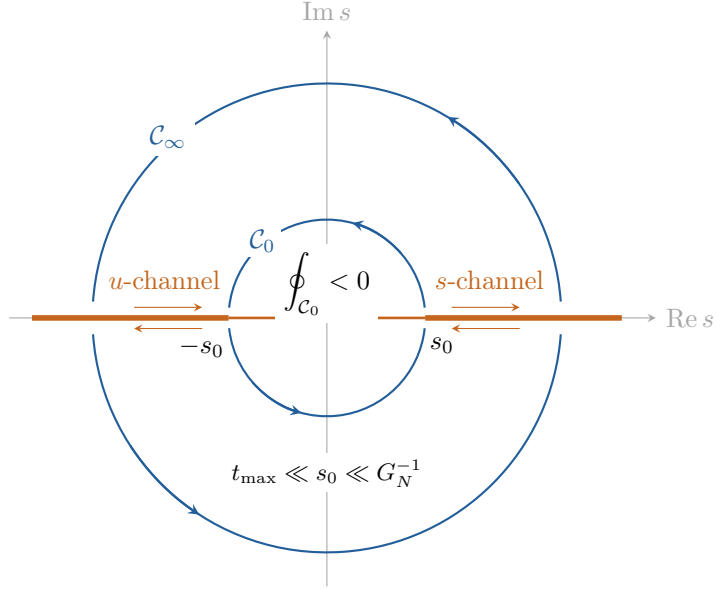

Whatever intermediate absorptive spectrum the geometry allows,
the physical-channel discontinuities are nonnegative by the optical
theorem \eqref{eq:single-packet-optical-theorem}.
For the packets above, however, the finite low-energy arc is
$\langle2\delta/s\rangle_\phi+\Ocal(G_N^2s_0)<0$.
Keeping the large contour explicitly gives
\begin{equation}
 \oint_{\mathcal C_0}\frac{\dd s\,\mathcal T_\phi(s)}{2\pi\ii s^2}
 =\int_{\Lambda_s^2}^{\infty}\frac{\dd s\,\Disc\mathcal T_\phi(s)}{2\pi\ii s^2}
 +\int_{-\infty}^{-\Lambda_u^2}\frac{\dd s\,\Disc\mathcal T_\phi(s)}{2\pi\ii s^2}
 +\oint_{\mathcal C_\infty}\frac{\dd s\,\mathcal T_\phi(s)}{2\pi\ii s^2}.
 \label{eq:compact-zero-mode-contour-diagnosis}
\end{equation}
The endpoints $\Lambda_s^2$ and $-\Lambda_u^2$ are the intersections
of the cuts with $\mathcal C_0$. For the circular contour in
Figure~\ref{fig:compact-dispersion-contours}, $\Lambda_s^2=\Lambda_u^2=s_0$.
The physical cuts extend inside $\mathcal C_0$ as well as outside it.
In the eikonal model of GR their discontinuities become order one
at $G_Ns\sim1$. For $G_Ns_0\ll1$, the weighted cut contribution inside
$\mathcal C_0$ is only $\Ocal(G_N^2s_0)$. The leading $\Ocal(G_N)$
contribution comes from outside, as indicated in
Figure~\ref{fig:compact-dispersion-contours}.

The contour $\mathcal C_\infty$ denotes the large-circle limit.
Its contribution is negative and nonzero. The zero-mode-subtracted
eikonal response considered here does not admit the extension of its
real-axis UV bound to arbitrary complex directions available for an
ordinary two-particle Fock amplitude under the standard analyticity
and growth assumptions\footnote{\(\mathcal T_\phi(s)=-\ii\int\dd^2b\,\widetilde\phi(b)[S(s,b)-1]
 =(8\pi s)^{-1}\int\dd t\,\phi(t)\M(s,t)
 =(8\pi s)^{-1}\M_\phi(s)\), where
$\M_\phi(s)$ denotes the momentum-space transfer smearing. The bound
$|\M_\phi(s)|\leq C|s|$ therefore becomes
$|\mathcal T_\phi(s)|\leq C/(8\pi)$.}~\cite{Caron-Huot:2022ugt}.
The real-axis UV bound alone does not establish this complex-energy bound.

An IR regulator that retains the zero mode can supply a sufficiently
large positive constant to make the regulated Green function
nonnegative throughout the transverse space. Removing this regulator
dependence from exclusive scattering requires a four-dimensional
FK construction using the modes admitted by the geometry.
The soft cloud must satisfy $R^{-1}\ll\omega\ll\sqrt{|s|}$.
Its effective IR cutoff $\omega$ then lies above the geometric
resolution scale $R^{-1}$. At this resolution, its soft sector is
effectively that of a flat-space dressed amplitude. In flat space, FK dressing specifies a
soft profile in momentum space and preserves translation symmetry.
The soft cloud acts as a background accompanying the hard particles.
Unitarity, analyticity and crossing symmetry can still hold when the
reference-frame dependence introduced by the dressing profile is
explicitly accounted for. The asymptotic states then contain arbitrarily
many soft quanta and are not ordinary two-particle Fock states.
The real-axis UV bound supplied by unitarity does not thereby extend
automatically to arbitrary complex directions, paralleling the
obstruction indicated above by the negative zero-mode-subtracted
Green function and the nonzero $\mathcal C_\infty$ contour
contribution~\cite{Dong:ProjectA}.

Within the weak-gravity region $s\ll(G_N^2t_{\max})^{-1}$, we use the
gravitational eikonal expression \eqref{eq:compact-eikonal-cut-positivity}.
The Schwarzschild radius is then smaller than the packet length scales,
and complex-energy growth is already visible in this region.
For a narrow packet pair sampling positive $G_{\cancel{0}}$, the
low-energy arc and the subtracted physical-cut contribution cancel.
For negative $G_{\cancel{0}}$, they reinforce. This follows from the
negative low-energy residue together with the nonnegative physical
$s$- and $u$-channel discontinuities. Their contour balance in
\eqref{eq:compact-zero-mode-contour-diagnosis} requires a nonzero
contribution from $\mathcal C_\infty$.
At fixed positions with $G_{\cancel{0}}(b,b')<0$, continuation to
$s+\ii y$ gives
\begin{equation}
 |S(s+\ii y,b,b')|
 =\exp\!\left[-4\pi G_NG_{\cancel{0}}(b,b')\,y\right],
 \qquad y>0.
 \label{eq:compact-eikonal-complex-growth}
\end{equation}
The scale $y\sim G_N^{-1}$ already displays exponential growth
within the weak-gravity regime above. This is an indication of a
nonzero $\mathcal C_\infty$ contour contribution in
\eqref{eq:compact-zero-mode-contour-diagnosis}.
Even if black-hole formation changes the ladder approximation at
$s\gtrsim(G_N^2t_{\max})^{-1}$, its contribution to
\eqref{eq:compact-zero-mode-contour-diagnosis} is subleading.
For normalized packets, unitarity gives
$0\leq\operatorname{Im}\mathcal T_\phi\leq2$, so the combined $s$- and
$u$-channel contribution above that scale is at most
$4G_N^2t_{\max}/\pi$, consistent with the eikonal black-hole tail
estimate of~\cite{Haring:2024wyz}.
This estimate concerns the breakdown of the classical post-Minkowskian
description at the black-hole scale. Quantum corrections of relative
order $\Ocal(G_Nt_{\max})\ll1$ are also omitted at the classical order
retained here~\cite{DiVecchia:2023frv}.
The negative low-energy arc, the nonzero $\mathcal C_\infty$ contour
contribution and complex-direction exponential growth are different
manifestations of the same obstruction.

We next consider transverse spaces without a zero mode, beginning with
the AdS$_4$-inspired eikonal toy model $\mathbb M_{1,1}\times H_2$.
The AdS eikonal construction in~\cite{Cornalba:2007zb}
also reduces the exchange to a
hyperbolic transverse space. Its graviton propagator on $H_2$ inverts
$-\Delta_{H_2}+2/R^2$, however, whereas the massless transverse model
here inverts $-\Delta_{H_2}$. The two share geometric suppression of
long-range exchange, but their Green functions and scattering problems
are different.

On the complete $H_2$, decay at infinity excludes the zero mode.
The continuous spectrum starts at $1/(4R^2)$, where $R$ is the curvature
radius. In geodesic polar coordinates, an $SO(2)$-adapted eigenbasis is
\begin{equation}
 \phi_{\kappa m}(r,\theta)
 =N_{\kappa m}\e^{\ii m\theta}
 P^{m}_{-\frac12+\ii\kappa}\!\left(\cosh\frac rR\right),
 \qquad
 \lambda_\kappa=\frac{\kappa^2+\frac14}{R^2}.
 \label{eq:H2-mode-basis}
\end{equation}
Here $\kappa\geq0$ is continuous and $m\in\mathbb Z$.
As for the inverse Laplacian above, the Green function solves
$-\Delta_{H_2,b}G_{H_2}(b,b')=\delta_{H_2}(b,b')$ and vanishes at
infinity. In terms of geodesic separation $r=r(b,b')$, its solution
is~\cite{Cohl:2012hyperbolic}
\begin{equation}
 G_{H_2}(r)=\frac1{2\pi}\log\coth\frac r{2R}.
 \label{eq:H2-massless-green}
\end{equation}
It is positive at distinct points and decays exponentially at large
separation. At fixed physical $r$, the flat logarithm is recovered as
$R\to\infty$,
\begin{equation}
 G_{H_2}(r)\sim\frac1\pi\e^{-r/R}\quad(r\gg R),
 \qquad
 G_{H_2}(r)=\frac1{2\pi}\log\frac{2R}{r}
             +\Ocal\!\left(\frac{r^2}{R^2}\right)\quad(R\to\infty).
 \label{eq:H2-large-distance}
\end{equation}
Take matched elastic packets $\psi_1\sim\psi_3$ and $\psi_2\sim\psi_4$,
with the normalized nonnegative densities of
\eqref{eq:transverse-packet-densities}. They need not have the same profile.
For smooth localized packets,
\begin{equation}
 \int_{H_2}\dd^2b\,\dd^2b'\,
 \rho_{13}(b)G_{H_2}\!\bigl(r(b,b')\bigr)\rho_{24}(b')>0.
 \label{eq:H2-positive-pairing}
\end{equation}
This is the genuine inverse of the transverse Laplacian, with no zero-mode
subtraction. Together with the packet positivity established in
\eqref{eq:eikonal-density-contraction}--\eqref{eq:single-packet-optical-theorem},
it gives an IR-finite positive smearing in the leading eikonal model and
removes the zero-mode obstruction to using Fock asymptotic states.
The remaining drawback is local curvature. The bulk derivative expansion
still carries power corrections, such as $\Ocal((R\Lambda_{\rm EFT})^{-2})$
in \eqref{eq:ads-eft-control}.

\begin{figure}[tbp]

 \centering
 \includegraphics[width=.97\linewidth]{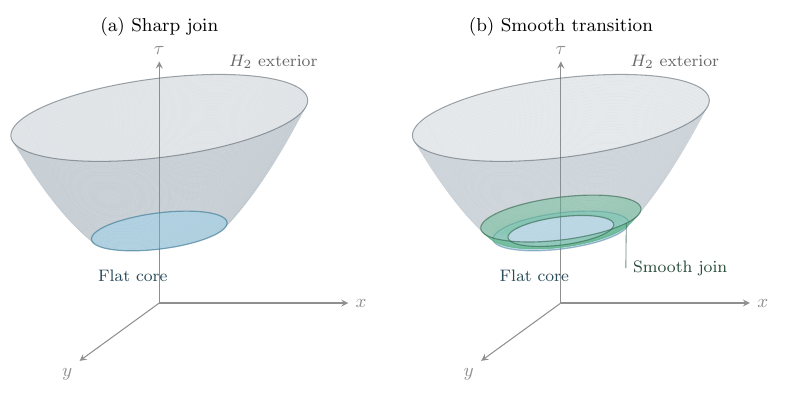}
 \caption{A flat core with a hyperbolic exterior, shown in the auxiliary
 embedding $\mathbb R^{1,2}$. The left panel replaces the hyperbolic cap by a flat
 disk, retaining a sharp circular join. The right panel smooths a finite annulus
 while preserving a strictly flat interior and the hyperbolic end.
 Light blue, green and gray indicate the flat, transition and hyperbolic regions.
 The right panel retains a translucent copy of the sharp join beneath the
 concentric, rotationally symmetric transition annulus.
 The coordinate $\tau$ is an embedding coordinate, not physical time.
 The outer rim only truncates the drawing.}
 \label{fig:flat-core-hyperbolic}
\end{figure}

An exactly flat scattering core can instead be joined to a hyperbolic
exterior. To distinguish their independent sizes, let $R$ now denote the
flat-disk radius and $R_{\rm ext}$ the exterior curvature radius.
In the auxiliary embedding $\mathbb R^{1,2}$, cut the upper hyperboloid
$\tau^2-x^2-y^2=R_{\rm ext}^2$ at
$\tau=\sqrt{R_{\rm ext}^2+R^2}$ and replace its cap by the flat disk
in that plane. The resulting surface has a sharp join. Smoothing an
annulus around the join leaves a smaller strictly flat core and an
exactly hyperbolic exterior, producing a smooth asymptotically hyperbolic
surface~\cite{Mazzeo:1987resolvent,Guillarmou:2005resolvent}. See
figure~\ref{fig:flat-core-hyperbolic}.

For the sharp join, the intrinsic metric in proper radial distance is
\begin{equation}
 \dd s_{\Sigma_2}^2=\dd r^2+f(r)^2\dd\theta^2,
 \qquad
 f(r)=\begin{cases}
 r,&0\leq r\leq R,\\[2pt]
 R_{\rm ext}\sinh\!\left[\dfrac{r-R}{R_{\rm ext}}
             +\operatorname{arsinh}\dfrac{R}{R_{\rm ext}}\right],&r\geq R.
 \end{cases}
 \label{eq:flat-core-hyperbolic-metric}
\end{equation}
\begin{samepage}
For the massless scalar Laplacian, a radial conformal map that leaves
the flat core unchanged sends the full surface to a disk of radius
\begin{equation}
 R_{\rm eff}
 =R\exp\!\left[\int_R^\infty\frac{\dd r}{f(r)}\right]
 =R_{\rm ext}+\sqrt{R^2+R_{\rm ext}^2}.
 \label{eq:flat-core-effective-radius}
\end{equation}
\end{samepage}
The integral measures the exterior's contribution to the conformal
radius. The interior is already represented by the unchanged coordinate
$r$. In two dimensions the conformal factors of the Laplacian and its
delta-function source cancel. With decay at infinity, the Green function
for two points in the core is therefore exactly
\eqref{eq:disk-massless-center-green} with $R$ replaced by $R_{\rm eff}$.
A rotationally symmetric smoothing changes the effective radius but
keeps it finite, and preserves this positive, IR-finite Green function.

For packets supported entirely inside the flat core, local vertex
overlaps coincide exactly with their flat-space values. A sharp
position-space cutoff produces power-law momentum tails, which are
unsuitable for the smearing needed to bound the scale of new physics.
To balance this bootstrap requirement with suppressing the geometric
corrections to local contact interactions, a good choice is a packet
that is Gaussian in both position and momentum in the flat core, centered
inside the flat core and away from the transition.
Only its exponentially small tails reach the
transition, so local three- and four-point overlaps receive exponentially
suppressed corrections by the local-vertex analysis above.
Long-range exchange can change by order one. This realizes the desired
separation on $\mathbb M_{1,1}\times\Sigma_2$ with an asymptotically
hyperbolic $\Sigma_2$. The full mode overlaps depend on the smoothing.
We do not compute them here.

Equivalently, eliminating the exterior wavefunction with the spectral
Dirichlet-to-Neumann (DtN) operator supplies the boundary condition for the
interior mode equation. For the massless Green problem, at fixed $R$
and fixed $SO(2)$ angular quantum number $m$, the DtN eigenvalue grows
as $1/R_{\rm ext}$ when $R_{\rm ext}\to0$, forcing the wavefunction
at $r=R$ to vanish. Equivalently,
\eqref{eq:flat-core-effective-radius} gives $R_{\rm eff}\to R$.
Thus increasing the magnitude of the exterior negative curvature
$-1/R_{\rm ext}^2$ induces the Dirichlet limit.

The resulting Dirichlet disk has the explicit mode basis
\begin{equation}
 \phi_{mn}(r,\theta)
 =N_{mn}J_{|m|}\!\left(\frac{j_{|m|,n}r}{R}\right)\e^{\ii m\theta},
 \qquad
 \lambda_{mn}=\frac{j_{|m|,n}^2}{R^2}>0.
 \label{eq:disk-mode-basis}
\end{equation}
Here $m\in\mathbb Z$, $n=1,2,\ldots$, and $j_{|m|,n}$ is the $n$th
positive zero of $J_{|m|}$. All eigenvalues are strictly positive, so
there is no zero mode. Orthogonality follows from
\begin{equation}
 \int_0^R r\dd r\,
 J_{|m|}\!\left(\frac{j_{|m|,n}r}{R}\right)
 J_{|m|}\!\left(\frac{j_{|m|,n'}r}{R}\right)
 =\frac{R^2}{2}J_{|m|+1}(j_{|m|,n})^2\delta_{nn'}.
 \label{eq:bessel-orthogonality}
\end{equation}
which gives $|N_{mn}|^{-2}=\pi R^2J_{|m|+1}(j_{|m|,n})^2$ in the complex
angular basis. The Green function for two arbitrary interior points,
and its specialization to a source at the center, are
\begin{equation}
 \begin{aligned}
 G_D(b,b')
 &=\frac1{4\pi}\log
 \frac{R^4+|b|^2|b'|^2-2R^2b\cdot b'}{R^2|b-b'|^2},\\
 G_D(r,0)&=\frac1{2\pi}\log\frac Rr,\qquad 0<r<R.
 \end{aligned}
 \label{eq:disk-massless-center-green}
\end{equation}
Both are positive in the interior and vanish when either point reaches
the boundary. Consequently, the normalized nonnegative densities
$\rho_{13}$ and $\rho_{24}$ give a positive pairing of the form
\eqref{eq:H2-positive-pairing}, with $H_2$ replaced by the disk.
The resulting eikonal smearing is positive and IR finite. For Gaussian
packets centered away from the wall, local EFT overlaps retain only
exponentially suppressed boundary corrections, while the long-range
Green function changes by order one.

These constructions establish feasibility in the leading eikonal regime.
Optimizing a bootstrap bound requires going beyond it. The present
low-energy contour lies at $s_0\gg t_{\max}$, restricting the new-physics
scales that can be resolved by this construction. To push the contour
toward $t_{\max}$, finite-angle scattering must be included.
In the eikonal limit, each free eigenmode has the separated form in
\eqref{eq:product-mode}. What fails at finite angle is the common
longitudinal plane-wave approximation for a localized transverse packet.
When $\lambda_\alpha/|s|$ is not small, its different transverse components
have different longitudinal kinematics, and the full mode overlaps enter.

Even the Dirichlet disk then lacks transverse translations and the full
spatial rotation group underlying flat-space partial waves. Its
transverse spectrum is discrete. A packet description must account for
repeated wall reflections and possible repeated encounters between the
beams. These features make finite-angle scattering substantially harder
than the leading straight-ray problem. The half-space considered next
keeps the bulk flat while retaining translations parallel to the wall
and a continuous transverse spectrum, providing a simpler setting for
the bootstrap beyond this eikonal test.

\subsection{Half-space from a curved exterior}
\label{sec:halfspace-geometry}

The minimal model is a flat half-space spacetime with a Dirichlet
boundary at $y=0$. Its transverse geometry is
$\mathbb H_2=\{(x,y):y\geq0\}$. It preserves $ISO(1,2)$ acting on
$(t,x,z)$, extending the longitudinal $ISO(1,1)$ of the preceding
product geometries. Of the four-dimensional Poincar\'e generators,
only the translation in $y$ and the three Lorentz generators mixing
$y$ with $t,x,z$ are broken. Its transverse massless Green function
follows directly from the image method. Writing $X=(x,y)$,
\begin{equation}
 G_h(X,X')=\frac1{2\pi}\log\frac{|X-\bar X'|}{|X-X'|},
 \qquad \bar X'=(x',-y').
 \label{eq:half-plane-green-mode-chapter}
\end{equation}
For $y,y'\geq0$, $|X-\bar X'|\geq|X-X'|$, so $G_h\geq0$.
It is strictly positive at distinct interior points.
Fourier transformation in $x-x'$ gives the $(p_x,y)$ momentum--position
mixed representation,
\begin{equation}
 G_h(p_x,y,y')
 =\frac{\e^{-|p_x||y-y'|}-\e^{-|p_x|(y+y')}}{2|p_x|},
 \label{eq:mixed-dirichlet-green}
\end{equation}
which is likewise nonnegative and extends continuously to $p_x=0$, with
\begin{equation}
 \lim_{p_x\to0}G_h(p_x,y,y')=\min(y,y').
 \label{eq:mixed-dirichlet-zero-limit}
\end{equation}
The $1/(2|p_x|)$ IR divergence thus cancels between the direct and image
terms. The eigenfunctions in the $y$ direction are
\begin{equation}
 \phi_{k_y}(y)=\sqrt{\frac2\pi}\sin(k_y y),\qquad k_y>0.
 \label{eq:halfline-sine-basis}
\end{equation}
Here $k_y$ labels the independent sine modes. Opposite signs give the
same mode up to a sign, and $k_y=0$ gives the zero function.
The spectrum has no hard gap. The boundary condition excludes a zero
mode, while arbitrarily soft sine modes remain. The half-space therefore
keeps the bulk spacetime flat and the massless spectrum continuous down
to zero. Its single-wall image cancellation removes the zero-mode
ambiguity without the repeated reflections of a finite cavity.

As in the preceding construction, spacetime curvature can induce the
Dirichlet boundary at $y=0$. A concrete metric realization keeps the
scattering region $y>0$ exactly flat and joins it to a strongly warped
exterior at $y<0$,
\begin{equation}
 \dd s^2=
 \begin{cases}
  \eta_{ab}\dd x^a\dd x^b-\dd y^2, & y>0,\\[2mm]
  \e^{-2y/R}\eta_{ab}\dd x^a\dd x^b-\dd y^2, & y<0,
 \end{cases}
 \qquad a,b=t,x,z.
 \label{eq:warped-halfspace-metric}
\end{equation}
Introduce a single conformal coordinate,
\begin{equation}
 \widetilde y=
 \begin{cases}
  y+R, & y\geq0,\\
  R\e^{y/R}, & y<0,
 \end{cases}
 \qquad
 \Omega(\widetilde y)=
 \begin{cases}
  1, & \widetilde y\geq R,\\
  R/\widetilde y, & 0<\widetilde y<R.
 \end{cases}
 \label{eq:warped-wall-coordinates}
\end{equation}
The geometry is a Poincar\'e $AdS_4$ region joined to a Minkowski
half-space at $y=0$. Its metric on the full manifold is
\begin{equation}
 \dd s^2=\Omega^2(\widetilde y)
 (\eta_{ab}\dd x^a\dd x^b-\dd\widetilde y^2),
 \qquad \Rcal_{\rm ext}=-\frac{12}{R^2}.
 \label{eq:warped-wall-curvature}
\end{equation}
Each fixed-$\widetilde y$ slice is $\mathrm{Minkowski}_{1,2}$.
The fourth coordinate $\widetilde y$ is spatial.
The join is at $\widetilde y=R$, or $y=0$, where both induced metrics
equal $\eta_{ab}$. The AdS conformal boundary is at $\widetilde y=0$,
equivalently $y\to-\infty$.

For a minimally coupled massless scalar, the standard no-source
AdS boundary condition sets the constant boundary value to zero.
The remaining branch vanishes as $\widetilde y^3$~\cite{Witten:1998qj}.
Write the exterior and flat-side profiles as $\chi_-(\widetilde y)$
and $\chi_+(\widetilde y)$. At zero tangential momentum the field
equation is $\partial_{\widetilde y}
(\Omega^2\partial_{\widetilde y}\chi_-)=0$.
Integrating from the no-source endpoint to the interface gives
\begin{equation}
 \chi_-(R)=\partial_{\widetilde y}\chi_-(R)
 \int_{0}^{R}\frac{\dd\widetilde y}{\Omega^2(\widetilde y)}
 =\frac R3\,\partial_{\widetilde y}\chi_-(R).
 \label{eq:warped-wall-endpoint}
\end{equation}
Here $\Omega^2$ is retained. Unlike the hyperbolic exterior joined
to a flat disk in the preceding subsection, the present problem
uses the four-dimensional scalar Laplacian, for which the
two-dimensional massless Green equation's conformal cancellation
does not apply.

For nonzero normal momentum $k_y$, the no-source exterior solution is
$\chi_-(\widetilde y)\propto\sin(k_y\widetilde y)
-k_y\widetilde y\cos(k_y\widetilde y)$.
Matching its value and normal derivative to the flat-side waves at
$\widetilde y=R$ determines the reflection coefficient exactly,
\begin{equation}
 \begin{aligned}
 \chi_+(\widetilde y)
 &=\e^{-\ii k_y(\widetilde y-R)}
   +\mathcal C\e^{\ii k_y(\widetilde y-R)},\\
 \mathcal C
 &=\frac{\sin(k_yR)-k_yR\e^{\ii k_yR}}
        {\sin(k_yR)-k_yR\e^{-\ii k_yR}}
   \xrightarrow{R\to0}-1.
 \end{aligned}
 \label{eq:warped-wall-impedance}
\end{equation}
Consequently $\chi_+(R)=1+\mathcal C\to0$ as $R\to0$.
The no-source AdS exterior thus induces a Dirichlet boundary at
$y=0$. Eliminating the exterior through its Dirichlet-to-Neumann
map gives the same reflection coefficient.

The same geometry also fixes reflection of Maxwell waves and linearized
gravitational waves. Use axial gauge, $A_{\widetilde y}=h_{\widetilde y\mu}=0$,
with the no-source AdS endpoint and no additional interface current or
transverse-traceless stress source. In four dimensions, conformal
invariance gives the Maxwell profile
$\chi_-(\widetilde y)\propto\sin(k_y\widetilde y)$.
Matching at $\widetilde y=R$ gives
$\mathcal C_\gamma=-\e^{2\ii k_yR}\to-1$.
The dimensionless transverse-traceless metric perturbation has radial
kinetic weight $\Omega^2$, so its profile obeys the same equation as the
massless scalar~\cite{DeWolfe:1999cp}. Continuity of the field and its flux
derivative then gives exactly \eqref{eq:warped-wall-impedance}.
The coordinate reflection has Jacobian
$r_y=\operatorname{diag}(1,1,-1,1)$ in $(t,x,y,z)$ order and acts on every
Lorentz index. In the $R\to0$ limit, we encode these matched
physical-wave reflections in the tensor parity extension
\begin{equation}
 \Phi^{\rm refl}_{\mu_1\cdots\mu_J}(x)
 =-(r_y)_{\mu_1}{}^{\nu_1}\cdots(r_y)_{\mu_J}{}^{\nu_J}
   \Phi^{\rm orig}_{\nu_1\cdots\nu_J}(r_yx).
 \label{eq:warped-spin-reflection}
\end{equation}
The superscript $\mathrm{orig}$ denotes the original, unreflected
wavefunction. The minus sign is the matched reflection coefficient,
while each $r_y$ factor is the geometric action on one tensor index.

\Needspace{12\baselineskip}
\section{Scattering in half-space}
\label{sec:halfspace-scattering}

The preceding section showed that IR-finite eikonal scattering in
half-space admits smearing that satisfies positivity and crossing.
In impact-parameter space, the construction uses matched elastic wave
packets, $\psi_1\sim\psi_3$ and $\psi_2\sim\psi_4$, with the densities in
\eqref{eq:transverse-packet-densities}. Their Fourier transform gives the
momentum-space smearing kernel. We now develop the half-space
wavefunctions and scattering amplitudes, leaving the explicit construction
of positive smearing functionals to the next section.

We use the $R\to0$ limit of the flat/$AdS_4$ joined geometry in
\eqref{eq:warped-halfspace-metric}. The exterior curvature magnitude
diverges, and the scalar reflection coefficient in
\eqref{eq:warped-wall-impedance} tends to $-1$. The scattering region is
\begin{equation}
  \mathcal X_h=\{(t,x,y,z)\in\mathbb M^{1,3}:y>0\}.
  \label{eq:halfspace-domain}
\end{equation}
For a scalar, the induced boundary condition is $\phi|_{y=0}=0$.
Translations and $SO(1,2)$ Lorentz transformations parallel to the wall
remain exact.

For scalars, vectors and tensors, reflection is determined by the field
equations, the no-source AdS boundary condition and interface matching.
The soft Green function in the scalar problem above remains IR finite at
finite $R$, including a smooth transition with the same no-source AdS
asymptotics. These geometries have more complicated eigenmodes,
wavefunctions and amplitudes, so we use the sharp join and its $R\to0$
limit to simplify the calculation. The no-source condition is at the AdS
end, $y\to-\infty$. The interface $y=0$ is a timelike slice of the connected
four-dimensional spacetime, and the field values on both sides are fixed
by the matched bulk solution. It carries no independent boundary modes.
The modes considered here are bulk modes of the joined AdS--Minkowski
spacetime.
\Cref{sec:halfspace-packets-properties,sec:halfspace-exchange-kinematics,sec:halfspace-s-matrix-properties}
first discuss scalar wavefunctions and amplitudes, and
\cref{sec:spin-frames-halfspace} extends the discussion to spinning states.

\subsection{Asymptotic states and kinematics in half-space}
\label{sec:halfspace-packets-properties}

In the leading eikonal regime of the preceding section, both the tree
exchange and its exponentiation are IR finite. The long-distance
divergence in \eqref{eq:worldline-log} therefore no longer obstructs Fock
asymptotic states in that regime. For finite-angle scattering, the
corresponding Wilson-line integrals are examined below. We first construct
the half-space Fock modes and parameterize their scattering amplitudes.

Half-space modes have a simple relation to flat-space plane waves.
The normal waves $\e^{\ii ky}$ combine with their reflected images into
the Dirichlet modes
\begin{equation}
  u_k(y)=2\ii\sin(ky),\qquad k>0,
  \label{eq:normal-sine-modes-ch4}
\end{equation}
which obey the same $2\pi\delta(k-k')$ normalization as the flat waves,
and the corresponding completeness relation,
\begin{equation}
  \int_0^\infty\!\dd y\,u_k^\dagger(y)u_{k'}(y)=2\pi\delta(k-k'),
  \qquad
  \int_0^\infty\!\frac{\dd k}{2\pi}\,u_k(y)u_k^\dagger(y')=\delta(y-y').
  \label{eq:normal-sine-completeness}
\end{equation}
The corresponding one-particle Fock state is $|p_\parallel,k\rangle_h$,
where $p_\parallel=(E,p_x,p_z)$. For a massless particle the on-shell
condition is $p_\parallel^2-k^2=0$.

The direct-minus-image construction also applies to general wave packets.
Start with flat-space Fock states and let
$R_y|p_\parallel,k\rangle=|p_\parallel,-k\rangle$. The odd projector is
\begin{equation}
  \Pi_h=\frac{1-R_y}{2},\qquad \Pi_h^2=\Pi_h=\Pi_h^\dagger,
  \qquad u_k(y)=2\Pi_h\e^{\ii ky}.
  \label{eq:halfspace-odd-map-ch4}
\end{equation}
The factor $2$ preserves the half-space mode normalization after
projection. The bulk field equations remain the ordinary flat-space
equations. Projection imposes the wall boundary condition. On an
$n$-particle state the projector is $\Pi_h^{\otimes n}$.
For state crossing from an incoming to an outgoing state, continue only
the longitudinal components $p_\parallel=(E,p_x,p_z)$ in the $tzx$ plane,
keeping the normal momentum $k$ fixed. The action of $\Pi_h$ is therefore
unchanged along this continuation. For four-point scattering, each leg
uses the same projection in the physical $2\to2$ and all-incoming
conventions, and the projected external wavefunctions inherit the
corresponding crossing relation.

A general flat-space one-particle wave packet, formed from plane waves
of a massive or massless particle of mass $m$, can be written as
\begin{equation}
  |\psi\rangle
  =\int_{\mathbb R^3}\!\frac{\dd p_x\,\dd p_z\,\dd k}{(2\pi)^3\,2E}\,
    \phi(p_\parallel,k)|p_\parallel,k\rangle,
  \qquad E=\sqrt{m^2+p_x^2+p_z^2+k^2}.
  \label{eq:flat-general-packet}
\end{equation}
Applying the projector in \eqref{eq:halfspace-odd-map-ch4} and taking
the resulting wavefunction in half-space gives the direct-minus-image packet,
\begin{equation}
 \begin{aligned}
  |\psi\rangle_h
  &=2\Pi_h|\psi\rangle
    =\int_{\mathbb R^3}\!\frac{\dd p_x\,\dd p_z\,\dd k}{(2\pi)^3\,2E}\,
      \phi(p_\parallel,k)|p_\parallel,k\rangle_h\\
  &=\int_{\mathbb R^2}\!\dd p_x\,\dd p_z
    \int_0^\infty\!\frac{\dd k}{(2\pi)^3\,2E}\,
      [\phi(p_\parallel,k)-\phi(p_\parallel,-k)]|p_\parallel,k\rangle_h.
 \end{aligned}
 \label{eq:normal-packet-state}
\end{equation}
The half-space momentum profile is the difference of the flat profiles
at $k$ and $-k$. Smearing on-shell momenta with $\phi(p_\parallel,k)$ can
localize the packet in one, two or all three spatial directions.
Using covariantly normalized mode kets, take the flat packet to satisfy
$\langle\psi|\psi\rangle=1$. Its half-space norm is
\begin{equation}
 \begin{aligned}
  {}_h\langle\psi|\psi\rangle_h
  &=\int_{\mathbb R^2}\!\dd p_x\,\dd p_z
    \int_0^\infty\!\frac{\dd k}{(2\pi)^3\,2E}\,
      |\phi(p_\parallel,k)-\phi(p_\parallel,-k)|^2\\
  &=\int_{\mathbb R^3}\!\frac{\dd p_x\,\dd p_z\,\dd k}{(2\pi)^3\,2E}\,
      \bigl[|\phi(p_\parallel,k)|^2
      -\phi^*(p_\parallel,k)\phi(p_\parallel,-k)\bigr].
 \end{aligned}
 \label{eq:packet-norms-halfspace}
\end{equation}
This need not equal one for a general packet. A nonzero half-space
packet is normalized by dividing by the square root of
\eqref{eq:packet-norms-halfspace}.

For example, a massive Lorentz-covariant packet is labeled by a central
on-shell momentum $\bar p$ and a center $\bar x$ on a chosen spacelike
hypersurface. The construction of \cite{Oda:2021tiv} gives
\begin{equation}
  |\bar p,\bar x;\sigma\rangle
  =\mathcal N\int\!\frac{\dd^3\boldsymbol p}{(2\pi)^3\,2E_{\boldsymbol p}}\,
  \exp\!\left[-\frac{\bar p\cdot p-m^2}{2\sigma^2}
                +\ii p\cdot\bar x\right]|p\rangle,
  \label{eq:oda-packet-short}
\end{equation}
Different spacelike hyperplanes give different but equivalent complete packet families.
These packets are localized in the three spatial directions on the chosen
surface and evolve in its normal time direction. For example, we may choose
the constant-time hypersurface $t=0$ and $\bar x=(0,0,y_0,0)$ with $y_0>0$.
The direct packet is centered at $y=y_0$, and reflection adds its image at
$-y_0$ with relative coefficient $-1$. The central momentum $\bar p$ may
have a normal component.

For a Gaussian profile in the delta-normalized momentum basis with normal
momentum width $\sigma$ and $y_0>10\sigma^{-1}$, the additional normalization
factor is $1+\mathcal{O}(\e^{-2\sigma^2y_0^2})$. The direct--image inner product is
exponentially suppressed, so this correction can be neglected.
Even when the central momentum $\bar p$ points toward the wall, so that
the direct and image wavefunctions overlap in position at the collision,
their opposite normal momenta produce an oscillating relative phase that
suppresses their integrated interference by the same amount. The complete packet family on $t=0$
and its free evolution to the collision hypersurface $t=t_0$ are equivalent
and have identical norms. Thus the correction remains exponentially small
whether or not $\bar p_y$ vanishes.

For fixed packet data, or equivalently
the fixed smearing profile used below, the normalization factor is a constant
absorbed into the smearing functional and does not affect the bootstrap or
dispersion relation. The normalization is specified here to define
normalized one-particle wave packets in half-space.

The four-point smearing functional constructed below correlates all four
external momenta in the all-incoming convention. Within a pure initial or
final two-particle state, a momentum profile that cannot be factorized into
one-particle profiles describes momentum entanglement. For correlations
between initial and final legs we use the broader term correlated packets.
Products of coherent Gaussian packets are particularly simple for real-state
positivity. Choosing identical incoming and outgoing profiles,
$\phi_1=\phi_3$ and $\phi_2=\phi_4$ in the physical convention, gives a
positive diagonal absorptive matrix element even away from the eikonal
limit. This construction can be applied separately in the physical $s$
and $u$ channels.

The smearing induced by these coherent packets acts on momentum components,
rather than Lorentz-invariant variables. Crossing of the amplitude in
Mandelstam variables need not survive smearing with momentum profiles.
Reconstructing momentum components from quadratic invariants requires a
square-root branch choice. The values $k$ and $-k$ give the same $k^2$
but lie on different branches of the inverse map. Along a complex crossing path, Gaussian decay can
also turn into exponential growth, making the momentum-smearing integral
diverge. The separate positive constructions in the two channels therefore
need not be analytic continuations of one another. To construct a positive
half-space S-matrix functional, we follow the invariant smearing used for
flat-space partial waves. There, one starts with the $4\times3$-dimensional
phase space of four independent on-shell particles. Imposing four-momentum
conservation and quotienting by the six-dimensional $SO(1,3)$ Lorentz group
leaves, for generic kinematics at fixed masses, two independent
Lorentz invariants, which may be chosen as $s$ and $t$. We likewise
formulate the half-space smearing directly in its Lorentz-invariant
kinematic variables, rather than in the individual momentum components.
The packets in \eqref{eq:oda-packet-short} are coherent packets,
whereas correlated packets entangle different coherent components.

\Needspace{5\baselineskip}
\label{sec:halfspace-six-variable-kinematics}

For each external leg, $p_i=(p_{i\parallel},k_i)$. In half-space, the
$4\times3$ on-shell degrees of freedom are reduced only by the three
wall-parallel momentum-conservation conditions and the three-dimensional
$SO(1,2)$ Lorentz group, leaving six independent variables. The normal
mode labels have a $\mathbb Z_2$ identification, since the Dirichlet modes
at $k_i$ and $-k_i$ are equivalent. Their complexified kinematics obey
\begin{equation}
  \sum_{i=1}^4p_{i\parallel}=0,
  \qquad k_i\in\mathbb C/\mathbb Z_2.
  \label{eq:halfspace-momentum-decomposition-ch4}
\end{equation}
A general amplitude therefore depends on six Lorentz invariants admitting
complex continuation, formed from Lorentz products among the external
momenta and with the unit spacelike wall normal $\hat y$. We choose
\begin{equation}
  (s_\parallel,t_\parallel;p_1\cdot\hat y,p_2\cdot\hat y,p_3\cdot\hat y,p_4\cdot\hat y)
  =(s_\parallel,t_\parallel;k_1,k_2,k_3,k_4).
  \label{eq:halfspace-six-variables-ch4}
\end{equation}
Here $s_\parallel=(p_{1\parallel}+p_{2\parallel})^2$ and
$t_\parallel=(p_{1\parallel}+p_{3\parallel})^2$ are Mandelstam invariants
formed from the wall-parallel momenta, while $k_i=p_i\cdot\hat y$ is the
Lorentz-invariant contraction with the wall normal $\hat y$.
There is no normal-momentum conservation because translations normal
to the wall are broken. Just as flat-space four-point
scattering leaves one of its two invariant variables to be smeared for
a one-variable dispersion relation, here we smear five. The positive
smearing functional is constructed in the next section.

A useful orthogonal change of normal variables is
\begin{align}
  k_1&=\frac{k_s+k_0+k_t+k_u}{2},&
  k_2&=\frac{k_s+k_0-k_t-k_u}{2},\nonumber\\
  k_3&=\frac{-k_s+k_0+k_t-k_u}{2},&
  k_4&=\frac{-k_s+k_0-k_t+k_u}{2},
  \label{eq:normal-hadamard-map-ch4}
\end{align}
with inverse
\begin{align}
  k_0&=\frac{k_1+k_2+k_3+k_4}{2},&
  k_s&=\frac{k_1+k_2-k_3-k_4}{2},\nonumber\\
  k_t&=\frac{k_1-k_2+k_3-k_4}{2},&
  k_u&=\frac{k_1-k_2-k_3+k_4}{2}.
  \label{eq:normal-hadamard-inverse-ch4}
\end{align}
This transformation has unit Jacobian.
In the phase $\sum_i k_i y_i$ of the joint Fourier transform of all four
particles, $k_0$ is conjugate to
$(y_1+y_2+y_3+y_4)/2$, twice the common normal center. The variables
$k_s,k_t,k_u$ are conjugate to channel-relative separations. For example,
$k_t$ is conjugate to $(y_1+y_3-y_2-y_4)/2$, the separation of the
centers of pairs $13$ and $24$. As we will see, localizing the common
center at a finite distance from the wall is essential to removing
the IR divergence.

The four-dimensional massless half-space Feynman Green function is
analogous to the two-dimensional transverse Green function in
\eqref{eq:mixed-dirichlet-green}. We use momentum space parallel to the
wall and position space in the normal direction. For the transfer,
write $q_\parallel^2=q_x^2+q_z^2-q_t^2$.
With the Feynman $\epsilon$ prescription of \eqref{eq:flat-double-worldline},
and its overall factor assigned to the parallel inverse Fourier transform,
\begin{equation}
 \begin{aligned}
  G_h^{\mathrm{4D}}(q_\parallel;y,y')
  &=\int_{-\infty}^{\infty}\!\frac{\dd q_y}{2\pi}\,
    \frac{\e^{\ii q_y(y-y')}-\e^{\ii q_y(y+y')}}
         {q_y^2+q_\parallel^2-\ii0}\\
  &=\frac{
    \e^{-\sqrt{q_\parallel^2-\ii0}\,|y-y'|}
    -\e^{-\sqrt{q_\parallel^2-\ii0}\,(y+y')}}
    {2\sqrt{q_\parallel^2-\ii0}}.
 \end{aligned}
  \label{eq:flat-limit-mixed-green}
\end{equation}
The square root is continued from its positive real branch. As the
wall-parallel momentum transfer tends to zero, the Green function has
an IR-finite limit,
\begin{equation}
  \lim_{q_\parallel\to0}G_h^{\mathrm{4D}}(q_\parallel;y,y')
  =\min(y,y').
  \label{eq:nonuniform-flat-limit}
\end{equation}
Inverse Fourier transforming the parallel momentum $q_\parallel$ to
position space, including the overall factor $-\ii$, gives the
position-space Green function,
\begin{equation}
  G_h^{\mathrm{4D}}(x,x')=\frac{1}{4\pi^2}
  \left[\frac{1}{-(x-x')^2+\ii0}
       -\frac{1}{-(x-R_yx')^2+\ii0}\right].
  \label{eq:flat-log-return}
\end{equation}
Translate all four packet centers together along $y$, keeping their
relative positions fixed. As the common center moves infinitely far from
the wall, the image separation in \eqref{eq:flat-log-return} diverges and
the image term vanishes. The flat-space propagator is recovered, and the
long-distance scattering integrals regain their flat-space IR divergence.
This approaches the flat-space amplitude by changing only the external
states, without changing the spacetime geometry. In momentum space,
smearing $k_s,k_t,k_u$ while fixing $k_0$ with a delta function
localizes relative positions but leaves the common center delocalized
over the half-space, admitting the same flat IR region. We therefore
smear the normal momenta $k_1,k_2,k_3,k_4$ jointly, with nonzero momentum-space width also in $k_0$, to
localize the common normal center. The momentum width is Fourier conjugate to the extent of the wave
packet along $y$, so a narrower momentum profile gives a broader position-space packet.

\par\medskip

\subsection{Infrared control in half-space}
\label{sec:halfspace-exchange-kinematics}
\label{sec:halfspace-beam-completion}

Four-point momentum conservation leaves at most three linearly independent
external four-momenta. We may therefore orient the flat-space scattering
plane parallel to the wall, allowing finite-angle scattering in the $tzx$
directions. Localizing the plane waves into packets in the normal direction
then approximately preserves their flat-space angular information while
making the wall-parallel ray exchange IR finite. The spreading caused by this normal-momentum width
and its distant-boundary regulation are discussed below.
We compute the long-distance exchange between any two external legs in
the Wilson-line description of Eqs.~(\ref{eq:virtual-loop-rays}--\ref{eq:worldline-log})
\cite{Melville:2013qca}. A packet with central momentum $\bar p_i$ has
spatial velocity $\boldsymbol v_i=\bar{\boldsymbol p}_i/\bar p_i^0$.
Parametrize its ray by $x_i(\tau_i)=b_i+\bar p_i\tau_i$. In the all-incoming
convention the external rays have $\tau_i\leq0$.
The propagator integral in the flat-space virtual-graviton exchange diagram is
\begin{equation}
 \begin{aligned}
  I_0(b_1,b_2;\bar p_1,\bar p_2)
  &=\int_{-\infty}^0\!\dd\tau_1\int_{-\infty}^0\!\dd\tau_2\,
    G_0\!\left(x_1(\tau_1)-x_2(\tau_2)\right)\\
  &=\int\!\frac{\dd^4\ell}{(2\pi)^4}\,
    \frac{\ii\e^{-\ii\ell\cdot(b_1-b_2)}}
    {(\ell^2+\ii0)(\bar p_1\cdot\ell+\ii0)(\bar p_2\cdot\ell-\ii0)}.
 \end{aligned}
 \label{eq:flat-double-worldline}
\end{equation}
Here $G_0$ is the massless Feynman propagator, with the Wilson-line
time ordering already included. Each projected ray has a direct and an
image component. Their pairwise exchanges give four sectors,
\begin{equation}
  I_h=\sum_{\alpha,\beta=0}^{1}(-1)^{\alpha+\beta}
  \int_{-\infty}^0\!\dd\tau_1\int_{-\infty}^0\!\dd\tau_2\,
  G_0\!\left(
    R_y{}^{\alpha}x_1(\tau_1)-R_y{}^{\beta}x_2(\tau_2)
  \right).
  \label{eq:four-image-worldline}
\end{equation}
These are direct--direct, image--direct, direct--image and image--image.
First let both central velocities be parallel to the wall, at heights
$y_1,y_2>0$. For massive rays with distinct velocities, choose their parallel
separation orthogonal to both momenta and denote its spatial length
by $\Delta x$. Integrating over $\tau_1,\tau_2$ cancels the common
long-distance logarithm and gives
\begin{equation}
 \begin{aligned}
  I_h
  &=-\frac{1}{4\pi^2m_1m_2}
    \log\frac{(\Delta x)^2+(y_1+y_2)^2}{(\Delta x)^2+(y_1-y_2)^2}
    \int_0^\infty\!\frac{\dd r}
    {r^2-2r\,\bar p_1\cdot\bar p_2/(m_1m_2)+1-\ii0}\\
  &=-\frac{\chi(\bar p_1,-\bar p_2)}
    {4\pi^2m_1m_2\sinh\chi(\bar p_1,-\bar p_2)}
    \log\frac{(\Delta x)^2+(y_1+y_2)^2}{(\Delta x)^2+(y_1-y_2)^2}.
 \end{aligned}
 \label{eq:asymptotic-image-factor}
\end{equation}
The angle integral is expressed in terms of the analytically continued
rapidity $\chi$. Contracting the two gravitational Wilson-line vertices
with the de Donder propagator gives the Lorentz-invariant logarithmic
coefficient $\gamma$ \cite{Melville:2013qca},
\begin{equation}
 \begin{aligned}
  \cosh\chi(p,q)&=\frac{p\cdot q}{m_pm_q}-\ii0,\\
  \gamma(p,q)&\equiv\frac{2G_N}{\pi}
    \frac{(p\cdot q)^2-\tfrac12m_p^2m_q^2}{m_pm_q}
    \frac{\chi(p,q)}{\sinh\chi(p,q)}.
 \end{aligned}
 \label{eq:soft-cusp-rapidity}
\end{equation}
The pair has rapidity $\chi(\bar p_i,-\bar p_j)$. On the real boundary,
the coefficient inherits the cut of $\chi/\sinh\chi$ only for
$\cosh\chi<-1$. The ratio has no discontinuity for $\cosh\chi>1$.
Eq.~\eqref{eq:asymptotic-image-factor} is IR finite. The separations
$\Delta x,y_1,y_2$ specify the endpoint geometry and hence the short-distance
scales of the Wilson-line integral. This expression treats packets whose
overlap is negligible as localized rays. When their endpoints coincide,
the pointlike approximation fails and the finite packet profiles must be
convolved with the exchange kernel.

For general central velocities, the flat-space pair contributes to the
virtual factor in the exclusive amplitude as follows. In this and the
next formula we retain only the logarithmically IR-divergent terms,
\begin{equation}
 \begin{aligned}
  \log W_{ij,0}
  &=-8\pi G_N
    \bigl[(\bar p_i\cdot\bar p_j)^2-\tfrac12m_i^2m_j^2\bigr]
    I_0(b_i,b_j;\bar p_i,\bar p_j)\\
  &=\gamma(\bar p_i,-\bar p_j)\log\lambda_{\rm IR}^{-1}.
 \end{aligned}
 \label{eq:sahoo-sen-triangle-rapidity}
\end{equation}
Here $\lambda_{\rm IR}$ is an IR cutoff taken to zero. For generic
scattering, the real part of the full soft exponent suppresses the
exclusive amplitude to zero in this limit. The positions $b_i,b_j$
affect only the IR-finite part, while the coefficient of
$\log\lambda_{\rm IR}^{-1}$ agrees with that of $\log\omega^{-1}$ in
\cite{Sahoo:2018lxl}. Both come from the same leading soft triangle,
with a lower IR cutoff here and the soft-energy scale $\omega$ there.
For the four direct/image current terms in \eqref{eq:four-image-worldline},
reflection acts on their energy-momentum tensors through the Wilson-line
trajectories, giving
\begin{equation}
 \begin{aligned}
  \log W_{12,h}
  =\bigl[&\gamma(\bar p_1,-\bar p_2)
  -\gamma(R_y\bar p_1,-\bar p_2)
  -\gamma(\bar p_1,-R_y\bar p_2)\\
  &+\gamma(R_y\bar p_1,-R_y\bar p_2)\bigr]
    \log\lambda_{\rm IR}^{-1}.
 \end{aligned}
 \label{eq:four-image-rapidity-coefficient}
\end{equation}
If either central ray is parallel to the wall, for example
$R_y\bar p_2=\bar p_2$, the first and third terms cancel, as do the second
and fourth. The long-distance logarithm therefore cancels in this ray
sector. The other ray may have a nonzero normal component, whereas if
both normal components are nonzero the four terms need not cancel.
We therefore choose all packet central velocities parallel to the wall.

The preceding cancellation treats each packet as a beam with a frozen
position-space width along $\hat y$. Large scattering angles are allowed
within the wall. Only the angular spread along $\hat y$ is small, of order
$\sigma/\sqrt{s}$ in the $s$-channel center-of-mass frame.
The frozen-profile calculation retains the leading order in $\sigma^2/s$.
Corrections are analogous in power counting to the next-to-eikonal
terms in $|t|/s$ for rotationally invariant flat-space scattering,
but here the expansion concerns only the $\hat y$ direction.
Equation~\eqref{eq:four-image-rapidity-coefficient} tests the central
momenta. A finite momentum width produces diffraction spreading.
A small angular spread keeps the width nearly frozen over a finite
propagation interval, but does not prevent late-time spreading.
The fixed-height limits in
Eqs.~\begingroup
\def\BLastRefPart#1.#2.#3\relax{#2}%
\edef\BEndRef{\getrefnumber{eq:flat-log-return}..}%
\hyperref[eq:nonuniform-flat-limit]{(\ref*{eq:nonuniform-flat-limit}--\expandafter\BLastRefPart\BEndRef\relax)}%
\endgroup{} do not control
that late-time behavior.

\begin{figure}[H]
 \centering
 \includegraphics[width=\linewidth]{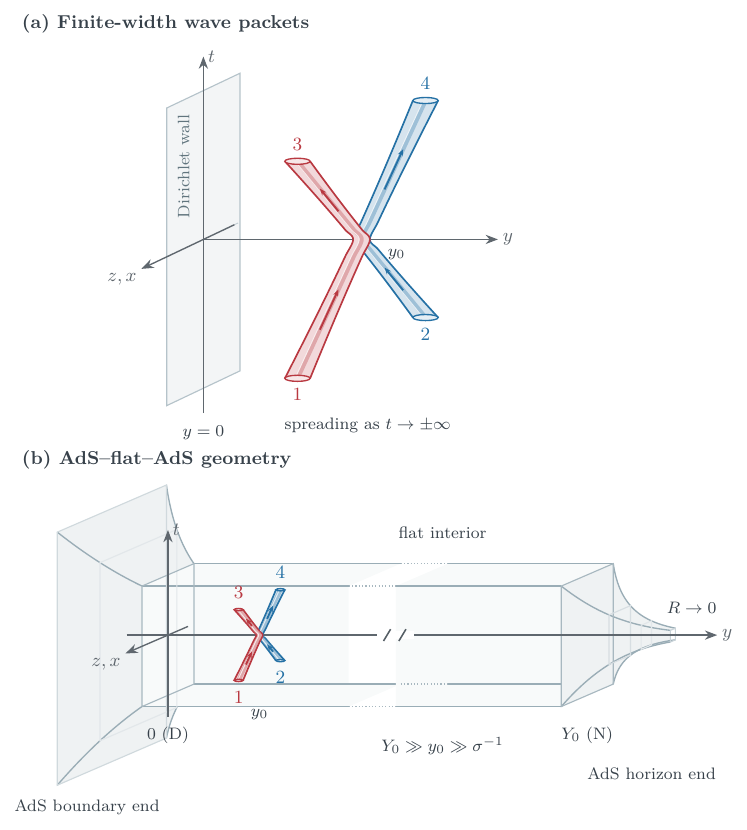}
 \caption{(a) Wave packets spread during propagation towards
 $t\to\pm\infty$. Unlike Fig.~\ref{fig:eikonal-cylinder}, scattering
 may occur at general angles in $tzx$. All four central trajectories
 remain parallel to the Dirichlet wall, with future-directed arrows.
 (b) AdS--flat--AdS geometry, with a long flat interval compressed across
 the axis break. The left side joins the AdS boundary end and the right
 side the Poincar\'e horizon end. Matching gives Dirichlet and Neumann
 conditions, respectively, as $R\to0$. The changing cross section
 represents the metric warp factor as a function of $y$.}
 \label{fig:halfspace-beam-completion}
\end{figure}

For example, set $u=t-z$, $v=t+z$, and take a massless on-shell beam
with $p_u=P$, $p_v=k^2/(4P)$ and $p_x=0$.
The momentum profile along $\hat y$ $\exp[-k^2/(4\sigma^2)-\ii ky_0]$
localizes it at $y_0$, with initial position standard deviation
$1/(2\sigma)$. Subtracting the reflected wave in half-space gives
\begin{equation}
 \begin{aligned}
  \phi_h(u,v,y)&\propto
  \frac{\e^{-\ii Pu}}{\sqrt{1+\ii\sigma^2v/P}}
  \left[\e^{-\sigma^2(y-y_0)^2/(1+\ii\sigma^2v/P)}
       -\e^{-\sigma^2(y+y_0)^2/(1+\ii\sigma^2v/P)}\right],\\
  (\Delta y)^2(v)&=
  \frac{1}{4\sigma^2}\left(1+\frac{\sigma^4v^2}{P^2}\right).
 \end{aligned}
 \label{eq:halfspace-diffracting-packet}
\end{equation}
The second line is the variance of the direct Gaussian before reflection.
The opposite beam follows by interchanging $u$ and $v$. A narrow smear
of $P$ can also localize the longitudinal profile, as illustrated in
Fig.~\ref{fig:halfspace-beam-completion}. For simplicity, the calculation
here localizes only $y$. The width is nearly constant for
$|v|\ll P/\sigma^2$ and grows linearly at large $|v|$.
This late-time spreading modifies the virtual soft factor $W$
multiplying the low-energy $S$ matrix. The leading IR-divergent
contribution to $\log W$ is quadratic in the normal momenta, with
characteristic size $G_N\sigma^2$ and logarithmic enhancements from
the soft and collinear regions. The flat-space collinear cancellation
uses momentum conservation, whereas the boundary breaks conservation
of the bulk legs' momentum along $\hat y$. That cancellation therefore
does not remove the reflected bulk contribution. Once packet
spreading is included, its soft integral diverges as
$\lambda_{\rm IR}\to0$. The resulting dimensionless soft-loop factor
gives a relative correction of order
$G_N\sigma^2\log^2(Y_0q_{\max})$ when the geometry regulates both
regions as described below.

Regulating this residual soft region requires a finite four-dimensional
flat strip of width $Y_0$ inserted into Poincar\'e $AdS_4$, rather than
an AdS exterior attached only to the left of a semi-infinite flat region.
Take $Y_0\gg y_0>10\sigma^{-1}$. As before, the left side joins the
exterior leading to the AdS boundary. The right interface at $y=Y_0$
now joins the exterior leading to the Poincar\'e horizon, as in
Fig.~\ref{fig:halfspace-beam-completion}.
Using the same field and
normal-flux matching as in \eqref{eq:warped-wall-endpoint}, the
horizon-regular profile induces the boundary condition
\begin{equation}
 \partial_y\phi(Y_0)=-\frac{R\,q_{\parallel}^{2}}{1+R\,q_\parallel}\phi(Y_0).
 \label{eq:far-horizon-neumann}
\end{equation}
Here $q_\parallel$ is the magnitude of the spacelike wall-parallel
momentum transfer. As the exterior curvature grows without bound,
$R\to0$, the right boundary becomes Neumann, just as the left becomes
Dirichlet in \eqref{eq:warped-wall-endpoint}. The flat strip therefore obeys
\begin{equation}
  \phi(0)=0,\qquad \partial_y\phi(Y_0)=0,
  \qquad k_n=\frac{(n+\tfrac12)\pi}{Y_0},\quad n=0,1,\ldots .
  \label{eq:far-dn-spectrum}
\end{equation}
The spectrum has no zero mode and starts at $\pi/(2Y_0)$.
Let $q_{\max}=\sqrt{t_{\max}}$ denote the characteristic
wall-parallel momentum transfer selected by the smearing, with
$q_{\max}^{-1}\ll y_0$. For the low-energy $S$ matrix containing
graviton exchange and local EFT contact interactions, write
\begin{equation}
 \begin{gathered}
 S_{\rm wp}(s)\simeq S_{\rm frozen}(s)\,W,\\
 S_{\rm frozen}(s)\sim
  1+\ii\bigl[G_Ns\log(y_0q_{\max})
                   +\text{local EFT contact terms}\bigr],\\
 \log W=\mathcal{O}\!\left(G_N\sigma^2\log^2(Y_0q_{\max})\right).
 \end{gathered}
 \label{eq:beam-logarithmic-scales}
\end{equation}
Here $S_{\rm frozen}$ is the IR-finite $S$ matrix evaluated in the
frozen-packet limit. Its second line gives the characteristic size of
a normalized packet matrix element, including local contact terms.
The factor $W$ is obtained from the Wilson-line dressing of the four
external packets, averaged over their momentum profiles at leading
logarithmic accuracy. The corresponding boundary contribution is
discussed in App.~\ref{app:beam-log-estimates}.

The logarithmic coefficient is most simply exhibited for half-space
momentum eigenstates, obtained by applying $\Pi_h$ to flat-space
plane waves. Reflect the gravitational tensor field as in
\eqref{eq:warped-spin-reflection}. The four original external lines
and their four images give 28 pair exchanges. There are six pairs
of distinct particles, each with four original/image combinations,
and four exchanges between a particle and its own image. Using
the massless logarithmic limit of \eqref{eq:soft-cusp-rapidity},
their sum is
\begin{equation}
 \begin{gathered}
 \sum_{i<j}\bigl(p_i\cdot p_j-p_i\cdot R_yp_j\bigr)
 -\frac12\sum_i p_i\cdot R_yp_i
 =-\Bigl(\sum_i k_i\Bigr)^2=-4k_0^2,\\
 \log W
 \simeq\frac{16G_N}{\pi}\,k_0^2
       \log\frac{q_{\max}}{\lambda_{\rm IR}}
       \log\frac{q_{\max}}{\mu_{\rm col}}.
 \end{gathered}
 \label{eq:beam-soft-collinear-coefficient}
\end{equation}
Here $\mu_{\rm col}$ regulates the collinear divergence, and $k_0$
is the normal variable of \eqref{eq:normal-hadamard-inverse-ch4}.
Taking its characteristic size $k_0^2\sim\sigma^2$ gives the relative
correction in \eqref{eq:beam-logarithmic-scales}.

For the forward scattering represented by the identity in
\eqref{eq:beam-logarithmic-scales}, each incoming momentum is paired
with its matching outgoing momentum. In the all-incoming convention,
$k_3=-k_1$ and $k_4=-k_2$, so $k_0=0$. The matching remains after
the half-space projection, and the 28 reflected pairs therefore give
zero for the logarithmic term in
\eqref{eq:beam-soft-collinear-coefficient}. The spreading-induced
IR contribution to scattering first appears at loop order, for both
local EFT contact terms and gravitational tree exchange. Its relative
size is $\mathcal{O}(G_N\sigma^2\log^2(Y_0q_{\max}))$, which can be kept small
by the wave-packet and accuracy conditions below. The half-space
tree-level packet amplitude, including contact interactions and
graviton exchange, is explicitly shown to be IR finite in
Sec.~\ref{sec:gravity-tree-functional}.

The finite DN strip regulates both logarithmic regions. Its soft gap
is $\lambda_{\rm IR}=\pi/(2Y_0)$. In the wall-parallel $2+1$-dimensional
kinematics, the DN eigenvalues $k_n$ are effective masses. Three
on-shell lines with collinear wall-parallel velocities would require
$k_a=k_b+k_c$, which is impossible for three half-integer eigenvalues
in \eqref{eq:far-dn-spectrum}. The collinear region is therefore also
regulated by the spectrum set by $Y_0$. At the level of the
characteristic logarithmic scales, the two logarithms combine into
$\log^2(Y_0q_{\max})$. The scale $q_{\max}$ is the characteristic
$t$-channel exchange scale and sets the hard end of the soft
Wilson-line integral.

To bootstrap UV physics above $\Lambda_{\rm EFT}$, take $q_{\max}$
as close to $\Lambda_{\rm EFT}$ as possible to obtain the strongest
bound. Prepare the correlated wave function and its background with
\begin{equation}
 10Y_0^{-1}\ll10y_0^{-1}<\sigma
 < q_{\max}\sim\Lambda_{\rm EFT}.
 \label{eq:beam-design-hierarchy}
\end{equation}
For an EFT bound with required relative accuracy $\epsilon$, the
relevant sizes must satisfy
\begin{equation}
 G_N\sigma^2\log^2(Y_0q_{\max})\ll\epsilon,\qquad
 \frac{y_0}{Y_0}\ll\epsilon,\qquad
 \e^{-\sigma^2y_0^2}\ll\epsilon.
 \label{eq:beam-accuracy-conditions}
\end{equation}
The first controls the relative soft-loop effect, the second makes
the DN exchange approximate half-space, and the third suppresses
the tails of the external wavefunction overlaps at local three- and
four-point contact vertices near the Dirichlet wall. The distant
Neumann boundary can be moved outward to improve the geometric
approximation while keeping the Wilson-line logarithmic enhancement
within the required accuracy. The choice of $\sigma/q_{\max}$ is
separately controlled by positivity of the correlated wave-packet
construction in Sec.~\ref{sec:halfspace-positive-functionals}.

The scales $q_{\max}$ and $\sigma$ set the momentum-profile sizes
parallel to the wall and along $\hat y$, respectively, while $y_0$
is the packet's distance from the Dirichlet wall. These parameters specify the external wave
packets. The spacetime background metric fixes
the flat-strip width $Y_0$, which can be adjusted by changing that
background geometry. At fixed
dimensionless ratios, changing $\Lambda_{\rm EFT}$ rescales $q_{\max}$
and $\sigma$ proportionally and $y_0,Y_0$ inversely. As demonstrated
in Sec.~\ref{sec:gravity-tree-functional}, the gravitational contribution
to the dispersive bound is then $G_N$ times a finite dimensionless
function of these ratios, independent of the
$\Lambda_{\rm EFT}$ scale. These choices of wave packets and geometry can therefore meet the
required accuracy while making gravity's effect on the EFT positivity
bound IR finite.

The right boundary is Neumann rather than Dirichlet because the geometry
changes the curvature of Poincar\'e $AdS_4$ over a finite interval while
retaining its boundary and horizon ends. At finite $R$, this also
suggests another geometric route towards flat-space scattering. One can hold the exterior AdS radius
fixed, widen the flat strip, and keep the
scattering region far from both ends. It may provide a setting to study
the recovery of scattering states and their Fock space, to be compared
with the usual large-radius construction~\cite{Fitzpatrick:2011jn}.
We leave this potential application for future work.
More concretely, with $y_<=\min(y,y')$ and $y_>=\max(y,y')$, the Dirichlet--Neumann and
Dirichlet--Dirichlet (DD)
strip Green functions are
\begin{equation}
 \begin{aligned}
 G_{DN}^{\mathrm{4D}}(q_\parallel;y,y')
 &=\frac{\sinh(q_\parallel y_<)\cosh[q_\parallel(Y_0-y_>)]}
 {q_\parallel\cosh(q_\parallel Y_0)}
 \xrightarrow{q_\parallel\to0}y_<,\\
 G_{DD}^{\mathrm{4D}}(q_\parallel;y,y')
 &=\frac{\sinh(q_\parallel y_<)\sinh[q_\parallel(Y_0-y_>)]}
 {q_\parallel\sinh(q_\parallel Y_0)}
 \xrightarrow{q_\parallel\to0}y_<\!\left(1-\frac{y_>}{Y_0}\right).
 \end{aligned}
 \label{eq:far-dn-green-comparison}
\end{equation}
Here $q_\parallel$ is the magnitude of real spacelike momentum transfer.
The DN zero-transfer limit agrees with the half-space result
\eqref{eq:nonuniform-flat-limit}, whereas DD retains dependence on the
distant boundary. For correlated packets at $y,y'\sim y_0\ll Y_0$,
the relative DN correction to the half-space Green function is uniformly
$\mathcal{O}(y_0/Y_0)$ over the real spacelike $q_\parallel$ range.
For $q_\parallel Y_0\gg1$, this relative correction becomes exponentially
suppressed as $\mathcal{O}(\e^{-2q_\parallel(Y_0-y_>)})$, beyond the uniform
power-law bound. The zero-transfer behavior preserves $y_0$ as the
effective IR length of frozen-packet scattering.

Spreading along
$\hat y$ produces a $\sigma^2/s$ subleading correction in soft virtual
exchange, whose IR cutoff is supplied by the more distant $Y_0$.
As in half-space, packet integrals of local three- and four-point EFT
vertices receive exponentially suppressed Gaussian image-tail corrections.
Choosing finite $Y_0$ and the packet scales within the required accuracy
allows the bootstrap below to use the half-space approximation.

\newpage

\subsection{Locality, analyticity, unitarity, and high-energy bounds}
\label{sec:halfspace-s-matrix-properties}

\textbf{Locality and microcausality.}
In flat spacetime, locality requires field operators at spacelike
separation to commute. We illustrate this with a free scalar field
$\Phi$. Its Pauli--Jordan commutator function is defined by
$[\Phi(x),\Phi(x')]=\ii\Delta(x-x')$ and vanishes at spacelike
separation. This is distinct from the time-ordered Feynman propagator.
For $x=(x_\parallel,y)$ and $x'=(x'_\parallel,y')$ in half-space,
the Dirichlet commutator is
\begin{equation}
 [\Phi_h(x),\Phi_h(x')]
 =\ii\bigl[\Delta(x_\parallel-x'_\parallel,y-y')
  -\Delta(x_\parallel-x'_\parallel,y+y')\bigr].
 \label{eq:halfspace-image-commutator}
\end{equation}
The second term connects $x$ to the mirror of $x'$ across the wall.
Reflection keeps the time difference and the wall-parallel spatial
separation fixed, while replacing the normal distance $|y-y'|$ by
$y+y'$. If the original separation is spacelike, the reflected
separation is also spacelike, so both commutators vanish.
The retarded Green function
$G_{\rm ret}(x,x')=\ii\theta(t-t')[\Phi(x),\Phi(x')]$
describes propagation from a source at $x'$ towards the future.
The image term cannot carry a signal outside the direct light cone.

For the DN strip $0<y<Y_0$, flat-space wavefunctions can likewise be
folded into admissible strip wavefunctions without first expanding in
DN eigenmodes. The extension is odd about $y=0$ and even about $y=Y_0$.
Composing these reflections translates by $2Y_0$ with a minus sign,
so the extended field satisfies
\begin{equation}
 \begin{gathered}
 \Phi_{DN}(-y)=-\Phi_{DN}(y),\qquad
 \Phi_{DN}(2Y_0-y)=\Phi_{DN}(y),\\
 \Phi_{DN}(y+2nY_0)=(-1)^n\Phi_{DN}(y),\qquad n\in\mathbb Z.
 \end{gathered}
 \label{eq:dn-image-periodicity}
\end{equation}
The period is therefore $4Y_0$. Starting with a smooth localized
flat-space profile $\phi(y)$, first periodize and then impose the
Dirichlet and Neumann reflection parities,
\begin{equation}
 \begin{aligned}
 \phi_{\rm per}(y)&=\Pi_{\rm per}\phi(y)
   =\sum_{n\in\mathbb Z}\phi(y+4nY_0),\\
 \phi_{DN}(y)&=(1-R_y)(1+R_{Y_0})\phi_{\rm per}(y)
   =\Pi_{DN}\phi(y),\\
 \Pi_{DN}&\equiv(1-R_y)(1+R_{Y_0})\Pi_{\rm per}.
 \end{aligned}
 \label{eq:scalar-dn-fold-projector}
\end{equation}
Here $R_yf(y)=f(-y)$ and $R_{Y_0}f(y)=f(2Y_0-y)$.
Both reflections square to the identity and commute on the periodized
input. The minus sign enforces Dirichlet parity and the plus sign
enforces Neumann parity. Including periodization, $\Pi_{DN}$ is a
linear map from flat-space profiles to strip profiles. Its symmetric
image kernel gives the formal-adjoint relation $\Pi_{DN}^\dagger=\Pi_{DN}$.
It is not an idempotent operator on the original flat-space function
space, so we call it a folding map rather than an orthogonal projector.
The resulting profile satisfies $\phi_{DN}(0)=0$ and
$\partial_y\phi_{DN}(Y_0)=0$ and can be normalized on the strip.
Applying the same map to the first endpoint of the flat-space
commutator and Feynman kernel gives
\begin{equation}
 \begin{aligned}
 [\Phi_{DN}(x),\Phi_{DN}(x')]
 &=\ii\Pi_{DN}\Delta(x_\parallel-x'_\parallel,y-y'),\\
 G_{DN}^{\mathrm{4D}}(x,x')
 &=\Pi_{DN}G_{\rm flat}^{\mathrm{4D}}(x-x').
 \end{aligned}
 \label{eq:dn-image-commutator}
\end{equation}
The Feynman kernel retains the prescription used above.
For two interior points, both direct-image distances
$|y-y'+2nY_0|$ and reflected-image distances $|y+y'+2nY_0|$
are at least the original normal separation $|y-y'|$.
Every image commutator therefore vanishes when the original points
are spacelike separated. At finite time separation only finitely many
image points can lie in the receiving point's past light cone.
Equivalently, only finitely many reflected copies of the receiver lie
in the source's future light cone. The second, Neumann wall therefore
preserves microcausality.

\textbf{Analyticity and crossing symmetry.}
Four momentum-conserving flat-space external momenta span at most
a three-dimensional vector subspace,
\begin{equation}
 p_4=-p_1-p_2-p_3,\qquad
 \epsilon_{\mu\nu\rho\sigma}p_1^\mu p_2^\nu p_3^\rho p_4^\sigma=0.
 \label{eq:four-momentum-span}
\end{equation}
For real $2\to2$ scattering this subspace consists of time and the
scattering plane, which may be oriented within any timelike
$2+1$-dimensional hyperplane. Generic complexified
Mandelstam data can likewise be realized by three-component complex
vectors $p_1,p_2,p_3\in\mathbb C^{1,2}$, without a fourth momentum
direction. Half-space wavefunctions can be obtained from suitable
flat-space inputs through $\Pi_h$. For general half-space modes,
we keep $k_1,k_2,k_3,k_4$ real and complexify only the wall-parallel
momenta. In the mostly-minus convention,
\begin{equation}
 p_{i\parallel}^2=m_{i\parallel}^2,\qquad
 m_{i\parallel}^2=m_i^2+k_i^2,\qquad
 \sum_i p_{i\parallel}=0,\qquad
 s_\parallel+t_\parallel+u_\parallel=\sum_i m_{i\parallel}^2.
 \label{eq:parallel-effective-masses}
\end{equation}
The wall preserves the $SO(1,2)$ transformations of the parallel
components, with $m_{i\parallel}$ acting as an effective mass at fixed
real $k_i$. The parallel Mandelstam invariants are
\begin{equation}
 \begin{aligned}
 s_\parallel&=2p_{1\parallel}\cdot p_{2\parallel}+m_{1\parallel}^2+m_{2\parallel}^2,\\
 t_\parallel&=2p_{1\parallel}\cdot p_{3\parallel}+m_{1\parallel}^2+m_{3\parallel}^2,\\
 u_\parallel&=2p_{2\parallel}\cdot p_{3\parallel}+m_{2\parallel}^2+m_{3\parallel}^2.
 \end{aligned}
 \label{eq:parallel-complex-gram}
\end{equation}
A momentum-conserving flat-space realization also requires
$\sum k_i=0$. The wall can exchange momentum along $\hat y$, so a
half-space amplitude does not impose this additional condition.

Keeping $k_i$ real is analogous to transfer smearing with $t=-q^2$
and real $q\geq0$, which keeps the spacelike transfer momentum on a
real integration domain. Our $\hat y$ smearing also integrates over
real $k$, while analytic continuation acts on the parallel
$SO(1,2)$ invariants. At fixed $k_i$, the channel poles and cuts are thus studied
in $s_\parallel,t_\parallel,u_\parallel$, with their locations allowed
to depend on these real $k_i$. Expanding the wavefunctions in normal
modes preserves locality in $(t,z,x)$. The propagators have the usual
parallel-momentum poles, and cubic and higher couplings contain
normal-mode overlap integrals, analogous to integration over the
radial coordinate $z$ in AdS/CFT.

Here the relevant statement is compatibility of four-dimensional
crossing with the external-state maps. Mandelstam crossing can be
realized by complex on-shell four-momenta. For a momentum-conserving
flat-space quartet, complexified $SO(1,3)$ transformations allow its
$\hat y$ components to take prescribed real values $k_i$ with
$\sum k_i=0$ on regular portions of the continuation.
In a common analytic domain connecting the crossed channels, we
therefore realize the Mandelstam continuation by complex parallel
momenta while keeping the real $k_i$ fixed. Both $\Pi_h$ and $\Pi_{DN}$
act only on the $\hat y$ profile. One may first fold the external
profiles and continue their parallel components, or first continue
those components and then fold, with the same result. This
compatibility uses the four-dimensional momentum realization of
crossing and does not require a separate $2+1$-dimensional crossing
proof. Parameterizing the half-space or DN amplitude
by the parallel invariants, and including charge conjugation and
polarization crossing in the particle labels when needed, the
crossing-related boundary values obey
\begin{equation}
 \M_{h/DN}^{12\to34}(s_\parallel,t_\parallel;k_1,k_2,k_3,k_4)
 =\M_{h/DN}^{32\to14}(u_\parallel,t_\parallel;k_3,k_2,k_1,k_4).
 \label{eq:halfspace-crossing-kernel}
\end{equation}
The exchange $1\leftrightarrow3$ interchanges $s_\parallel$ and
$u_\parallel$ while keeping $t_\parallel$ fixed. The normal momenta
move with their external legs and remain real throughout the
continuation and subsequent smearing. For the DN strip, replace
$k_i$ by the discrete $k_{n_i}$ in \eqref{eq:far-dn-spectrum}.
This compatibility concerns continuation between the $s$ and $u$
channels. It does not identify the full boundary amplitude or its
discontinuity with a linear image of the flat-space amplitude.
Internal exchanged particles, including those in loops, must also
obey the boundary conditions and use the corresponding Green functions.

\textbf{Unitarity and asymptotic states.}
The reflecting boundaries conserve probability. For a positive-frequency
scalar wavefunction, the Klein--Gordon current is
$j^\mu=\ii(\Phi^*\partial^\mu\Phi-\Phi\partial^\mu\Phi^*)$.
In the strict $R\to0$ limit the two boundary conditions give
\begin{equation}
 \begin{aligned}
 \Phi(0)=0&\quad\Longrightarrow\quad j^y(0)=0
       &&\text{(Dirichlet)},\\
 \partial_y\Phi(Y_0)=0&\quad\Longrightarrow\quad j^y(Y_0)=0
       &&\text{(Neumann)}.
 \end{aligned}
 \label{eq:zero-boundary-flux}
\end{equation}
Neither wall leaks probability. The same conditions remove the
integration-by-parts boundary term for every pair of allowed
wavefunctions and define the standard self-adjoint boundary domain
of $-\partial_y^2$~\cite{Asorey:2006ij}.
For a local interacting Hermitian bulk Hamiltonian on this domain, the half-space or DN Hamiltonian
therefore has no
additional boundary obstruction to self-adjointness and unitary
evolution. Equation~\eqref{eq:zero-boundary-flux} applies to the
$R\to0$ limit of the AdS--flat gluing. At finite $R$, the flat region
must be treated together with its connected $AdS_4$ exterior.

The modes in \eqref{eq:normal-sine-modes-ch4}, with the discrete
momenta of \eqref{eq:far-dn-spectrum} for DN, give complete one-particle
expansions and hence Fock bases. General half-space plane waves with
$k>0$ can still have IR-divergent scattering and require dressed
asymptotic states. But for packets localized along $\hat y$ with
central velocities parallel to the wall, the frozen-profile limit
has the IR cancellation in \eqref{eq:four-image-rapidity-coefficient}.
In a finite DN strip, a bulk massless field has wall-parallel effective
masses $k_n\geq\pi/(2Y_0)>0$. No mode can simultaneously have vanishing
energy and momentum, so the spreading correction is also free of
soft IR divergence. Equations~\eqref{eq:nonuniform-flat-limit} and
\eqref{eq:far-dn-green-comparison} show that the tree-exchange kernels
in both half-space and DN are IR finite. Keeping the finite DN
spectrum in the $\sigma^2/s$ expansion makes the long-distance
contributions illustrated by \eqref{eq:worldline-log} IR finite at
each order. For the long-distance contributions considered here, the DN modes
and their wave packets require no soft-graviton dressing within
this expansion.

\Needspace{5\baselineskip}
\textbf{UV bounds.}
Consider a flat-space profile with fixed transfer, or transfer smearing
over $-t_{\max}\leq t\leq0$, followed by $\Pi_h$ or $\Pi_{DN}$.
At $s_\parallel\gg t_{\max},m_i^2$ the scattering returns to the
eikonal regime. The hierarchy $\sigma^2\ll t_{\max}$ makes the
characteristic $|t_\parallel|$ of order $t_{\max}$. The high-energy
motion is along $z$ within $(t,z,x)$, while $x$ and $y$ form the
transverse impact space of \cref{sec:transverse-modes}.
The full $S$ matrix, including elastic and inelastic channels, is
unitary. Its eigenvalues have unit modulus in an eigenchannel basis,
so every elastic impact matrix element obeys
$|S_{h/DN}(s_\parallel,\boldsymbol b)|\leq1$.
Here $\boldsymbol b$ denotes the transverse configuration, written
below as the relative wall-parallel separation $x$ and the two
normal positions $y_1,y_2$. Define the impact transition response by
$\mathcal T_{h/DN}=-\ii(S_{h/DN}-1)$. After transfer smearing, the
momentum-space-normalized amplitude between the initial and outgoing
profiles is a function of $s_\parallel$ alone,
\begin{equation}
 \begin{aligned}
 \M_{h/DN}^{\rm wp}(s_\parallel)
 &=2s_\parallel\int_{\mathbb R}\!\dd x\int_{I^2}\!\dd y_1\dd y_2\,
 \psi_{\rm in}(x,y_1,y_2)\\
 &\qquad\times\mathcal T_{h/DN}(s_\parallel,x,y_1,y_2)
 \psi_{\rm out}^*(x,y_1,y_2).
 \end{aligned}
 \label{eq:real-impact-unitarity-bound}
\end{equation}
The factor $2s_\parallel$ restores the momentum-space amplitude
normalization. A correlated functional integrates such initial and outgoing profiles
against a joint state density. Correlation alone does not imply
positivity. The positive densities in \cref{sec:halfspace-positive-functionals}
resolve into nonnegative weights pairing the same state on the two
sides. Here $I=(0,\infty)$ for half-space and $I=(0,Y_0)$ for DN.
The external profiles can be expanded in the modes of
\eqref{eq:normal-sine-modes-ch4} and \eqref{eq:far-dn-spectrum},
or equivalently obtained from flat-space profiles by $\Pi_h$ and
$\Pi_{DN}$. The latter form makes contact with flat-space
state constructions directly.

For normalized packets, two-particle scattering unitarity bounds the
amplitude on the physical high-energy ray. With crossing, analyticity
and subexponential growth, the two-particle Fock amplitude has the
same linear growth bound along complex high-energy rays in its
physical-sheet analytic domain~\cite{Caron-Huot:2022ugt}.
On the physical ray, using the $2s_\parallel$ normalization in
\eqref{eq:real-impact-unitarity-bound},
$|\mathcal T_{h/DN}|=|1-S_{h/DN}|\leq2$ gives
\begin{equation}
 |\M_{h/DN}^{\rm wp}(s_\parallel)|
 \leq4|s_\parallel|
 \int_{\mathbb R}\!\dd x\int_{I^2}\!\dd y_1\dd y_2\,
 |\psi_{\rm in}\psi_{\rm out}^*|
 \leq C_{\rm wp}|s_\parallel|.
 \label{eq:halfspace-linear-real-bound}
\end{equation}
Cauchy--Schwarz bounds the overlap integral by the product
$\|\psi_{\rm in}\|_2\|\psi_{\rm out}\|_2=1$, so one may take
$C_{\rm wp}=4$ independently of energy. The expression above uses
pure initial and outgoing states. The correlated wave packets in
the next section instead average against a positive semidefinite
state density. A normalized density obeys the same UV bound.
The normalized-packet argument applies equally in flat space,
half-space and the DN strip. Folding changes the eikonal profiles
and their normalization factors. For the localized Gaussian packets
used here, with the same single-packet normalization convention and
centers far from the walls, these factors differ only by exponentially
small relative corrections.

\subsection{Spinning wavefunctions and boundary reflection}
\label{sec:spin-frames-halfspace}

The preceding constructions used scalars to study scattering in
half-space and the DN strip. A wave-packet profile specifies a momentum
distribution on the mass shell. Integrating flat-space plane waves
against this distribution gives the packet wavefunction, and applying
$\Pi_h$ or $\Pi_{DN}$ gives the corresponding boundary wavefunction.
We now complete this construction for spinning coherent wave packets
by specifying the polarization of each momentum component. The
subsequent reflection acts on the complete spinning wavefunction,
including its momentum and every open Lorentz index, to implement
the Dirichlet and Neumann boundary conditions.

We first specify the polarization at the packet's central momentum
$\bar p$, using spin one as an example. Choose two real orthonormal
transverse polarizations $\bar\epsilon_x,\bar\epsilon_y$ and define
\begin{equation}
 \bar\epsilon_\pm=\frac{\bar\epsilon_x\pm\ii\bar\epsilon_y}{\sqrt2},
 \qquad \bar p\cdot\bar\epsilon_a=0,\qquad
 \bar\epsilon_a^*\cdot\bar\epsilon_b=-\delta_{ab}.
 \label{eq:spin-helicity-polarizations}
\end{equation}
The subscripts $x,y$ label the chosen polarization axes. For a massive
vector, a third polarization $\bar\epsilon_0$ completes the three
spin projections $a=+,0,-$ along the chosen quantization axis.
They form a spin-one representation of the $SU(2)$ little group and
can mix under Lorentz-induced Wigner rotations. For a massless vector,
$a=+,-$ labels helicities $\pm1$. Under $U(1)$ little-group rotations,
these states acquire phases without mixing.
The polarizations also admit gauge shifts proportional to $\bar p$.
For a massive particle, rotations about the quantization axis likewise
act by phases on the spin projections. Higher integer spins use
symmetric traceless polarization tensors. In particular, the massless
helicity $\pm J$ tensor is $\bar\epsilon_\pm^{\otimes J}$.

For a massive particle with $p^2=\bar p^2=m^2$, the canonical boost
depends only on $p$ and $\bar p$. It is generated in the
$\bar p\wedge p$ plane and has the covariant form
\begin{equation}
 \Lambda^\mu{}_{\nu}(p,\bar p)
 =\delta^\mu{}_{\nu}
  +\frac{2p^\mu\bar p_\nu}{m^2}
  -\frac{(p+\bar p)^\mu(p+\bar p)_\nu}{m^2+p\cdot\bar p}.
 \label{eq:spin-canonical-boost}
\end{equation}
It maps $\bar p$ to $p$, fixes the orthogonal complement of their
plane, and defines the polarization at $p$ by the same Lorentz
transformation.

For a massless particle with $p^2=\bar p^2=0$, the little-group
freedom prevents $p$ and $\bar p$ alone from specifying a unique
canonical boost. We choose a null reference vector $\xi^\mu$ and
fix the gauge by $\xi\cdot\bar\epsilon_\pm=0$.
For a given helicity, different reference choices can change the
resulting polarization by a gauge shift and a little-group phase.
The gauge shift drops out of the field strength and gauge-invariant
amplitudes, whereas the phase remains. The reference-dependent Lorentz
transformation is
\begin{equation}
 \begin{aligned}
 \Lambda^\mu{}_{\nu}(p,\bar p;\xi)
 &=\delta^\mu{}_{\nu}
  +\frac{(p-\bar p)^\mu\xi_\nu}{\bar p\cdot\xi}
  +\frac{\xi^\mu(\bar p-p)_\nu}{p\cdot\xi}
  +\frac{p\cdot\bar p}{(p\cdot\xi)(\bar p\cdot\xi)}\xi^\mu\xi_\nu.
 \end{aligned}
 \label{eq:spin-standard-null-boost}
\end{equation}
The massive and massless constructions are Lorentz covariant and give
\begin{equation}
 \Lambda\bar p=p,\qquad
 \epsilon_a^\mu(p)=\Lambda^\mu{}_{\nu}\bar\epsilon_a^\nu,\qquad
 p\cdot\epsilon_a(p)=0,\qquad
 \epsilon_a(p)^*\cdot\epsilon_b(p)=-\delta_{ab}.
 \label{eq:spin-screen-orthogonality}
\end{equation}
The massless polarizations also satisfy $\xi\cdot\epsilon_\pm(p)=0$.
For integer spin $J$, the same transformation acts through
$\Lambda^{\otimes J}$. Half-integer spins use the corresponding
left- and right-handed Weyl representations of $SL(2,\mathbb C)$,
which we do not develop here. For a chosen polarization $a$, the
rank-$J$ coherent spinning packet is
\begin{equation}
 \Phi_a^{\mu_1\cdots\mu_J}(x)
 =\int\frac{\dd^3\mathbf p}{(2\pi)^3\,2E_p}\,
 \phi(p)
 (\Lambda^{\otimes J})^{\mu_1\cdots\mu_J}{}_{\nu_1\cdots\nu_J}
 \bar\epsilon_a^{\nu_1\cdots\nu_J}
 \e^{-\ii p\cdot x}.
 \label{eq:single-spinning-coherent-packet}
\end{equation}
Here $\Lambda$ is the massive transformation or the massless one
with reference vector $\xi$, evaluated at $p$.
The polarization $\bar\epsilon_a$ is specified at $\bar p$.
It may be any chosen linear combination of spin projections for a
massive particle, while $a=+$ or $-$ selects a definite massless
helicity. The $J$ Lorentz factors act on the $J$ polarization indices.
Integrating the given momentum profile $\phi(p)$ then produces the
spinning wave packet, just as for scalar particles.

The spinning smearing functionals in
Sec.~\ref{sec:congruence-construction} use correlated wave packets,
rather than independent combinations of the single-particle coherent
packets above. Conventional partial-wave functionals are defined on
momentum-conserving Mandelstam kinematics, so the four external momenta
are already correlated. For massless helicities, comparing elastic
and inelastic channels through real Wigner
$d^J_{h_1-h_2,h_4-h_3}(\theta)$ kernels also uses a common convention
for the four external little-group phases. Fixing these phases to zero
in the scattering convention removes the phase factors of the full
Wigner $D$ function. The polarization frame of each leg is then chosen
relative to the complete momentum configuration. For example,
$\epsilon_1$ can depend on $p_2,p_3,p_4$ as well as $p_1$.
Relations such as $[12]/\langle12\rangle=1$ express this correlated
choice, which is described in detail in that subsection.

The canonical boosts in
\crefrange{eq:spin-canonical-boost}{eq:spin-standard-null-boost}
specify how polarization changes under on-shell momentum smearing,
extending a momentum profile to a spinning wavefunction.
This prescription applies both to coherent wave packets and to momentum
shaking in the congruence construction of
Sec.~\ref{sec:congruence-construction}, where the shaking achieves the
$\hat y$ localization required for IR safety by giving
$k_1,k_2,k_3,k_4$ finite-width distributions.
Shaking the momenta in this congruence also requires changing the
polarizations, which is exactly described by the canonical boosts here,
preserving the transversality and inner products in
\eqref{eq:spin-screen-orthogonality}.
Note that the external wavefunction after this congruence contains
components both satisfying and violating $\hat y$ momentum conservation
and is parameterized by $s_\parallel,t_\parallel,k_1,k_2,k_3,k_4$.
Even in flat space, the external wavefunction's $\hat y$ weight need not
impose momentum conservation. This is enforced by the delta function
carried by the amplitude on which the functional acts.
With parallel momentum conservation imposed, this amplitude is
\begin{equation}
 2\pi\delta\!\left(\sum k_i\right)\M(s,t)
 =\pi\delta(k_0)\M(s,t),\qquad k_0=\frac12\sum k_i.
 \label{eq:spin-flat-smearing-conserved-slice}
\end{equation}
In the trace pairing of the correlated state density with the S-matrix,
this delta function selects the momentum-conserving slice of the external
wavefunction. Integrating over that slice with the induced measure
turns the functional of $s_\parallel,t_\parallel,k_1,k_2,k_3,k_4$
into a flat-space smearing functional of $s,t$.

To obtain half-space or DN wavefunctions, we apply
reflection to the complete wavefunction. This acts on every particle
argument and every open Lorentz index, independently of how the
wavefunction was prepared or whether its momenta are correlated.
Let $(r_y)^\mu{}_{\nu}$ denote the Lorentz-index matrix of the geometric
reflection $R_y$ in \eqref{eq:halfspace-odd-map-ch4}. It flips the
$y$ component of each Lorentz index, so that
$p=(p_\parallel,k)$ becomes $r_yp=(p_\parallel,-k)$.
For a rank-$J$ tensor, the complete reflection is
\begin{equation}
 (R_y\Phi)_{\mu_1\cdots\mu_J}(x)
 =(r_y)_{\mu_1}{}^{\nu_1}\cdots(r_y)_{\mu_J}{}^{\nu_J}
   \Phi_{\nu_1\cdots\nu_J}(r_yx).
 \label{eq:tensor-reflection-ch4}
\end{equation}
This is the tensor reflection entering \eqref{eq:warped-spin-reflection}.
The operator obeys $R_y^\dagger=R_y$ and $R_y^2=1$.
For a massless particle, helicity is the projection of angular momentum
$\boldsymbol J$ onto momentum $\boldsymbol p$, both understood here as
operators on wavefunctions. Momentum is a polar vector and angular
momentum an axial vector, so the reflection with $\det r_y=-1$ gives
\begin{equation}
 R_y^\dagger\frac{\boldsymbol J\cdot\boldsymbol p}{|\boldsymbol p|}R_y
 =-\frac{\boldsymbol J\cdot\boldsymbol p}{|\boldsymbol p|},
 \qquad h\longmapsto-h.
 \label{eq:spin-helicity-reflection}
\end{equation}
Thus $r_y^{\otimes J}\epsilon_h(p)$ has helicity $-h$ at momentum
$r_yp$, independently of the momentum direction or polarization convention.
Momentum and polarization undergo the same Lorentz
reflection, preserving the on-shell condition and the orthogonality
relations in \eqref{eq:spin-screen-orthogonality}.
Replacing the scalar reflection in \eqref{eq:halfspace-odd-map-ch4}
by \eqref{eq:tensor-reflection-ch4} gives the linear half-space projector
$\Pi_h$, which maps flat-space spinning wavefunctions to half-space
spinning wavefunctions. For a multiparticle wavefunction, the same
operation acts on each particle argument and its spin indices.

For the DN strip, use the same complete reflection about $y=Y_0$
and substitute the two reflections into \eqref{eq:scalar-dn-fold-projector}.
The resulting $\Pi_{DN}$ folds a flat-space spinning wavefunction into
one satisfying the tensor DN conditions. Lorentz indices are reflected
along with the arguments and do not alter the periodization and folding
operations. Suppressing the unchanged parallel coordinates, the folded
wavefunction is
\begin{equation}
 \begin{aligned}
 \Phi_{DN}(y)&=\Pi_{DN}\Phi_0(y)\\
 &=\sum_{n\in\mathbb Z}(-1)^n
 \left[\Phi_0(y+2nY_0)
       -r_y^{\otimes J}\Phi_0(-y+2nY_0)\right],
 \qquad 0<y<Y_0.
 \end{aligned}
 \label{eq:spin-dn-fold}
\end{equation}
At the two reflection planes, the coordinate argument is unchanged,
and the boundary values reduce to the action on tensor indices,
\begin{equation}
 \begin{aligned}
 [(1-R_y)\Phi]_{\mu_1\cdots\mu_J}(x_\parallel,0)
 &=[(1-r_y^{\otimes J})\Phi]_{\mu_1\cdots\mu_J}(x_\parallel,0),\\
 [(1+R_{Y_0})\Phi]_{\mu_1\cdots\mu_J}(x_\parallel,Y_0)
 &=[(1+r_{Y_0}^{\otimes J})\Phi]_{\mu_1\cdots\mu_J}(x_\parallel,Y_0).
 \end{aligned}
 \label{eq:spin-boundary-reflection-values}
\end{equation}
Here $R_{Y_0}$ reflects about $y=Y_0$. Its Lorentz-index matrix
$r_{Y_0}$, like $r_y$, flips only the $y$ component.
A component with $N_y$ indices along $\hat y$
acquires the index factor $(-1)^{N_y}$. Its values and derivatives
therefore satisfy
\begin{equation}
 \renewcommand{\arraystretch}{1.3}
 \setlength{\aboverulesep}{0.75ex}
 \setlength{\belowrulesep}{0.3ex}
 \begin{array}{@{}l@{\qquad}l@{\qquad}l@{}}
 \toprule
 &y=0&y=Y_0\\
 \midrule
 N_y\ \hbox{even}&\Phi_{DN}=0&\partial_y\Phi_{DN}=0\\
 N_y\ \hbox{odd}&\partial_y\Phi_{DN}=0&\Phi_{DN}=0\\
 \bottomrule
 \end{array}
 \label{eq:spin-tensor-boundary-components}
\end{equation}
In particular, at the Dirichlet wall a vector has
$A_t=A_x=A_z=0$ and $\partial_yA_y=0$, while $A_y$ may be nonzero.
A rank-two field has $T_{ab}=T_{yy}=0$ and $\partial_yT_{ay}=0$,
while $T_{ay}$ may be nonzero, with $a,b$ tangent to the wall.
These are the exact boundary conditions of the specified tensor parity
extension. The rank-two field may be the electromagnetic field strength
$F_{\mu\nu}$ or a linearized graviton perturbation $h_{\mu\nu}$.

\section{The half-space positivity bootstrap}
\label{sec:halfspace}
\label{sec:halfspace-positive-functionals}

The preceding section established the scattering properties in half-space.
We now construct positive state densities localized perpendicular to the
wall and use them to develop the half-space positivity bootstrap. We start from two
momentum-conserving flat-space seeds, forward scattering at zero momentum
transfer and a general partial-wave smearing functional. A congruence
transformation applies paired normal-momentum shaking to the initial and
final states, producing correlated flat-space wave packets localized along
$\hat y$. Applying $\Pi_h$ to each particle wavefunction gives packets
satisfying the half-space boundary conditions while preserving their
positive semidefinite (PSD) state density. The trace of this density with
the physical absorptive operator is nonnegative. The exact $s$-channel
construction also identifies the transfer measure needed for half-space
smearing. Using this structure, we separately specify a crossing-symmetric
half-space functional and numerically test its trace pairing with both
momentum-conserving and nonconserving absorptive kernels. This directly defined functional is then
applied to the half-space
dispersion relation to obtain an IR-finite EFT bound.

We adopt the scale relation in \eqref{eq:beam-design-hierarchy} and the
accuracy conditions in \eqref{eq:beam-accuracy-conditions}. As shown in
\eqref{eq:beam-logarithmic-scales}, packet spreading produces a relative
soft-loop effect of order $G_N\sigma^2\log^2(Y_0q_{\max})$ that carries
the $Y_0$ dependence. The DN strip regulates the IR at every
loop order, while these conditions allow its leading tree-level amplitude
to be evaluated with the half-space kernel at the prescribed accuracy.
The positivity argument also applies on the strip after replacing
$\Pi_h$ by $\Pi_{DN}$, since both are linear maps.
Here $\sigma$ is the width of the external wave packets in normal momentum, with a conjugate
position-space width of order $1/\sigma$ along $y$. The ratio $\sigma/\Lambda_{\rm EFT}$ is chosen
to obtain positive responses. In this section, finite-width responses retain this nonzero
normal-momentum width. We impose a small-momentum-width
hierarchy only when displaying a limiting expansion, not as a general
condition on the bootstrap functional. In this section we consider elastic scattering with
$t$-channel smearing
and a twice-subtracted dispersion relation (2SDR), without using improved or
crossing-symmetric dispersion relations (CSDRs).
The state construction can likewise be combined with other subtraction
orders and with these more general dispersion relations.

\subsection{Positivity and congruence}
\label{sec:dirichlet-packets}

We first interpret smearing with the flat-space profile $\phi(t)$ in terms of
correlated wave packets. General incoming and outgoing two-particle
Fock states may be entangled, meaning that their joint wavefunctions
need not factor into two one-particle coherent wavefunctions. Write
\begin{equation}
 \begin{aligned}
 |\Phi_{12}\rangle&=\int\dd\nu_{12}\,
     \Phi_{12}(p_1,p_2)|p_1,p_2\rangle,\\
 |\Phi_{34}\rangle&=\int\dd\nu_{34}\,
     \Phi_{34}(p_3,p_4)|-p_3,-p_4\rangle.
 \end{aligned}
 \label{eq:joint-two-particle-packets}
\end{equation}
The measures contain the on-shell phase-space integrals for the two
particles and sums over their particle species and helicities.
The all-incoming variables $p_3,p_4$ represent the physical outgoing
momenta $-p_3,-p_4$.

The optical theorem gives a nonnegative absorptive part when the same
pure state is used on both sides of a physical channel, or when such
matched states are combined with nonnegative weights. With $S=1+\ii T$
and $\operatorname{Im}T=(T-T^\dagger)/(2\ii)$, unitarity gives
\begin{equation}
 \begin{aligned}
 \langle\lambda;{\rm in}|\operatorname{Im}T
                  |\lambda';{\rm out}\rangle
 &=\frac12\sumint_X
       \Acal_X(\lambda)\Acal_X(\lambda')^*,\\
 \Acal_X(\lambda)&=\langle\lambda;{\rm in}|T|X\rangle .
 \end{aligned}
 \label{eq:halfspace-abs-kernel}
\end{equation}
Here $\lambda,\lambda'$ specify the momenta, helicities and particle
species of the two-particle external states. The sum and integral over
$X$ include all allowed intermediate channels, their phase spaces and
helicities. For any nonnegative measure $\dd\rho(\lambda)$, the correlated
state density and the absorptive part of its scattering are
\begin{equation}
 \begin{aligned}
 K&=\int\dd\rho(\lambda)\,|\Phi_\lambda\rangle
       \langle\Phi_\lambda|\succeq0,\\
 \operatorname{Im}\Tr(KT)
 &=\Tr(K\operatorname{Im}T)
 =\frac12\int\dd\rho(\lambda)\,
       \sumint_X
       |\langle\Phi_\lambda|T|X\rangle|^2\geq0.
 \end{aligned}
 \label{eq:halfspace-gram}
\end{equation}
The smearing kernel $K$ is a positive state density specified by a nonnegative sum of projectors
onto correlated two-particle wave packets. The absorptive operator $\operatorname{Im}T$ is
determined by
the theory's interactions and UV completion. These are distinct PSD
operators, whose trace pairing gives the nonnegative observable.

For a general pair of initial and final states, a linear map $V$ acts by
\begin{equation}
 V|\Phi_{\rm in}\rangle\langle\Phi_{\rm out}|V^\dagger
 =|(V\Phi)_{\rm in}\rangle\langle(V\Phi)_{\rm out}|.
 \label{eq:one-side-state-map}
\end{equation}
Its action on each side depends only on that side's state, not on the
state across the contraction. It need not be invertible or preserve
norms. We use $V$ for localization along $\hat y$ in flat space, so that
$V\Phi$ is still a flat-space wavefunction. Boundary projection or folding
is a separate step. For the density in \eqref{eq:halfspace-gram}, the map
sends each $\Phi_\lambda$ to $V\Phi_\lambda$ while leaving the nonnegative
measure $\dd\rho(\lambda)$ unchanged. This congruence preserves positivity,
\begin{equation}
 VKV^\dagger\succeq0,\qquad
 \Tr\!\left(VKV^\dagger\operatorname{Im}T\right)\geq0.
 \label{eq:common-wall-congruence}
\end{equation}
where $T$ is the flat-space scattering operator.

\phantomsection
\label{sec:spinning-external-functionals}
\label{app:spin-reflection}
\label{app:spin}
For spinning external particles, $K$ and $\operatorname{Im}T$ act on the
Hilbert space of spinning wavefunctions. The $\Phi_\lambda$ in
\eqref{eq:halfspace-gram} therefore already contain their polarization
information before localization and boundary projection. For the massless
particles considered below, they may be linear combinations of helicity
wavefunctions.
Before the congruence, the usual partial-wave functional smears the
momentum-conserving Mandelstam variables $s,t$. Numerical positivity
tests use Wigner-$d$ functions with fixed external-state phase conventions,
rather than the full Wigner-$D$ matrices~\cite{Caron-Huot:2022ugt}.
In spinor variables, the corresponding little-group phase factors,
such as $[12]/\langle12\rangle$, are fixed to constants as described in
\crefrange{eq:four-leg-little-group-phase}{eq:redundant-spinor-phase-fixing}.
The seed $K$ thus specifies both momenta and polarizations, including their
little-group phases. For massive particles the little group is $SU(2)$,
so the amplitude and state density carry $SU(2)$ little-group indices
whose contraction in the trace gives a basis-independent observable.

Starting from this fully specified seed, we first apply $V$ to the
two-particle state on each side. In the $s$ channel it shakes the normal
momenta of particles $1,2$, keeping each particle on shell. The canonical
boost in \eqref{eq:spin-standard-null-boost} simultaneously acts on their
polarization tensors with the phases fixed above. The resulting
correlated flat-space density includes components both with and without
normal momentum conservation, with the conserved slice selected by
$\delta(k_0)$. A finite width in
$k_0=\tfrac12\sum_i k_i$ localizes the common normal position, as in
\eqref{eq:normal-hadamard-map-ch4}. We then apply $\Pi_h$ to each particle,
obtaining a positive half-space density with the normal localization
required for the IR safety specified above,
\begin{equation}
 \begin{gathered}
 K_h=\Pi_h^{\otimes2}\,VKV^\dagger\,
       (\Pi_h^{\otimes2})^\dagger\succeq0,\\[3pt]
 \begin{aligned}
 \Tr\!\biggl[\int_0^\infty\!\dd k_i\,
       K_h\operatorname{Im}T_h\biggr]
 &=\Tr\!\biggl[\int_{-\infty}^\infty\!\dd k_i\,
       VKV^\dagger(\Pi_h^{\otimes2})^\dagger
       \operatorname{Im}T_h\,\Pi_h^{\otimes2}\biggr]\\[3pt]
 &=\Tr\!\biggl[\int_{-\infty}^\infty\!\dd k_i\,
       VKV^\dagger\operatorname{Im}T_h\biggr]\geq0.
 \end{aligned}
 \end{gathered}
 \label{eq:spin-projector}
\end{equation}
Here $\Pi_h$ reflects the momentum and every open Lorentz index.
Since $T_h$ describes scattering between half-space eigenmodes, applying
the idempotent projector to its external states leaves them unchanged,
so $(\Pi_h^{\otimes2})^\dagger T_h\Pi_h^{\otimes2}=T_h$.
We can therefore smear the half-space amplitude directly with the
flat-space PSD density $VKV^\dagger$.
For example, for scattering between the $\sin(k_i y_i)$ modes entering
$T_h$, the contributions from $k_i$ and $-k_i$ are automatically
antisymmetrized, reproducing the original smearing by $K_h$.
The operator $T_h$ is always that of the boundary theory.
Unlike the external-state smearing kernel $K$, it cannot in general be
obtained from its flat-space counterpart $T$ through these linear maps
alone, since internal propagation, including loops and multiparticle
channels, must also obey the boundary conditions.

For the DN strip, $\Pi_{DN}$ in \eqref{eq:scalar-dn-fold-projector}
is the unnormalized folding map, and the analogous trace is
\begin{equation}
 \begin{gathered}
 K_{DN}=\Pi_{DN}^{\otimes2}\,VKV^\dagger
          (\Pi_{DN}^{\otimes2})^\dagger,\\[3pt]
 \Tr\!\biggl[\sum_{n_i\geq0}
       K_{DN}\operatorname{Im}T_{DN}\biggr]
 =\Tr\!\biggl[\int_{-\infty}^\infty\!\dd k_i\,
       VKV^\dagger\operatorname{Im}T_{DN}\biggr]\geq0.
 \end{gathered}
 \label{eq:dn-folded-trace}
\end{equation}
The sum includes the normal-mode weight $(2Y_0)^{-4}$ for the four
external legs, with $k_n=(n+\tfrac12)\pi/Y_0$.
The reflected and periodized boundary eigenfunctions on each external
leg of $T_{DN}$ supply the spectral selection $\sum_n\delta(k-k_n)$.
Flat-space and half-space normal modes have Dirac-delta normalization,
whereas DN normal modes are normalized on the finite interval of length
$Y_0$. The unnormalized folding map $\Pi_{DN}$ between these mode spaces
is not idempotent. With these mode-normalization weights, however, the
spectral delta functions exactly reproduce the discrete sum, transferring
the $\Pi_{DN}$ folding to the amplitude for direct pairing with $VKV^\dagger$.

\subsection{Half-space positivity from forward scattering}
\label{sec:collinear-functional}

Forward scattering provides the simplest momentum-conserving positive
seed $K$. We apply a congruence transformation to its initial and final
states, followed by boundary projection, to obtain a positive half-space
kernel. For forward scattering in the $tz$ plane, the all-incoming
momenta obey
\begin{equation}
 \bar p_1+\bar p_3=0,\qquad
 \bar p_2+\bar p_4=0,\qquad t=0.
 \label{eq:forward-seed-pairing}
\end{equation}
The bars denote the seed momenta before shaking, which set the centers
of the ensuing packets. These relations identify the two sides of the
forward density. The incoming state is the product of the plane waves
of particles $1,2$, and the outgoing state is the same product, labeled
by $3,4$. This is a single unentangled pure state on each side, with no
mixture over states and no localization in any direction.

We now give each leg a momentum component $k_i$ along $\hat y$.
For massless external particles we preserve the on-shell conditions
$p_{i\parallel}^2=k_i^2$ and fix the total parallel momentum
$P_{s\parallel}=p_{1\parallel}+p_{2\parallel}=\bar p_1+\bar p_2$.
Here $\parallel$ denotes the components in $tzx$.
On the physical on-shell domain, the one-side map is
\begin{equation}
 V|\bar p_1,\bar p_2\rangle
 =\int\dd k_1\,\dd k_2\,C(k_1)C'(k_2)
       |p_{1\parallel},k_1;p_{2\parallel},k_2\rangle.
 \label{eq:collinear-momenta}
\end{equation}
At fixed $k_i$, $p_{i\parallel}$ lies on the $2+1$-dimensional mass shell
with effective mass squared $k_i^2$, leaving two momentum degrees of
freedom for each particle. Fixing $P_{s\parallel}$ leaves one shared
degree of freedom, an $SO(2)$ rotation of the pair about $P_{s\parallel}$.
To specify $V$ using only the state on which it acts, we require
$p_{1\parallel}\wedge p_{2\parallel}$ to span the same oriented plane
as $\bar p_1\wedge\bar p_2$. Their magnitudes need not agree.
For each allowed pair $k_1,k_2$, this condition, together with the
on-shell constraints and fixed $P_{s\parallel}$, uniquely determines
$p_{1\parallel},p_{2\parallel}$. Thus $V$ depends only on
$\bar p_1,\bar p_2$ and the two shaking profiles $C,C'$.
Once these profiles are given, \eqref{eq:collinear-momenta} defines a
two-particle state localized along $\hat y$. The two particles are
generally entangled because their parallel momenta depend jointly on
$k_1,k_2$.

Independent $\hat y$ shaking gives a positive density $VKV^\dagger$
localized along $\hat y$, whose momentum distribution extends away from
the conserving slice $k_0=0$. Applying the projection in
\eqref{eq:spin-projector} then gives the positive half-space kernel
generated from the forward seed. A simple choice is the normalized
Gaussian profiles
\begin{equation}
 \begin{gathered}
 C(k)=\frac{\e^{-k^2/(2\sigma^2)-\ii k y_1}}
                 {(\pi\sigma^2)^{1/4}},\qquad
 C'(k)=\frac{\e^{-k^2/(2\sigma^2)-\ii k y_2}}
                  {(\pi\sigma^2)^{1/4}},\\
 \int_{\mathbb R}\dd k\,|C(k)|^2
 =\int_{\mathbb R}\dd k\,|C'(k)|^2=1.
 \end{gathered}
 \label{eq:forward-gaussian-profile}
\end{equation}
The positions $y_1,y_2$ are their respective localization centers along
$\hat y$, with widths of order $\sigma^{-1}$.
Elastic matching is $1\to3$ and $2\to4$.
For these matched profiles, the smearing weight in the all-incoming
convention is
\begin{equation}
 \begin{aligned}
 W(k_i)=C(k_1)C'(k_2)C(k_3)C'(k_4)
 &=\bigl[C(k_1)C'(k_2)\bigr]
   \bigl[C(-k_3^*)C'(-k_4^*)\bigr]^*,\\
 &=\bigl[C(k_3)C'(k_2)\bigr]
   \bigl[C(-k_1^*)C'(-k_4^*)\bigr]^*.
 \end{aligned}
 \label{eq:common-su-rank-one-weight}
\end{equation}
The outgoing factors are conjugated wavefunctions evaluated at the
sign-reversed conjugate momenta. The inner stars do not affect the real
$k_i$ integrations, but specify the analytic continuation of the bra
when parallel momenta become complex. The weight therefore respects
$s$--$u$ crossing, with the physical pairings $(12|34)$ and $(32|14)$
and their respective positive channel measures.

The planar restriction keeps all four shaken $p_{i\parallel}$ in the
two-plane of the seed forward scattering at the orientation fixed above,
so this oriented plane and the total parallel momentum $P_{s\parallel}$
are conserved in the $s$ channel.
Each on-shell momentum then has two continuous degrees of freedom.
The two parallel momentum-conservation conditions and one $SO(1,1)$
boost remove three of the eight degrees of freedom, leaving the five
variables $(s_\parallel;k_1,k_2,k_3,k_4)$.
Without fixing orientation, coincidence of the planes leaves a discrete
$\mathbb Z_2$ choice. At small $k_i$, its two relative orientations give
the forward pairing $p_1\simeq-p_3$ and the backward pairing
$p_1\simeq-p_4$. As \eqref{eq:common-su-rank-one-weight} shows, the
additional backward pairing is $1\to4$ in the $s$ channel and $1\to2$
in the $u$ channel. Including both $s$-channel $\mathbb Z_2$ sheets therefore
breaks the prescribed $s$--$u$ crossing.
We denote the forward sheet selected by the orientation condition by $+$.
Solving the on-shell constraints gives its square-root structure,
\begin{equation}
 \begin{aligned}
 t_\parallel^+(s_\parallel;k_i)
 &=\frac{\sum k_i^2-s_\parallel}{2}
 +\frac{\sqrt{R(s_\parallel;k_i)}
 -(k_1^2-k_2^2)(k_3^2-k_4^2)}{2s_\parallel},\\
 u_\parallel^+&=\sum k_i^2-s_\parallel-t_\parallel^+,\\
 R(s_\parallel;k_i)
 &=\bigl[(s_\parallel-k_1^2-k_2^2)^2-4k_1^2k_2^2\bigr]
   \bigl[(s_\parallel-k_3^2-k_4^2)^2-4k_3^2k_4^2\bigr].
 \end{aligned}
 \label{eq:R-curve}
\end{equation}
In the high-energy eikonal limit this sheet must approach
$p_1\simeq-p_3$, so the squared parallel momentum transfer
$t_\parallel^+$ must remain finite. The square root must therefore satisfy
\begin{equation}
 \lim_{|s_\parallel|\to\infty}
 \frac{\sqrt{R(s_\parallel;k_i)}}{s_\parallel^2}=1.
 \label{eq:forward-sheet-infinity}
\end{equation}
This condition fixes the branch at infinity in every direction of the
complex $s_\parallel$ plane. Continuing $t_\parallel^+$ inward from
$s_\parallel=\infty$ encounters the four real branch points
$(k_1\pm k_2)^2$ and $(k_3\pm k_4)^2$.
We order them along the real axis and take finite cuts between the first
and second and between the third and fourth. Away from these cuts,
$t_\parallel^+$ is analytic except for a possible pole at $s_\parallel=0$.
These unphysical cuts arise from the sheet choice of $\sqrt{R}$,
rather than physical on-shell intermediate states.
For the Gaussian profiles the cuts have a characteristic extent of
$\mathcal{O}(\sigma^2)$ along the real axis. Replacing
\eqref{eq:forward-gaussian-profile} by the compact-momentum $\mathrm{C}^{\infty}$
profiles of \eqref{eq:compact-bump-tail}, with support in $|k_i|\leq\sigma$,
confines these cuts strictly to $0\leq s_\parallel\leq4\sigma^2$.
For the bootstrap we choose $4\sigma^2<\Lambda_{\rm EFT}^2$, placing this
region below the lightest UV resonance of mass $\Lambda_{\rm EFT}$.

By \eqref{eq:spin-projector}, the boundary projections can be transferred
to $T_h$, so we directly smear the half-space amplitude $\M_h$ with the
flat-space density $VKV^\dagger$, represented by the weight $W(k_i)$.
For spinning wavefunctions, the polarization data are supplied by the
seed and transformed by the canonical boosts above.
For massless particle $1$, a reference vector in the oriented seed plane
is $\xi=\bar p_2$, equivalently $\bar p_4=-\bar p_2$, since
\eqref{eq:spin-standard-null-boost} depends only on the reference null
direction, not its sign.
Here $\M_h$ is the half-space scattering matrix kernel in the momentum
parameterization. For the $tz$ seed, the plane condition sets every
$p_{ix}=0$, so shaking produces momenta only in $tzy$. At fixed
$s_\parallel$ and $k_i$, the on-shell map fixes the parallel components.
The four $\dd k_i$ below are precisely the integration measures defined
on the two sides of the state density in \eqref{eq:collinear-momenta}.
By \eqref{eq:spin-projector}, the smeared amplitude at fixed $s_\parallel$ is
\begin{equation}
 \begin{aligned}
 \M_W(s_\parallel)
 &=\Tr\!\left[K_hT_h(s_\parallel)\right]\\
 &=\int_{-\infty}^\infty\prod_i\dd k_i\,W(k_i)\,
 \M_h\bigl(s_\parallel,t_\parallel^+(s_\parallel;k_i);k_i\bigr).
 \end{aligned}
 \label{eq:construction-I-smeared-function}
\end{equation}
For compact profiles supported in $|k_i|\leq\sigma$, the discontinuity
at $s_\parallel\geq4\sigma^2$ comes entirely from physical intermediate
states. For example, tree-level exchange of a resonance of mass $m$
contributes at $s_\parallel=s_{\parallel,m}$, where
$s_{\parallel,m}=m^2+k_m^2$.
Products of external modes expand in the complete half-space eigenmodes
with real normal momentum $k_m$. Since normal momentum is not conserved,
$k_m^2$ ranges over nonnegative values, producing a cut supported on
$s_\parallel\geq m^2$.
The same real spectral decomposition places loop cuts with mass
threshold $m$ at $s_\parallel\geq m^2$.

For the bootstrap we apply a 2SDR to
$\M_W$,
\begin{equation}
 \oint_{\mathcal C_{\rm EFT}}\frac{\dd s_\parallel}{(2\pi\ii)s_\parallel^3}
 \M_W(s_\parallel)
 =\left(\int_{s\text{-cut}}+\int_{u\text{-cut}}\right)
 \frac{\dd s_\parallel}{(2\pi\ii)s_\parallel^3}
 \Disc_s\M_W+B_\infty .
 \label{eq:forward-dispersion}
\end{equation}
The linear UV bound of \eqref{eq:halfspace-linear-real-bound} gives
$B_\infty=0$. We choose a counterclockwise contour $\mathcal C_{\rm EFT}$
enclosing all low-energy singularities, including the unphysical poles
and branch cuts induced by the momentum parameterization. Provided
every point on it lies within the EFT regime, its integral can be
computed by substituting the EFT amplitude into
\eqref{eq:construction-I-smeared-function}. Only the physical $s$- and
$u$-channel UV cuts lie outside it. A positivity bound on the EFT
coefficients then requires nonnegative contributions from both channels.

The $s$-channel contribution is positive by \eqref{eq:spin-projector}.
Here the conserved invariant is $s_\parallel$, rather than the flat-space
$s$, so the absorptive spectrum takes the form
$\delta(s_\parallel-s_{\parallel,m})\rho_s(k_i)$.
At fixed $s_{\parallel,m}$, $\rho_s$ is the PSD sum and phase-space
integral of intermediate-state production amplitudes paired with their
conjugates in half-space $2\to2$ scattering. It replaces the
Wigner $d$-functions of the flat-space partial-wave expansion.
Its contribution is
\begin{equation}
 \begin{aligned}
 &\int_{s\text{-cut}}\frac{\dd s_\parallel}{s_\parallel^3}
 \int_{-\infty}^\infty\prod_i\dd k_i\,
 W(k_i)\,\delta(s_\parallel-s_{\parallel,m})\,\rho_s(k_i)\\
 &\qquad=\frac{1}{s_{\parallel,m}^3}
 \int_{-\infty}^\infty\prod_i\dd k_i\,W(k_i)\rho_s(k_i)\geq0 .
 \end{aligned}
 \label{eq:forward-spectral-s}
\end{equation}
The remaining integral is the trace pairing of two PSD kernels at fixed
$s_{\parallel,m}$. Integration over the intermediate-state spectral labels
in $\rho_s$ preserves this PSD property.

For the $u$ channel, the dispersion denominator is $s_\parallel^3$,
not $u_\parallel^3$. In the flat-space S-matrix, for example, the crossed
denominator modifies the smearing weight, whose action on the crossed
Wigner $d$-functions must be checked numerically.
Here there is an additional correction, since at fixed $k_i$ the
derivative $\partial u_\parallel^+/\partial s_\parallel$ depends on
$s_\parallel$ and $k_i$, whereas in flat space at fixed transfer $t$
one has $\partial u/\partial s=-1$.
On the outer physical $u$ cut, where $\partial_{s_\parallel}u_\parallel^+<0$,
a PSD spectral kernel $\rho_u(k_i)$ at $u_{\parallel,m}=m^2+k_m^2$ contributes
\begin{equation}
 \begin{aligned}
 &-\int_{u\text{-cut}}\frac{\dd s_\parallel}{s_\parallel^3}
 \int_{-\infty}^\infty\prod_i\dd k_i\,
 W(k_i)\,\delta\bigl(u_\parallel^+(s_\parallel;k_i)-u_{\parallel,m}\bigr)
 \rho_u(k_i)\\
 ={}&\int_{-\infty}^\infty\prod_i\dd k_i\,
 \frac{W(k_i)}{s_\parallel^3\,(\partial_{s_\parallel}u_\parallel^+)}
 \bigg|_{u_\parallel^+=u_{\parallel,m}}\rho_u(k_i).
 \end{aligned}
 \label{eq:forward-spectral-u}
\end{equation}
The inverse Jacobian supplied by the delta function and the dispersion
denominator both multiply $W$, so its original $u$-channel PSD property
in \eqref{eq:common-su-rank-one-weight} does not by itself establish
positivity of the modified kernel.
By \eqref{eq:spin-projector}, its positivity can be tested numerically
in the flat-space two-particle state space before boundary projection.
A rigorous test must include matrix elements that do not conserve
normal momentum. Partial-wave tests on the conserving flat-space slice
can estimate the size of the corrections.

To exhibit the small-width behavior, consider
$k_i^2\sim\sigma^2\ll u_{\parallel,m}$. The weight then admits
a controlled expansion on the $u$-channel branch. This hierarchy is
used for the expansion below and is not a condition imposed on the
finite-width numerical functional constructed later,
\begin{equation}
 \begin{aligned}
 \frac{W(k_i)}{s_\parallel^3\,(\partial_{s_\parallel}u_\parallel^+)}
 \bigg|_{u_\parallel^+=u_{\parallel,m}}
 &=\frac{1}{u_{\parallel,m}^3}
 \biggl[1+\frac{3(k_3^2+k_2^2)}{u_{\parallel,m}}\biggr]
 W(k_i)
 \biggl[1+\frac{3(k_1^2+k_4^2)}{u_{\parallel,m}}\biggr]\\
 &\quad+\frac{W(k_i)}{u_{\parallel,m}^3}
 \mathcal{O}\!\left(\frac{\sigma^4}{u_{\parallel,m}^2}\right).
 \end{aligned}
 \label{eq:forward-crossed-nlo}
\end{equation}
The two factors act on the two sides of the $u$-channel pairing
$(32|14)$. Their product preserves the PSD congruence form while
retaining the leading term and next-to-leading correction.

For UV spectral scales $s_{\parallel,m},u_{\parallel,m}\gtrsim
\Lambda_{\rm EFT}^2$, the small-width expansion in
\eqref{eq:forward-crossed-nlo} applies near the lowest threshold when
$\sigma^2\ll\Lambda_{\rm EFT}^2$. In this limit, \eqref{eq:R-curve} gives
$t_\parallel^+\sim\sigma^4/s_\parallel$, while $k_t^2$ is of order
$\sigma^2$, so the transfer smearing probes only squared momenta of order
$\sigma^2$, with the corresponding momenta far below the UV scale
$\Lambda_{\rm EFT}$.
It therefore yields weak EFT bounds similar to those in
\cref{sec:transverse-modes}. The partial-wave construction below instead
allows the tangential transfer to reach $q_{\max}\sim\Lambda_{\rm EFT}$
while the normal-momentum width is chosen separately.

After fixing the normal smearing, the amplitude in
\eqref{eq:construction-I-smeared-function} has no freely variable $t$.
One therefore cannot perform the usual forward-limit power expansion in
$t$ to obtain a sequence of sum rules. In addition, a gravitational
amplitude with a $1/t$ pole has no regular forward power series in $t$.
Half-space smearing makes this contribution IR finite but retains
logarithmic dependence on the distance $y_0$ from the wall.

\subsection{Half-space positivity from partial-wave smearing}
\label{sec:congruence-construction}
\label{sec:pair-basis-exact-map}
\label{app:localized-middle-kernel-proof}

The forward seed has no independent wall-parallel transfer.
Its tangential transfer is induced by normal momentum shaking through
the on-shell conditions. A partial-wave seed instead allows independent
wall-parallel transfer up to the EFT scale, while a separate scale
$\sigma$ controls normal localization.
We construct an $s$-channel positive kernel by restricting a positive
spherical kernel to one fixed circle and applying independent
one-side state maps.
The resulting trace measure also determines
the forward weighting required of a general smearing functional.

Partial-wave positivity is positivity of the kernel relating the
relative wavefunctions of the incoming pair $(1,2)$ and outgoing pair
$(3,4)$. Momentum conservation, $p_1+p_2=-p_3-p_4$, gives the same
center-of-mass plane wave on both sides, multiplied by each pair's
relative wavefunction. Thus positivity reduces to the relative-state
kernel. In the common $s$-channel center-of-mass frame, $SO(3)$ rotations
allow the relative plane waves of $(1,2)$ or $(3,4)$ to be expanded in
spherical harmonics. For spinning external particles, the relative
rotation between $p_1-p_2$ and $p_4-p_3$ is described by Wigner
$D^J_{h_1-h_2,\,h_4-h_3}$ functions. Fixing the $U(1)$ little-group
phase convention for massless states gives the reduced Wigner $d$
functions, with Legendre polynomials for scalar external particles. For example,
the flat-space functional associated with the 2SDR satisfies
\begin{equation}
 \begin{alignedat}{1}
 &-\oint_{\mathcal C_{\rm EFT}}\frac{\dd s}{2\pi\ii}
 \int_{-t_{\max}}^0\dd t\,\phi(t)\frac{\M(s,t)}{s^2u}\\
 ={}&\left(\int_{s\text{-cut}}+\int_{u\text{-cut}}\right)
 \frac{\dd s}{2\pi\ii}\int_{-t_{\max}}^0\dd t\,\phi(t)
 \frac{\operatorname{Disc}_s\M(s,t)}{s^2(s+t)}.
 \end{alignedat}
 \label{eq:partial-wave-seed-dispersion}
\end{equation}
The flat-space high-energy bound makes the contour at infinity in
the 2SDR vanish. The partial-wave test requires the complete
weight on each physical cut to give nonnegative partial-wave eigenvalues
for every allowed mass and spin, or PSD helicity matrices when inelastic
transitions are included \cite{Caron-Huot:2022ugt}.
At each spectral mass, the positive kernel is the corresponding sum of angular
partial-wave projectors with nonnegative weights, or PSD matrix weights
in the spinning case.

For the half-space construction we continue to use the normal-momentum
parameterization of \eqref{eq:normal-hadamard-inverse-ch4}.
The localized state density includes components away from normal
momentum conservation, so the incoming and outgoing pair-center
wavefunctions need not coincide. To implement the congruence map, we
separate the pair-center and relative momenta on each side. For the
incoming pair, choose the conjugate variables
\begin{equation}
 \begin{gathered}
 P=p_1+p_2,\qquad q^\mu=(p_1-p_2)^\mu,\qquad
 X=\frac{x_1+x_2}{2},\qquad r=\frac{x_1-x_2}{2},\\
 p_1\cdot x_1+p_2\cdot x_2=P\cdot X+q\cdot r.
 \end{gathered}
 \label{eq:scheme-II-pair-conjugacy}
\end{equation}
Here $X$ is the pair center, and $r$ is the position of particle $1$
relative to it. The normal components $K,q_y$ use the incoming convention, whereas
$K',q_y'$ use the outgoing convention,
\begin{equation}
\begin{aligned}
 K&=k_1+k_2=k_s+k_0,&
 K'&=-(k_3+k_4)=k_s-k_0,\\
 q_y&=k_1-k_2=k_t+k_u,&
 q_y'&=k_4-k_3=k_u-k_t.
\end{aligned}
\label{eq:scheme-II-s-pair-variables}
\end{equation}
Here $q_y,q_y'$ describe relative normal motion, while $K,K'$ describe
the normal motion of the two pair centers.

We denote the positive flat-space angular kernel associated with
\eqref{eq:partial-wave-seed-dispersion} by $K_{\text{pw-2}}$
\cite{Caron-Huot:2021rmr,Caron-Huot:2022ugt,Dong:2024omo,Beadle:2025cdx},
where the suffix denotes the two subtractions of the 2SDR. This section aims at the
bootstrap, so the kernel includes the subtracted dispersion denominator
from the outset, rather than first constructing a smeared amplitude.
Let $P_s$ be the pair-center momentum, with $P_s^2=s$.
In the physical region $s>0$, an $SO(1,3)$ transformation places
$P_s$, which is shared by the initial and final states, in the
wall-parallel $(t,z,x)$ slice, so that $P_{s\parallel}=P_s$ and $s_\parallel=s$.
In the pair center-of-mass frame, the massless relative reference
directions can then be parameterized as
\begin{equation}
 \begin{gathered}
 |\mathbf q|=|\mathbf q'|=\sqrt{s},\qquad
 \Omega=(\theta,\phi)=\widehat{\mathbf q},\quad
 \Omega'=(\theta',\phi')=\widehat{\mathbf q}',\\
 q_y=\sqrt{s}\,\Omega_y,\qquad q_y'=\sqrt{s}\,\Omega_y'.
 \end{gathered}
 \label{eq:scheme-II-relative-sphere}
\end{equation}
The sphere lies in the three-space orthogonal to $P_{s\parallel}$,
with tangential axes $z,x$ and normal axis $y$.
For the physical $s$-cut weight, its angular kernel and spherical
eigenvalues are
\begin{equation}
 \begin{gathered}
 K_{\text{pw-2}}(s,\Omega\cdot\Omega')
 =\frac{\phi(t)}{4\pi s(s+t)}
 =\sum_{J,m}a_J(s)
   \bigl[\e^{\ii P_s\cdot X}Y_{Jm}(\Omega)\bigr]^*
   \bigl[\e^{\ii P_s\cdot X}Y_{Jm}(\Omega')\bigr],\\
 a_J(s)=\int_{-t_{\max}}^{0}\!\dd t\,
 \frac{\phi(t)}{s^2(s+t)}\,
 P_J\!\left(1+\frac{2t}{s}\right)\geq0.
 \end{gathered}
 \label{eq:Kpw}
\end{equation}
Here $t=-s(1-\Omega\cdot\Omega')/2$.
The factor $s/(4\pi)$ converting $\phi(t)/[s^2(s+t)]$ into the angular
kernel follows from $\dd\Omega'=4\pi\,\dd t/s$ after the common
azimuthal integral. The center plane waves cancel in the matched
pairing, but are displayed because shaking also acts on the pair-center
state. All mode variables in this physical pairing are real.
Note that $a_J(s)$ specify the density of correlated external wave packets,
rather than the absorptive part of the scattering amplitude.

\begin{figure}[H]
 \centering
 \includegraphics[width=.64\linewidth]{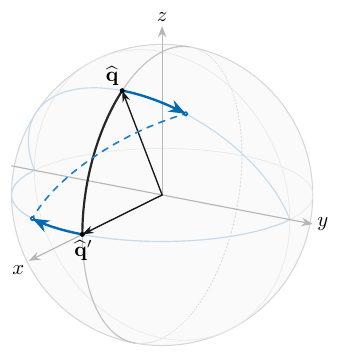}
 \caption{Schematic illustration of normal shaking of the incoming
 and outgoing relative momenta. The black reference directions lie
 on the fixed $zx$ great circle. Blue arrows show independent normal
 shaking, and the dashed blue arc joins the displaced directions.
 The common rotation about $y$ remains free. Shaking of the incoming
 and outgoing center-of-mass momenta is not shown.}
 \label{fig:relative-angular-shaking}
\end{figure}

Now restrict both directions to the same wall-parallel great circle,
$\Omega_y=\Omega_y'=0$, so that
$\Omega\cdot\Omega'=\cos(\phi-\phi')$.
With $y$ as the magnetic quantization axis, the addition theorem gives
\begin{equation}
 \begin{gathered}
 K_{\text{pw-2}}\bigl(s,\cos(\phi-\phi')\bigr)
 =\sum_{m\in\mathbb Z}\lambda_m(s)
   \bigl[\e^{\ii P_s\cdot X}\e^{\ii m\phi}\bigr]^*
   \bigl[\e^{\ii P_s\cdot X}\e^{\ii m\phi'}\bigr],\\
 \lambda_m(s)=\sum_{J\geq|m|}a_J(s)
       |Y_{Jm}(\pi/2,0)|^2\geq0.
 \end{gathered}
 \label{eq:fixed-circle-restriction}
\end{equation}
The functions $\e^{\ii m\phi}$ are Fourier harmonics on the circle
$S^1$, orthonormal with measure $\dd\phi/(2\pi)$, so $\lambda_m(s)$ are
the circle eigenvalues, while the $a_J(s)$ in \eqref{eq:Kpw} remain
the spherical eigenvalues.
 The center plane wave is the same in both expansions.
Before shaking, $P_s$ is a timelike vector in the wall-parallel
$(t,z,x)$ space. The relative direction runs around the circle in the
two-dimensional plane perpendicular to $P_s$ within that space,
which is the $zx$ circle in the pair rest frame.
$V_s$ acts on the complete state, redistributing both pair-center
and relative normal momenta without changing the angular integration
measure. Equation~\eqref{eq:fixed-circle-restriction} takes the equatorial
values of the same kernel in \eqref{eq:Kpw}, and the addition theorem
preserves its positivity. This does not rotate all four momenta of
off-equator components into the $(t,z,x)$ slice by an $SO(1,3)$
transformation. Such a choice based on both the initial and final
states would spoil the one-side congruence construction of positivity.
The seed momenta in this realization, denoted $p_i^{\rm sd}$, obey
$p_i^{\rm sd}\cdot\hat y=0$. The reference directions and their
independent shaking are shown in Fig.~\ref{fig:relative-angular-shaking}.

The one-side on-shell map \eqref{eq:collinear-momenta} shakes the
$\hat y$ components of the incoming pair while keeping $P_{s\parallel}$
and the oriented plane $p_1^{\rm sd}\wedge p_2^{\rm sd}$ fixed.
The outgoing pair is treated in the same way. For spinning states,
the canonical boost transports the seed polarizations with the momenta,
using $p_2^{\rm sd}$ as the reference vector for particle $1$.
Unlike the forward seed, within the wall-parallel $(t,z,x)$ space the
projections of $q_\parallel,q_\parallel'$ perpendicular to
$P_{s\parallel}$ can make a nonzero angle already before shaking.
The tangential transfer scale is therefore independent of the normal
shaking scale.
Applying $V$ to both sides of the positive seed and then projecting
each particle with $\Pi_h$ gives the half-space PSD kernel
$K_h=\Pi_h^{\otimes2}VK_{\text{pw-2}}V^\dagger
(\Pi_h^{\otimes2})^\dagger$.
This separates the normal localization that regulates the IR divergence
from the tangential smearing range, which can reach
$q_{\max}\sim\Lambda_{\rm EFT}$. These scales control different parts of the external wave-packet profiles.

All normal-momentum shaking is contained in $V$, including
$k_1,k_2$, or equivalently $K,q_y$, on the incoming side.
Write the invariant smearing action of the seed in
\eqref{eq:partial-wave-seed-dispersion} as
\begin{equation}
 K_{\text{pw-2}}[\M]
 =-\int\!\frac{\dd s^{\rm sd}}{2\pi\ii}
 \int_{-t_{\max}}^0\!\dd t^{\rm sd}\,
 \frac{\phi(t^{\rm sd})}{(s^{\rm sd})^2u^{\rm sd}}\,
 \M(s^{\rm sd},t^{\rm sd}).
 \label{eq:planar-seed-smearing-action}
\end{equation}
Here $K_{\text{pw-2}}[\M]$ is the invariant form of the flat-space
trace. At fixed $s^{\rm sd}$, the two directions have four
angular coordinates. For an $SO(3)$-invariant integrand, integrating the
three common rotation coordinates leaves $\dd\cos\theta_s$.
Equivalently, the transfer integral from $-t_{\max}$ to zero becomes
an integral from zero to $q_{\max}$ with measure
$2q^{\rm sd}\,\dd q^{\rm sd}$, where $q^{\rm sd}=\sqrt{-t^{\rm sd}}$.
We retain the indefinite $\dd s^{\rm sd}$ integral because the eventual
dispersion variable need not be $s^{\rm sd}$, so changing to that variable
automatically supplies the corresponding Jacobian.
For the kernel in \eqref{eq:fixed-circle-restriction}, whose seed has only
$(t,z,x)$ momenta, shaking breaks the spatial $SO(3)$ symmetry of the
momentum distribution. Only one common $SO(2)$ rotation in the $zx$ plane
can be integrated out, so the relative angle $\Omega\cdot\Omega'$ and
both normal coordinates $\Omega_y,\Omega_y'$ must be retained.

For fixed shaken momenta, $p_{i\parallel}^2=k_i^2$ is the effective
mass shell in $(t,z,x)$. Keeping the oriented pair planes preserves
the angle $\theta_s$ between the relative directions orthogonal to
$P_{s\parallel}$. The seed has zero normal momentum, whereas the
shaken pairs have effective masses $|k_i|$. Their on-shell kinematics
therefore gives
\begin{equation}
 \begin{aligned}
 t^{\rm sd}&=-\frac{s^{\rm sd}}{2}(1-\cos\theta_s),\\
 t_\parallel&=\frac{\sum k_i^2-s_\parallel}{2}
 -\frac{(k_1^2-k_2^2)(k_3^2-k_4^2)}{2s_\parallel}
 +\frac{\sqrt{R(s_\parallel;k_i)}}{2s_\parallel}\cos\theta_s.
 \end{aligned}
 \label{eq:planar-seed-preserved-angle}
\end{equation}
The definition of $R$ and its sheet choice are given in
(\ref{eq:R-curve}--\ref{eq:forward-sheet-infinity}).
Since the map preserves $P_{s\parallel}$, eliminating $\theta_s$
reconstructs the seed invariants from the external kinematic variables
$s_\parallel,t_\parallel,u_\parallel$ and the four $k_i$,
\begin{equation}
 \begin{aligned}
 s^{\rm sd}&=s_\parallel,\\
 t^{\rm sd}&=-\frac{s_\parallel}{2}
 +\frac{s_\parallel^2}{\sqrt{R(s_\parallel;k_i)}}
 \left[t_\parallel+\frac{s_\parallel-\sum k_i^2}{2}
 +\frac{(k_1^2-k_2^2)(k_3^2-k_4^2)}{2s_\parallel}\right],\\
 u^{\rm sd}&=-\frac{s_\parallel}{2}
 +\frac{s_\parallel^2}{\sqrt{R(s_\parallel;k_i)}}
 \left[u_\parallel+\frac{s_\parallel-\sum k_i^2}{2}
 -\frac{(k_1^2-k_2^2)(k_3^2-k_4^2)}{2s_\parallel}\right].
 \end{aligned}
 \label{eq:planar-seed-inverse-map}
\end{equation}
Note that the seed invariants always obey
$s^{\rm sd}+t^{\rm sd}+u^{\rm sd}=0$, whereas the shaken external
variables obey $s_\parallel+t_\parallel+u_\parallel=\sum k_i^2$.

To make the exact map physically transparent, temporarily take
$k_i=\mathcal O(\sigma)$ and $\sigma^2\ll|s_\parallel|$. This expansion
is not required in the finite-width numerical positivity test.
Retaining the leading and next-to-leading shaking terms and factoring
out their common rescaling gives
\begin{equation}
 (s^{\rm sd},t^{\rm sd},u^{\rm sd})
 =\frac{s_\parallel}{s_\parallel-\sum k_i^2}
 \Bigl(s_\parallel-\sum k_i^2,\,
 t_\parallel+\mathcal O(\tfrac{k^4}{s_\parallel}),\,
 u_\parallel-\mathcal O(\tfrac{k^4}{s_\parallel})\Bigr).
 \label{eq:planar-seed-factorized-width}
\end{equation}
The common factor collects the shared next-to-leading rescaling, leaving
quadratic arguments that depend only on their respective parallel
Mandelstam variable and the normal momenta. This expansion requires no
hierarchy between $|t_\parallel|$ and
$\sigma^2$. 

The exact map \eqref{eq:planar-seed-inverse-map} respects $t$--$u$ crossing under
$t_\parallel\leftrightarrow u_\parallel$ and $k_3\leftrightarrow k_4$
(equivalently $k_t\leftrightarrow k_u$). Accordingly, the exact
$s$-channel congruence preserves $t$--$u$ symmetry of the S-matrix
when the normal weight $W(k_i)$ respects the same exchange.
For a $t$--$u$-symmetric seed, for example when particles $3,4$ are
identical, we choose $C=C'$ in \eqref{eq:common-su-rank-one-weight} so
that normal smearing does not break their exchange symmetry.
For the Gaussian profiles in \eqref{eq:forward-gaussian-profile},
this choice gives the four wave packets the same normal localization
center, $y_1=y_2$.
Exchanging $s_\parallel\leftrightarrow u_\parallel$ and
$k_s\leftrightarrow k_u$ (equivalently $k_2\leftrightarrow k_4$) in
this map directly gives the $u$-channel parameterization, which is
correspondingly covariant under $s$--$t$ crossing. The factorized
expansion inherits these crossing relations.

We now apply \eqref{eq:one-side-state-map} to the states while retaining
the angular integrations. With the normal weight $W(k_i)$, the resulting
smearing of the dispersion relation on the physical $s$ branch is
\begin{equation}
 \begin{alignedat}{1}
 &(V_sK_{\text{pw-2}}V_s^\dagger)[\M_h]\\
 ={}&\int\!\frac{\dd s_\parallel}{2\pi\ii}
   \int_0^{2\pi}\!\frac{\dd\phi\,\dd\phi'}{(2\pi)^2}
   \int_{\mathcal D_s}\!\prod\limits_{i=1}^{4}\dd k_i\,W(k_i)
   K_{\text{pw-2}}\bigl(s^{\rm sd},\cos(\phi-\phi')\bigr)
   \M_h(s_\parallel,t_\parallel;k_i)\\
 ={}&-\int\!\frac{\dd s_\parallel}{2\pi\ii}
   \int_{-t_{\max}}^0\!\dd t^{\rm sd}\,
   \frac{\phi(t^{\rm sd})}
        {4\pi^2s^{\rm sd}u^{\rm sd}\sqrt{t^{\rm sd}u^{\rm sd}}}
   \int_{\mathcal D_s}\!\prod\limits_{i=1}^{4}\dd k_i\,W(k_i)
   \M_h(s_\parallel,t_\parallel;k_i).
 \end{alignedat}
 \label{eq:planar-seed-sixfold-nlo}
\end{equation}
Here $s_\parallel=s^{\rm sd}$ and
$\cos\theta_s=\cos(\phi-\phi')$ determine $t_\parallel$ through
\eqref{eq:planar-seed-preserved-angle}. The amplitude is evaluated at
the reconstructed external variables $(s_\parallel,t_\parallel;k_i)$,
and the four normal integrations include pair-center and relative shaking.

 To keep these external states in the physical regime,
the normal-momentum integration domain $\mathcal D_s$ satisfies
\begin{equation}
 |K|,\ |q_y|,\ |K'|,\ |q_y'|<\sqrt{s^{\rm sd}}.
 \label{eq:fixed-circle-physical-support}
\end{equation}
To construct correlated wave packets supported strictly inside this physical momentum domain, choose
the $\mathrm{C}^{\infty}$
compact profiles of \eqref{eq:compact-bump-tail} on both the incoming and outgoing sides. They also
give stretched-exponential
localization tails in position space.
We continue to use Gaussian profiles in the numerical calculations.
The four external-line factors in $W(k_i)$ retain a factorized form
in the equivalent variables $(K,K',q_y,q_y')$ of
\eqref{eq:scheme-II-s-pair-variables}, which simplifies the integrations.
For a profile proportional to $\e^{-k^2/(2\sigma^2)}$, we retain
$-10\sigma\leq k\leq10\sigma$ and neglect the tail
whose magnitude is exponentially suppressed to $\e^{-50}$.
 The normal-momentum width is chosen so that
the full retained four-leg support lies inside
\eqref{eq:fixed-circle-physical-support}, keeping the numerical
integration on the physical branches.

Substituting the positive wavefunctions in
\eqref{eq:fixed-circle-restriction} gives the physical $s$-channel
absorptive trace
\begin{equation}
 \Tr[V_sK_{\text{pw-2}}V_s^\dagger\operatorname{Im}T_h]
 =\frac12\sum_{m\in\mathbb Z}\lambda_m(s)
   \sumint_{X_h}
   |\langle X_h|T_h\Pi_h^{\otimes2}V_s|P_s,m\rangle|^2\geq0.
 \label{eq:scheme-II-leading-Gram-square}
\end{equation}
Here $|P_s,m\rangle$ is the complete center and relative state displayed
in \eqref{eq:fixed-circle-restriction}. The sum and integral over $X_h$
run over all intermediate states in the half-space Hilbert space.
The boundary projectors can be moved to $T_h$ as in
\eqref{eq:spin-projector}. The linear congruence generated by $V_s$
therefore preserves this $s$-channel positivity exactly at every order.

The factor $s^{\rm sd}/\sqrt{t^{\rm sd}u^{\rm sd}}$ in
\eqref{eq:planar-seed-sixfold-nlo} comes from the relative circle angle.
After integrating the common azimuth,
$-\dd t^{\rm sd}/\dd\theta_s=\sqrt{t^{\rm sd}u^{\rm sd}}$,
so the two angular integrations directly give the last line of
\eqref{eq:planar-seed-sixfold-nlo}.
 Physically, the reference relative
directions $\mathbf q,\mathbf q'$ lie on the equator shown in
Fig.~\ref{fig:relative-angular-shaking}, and normal shaking distributes
them near it. The relative spherical-area measure
$\dd\cos\theta_s$ is replaced by the circle-length measure
$\dd\theta_s$. This is the geometric origin of the inverse square root.

However, this angular measure introduces an additional branch cut.
 When the
$s$-channel expression is continued to the $u$ channel, the square root
in the measure contributes a discontinuity as well as the physical
$u$-channel intermediate states, even if $t^{\rm sd}$ remains on the
negative real axis. A dispersion relation for
\eqref{eq:planar-seed-sixfold-nlo} therefore has a discontinuity on the
negative real $s^{\rm sd}$ axis that is not entirely due to physical
$u$-channel states, and its positivity is not guaranteed. The exact
congruence guarantees positivity in the physical $s$ channel.

What \eqref{eq:planar-seed-sixfold-nlo} identifies for a localized positive
smearing functional is the forward power of its transfer measure.
As $t^{\rm sd}\to0$, multiplying $\dd t^{\rm sd}$ by the angular factor
changes the positive radial measure from
$q^{\rm sd}\dd q^{\rm sd}$ to $\dd q^{\rm sd}$.
 The two-dimensional area integral over
the position of $\widehat{\mathbf q}'$ relative to
$\widehat{\mathbf q}$ on the direction sphere becomes a one-dimensional
equatorial line integral. The other angular degree of freedom,
perpendicular to the equator, is described by $\dd k_t$.
The remaining factor $s^{\rm sd}/\sqrt{-u^{\rm sd}}$ is smooth in the
forward region of the physical $s$ channel and is not essential to
the forward power required of a crossing-symmetric smearing prescription.
Near forward scattering, \eqref{eq:planar-seed-factorized-width} gives
$q^{\rm sd}\sim q_\parallel$. Together with $k_t$, these are two
coordinates for transverse momentum exchange, restoring the local
two-dimensional area integral.

The construction extracts the equatorial components of $K_{\text{pw-2}}$
through \eqref{eq:fixed-circle-restriction}. It does not rotate all four
momenta into the wall-parallel $(t,z,x)$ space by an $SO(1,3)$
transformation chosen from both sides of the scattering.
Compressing the original spherical-area distribution onto the equator
and then shaking it, while retaining its radial area weight, would
instead give $q_\parallel\dd q_\parallel\dd k_t$ times a smooth factor
near forward transfer. Relative to the two independent transverse
coordinates, the extra $q_\parallel$ forces that effective
momentum-space kernel to vanish at the origin. %
However, a nonzero PSD momentum-space kernel
has a nonnegative density in its Fourier-conjugate position space.
Its value at zero transfer is the integral of that density and therefore
cannot vanish.
 The spherical
version is the diagonal sum of nonnegative harmonic-mode densities.
Such a compression therefore cannot retain a nontrivial PSD kernel
with a regular remaining factor. This explains why a joint rotation
depending on the initial and final momenta, followed by normal shaking,
does not preserve the positivity argument.

Analogous congruences can be applied to seeds from improved dispersion
relations \cite{Caron-Huot:2021rmr,Caron-Huot:2022ugt}
and CSDRs \cite{Sinha:2020win,Pasiecznik:2025eqc}.
They guarantee positivity in the channel used for the state construction,
while the crossed-channel positivity and additional measure cuts require
separate checks. The useful common requirement is the forward behavior
of the smearing measure. For example, for a $t$--$u$-symmetric amplitude
one can smear the fixed symmetric combination used in
Ref.~\cite{Dong:2025dpy}. %
In the signed convention
$w^{-1}=t^{-1}+u^{-1}$, one has $w=-tu/s<0$ in the physical
$s$-channel regime, and the flat-space smearing uses
$\dd w\,\phi(w)$ from a negative cutoff to zero.
For a smearing functional constructed by equatorial restriction and shaking
as in \eqref{eq:planar-seed-sixfold-nlo}, the measure near
$w^{\rm sd}\to0$ requires
$\dd\sqrt{-w^{\rm sd}}\,\phi(w^{\rm sd})$ times a smooth factor.

The preceding construction generates positive smearing kernels from
states whose center and relative momenta initially lie in the
wall-parallel $(t,z,x)$ space. It therefore uses only the equatorial
values of $K_{\text{pw-2}}$, rewritten in the $SO(2)$ Fourier basis by
\eqref{eq:fixed-circle-restriction}, before an integral map redistributes
the states to produce normal-momentum components. The complementary
construction in Appendix~\ref{sec:archive531-relative-state-envelopes} retains the full spherical
harmonics and localizes
the relative states by multiplying them with an envelope, while leaving
the pair-center motion to momentum shaking.
These operations have the physical pictures of angular convolution and
pointwise multiplication. An integral map spreads a sharply supported
direction over nearby directions, whereas an envelope reweights the
wavefunction directly in angular coordinates. The angular and
angular-momentum representations play Fourier-related momentum-like
and position-like roles, respectively, without being literal momentum
and position spaces. A rotationally invariant convolution on the sphere
is diagonal in $(J,m)$ and multiplies each harmonic component by its
corresponding coefficient.

The complementary construction in
Appendix~\ref{sec:archive531-relative-state-envelopes} may help further
optimize the EFT bound obtained in
Sec.~\ref{sec:general-halfspace-dispersion}. There, the
gravitational exchange contribution to the dispersion relation is
enhanced by a factor $(q_{\max}/\sigma)^1$. For the wave-packet profiles
considered here, this factor is $100$ for Gaussian profiles and $10$
for compact $\mathrm{C}^{\infty}$ profiles. Bringing $\sigma$ closer to $q_{\max}$ could further suppress this
IR-finite gravitational contribution. A possible extension keeps
$k_s,k_t,k_u$ at the $q_{\max}$ scale while restricting only $k_0$ to a
separate width $\sigma_0\ll q_{\max}$. Each external normal momentum
$k_i$ can then remain of order $q_{\max}$, potentially improving the
bound. Whether this construction preserves positivity and crossing,
and how far it can improve the bound, are left to future work.

\subsubsection{Spinning smearing functionals}
\label{sec:spinning-smearing-functionals}

We next construct smearing functionals for spinning wavefunctions.
This requires specifying the spinning states of the unshaken seed
and their polarization conventions.
Once these are fixed, the canonical
boosts in Sec.~\ref{sec:spin-frames-halfspace} determine the polarizations
after shaking for both massive and massless particles. This prescription applies to $K_{\text{pw-2}}$ with a
momentum-dependent convention for the external helicity phases.

For spinning particles the spherical wavefunctions carry the
two-particle helicity difference $h=h_1-h_2$. With the external
polarization convention specified, their angular dependence is
\begin{equation}
 \mathcal Y_{Jm,h}(\Omega)
 =\sqrt{\frac{2J+1}{4\pi}}\,
 D^J_{mh}(\phi,\theta,0)^*,\qquad
 \int\!\dd\Omega\,\mathcal Y_{Jm,h}(\Omega)^*
 \mathcal Y_{J'm',h}(\Omega)=\delta_{JJ'}\delta_{mm'}.
 \label{eq:spinning-spherical-wavefunction}
\end{equation}
For a massive incoming pair, $h_1,h_2$ are the spin projections
along the respective particle momenta in the rest frame of $p_1+p_2$,
whose relative axis is $p_1-p_2$. They therefore refer to a chosen
helicity basis in the spin space carrying the massive $SU(2)$ little-group representation. For massless
particles they are the individual helicities, while the external
$U(1)$ little-group phases still require a convention. The outgoing
pair is treated in the same way. The quantization axis and angular
coordinates are those of \eqref{eq:scheme-II-relative-sphere}, and
$h=0$ gives the ordinary spherical harmonics. Conjugate contraction and
the sum over $m$ give Wigner $d^J_{hh'}$ in a common scattering plane,
reducing to $P_J$ when $h=h'=0$. The spinning version of
\eqref{eq:Kpw} consequently has a positive matrix of partial-wave
weights,
\begin{equation}
 [K_{\text{pw-2}}(\Omega,\Omega')]_{hh'}
 =\sum_{J,m}\mathcal Y_{Jm,h}(\Omega)^*
 [a_J]_{hh'}\mathcal Y_{Jm,h'}(\Omega'),
 \qquad a_J\succeq0.
 \label{eq:spinning-spherical-positive-seed}
\end{equation}
The matrix indices may also include particle species and other
polarizations. For example, particle $1$ in the initial pair may be
a superposition of the two graviton helicities $(h^+,h^-)$, or of a
graviton and a pion $(h^+,\pi)$, with particle $2$ fixed. The final
pair is treated in the same way, allowing particle $3$ to lie in the
corresponding superposition. The pion may be taken in the massless
limit to share the graviton's on-shell condition. At fixed $J$, the
eigenvectors of $[a_J]_{hh'}$ specify superpositions of these
two-particle states, and their nonnegative eigenvalues give the weights
in the selected correlated wave packet.

In flat space, $\Tr(K_{\text{pw-2}}T)$ is a singlet under the $U(1)$
little-group transformations of the four external legs, with the
phases carried by the external states in $K_{\text{pw-2}}$ canceling
those of the amplitude. Standard spinning positivity tests use
small Wigner $d$ functions and amplitude matrix elements in a fixed
helicity convention \cite{Caron-Huot:2022ugt}, so the correlated
wave packet must adopt the same $U(1)$ little-group phase convention.
Before that choice, the full helicity amplitude depends on the
external little-group phases as well as the Mandelstam variables.
In all-incoming notation,
\begin{equation}
 |i\rangle\longmapsto e^{-\ii\varphi_i/2}|i\rangle,\qquad
 |i]\longmapsto e^{\ii\varphi_i/2}|i],\qquad
 \M_{h_1h_2h_3h_4}\longmapsto
 e^{\ii\sum h_i\varphi_i}\M_{h_1h_2h_3h_4}.
 \label{eq:four-leg-little-group-phase}
\end{equation}
For the four-graviton amplitude in all-incoming $(++--)$ ordering,
rewrite the reduced function of Ref.~\cite{Caron-Huot:2022ugt} in
our Mandelstam variables and denote it by $f(s,t)$. Then
\begin{equation}
 \begin{gathered}[c]
 \M_{++--}=[12]^4\langle34\rangle^4 f(s,t)
 =\frac{[12]^2\langle34\rangle^2}{\langle12\rangle^2[34]^2}
   \M^{\bcancel{U(1)}}_{++--}(s,t),\\
 \M^{\bcancel{U(1)}}_{++--}(s,t)=s^4f(s,t).
 \end{gathered}
 \label{eq:graviton-phase-only-stripping}
\end{equation}
The dimensionless bracket ratio carries only helicity phase in the
physical real region. Removing it gives $\M^{\bcancel{U(1)}}$ with
the mass dimension of the full amplitude, whereas $f$ also removes
the dimensionful factor $s^4$. The phase-fixed, normalized wave-packet
amplitude therefore obeys the unitarity bound
\eqref{eq:halfspace-linear-real-bound} in both flat space and half-space.
The reduced function $f$ instead obeys the stronger UV bound underlying
the superconvergent dispersion relations.

In flat space, removing the phase factor is equivalent to fixing
the four external phases. In spinor-helicity variables we set the
ratio of square and angle brackets for each pair to one, equivalently
choosing the four continuous little-group phases to vanish relative
to this convention,
\begin{equation}
 \begin{aligned}
 &\frac{[12]}{\langle12\rangle}
 =\frac{[13]}{\langle13\rangle}
 =\frac{[14]}{\langle14\rangle}
 =\frac{[23]}{\langle23\rangle}
 =\frac{[24]}{\langle24\rangle}
 =\frac{[34]}{\langle34\rangle}=1,\\
 &\mathllap{\Longleftrightarrow\quad}\hspace{1.2pt}
 \varphi_1=\varphi_2=\varphi_3=\varphi_4=0.
 \end{aligned}
 \label{eq:redundant-spinor-phase-fixing}
\end{equation}
The six bracket ratios are not independent, with only four independent
little-group phases $\varphi_i$. In this convention
$[12]^4\mapsto s^2$ and $\langle34\rangle^4\mapsto s^2$, so the
helicity-fixed amplitude in \eqref{eq:graviton-phase-only-stripping}
is an analytic function of the Mandelstam variables $s,t$.

At a Mandelstam point $(s,t)$ in the support of the smearing profile, choose a
set of momenta realizing these invariants. Common Lorentz transformations
in $SO(1,3)$ relate equivalent choices of this momentum configuration.
For $stu\ne0$, the common normal to the four conserved momenta is
\begin{equation}
 n^\mu=\frac{2\varepsilon^{\mu\nu\rho\sigma}
 p_{1\nu}p_{2\rho}p_{3\sigma}}{\sqrt{stu}},\qquad
 n^2=-1,\qquad n\cdot p_i=0.
 \label{eq:conserved-seed-polarization-normal}
\end{equation}
At $t\to0$, this normal has a finite continuous limit along a fixed
scattering plane, which defines its value at the degenerate configuration.
For example, $n$ can be chosen as $\hat y$ for the planar realization
of $K_{\text{pw-2}}$, whereas for $K_{\text{pw-2}}'$ it follows the tilt of the
timelike three-plane spanned by the conserved momenta.
An explicit polarization basis for particle $1$ and its field strengths is
\begin{equation}
 \begin{gathered}
 \epsilon_{1y}^\mu=n^\mu,\qquad
 \epsilon_{1x}^\mu=
 \frac{\varepsilon^{\mu\nu\rho\sigma}p_{1\nu}p_{2\rho}n_\sigma}
 {p_1\cdot p_2},\qquad
 \epsilon_{1,\pm}^\mu=\frac{\epsilon_{1x}^\mu\pm\ii\epsilon_{1y}^\mu}{\sqrt2},\\
 F_{1,\pm}=p_1\wedge\epsilon_{1,\pm}
 =\frac{\widetilde{F}_{1y}\pm\ii F_{1y}}{\sqrt2},\qquad
 F_{1y}=p_1\wedge n.
 \end{gathered}
 \label{eq:conserved-seed-polarizations}
\end{equation}
The labels $x,y$ distinguish the two polarization components and need
not coincide with the coordinate axes in Fig.~\ref{fig:relative-angular-shaking}.
The circular combinations fix the $U(1)$ little-group phases, and the
other legs are treated in the same way. Although $\epsilon_{1x}$ uses
$p_2$ to choose a gauge representative, its field strength depends
only on $p_1$ and $n$. The same covariant expression therefore continues
to the crossed channels. These vectors specify the polarizations in
\eqref{eq:spin-helicity-polarizations} for each wave-packet component.

Here $\widetilde F^{\mu\nu}=\varepsilon^{\mu\nu\rho\sigma}F_{\rho\sigma}/2$,
with $\varepsilon^{0123}=-\varepsilon_{0123}=1$, and we use the bracket
orientation $2p_i\cdot p_j=\langle ij\rangle[ij]$.
Substituting \eqref{eq:conserved-seed-polarizations} gives explicitly
\begin{equation}
 \frac{[12]}{\langle12\rangle}
 =\frac{[12]^2}{s}
 =\frac{F_{1,+}^{\mu\nu}F_{2,+\mu\nu}}{s}
 =\frac{(\widetilde F_{1y}+\ii F_{1y})^{\mu\nu}
         (\widetilde F_{2y}+\ii F_{2y})_{\mu\nu}}{2s}
 =1.
 \label{eq:conserved-seed-field-strength-phase}
\end{equation}
In the last equality, $F_{1y}\cdot F_{2y}=-s$,
$\widetilde F_{1y}\cdot\widetilde F_{2y}=s$ and the mixed
contractions vanish because the two field strengths share $n$.
The same calculation applies to every pair, realizing
\eqref{eq:redundant-spinor-phase-fixing}.

Thus each Mandelstam point of the correlated wave packet, used above
as the seed for congruence, has a set of Lorentz-covariant polarization
vectors. The maps $V$ or $V_s'$ transport these vectors with the
canonical boosts of Sec.~\ref{sec:spin-frames-halfspace}, pairing
each shaken momentum with its shaken polarization. This completes
the spinning correlated wave packet and produces a PSD flat-space
kernel with components outside the normal-momentum-conserving slice.
Applying the boundary projection afterwards gives the positive
half-space functional through \eqref{eq:spin-projector}.

The projection changes the helicity content of the packet. Before
projection, the canonical boosts preserve the chosen helicity of each
massless leg, even when the momenta are shaken or correlated between
legs. The reflected component has the opposite helicity by
\eqref{eq:spin-helicity-reflection}. For example, when $n=\hat y$ in
\eqref{eq:conserved-seed-polarizations}, the reflection in
\eqref{eq:tensor-reflection-ch4} reverses $\epsilon_y$ and leaves
$\epsilon_x$ unchanged, exchanging $\epsilon_+$ and $\epsilon_-$.
Half-space eigenmodes and wavepackets therefore generally superpose
opposite helicities, so a single flat-space helicity wavefunction does
not describe the projected state.

This helicity mixing leaves the PSD property intact. The projector
acts on each leg's own momentum and polarization, so its action on
the initial pair uses only initial data, and likewise for the final
pair. Applying $\Pi_h^{\otimes2}$ on the initial side of $VKV^\dagger$
and its adjoint on the final side is a congruence, yielding a PSD
smearing kernel for a half-space spinning correlated wave packet.
By \eqref{eq:spin-projector}, pairing this kernel directly
with the half-space amplitude defines the corresponding functional.

\subsection{Crossing-symmetric smearing and numerical positivity}
\label{app:four-scalar-resonance-checks}
The state congruence of Sec.~\ref{sec:congruence-construction} gives
an exact positive $s$-channel trace, but its geometric square-root
measure also introduces an unphysical crossed-channel discontinuity.
We now use the $\dd q_\parallel$ measure in \eqref{eq:planar-seed-sixfold-nlo}
to define a smearing prescription shared by the physical $s$ and $u$
channels and test the positivity of both responses.

Choose $s_h=s_\parallel-k_s^2$ as the dispersion variable, with
$t_\parallel=-q_\parallel^2=t_h+k_t^2$ and
$u_h=k_0^2-s_h-t_h$. Using the smearing profile
$\phi(-q_\parallel^2)=(1-q_\parallel/q_{\max})^3$, define the
smeared dispersion integrand
\begin{equation}
 \mathcal B_{2,h}(s_h)
 =-\int_0^{q_{\max}}\dd q_\parallel\,\phi(-q_\parallel^2)
 \prod_{i=1}^{4}\int\dd k_i\,W(k_i)
 \frac{\M_h(s_h+k_s^2,-q_\parallel^2;k_i)}{s_h^2u_h}.
 \label{eq:common-dq-smeared-amplitude}
\end{equation}
The contour is counterclockwise, with measure $\dd s_h/(2\pi\ii)$.
The minus sign makes a positive contact coefficient multiplying $-su$
give a positive contour moment. The transfer measures $\dd q_\parallel$
and $\dd t$ have different dimensions. There are also four additional
normal-momentum integrals with weight $W$, so an extra dimensional
normalization factor is needed and will be given below.
By the idempotence of the boundary projection in \eqref{eq:spin-projector},
the external-state projections described by the smearing functional
are already included in $\M_h$.
This definition uses the full amplitude without pulling back
the seed-invariant functions through the state congruence. It therefore
does not introduce the additional branch cuts associated with the
square-root choice in \eqref{eq:R-curve} and
\eqref{eq:planar-seed-sixfold-nlo}. At fixed normal momenta,
$\dd s_h=\dd s_\parallel$, so the outer contour variable can be changed.
We use $s_h$ because the dispersion denominator is $s_h^2u_h$.

We use the Gaussian normal-momentum profile $W(k_i)$ defined in
\eqref{eq:common-su-rank-one-weight}, retaining the normal integration
band $[-10\sigma,10\sigma]$ and neglecting the
exponentially suppressed tails. An external wave packet with strictly compact momentum support can instead use
the compact-momentum $\mathrm{C}^{\infty}$ profiles of \eqref{eq:compact-bump-tail},
with compact support separated from kinematic singularities.
For the Gaussian we take $\sigma/q_{\max}\leq10^{-2}$.
A compact profile supported in $(-\sigma,\sigma)$ can allow
$\sigma/q_{\max}\sim10^{-1}$, subject to the same two-channel tests.
For simplicity, we do not adopt this compact-profile alternative here.
The normal profile $W(k_i)$ respects $s$--$u$ crossing and is used on both
physical channels. In the dispersion relation, the poles of the dispersion
denominator remain inside the low-energy contour, while the UV
discontinuities come from physical $s$- and $u$-channel intermediate states.

The physical $s$- and $u$-channel absorptive operators in half-space
allow nonconserving normal momenta and need not have full spacetime
$\mathrm{SO}(1,3)$ symmetry. Their spectral data require an
infinite-dimensional function space, beyond the channel energy and
partial-wave angular momentum of the conserving flat-space spectrum.
For example, an $s$-channel block is specified by the conserved invariant
$s_\parallel$, an angular harmonic $\ell$ of $\mathrm{SO}(2)$, and a
normal two-particle wavefunction with infinitely many functional degrees
of freedom. This $\mathrm{SO}(2)$ acts in the
plane perpendicular to the total wall momentum within the $(t,z,x)$ slice.
By \eqref{eq:spin-projector}, the idempotent boundary projection preserves
positivity. In other words, we can test \eqref{eq:common-dq-smeared-amplitude}
before $\Pi_h$ projection against positive absorptive kernels expressed
in the flat-space basis, requiring only the wall Poincar\'e symmetry
$\mathrm{ISO}(1,2)$. This group comprises translations and
$\mathrm{SO}(1,2)$ transformations along the $(t,z,x)$ slice.
Applying $\Pi_h$ gives the corresponding boundary-compatible half-space
kernels, preserving this symmetry and their positivity.

Accordingly, the positivity tests have two parts. The first uses the
$\mathrm{ISO}(1,3)$-invariant subset and requires nonnegative traces for
physical $s$- and $u$-channel resonances of mass $M\geq\Lambda_{\rm EFT}$
and arbitrary spin $J$. We scan their mass and spin numerically.
These tests are analogous to checking the momentum- and
angular-momentum-diagonal sectors of the state-density kernel underlying
the functional. The second uses general $\mathrm{ISO}(1,2)$-invariant
absorptive kernels. On either the incoming or outgoing side, the normal
two-particle wavefunction is a single function of two independent normal
momenta, represented by $f(k_1,k_2)$ on the incoming side, with infinitely
many degrees of freedom. We sample its shape and, for each fixed sample, scan the
conserved invariant $s_\parallel$ or $u_\parallel$ and the
$\mathrm{SO}(2)$ harmonic label $\ell$.
At the scanned widths $\sigma/q_{\max}\leq10^{-2}$, the full finite-width
responses are nonnegative. We also calculate the leading term as
$\sigma/q_{\max}\to0$ and prove its positivity analytically.
Together, these nonnegative spectral responses and leading-order
positivity support using the chosen small-width functional for the
half-space bootstrap and the resulting EFT bound.

For the conserving tests, directly insert the flat-space unit
discontinuity densities of the physical $s$ and $u$ channels,
expressed in plane-wave states. In the respective spectral integrals,
this replaces the amplitude in \eqref{eq:common-dq-smeared-amplitude} by
\begin{equation}
 \M_h(s_\parallel,t_\parallel;k_i)
 \xmapsto{\ \mathrm{spectral\ test}\ }
 \begin{cases}
  \pi\delta(k_0)\,\delta(s_h-M_s^2)
       P_J(1+2t_h/M_s^2),\\[2pt]
  \pi\delta(k_0)\,\delta(u_h-M_u^2)
       P_J(1+2t_h/M_u^2).
 \end{cases}
 \label{eq:common-dq-spectral-probes}
\end{equation}
\makeatletter%
\begingroup%
\protected@edef\@currentlabel{\theequation}%
\protected@edef\cref@currentlabel{[equation][\arabic{equation}][\arabic{section}]\theequation}%
\label{eq:relative-su-response-definitions}%
\label{eq:d3-unit-resonances}%
\endgroup%
\makeatother%
These are the physical $s$- and $u$-channel resonances. Their responses
to the same \eqref{eq:common-dq-smeared-amplitude} are denoted by
$\lambda_s(M_s,J)$ and $\lambda_u(M_u,J)$.
The spectral delta functions are integrated out of these responses.
For the numerical tests, set the packet centers of the pairs $1,3$ and
$2,4$ to $y_1=y_2=y_0$. At fixed channel mass, the spectral density
depends only on $s_h,t_h$, so we can integrate
$k_0,k_s,k_u$ first. Within the Gaussian tail accuracy already adopted,
the normal integrals reduce to
\begin{equation}
 W_\sigma(k_t)\equiv
 \int_{-10\sigma}^{10\sigma}\dd k_0\,\dd k_s\,\dd k_u\,
 W(k_i)\,\pi\delta(k_0)
 =2\pi\e^{-k_t^2/(2\sigma^2)}\bigl(1-\Ocal(\e^{-50})\bigr).
 \label{eq:d3-Gaussian-factorization}
\end{equation}
We neglect the exponential correction in this expression.
For the $s$ probe, integration over the dispersion variable $\dd s_h$
sets $s_h=M_s^2$ and $u_h=-M_s^2+q_\parallel^2+k_t^2$.
For the $u$ probe, it sets $u_h=M_u^2$ and
$s_h=-M_u^2+q_\parallel^2+k_t^2$.
The sign of the crossed-cut discontinuity compensates the Jacobian
$\dd s_h=-\dd u_h$.
The two finite-width responses are
\begin{equation}
 \begin{aligned}
 \lambda_s(M_s,J)
 &=\int_0^{q_{\max}}\dd q_\parallel\,\phi(-q_\parallel^2)
   \int_{-k_{\max}}^{k_{\max}}\dd k_t\,W_\sigma(k_t)
   \frac{P_J\!\left(1-2(q_\parallel^2+k_t^2)/M_s^2\right)}
        {M_s^4(M_s^2-q_\parallel^2-k_t^2)},\\[3pt]
 \lambda_u(M_u,J)
 &=\int_0^{q_{\max}}\dd q_\parallel\,\phi(-q_\parallel^2)
   \int_{-k_{\max}}^{k_{\max}}\dd k_t\,W_\sigma(k_t)
   \frac{P_J\!\left(1-2(q_\parallel^2+k_t^2)/M_u^2\right)}
        {M_u^2(M_u^2-q_\parallel^2-k_t^2)^2}.
 \end{aligned}
 \label{eq:common-dq-resonance-responses}
\end{equation}
\makeatletter%
\begingroup%
\protected@edef\@currentlabel{\theequation}%
\protected@edef\cref@currentlabel{[equation][\arabic{equation}][\arabic{section}]\theequation}%
\label{eq:relative-su-conserved-exact-pairings}%
\label{eq:d3-check-II-s}%
\label{eq:d3-check-II-u}%
\endgroup%
\makeatother%
These formulas retain the dependence of the absorptive part,
the dispersion denominator and the integration support on
$q_\parallel,k_t$, and do not introduce the unphysical square-root
cut in \eqref{eq:planar-seed-sixfold-nlo}. The different powers in the
two denominators come from the same $s_h^2u_h$ weight evaluated on
the two physical resonances.

We next give the normal-momentum support condition for
\eqref{eq:common-dq-smeared-amplitude}, including both conserving and
nonconserving half-space absorptive spectra. A resonance of mass $M$
contributes at $s_\parallel=M^2+k_m^2$, where the internal normal
momentum $k_m$ is independent of the external normal momenta.
Thus a resonance of given mass $M$ contributes a branch cut on
$s_\parallel\in[M^2,\infty)$.
A UV spectrum with $M_s,M_u\geq\Lambda_{\rm EFT}$ therefore produces
discontinuities starting at
$s_\parallel,u_\parallel\geq\Lambda_{\rm EFT}^2$.
For this spectrum, the transfer integral has $t_\parallel\leq0$ and
does not reach the $t$-channel singularities.
Let the maximum of $|k_{0,s,t,u}|$ on the retained normal support
be $k_{\max}$.
In the complex $s_h$ plane, the poles of the
dispersion denominator lie at $s_h=0$ and
$s_h=q_\parallel^2+k_0^2+k_t^2$.
The right UV cut from the $s$ channel starts at
$\Lambda_{\rm EFT}^2-k_s^2$, while the left UV cut from the $u$ channel
starts at $q_\parallel^2+k_0^2+k_t^2+k_u^2-\Lambda_{\rm EFT}^2$
and extends towards negative infinity.
Requiring the integration domain of
\eqref{eq:common-dq-smeared-amplitude} to keep these UV cuts away
from the poles of $s_h^2u_h$ for every
$0\leq q_\parallel\leq q_{\max}$ gives
\begin{equation}
 q_{\max}^2+3k_{\max}^2<\Lambda_{\rm EFT}^2.
 \label{eq:common-dq-compact-support}
\end{equation}
\makeatletter%
\begingroup%
\protected@edef\@currentlabel{\theequation}%
\protected@edef\cref@currentlabel{[equation][\arabic{equation}][\arabic{section}]\theequation}%
\label{eq:relative-s-u-real-support}%
\endgroup%
\makeatother%
To make the resulting EFT bound constrain the UV spectrum as
strongly as possible, we want $q_{\max}$ close to
$\Lambda_{\rm EFT}$, but finite normal support prevents exact
coincidence. This is one of the support conditions for numerically
testing \eqref{eq:common-dq-smeared-amplitude} on general half-space
absorption, and is not restricted to conserving probes.

The absorptive test spectrum in \eqref{eq:common-dq-spectral-probes} selects
only the $k_0=0$ slice, so the complete Gaussian normal integral
supplies a common factor proportional to $\sigma$ in the raw
responses of \eqref{eq:common-dq-resonance-responses}.
The transfer measure also changes its dimension relative to flat
space, replacing $2q_\parallel\dd q_\parallel$ by
$\dd q_\parallel$. To remove this common normal factor and restore
the mass dimension of the flat-space functional, we define
\begin{equation}
 \widehat\lambda_{s/u}(M_{s/u},J)
 =\frac{q_{\max}}{2\pi\sqrt{2\pi}\sigma}\lambda_{s/u}(M_{s/u},J).
 \label{eq:common-dq-normalization}
\end{equation}
We will see in Sec.~\ref{sec:general-halfspace-dispersion} that
applying this normalization factor to the half-space smearing in
\eqref{eq:common-dq-smeared-amplitude} gives an EFT contribution
in the smeared dispersion relation of the same order as the
flat-space functional, making the resulting EFT bounds easier to compare.

Before numerically checking the responses at finite widths, we first
examine whether the leading term of
\eqref{eq:common-dq-resonance-responses} is strictly positive when
$\sigma^2\ll q_{\max}^2,\Lambda_{\rm EFT}^2-q_{\max}^2$,
with $q_{\max}$ close to
$\Lambda_{\rm EFT}$. Applying the normalization in
\eqref{eq:common-dq-normalization} gives the leading responses
\begin{equation}
 \begin{aligned}
 \widehat\lambda_s^{(0)}(M_s,J)
 &=q_{\max}\int_0^{q_{\max}}\dd q_\parallel\,
 \frac{\phi(-q_\parallel^2)P_J(1-2q_\parallel^2/M_s^2)}
 {M_s^4(M_s^2-q_\parallel^2)},\\[3pt]
 \widehat\lambda_u^{(0)}(M_u,J)
 &=q_{\max}\int_0^{q_{\max}}\dd q_\parallel\,
 \frac{\phi(-q_\parallel^2)P_J(1-2q_\parallel^2/M_u^2)}
 {M_u^2(M_u^2-q_\parallel^2)^2}.
 \end{aligned}
 \label{eq:d3-unshaken-baseline}
\end{equation}
These are precisely the flat-space $s$- and $u$-channel partial-wave
responses of the dispersion relation smeared with
$q_{\max}\dd q_\parallel\,\phi(-q_\parallel^2)$. For the chosen profile,
their strict positivity over the physical mass range can be established
by the flat-space partial-wave positivity criterion
\cite{Caron-Huot:2021rmr}. Thus, as $\sigma/q_{\max}\to0$, the two UV
responses approach their respective flat-space limits.

\begin{figure}[H]
\centering
\includegraphics[width=\linewidth]{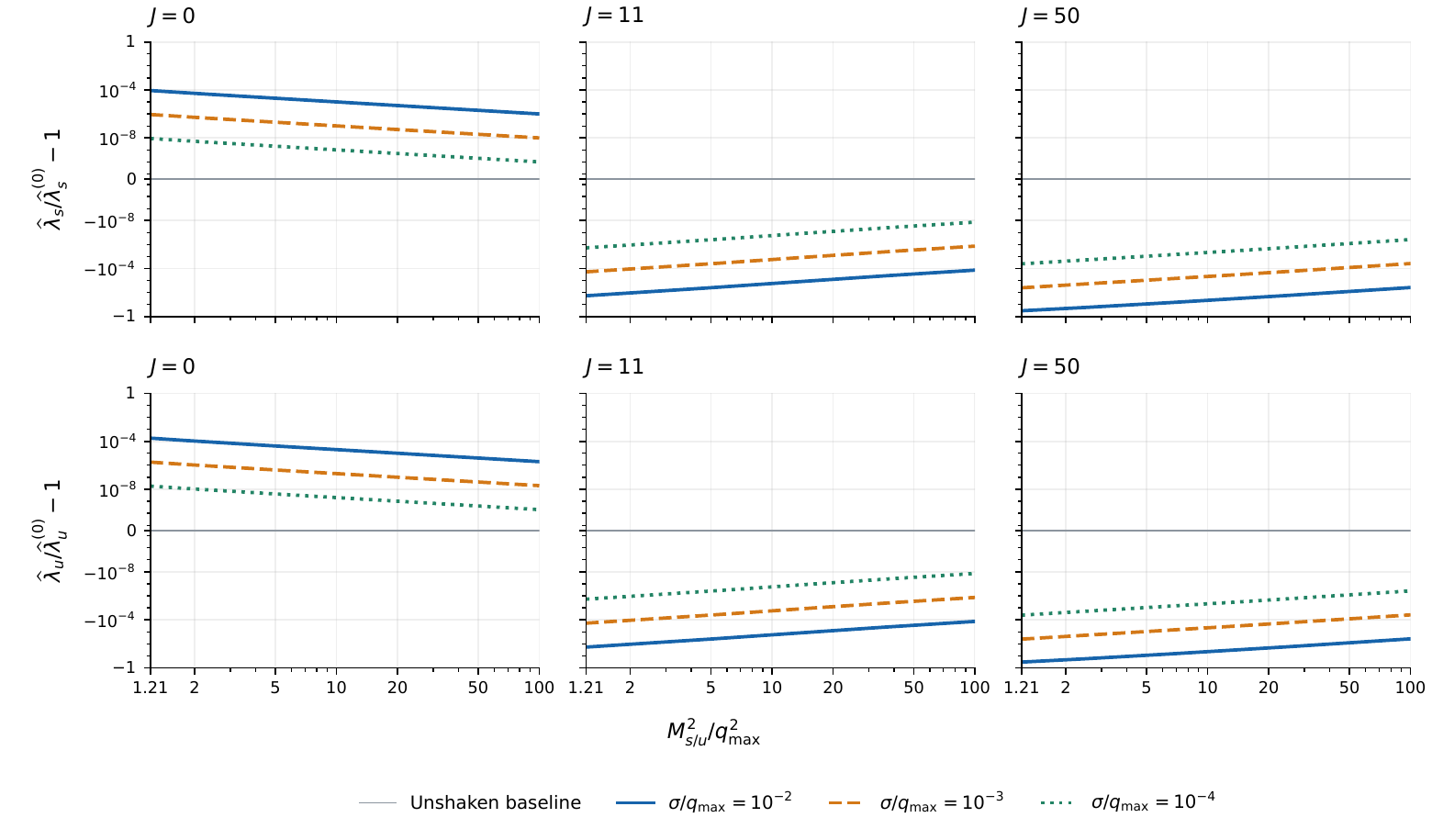}
\caption{Relative changes of the two physical-channel responses under one smearing prescription. Columns show $J=0,11,50$, with the $s$ channel above and the $u$ channel below. Symmetric-logarithmic tick labels give the relative changes.}
\label{fig:d3-construction2_s}
\label{fig:d3-construction2_u}
\label{fig:common-dq-relative-merged}
\end{figure}

However, in this limit the $t$-channel graviton exchange contributes
$\dd q_\parallel/q_\parallel^2$ to the dispersion relation, producing a
$\lambda_{\rm IR}^{-1}$ divergence, stronger than a logarithm.
The actual half-space bootstrap therefore retains a finite
$\sigma/q_{\max}$. A smaller ratio brings the UV responses closer to
their positive reference values, whereas a larger allowed width
improves the IR behavior. We choose the maximum $k_{\max}/q_{\max}$
to be $0.1$ and set $k_{\max}=10\sigma$, so that the omitted Gaussian
tails are exponentially suppressed and $\sigma/q_{\max}\leq10^{-2}$.
To satisfy the common support condition
\eqref{eq:common-dq-compact-support}, we set
$\Lambda_{\rm EFT}=1.1q_{\max}$ and test three Gaussian widths,
\begin{equation}
 \begin{gathered}
 M_s,M_u\geq\Lambda_{\rm EFT}=1.1q_{\max}
 >\sqrt{q_{\max}^2+300\sigma^2},\\[2pt]
 \frac{\sigma}{q_{\max}}\in\{10^{-2},10^{-3},10^{-4}\}.
 \end{gathered}
 \label{eq:d3-fixed-mass-widths}
\end{equation}
Substituting the three widths and parameter choices above into
\eqref{eq:common-dq-resonance-responses}, we calculate the $s$- and
$u$-channel responses of momentum-conserving absorptive parts.
We consider general $\phi_1\phi'_2\to\phi_3\phi'_4$ scattering,
allowing any integer spin in both channels. If $\phi$ and $\phi'$
are identical, only even spins contribute.
Figure~\ref{fig:common-dq-relative-merged} compares the relative changes
at the three widths against the zero-width responses in
\eqref{eq:d3-unshaken-baseline}. These changes tend to zero as
$\sigma/q_{\max}$ decreases, so the responses approach strictly positive
values and a small ratio is a safe choice for positivity.
The relative corrections are nevertheless not uniformly of order
$\sigma^2/\Lambda_{\rm EFT}^2$ for all spins $J$.
For example, at $J=50$, $M_s/q_{\max}=1.1$ and
$\sigma/q_{\max}=10^{-2}$,
$\widehat\lambda_s/\widehat\lambda_s^{(0)}\simeq0.67$.
As \eqref{eq:common-dq-large-spin} below shows, the finite-width
responses decay faster with angular momentum than their zero-width values.

\begin{figure}[H]
\centering
\includegraphics[width=\linewidth]{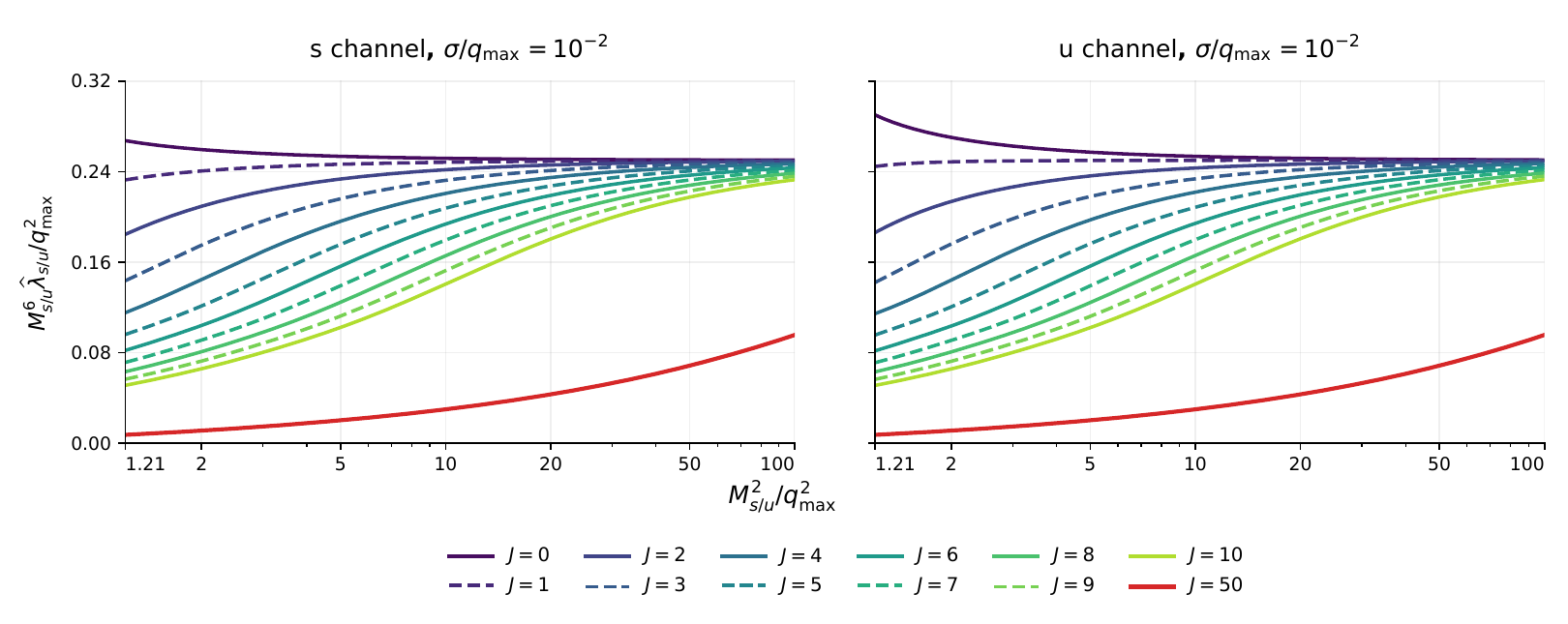}
\caption{Mass scans testing the positivity of the $s$- and $u$-channel responses $\widehat\lambda_s$ and $\widehat\lambda_u$ at fixed angular momenta.}
\label{fig:relative-pullback-weighted-responses}
\label{fig:d3-construction2_s_responses}
\label{fig:d3-construction2_u_responses}
\end{figure}
\begin{figure}[H]
\centering

\includegraphics[width=\linewidth]{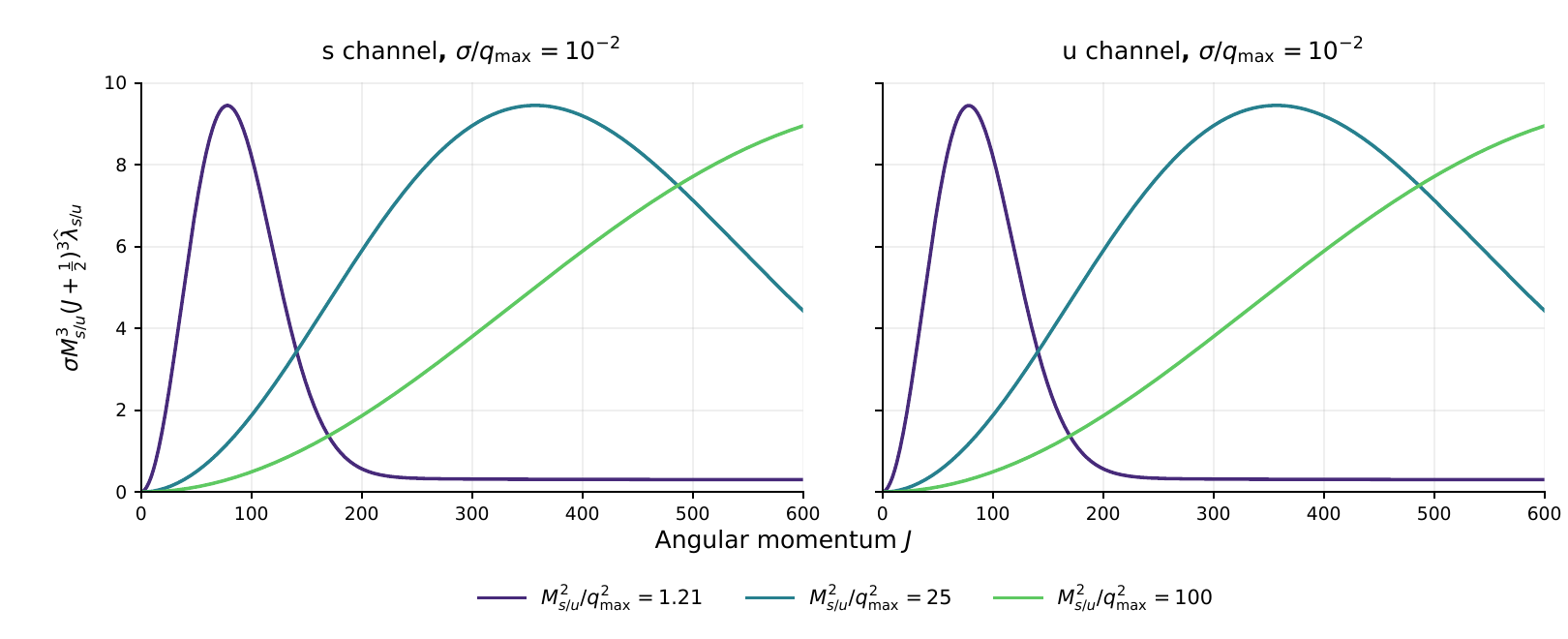}
\caption{Spin scans testing the positivity of the $s$- and $u$-channel responses $\widehat\lambda_s$ and $\widehat\lambda_u$ at fixed resonance masses.}
\label{fig:common-dq-fixed-mass-spin}
\end{figure}

Figure~\ref{fig:relative-pullback-weighted-responses} shows mass scans
of the finite-width $s$- and $u$-channel responses at
$\sigma/q_{\max}=10^{-2}$ for a series of fixed angular momenta.
For plotting, we multiply $\widehat\lambda$ by $M^6/q_{\max}^2$
to remove the overall mass decay and make the response dimensionless. At each of the displayed
spins, both responses remain
positive and decay as $M_{s/u}^{-6}$ at large mass, while the weighted
responses tend to $1/4$.
A complementary test fixes the resonance mass and varies integer
spin, including both even and odd spins. We take
$M_{s/u}^2/q_{\max}^2=1.21,25,100$ and evaluate $J=0,\ldots,600$ at the
same width $\sigma/q_{\max}=10^{-2}$, as shown in
Fig.~\ref{fig:common-dq-fixed-mass-spin}.

The large-spin behavior follows from the local forward-endpoint
behavior of \eqref{eq:common-dq-resonance-responses} and
\eqref{eq:d3-unshaken-baseline} at $(q_\parallel,k_t)\to(0,0)$,
using the endpoint expansion of Ref.~\cite{Dong:2025dpy}.
For fixed $M_{s/u}$, the two large-spin asymptotes are
\begin{equation}
 \begin{aligned}
 \widehat\lambda_{s/u}^{(0)}(M_{s/u},J)
 &\underset{J\to\infty}{\sim}
  \frac{q_{\max}}{2M_{s/u}^5(J+\tfrac12)},\\[2pt]
 \widehat\lambda_{s/u}(M_{s/u},J)
 &\underset{J\to\infty}{\sim}
  \frac{3}{4\sqrt{2\pi}\sigma M_{s/u}^3(J+\tfrac12)^3}.
 \end{aligned}
 \label{eq:common-dq-large-spin}
\end{equation}
The first line comes from the one-dimensional forward endpoint of
the zero-width $\dd q_\parallel$ integral. The second comes from
the two-dimensional forward endpoint of the finite-width
$\dd q_\parallel\dd k_t$ integral. The Gaussian is smooth there,
and the linear falloff of the selected weight gives the first
nonzero contribution after transverse angular integration.
The finite-width asymptote requires
$(J+\tfrac12)\sigma/M_{s/u}\gg1$, which becomes $J\gg110$ for the
lightest resonance in Fig.~\ref{fig:common-dq-fixed-mass-spin},
with $M_{s/u}/q_{\max}=1.1$ and $\sigma/q_{\max}=10^{-2}$.
For plotting, we multiply $\widehat\lambda$ by
$\sigma M^3(J+\tfrac12)^3$, so each weighted response tends to
the positive constant $3/(4\sqrt{2\pi})$ in this asymptotic regime.
Thus Fig.~\ref{fig:relative-pullback-weighted-responses} scans resonance
mass at fixed spin, while Fig.~\ref{fig:common-dq-fixed-mass-spin}
scans integer spin at fixed mass. Both scans give positive
$s$- and $u$-channel responses and approach positive asymptotic limits,
supporting the positivity of \eqref{eq:common-dq-resonance-responses}
on momentum-conserving absorptive parts.

A general half-space absorptive operator need only respect
$\mathrm{ISO}(1,2)$, rather than $\mathrm{ISO}(1,3)$, and therefore
also depends on the four external normal momenta. A PSD spectral
decomposition expresses this dependence through two-particle
profiles $f(k_1,k_2)$, using the same profile on the outgoing side
with complex conjugation. At fixed conserved timelike momentum
$P_{s\parallel}$ in the wall-parallel $(t,z,x)$ space, the relative
directions transform under $\mathrm{SO}(2)$. The Legendre polynomial
of the four-dimensional angular decomposition is then replaced by
the first-kind Chebyshev polynomial
$T_\ell(\cos\theta)=\cos(\ell\theta)$, combining the $\ell$ and
$-\ell$ modes. The corresponding positive absorptive blocks are
\begin{equation}
 \M_h(s_\parallel,t_\parallel;k_i)
 \xmapsto{\ \mathrm{spectral\ test}\ }
 \frac{\pi}{q_{\max}}
 \begin{cases}
  \delta(s_\parallel-\mu^2)T_\ell(\cos\theta_s)\,
   f(k_1,k_2)f(-k_3,-k_4)^*,\\[2pt]
  \delta(u_\parallel-\mu^2)T_\ell(\cos\theta_u)\,
   f(k_1,k_4)f(-k_3,-k_2)^*.
 \end{cases}
 \label{eq:common-dq-halfspace-probes}
\end{equation}
Here $\mu^2$ is the wall-parallel spectral energy invariant, fixing
$s_\parallel$ or $u_\parallel$. The profile $f$ is dimensionless,
and the factor $q_{\max}^{-1}$ gives the unit spectral test the same
mass dimension as \eqref{eq:common-dq-spectral-probes}.
The angles $\theta_s,\theta_u$ are those between the incoming and
outgoing pair relative momenta in the wall-parallel $(t,z,x)$ space.

Choose the analytic odd Gaussian profile
\begin{equation}
 f(k_1,k_2)=
 \Bigl(\e^{-\frac{(k_1-\sigma/2)^2}{2\tau^2}}
       -\e^{-\frac{(k_1+\sigma/2)^2}{2\tau^2}}\Bigr)
 \Bigl(\e^{-\frac{(k_2-\sigma/2)^2}{2\tau^2}}
       -\e^{-\frac{(k_2+\sigma/2)^2}{2\tau^2}}\Bigr).
 \label{eq:common-dq-halfspace-normal-basis}
\end{equation}
We take $\tau=0.05\sigma$, without an additional prefactor in $f$.
The profile is odd in both $k_1$ and $k_2$. Although written in the
flat-space plane-wave representation, it already lies in the image of
$\Pi_h$ on each leg and hence defines a boundary-compatible half-space
absorptive block. At fixed $k_i$ and $q_\parallel$, completing the
spectral-energy delta integral with $\dd s_h$ gives the finite-width
responses. The angular variables $\cos\theta_{s/u}$ follow from the
exact on-shell map \eqref{eq:planar-seed-preserved-angle}.
We use the same normalization
$q_{\max}/(2\pi\sqrt{2\pi}\sigma)$ as in
\eqref{eq:common-dq-normalization}. Its product with the unit spectral
factor $\pi/q_{\max}$ in \eqref{eq:common-dq-halfspace-probes} gives
$1/(2\sqrt{2\pi}\sigma)$, so the two responses are
\begin{equation}
 \begin{aligned}
 \widehat\lambda_s(s_\parallel,\ell;f)
 &=\int_0^{q_{\max}}\dd q_\parallel\,\phi(-q_\parallel^2)
 \prod_{i=1}^{4}\int\dd k_i\,
 \frac{W(k_i)f(k_1,k_2)f(-k_3,-k_4)^*T_\ell(\cos\theta_s)}
 {-2\sqrt{2\pi}\sigma s_h^2u_h},\\[3pt]
 \widehat\lambda_u(u_\parallel,\ell;f)
 &=\int_0^{q_{\max}}\dd q_\parallel\,\phi(-q_\parallel^2)
 \prod_{i=1}^{4}\int\dd k_i\,
 \frac{W(k_i)f(k_1,k_4)f(-k_3,-k_2)^*T_\ell(\cos\theta_u)}
 {2\sqrt{2\pi}\sigma s_h^2u_h}.
 \end{aligned}
 \label{eq:common-dq-halfspace-response-matrix}
\end{equation}
Here $s_h=s_\parallel-k_s^2$ and
$u_h=\sum_i k_i^2-s_\parallel-t_\parallel-k_u^2$.
Each normal-momentum integral runs over the Gaussian band
$[-10\sigma,10\sigma]$, independently of $q_\parallel$.
The parameter choices in \eqref{eq:d3-fixed-mass-widths} do not
touch the square-root singularities of
\eqref{eq:planar-seed-preserved-angle}.
The crossed functions follow from
$s_\parallel\leftrightarrow u_\parallel$ and $k_2\leftrightarrow k_4$,
so the $u$-channel integrand uses $(u_\parallel,t_\parallel;k_i)$.

The $s$- and $u$-channel discontinuities begin at
$s_\parallel,u_\parallel\geq\Lambda_{\rm EFT}^2>q_{\max}^2$.
As $\sigma/q_{\max}\to0$, \eqref{eq:planar-seed-preserved-angle}
gives $\cos\theta_s\to1-2q_\parallel^2/s_\parallel$,
and the $u$ channel is analogous. Denoting the leading responses
by $\widehat\lambda_{s/u}^{(0)}$, we have
\begin{equation}
 \begin{aligned}
 \widehat\lambda_s^{(0)}(s_\parallel,\ell;f)
 &=\int_0^{q_{\max}}\dd q_\parallel\,
 \frac{\phi(-q_\parallel^2)T_\ell(1-2q_\parallel^2/s_\parallel)}
 {2\sqrt{2\pi}\sigma s_\parallel^2(s_\parallel-q_\parallel^2)}
 \int\prod_{i=1}^{4}\dd k_i\,W(k_i)f(k_1,k_2)f(-k_3,-k_4)^*,\\[3pt]
 \widehat\lambda_u^{(0)}(u_\parallel,\ell;f)
 &=\int_0^{q_{\max}}\dd q_\parallel\,
 \frac{\phi(-q_\parallel^2)T_\ell(1-2q_\parallel^2/u_\parallel)}
 {2\sqrt{2\pi}\sigma u_\parallel(u_\parallel-q_\parallel^2)^2}
 \int\prod_{i=1}^{4}\dd k_i\,W(k_i)f(k_1,k_4)f(-k_3,-k_2)^*.
 \end{aligned}
 \label{eq:common-dq-halfspace-leading-limit}
\end{equation}
The weight $W(k_i)$ in \eqref{eq:common-su-rank-one-weight} is a
PSD normal-momentum kernel in both channel pairings. At leading order
in $\sigma/q_{\max}$, the normal integrations in
\eqref{eq:common-dq-halfspace-leading-limit} separate from the transfer
integral. Each is the squared modulus of the overlap between an
arbitrary two-particle normal profile $f$ and the chosen correlated two-particle
wavefunction, and is therefore nonnegative. The remaining transfer
integral gives the $\mathrm{SO}(2)$ angular coefficients. The chosen
weight produces a convex decreasing angular density whose value and
first derivative vanish at its cutoff, giving nonnegative
$T_\ell$ coefficients. Combining these two facts,
\eqref{eq:common-dq-halfspace-leading-limit} is PSD. For a fixed profile
with nonzero normal overlap, the leading response of each fixed angular
mode is strictly positive, analogous to the comparison in
Fig.~\ref{fig:common-dq-relative-merged}.

\begin{figure}[H]
\centering
\includegraphics[width=\linewidth]{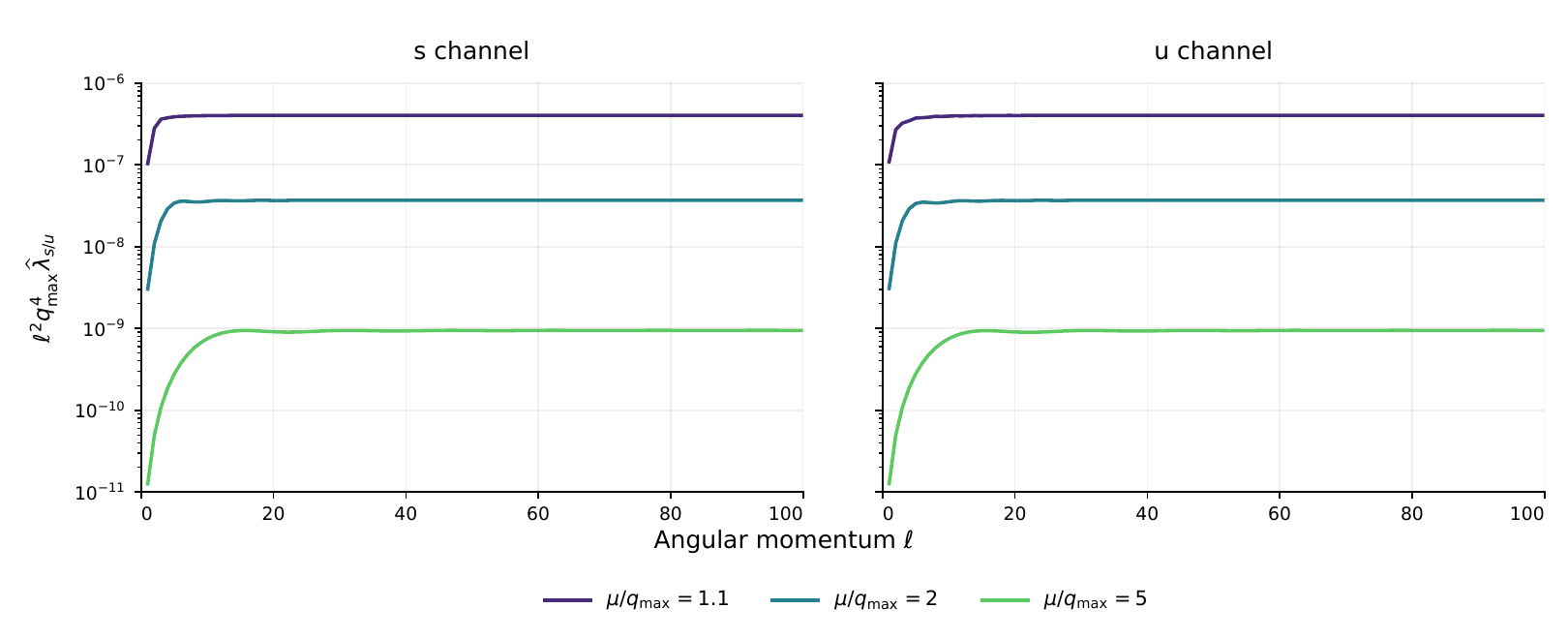}
\caption{Angular-momentum scans at fixed wall-parallel spectral energies for the same normal profile. Both panels use logarithmic vertical axes.}
\label{fig:common-dq-halfspace-normal-modes}
\end{figure}

We fix the wall-parallel spectral energy, $s_\parallel=\mu^2$ or
$u_\parallel=\mu^2$, and scan angular momentum with the same normal
profile $f$. Figure~\ref{fig:common-dq-halfspace-normal-modes} shows the
finite-width responses of \eqref{eq:common-dq-halfspace-response-matrix}
at $\mu/q_{\max}=1.1,2,5$. Both channels are positive at every integer
mode $0\leq\ell\leq1000$. In the scanned large-angular-momentum range,
the responses follow a positive $\ell^{-2}$ behavior.
We therefore plot $\ell^2q_{\max}^4\widehat\lambda$, retaining the dimensionless
response while making its positive plateau visible.

\begin{figure}[H]
\centering
\includegraphics[width=\linewidth]{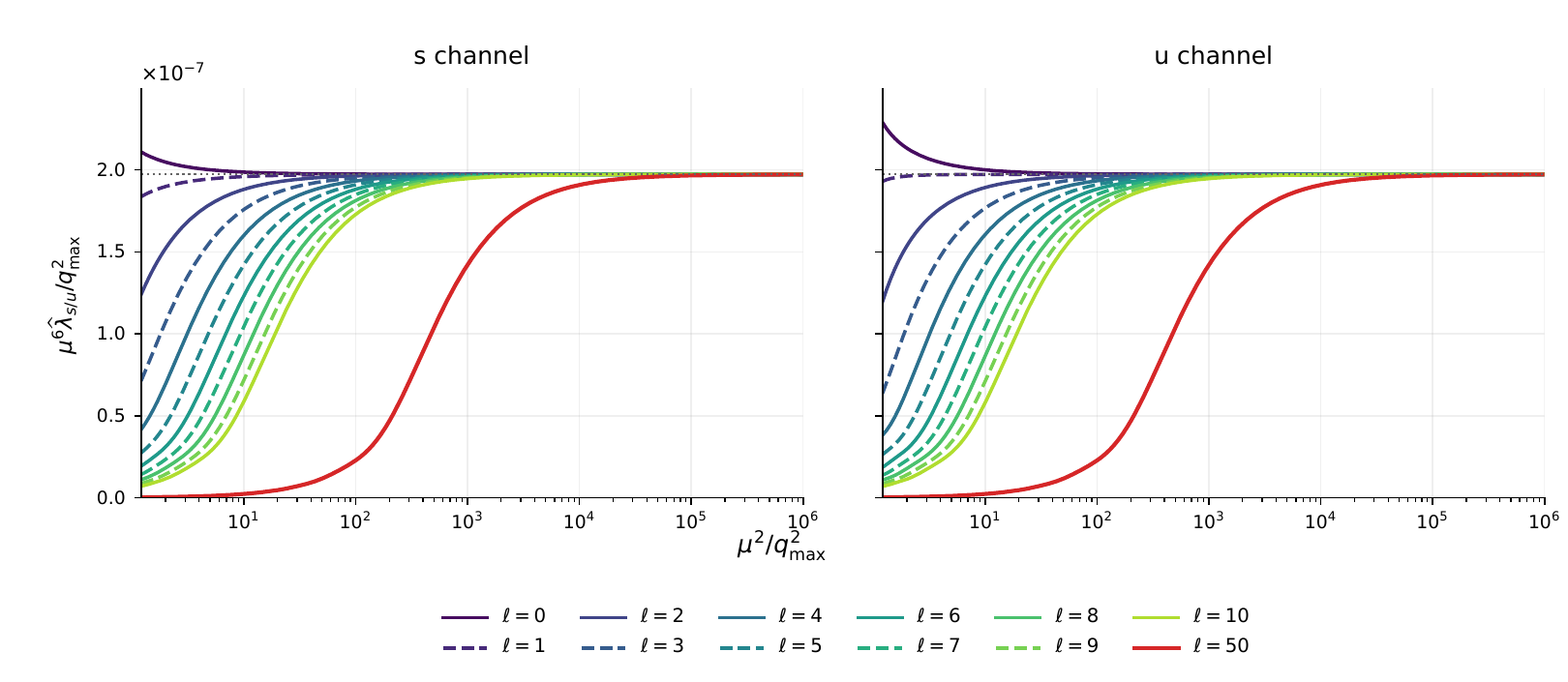}
\caption{Wall-parallel spectral-energy scans at fixed angular momenta for the same normal profile. The gray dotted line is the positive large-energy leading limit.}
\label{fig:common-dq-halfspace-mass-scan}
\end{figure}

To compare with the large-mass behavior in
Fig.~\ref{fig:relative-pullback-weighted-responses}, we also fix
$\ell=0,1,\ldots,10,50$ and scan the wall-parallel spectral energy
starting at $\mu^2/q_{\max}^2=1.21$.
Figure~\ref{fig:common-dq-halfspace-mass-scan} plots
$\mu^6\widehat\lambda/q_{\max}^2$, removing the overall large-energy
decay. Both channels remain positive and approach a positive common
plateau. At fixed angular momentum and normal profile, increasing
$\mu$ suppresses the relative normal-kinematic dependence in
\eqref{eq:common-dq-halfspace-response-matrix}, bringing it closer to
the leading expression \eqref{eq:common-dq-halfspace-leading-limit}.
The two scans thus test angular momentum at fixed spectral energy and
spectral energy at fixed angular momentum, including their positive
large-parameter behavior in the displayed ranges.

We additionally vary the shape of the normal profile $f(k_1,k_2)$.
For the Gaussian products in \eqref{eq:common-dq-halfspace-normal-basis},
we replace the centers $\pm\sigma/2$ by $\pm k_{\rm c}$ and sample
\begin{equation}
 \frac{k_{\rm c}}{\sigma}\in\{0.4,0.6,0.8\},\qquad
 \frac{\tau}{\sigma}\in\{0.025,0.05,0.1\},\qquad
 k_{\rm c}+10\tau<10\sigma.
 \label{eq:common-dq-profile-samples}
\end{equation}
The last condition keeps the ten-width neighborhoods of both
Gaussian centers inside the normal integration band
$[-10\sigma,10\sigma]$, with the profile suppressed by $\e^{-50}$
at ten widths from each center. We also replace the Gaussians by the
$\mathrm{C}^{\infty}$ compact shape of \eqref{eq:compact-bump-tail}, using the
same three centers and half-width $0.1\sigma$, with odd reflection on
each leg. For every sampled profile, both channels are positive at
every integer mode $0\leq\ell\leq1000$ and throughout the sampled
spectral-energy range of Fig.~\ref{fig:common-dq-halfspace-mass-scan}.
The sampled tails retain positive $\ell^{-2}$ and $\mu^{-6}$ behavior
in these ranges. These checks provide numerical support for a
half-space bootstrap.

\subsection{Low-energy amplitude and EFT bound}
\label{sec:sum-rules-examples}
\label{sec:general-halfspace-dispersion}
\label{sec:construction-two-example}
\label{sec:gravity-tree-functional}

We calculate the low-energy approximation to the 2SDR for four
massless scalars defined by
\eqref{eq:common-dq-smeared-amplitude}. The smearing lets us integrate
the wall-parallel smearing profile $\phi(-q_\parallel^2)$ and the
normal-momentum distribution $W(k_i)$ separately, greatly simplifying
the four-point EFT contact and graviton-exchange diagrams. For
$t$-channel graviton exchange, retaining momenta in $(t,z,x)$ and
positions in $y$ gives the half-space amplitude by convolving the
normal-direction graviton Green function with the wave-packet vertex
overlaps. This parallels an AdS$_5$/CFT$_4$ exchange calculation,
which retains momenta on the Minkowski slices while integrating
radial position-space mode overlaps. It is simpler than the full
momentum-space calculation because the wall breaks normal-momentum
conservation and removes the simple flat-space $1/t$ propagator form.
We thereby obtain the low-energy amplitude of the finite-width
correlated wave packet specified by the functional and its EFT bound.

Two corrections to this bound arise on its low-energy side. The first
comes from higher gravitational loops, with relative suppression
$G_N\Lambda_{\rm EFT}^2$. The Wilson coefficients contain the matching
contributions of heavy massive particles, including their loops,
but massless gravitational loops remain, with the soft part acting
as the multiplicative Wilson-line dressing of
Sec.~\ref{sec:halfspace-beam-completion}. As explained there,
wave-packet spreading reintroduces IR divergences in the half-space
loop diagrams even though the tree diagrams are strictly IR finite.
The DN strip cuts off this effect. We choose its width $Y_0$, subject
to \eqref{eq:beam-accuracy-conditions}, to keep the relative correction
at $\mathcal{O}(G_N\sigma^2\log^2(Y_0q_{\max}))\lesssim G_N\Lambda_{\rm EFT}^2$.
For $M_{\rm Pl}\gg\Lambda_{\rm EFT}$, neglecting this higher-order effect
allows us to bootstrap the IR-finite half-space tree amplitude
directly. For greater precision, the higher-loop gravitational
amplitudes can instead be calculated in the DN strip and included
in the bootstrap. Its absorptive spectrum still consists of
$\mathrm{ISO}(1,2)$ blocks, to which the same two-channel positivity
criterion of Sec.~\ref{app:four-scalar-resonance-checks} applies.

The second correction comes from the wall reflection's effect on
local three- and four-point EFT vertices. As the tree calculation
will show, the effective IR length is the scattering center's
distance $y_0$ from the Dirichlet wall. Reducing $y_0$ suppresses the
gravitational contribution and strengthens the bound, but local
EFT accuracy and external-state normalization limit how close the
interaction can approach the wall. The Dirichlet condition comes
from the $R\to0$ matching limit, where the exterior curvature diverges
and the low-energy expansion of local vertices fails. Reflecting
the external wavefunctions alone therefore cannot predict those
vertices from the flat bulk EFT. In addition, when a packet center
approaches the wall, destructive interference between its direct
and reflected parts suppresses the unrenormalized external
wavefunction, and its half-space norm
\eqref{eq:packet-norms-halfspace} departs from the far-wall value.
Localizing the vertex overlaps inside the flat region leaves only
exponentially suppressed sensitivity of $\mathcal{O}(\e^{-\sigma^2y_0^2})$,
with operator-dependent coefficients. These tails must remain in
the bound's error, so gravity cannot be suppressed arbitrarily at
fixed local accuracy. Long-range propagation is governed by the
boundary Green function \eqref{eq:flat-limit-mixed-green}, obtained
by solving the field equations in the metric-induced $R\to0$
matching limit, and can be retained in the IR-finite tree exchange.

Spinning extensions require the appropriate helicity kernels and
positivity tests. For four-graviton elastic scattering, the
$s$-channel absorptive partial waves carry $P_J$, while the crossed
$u$-channel ones carry $d^J_{44}$. An $\mathrm{ISO}(1,2)$-covariant
polarization prescription fixes $s_\parallel,t_\parallel,k_s,k_t,k_u$,
uses \eqref{eq:conserved-seed-polarizations} on the $k_0=0$ slice,
then determines the momenta for nonzero $k_0$ and transports the
polarizations with \eqref{eq:spin-standard-null-boost}. A functional
defined as in \eqref{eq:common-dq-smeared-amplitude} must subsequently
be tested on general $\mathrm{ISO}(1,2)$ absorptive blocks.
Here we discuss the possibility of a spinning bootstrap without
constructing a specific positive four-graviton wave packet.
A weak bound can be studied in the eikonal regime of
Sec.~\ref{sec:transverse-modes}. When
$M_{s/u}^2\gg q_{\max}^2,\sigma^2$, the fixed-helicity large-spin
forward limits of $P_J$ and $d^J_{44}$ share the same Bessel $J_0$
limit, and the physical polarizations are described by their
transverse components. A stronger bound requires moving $q_{\max}$
toward the lightest $s$- and $u$-channel resonance masses, adjusting
the smearing functional and testing its positivity. These
helicity-dependent tests and optimization are left to future work.

Applying \eqref{eq:common-dq-smeared-amplitude} with the normalization
of \eqref{eq:common-dq-normalization}, the low-energy smeared dispersion moment of the
2SDR~\cite{Caron-Huot:2022ugt} is
\begin{equation}
 \widehat B_{2,h}^{\rm EFT}
 =\oint_{\Gamma_{\rm EFT}}\frac{\dd s_h}{2\pi\ii}\,
   \widehat{\mathcal B}_{2,h}(s_h)
 \equiv\frac{q_{\max}}{2\pi\sqrt{2\pi}\sigma}
   \oint_{\Gamma_{\rm EFT}}\frac{\dd s_h}{2\pi\ii}\,
   \mathcal B_{2,h}(s_h).
\label{eq:general-eft-contour}
\end{equation}
The counterclockwise contour encloses the rational poles from
$1/(s_h^2u_h)$ and the low-energy singularities. The support condition
\eqref{eq:common-dq-compact-support} keeps the UV cuts outside it.
Under the complex-energy growth assumptions accompanying
\eqref{eq:halfspace-linear-real-bound}, the amplitude between
normalizable correlated wavepackets retains the linear UV bound.
The large arc in the 2SDR therefore vanishes, and contour
deformation gives the separate $s$- and $u$-cut integrals. With
both integrands evaluated on their upper rims $s_h+\ii0$,
\begin{equation}
 \widehat B_{2,h}^{\rm EFT}
 =\int_{\Lambda_{\rm EFT}^2-k_{\max}^2}^{\infty}
    \frac{\dd s_h}{\pi}\,\Im\widehat{\mathcal B}_{2,h}(s_h)
 +\int_{-\infty}^{q_{\max}^2+3k_{\max}^2-\Lambda_{\rm EFT}^2}
    \frac{\dd s_h}{\pi}\,\Im\widehat{\mathcal B}_{2,h}(s_h)\geq0.
\label{eq:channel-wise-sum-rule}
\end{equation}
The inequality uses the two-channel positivity condition of
Sec.~\ref{app:four-scalar-resonance-checks}, including the general
$\mathrm{ISO}(1,2)$ half-space absorptive blocks as well as their
normal-momentum-conserving $\mathrm{ISO}(1,3)$ subset. The following
tree calculation gives the low-energy side of this dispersion relation.

We first consider the contact term for four identical massless
scalars. In flat space, their crossing-symmetric local EFT amplitude is
\begin{equation}
\begin{aligned}
 \M_{\rm loc}(s,t)
 &=\frac{g_2}{\Lambda_{\rm EFT}^4}(s^2+t^2+u^2)
   +\frac{g_3}{\Lambda_{\rm EFT}^6}stu+\cdots\\
 &=-su\left[\frac{2g_2}{\Lambda_{\rm EFT}^4}
           -\frac{tg_3}{\Lambda_{\rm EFT}^6}+\cdots\right]
   +\frac{2t^2g_2}{\Lambda_{\rm EFT}^4}+\cdots.
\end{aligned}
\label{eq:concrete-scalar-crossing-basis}
\end{equation}
\makeatletter%
\begingroup%
\protected@edef\@currentlabel{\theequation}%
\protected@edef\cref@currentlabel{[equation][\arabic{equation}][\arabic{section}]\theequation}%
\label{eq:contact-crossing-paired-expansion}%
\endgroup%
\makeatother%
Integrating the overlap of four external wavefunctions at one
normal position gives a contact vertex an effective normal-momentum
conservation constraint even in half-space. For Dirichlet modes
with positive normal momenta $|k_i|$, the overlap is
\begin{equation}
 \int_0^\infty\dd y\,\prod_{i=1}^{4}\sin(|k_i|y)
 =\frac{\pi}{16}\sum_{r_1,r_2,r_3,r_4=\pm1}
    r_1r_2r_3r_4\,
    \delta\!\left(\sum_{i=1}^{4}r_i|k_i|\right).
\label{eq:contact-four-sine-overlap}
\end{equation}
Each $|k_i|$ runs over $(0,\infty)$. Unfolding the four ranges
to signed momenta incorporates the sixteen reflection coefficients
as the action of $\Pi_h^{\otimes4}$ on the correlated wave packet.
Their delta functions impose the normal-momentum constraint
$2\pi\delta(\sum k_i)=\pi\delta(k_0)$ on the smearing kernel.
The idempotence in \eqref{eq:spin-projector} leaves just one
projection on each external leg of the state.
The EFT contact term thus depends only on the pre-projection distribution
on the $k_0=0$ momentum-conserving slice. The same overlap argument
applies to any even number of scalar Dirichlet modes.

On the conserving slice, $(s,t,u)=(s_h,t_h,u_h)$.
The $s_h^2u_h$ denominator in \eqref{eq:common-dq-smeared-amplitude}
selects the coefficient of $-su$ in
\eqref{eq:contact-crossing-paired-expansion}, a power series in $t$.
Other powers of $su$ have zero residue on the low-energy contour
enclosing both $s_h=0$ and $u_h=0$. For the leading and next-to-leading
$g_2$ and $g_3$ terms, the EFT contribution to the normalized dispersion integrand is
\begin{equation}
 \widehat{\mathcal B}_{2,h}^{\rm loc}
 =-\frac{4q_{\max}}{\pi\sqrt{2\pi}\sigma}
 \int_0^{q_{\max}}\dd q_\parallel\,\phi(-q_\parallel^2)
 \prod_{i=1}^{4}\int\dd k_i\,W
 \frac{\Pi_h^{\otimes2}
       [\pi\delta(k_0)\M_{\rm loc}(s_h,t_h)]\Pi_h^{\otimes2}}
      {s_h^2u_h}.
\label{eq:contact-projected-dispersive-integrand}
\end{equation}
The projection leaves the parallel momenta unchanged and acts
on the normal momenta of the initial and final states.

Throughout the finite $q_\parallel,k_i$ integration domain, the
same $\Gamma_{\rm EFT}$ encloses all rational poles of $1/(s_h^2u_h)$,
allowing us to perform the $s_h$ contour integral first and extract
the coefficient of $-s_hu_h$ in the displayed $g_2,g_3$ vertex.
The Hermiticity of $\Pi_h$ then transfers the two external projections
from $\M_{\rm loc}$ to the external-state weight $W$, giving
\begin{equation}
 \widehat B_{2,h}^{\rm loc}
 =\frac{4q_{\max}}{\pi\sqrt{2\pi}\sigma}
 \int_0^{q_{\max}}\dd q_\parallel\,\phi
 \prod_{i=1}^{4}\int\dd k_i\,
 \Pi_h^{\otimes2}W\Pi_h^{\otimes2}\,
 \pi\delta(k_0)
 \left[\frac{2g_2}{\Lambda_{\rm EFT}^4}
       -\frac{g_3}{\Lambda_{\rm EFT}^6}t_h+\cdots\right].
\label{eq:contact-projected-contour-moment}
\end{equation}
The ellipsis retains higher coefficients in the $t$ series.
With the wave-packet centers $y_1=y_2=y_0$, the Fourier transform
of the Gaussian normal weight in \eqref{eq:common-su-rank-one-weight}
against the half-space eigenmodes is
\begin{equation}
\begin{gathered}
 \prod_{i=1}^{4}\int_{\mathbb R}\dd k_i\,W(k_i)
 \prod_{i=1}^{4}(\e^{\ii k_i y_i}-\e^{-\ii k_i y_i})
 =4\pi\sigma^2\prod_{i=1}^{4}\psi_h(y_i),\\
 \psi_h(y)=\e^{-\sigma^2(y-y_0)^2/2}
           -\e^{-\sigma^2(y+y_0)^2/2},\qquad y>0.
\end{gathered}
\label{eq:common-dq-mixed-gaussian}
\end{equation}
This product describes the normal-position part of the correlated
wave packet, while the parallel momenta retain their on-shell kinematics.

At the local four-point vertex, we act with the derivatives and set
$y_1=y_2=y_3=y_4=y$. On the conserving slice $k_t=k_1+k_3$, so
$-t_h=q_\parallel^2+k_t^2$ becomes
$q_\parallel^2-(\partial_{y_1}+\partial_{y_3})^2$ in the mixed representation.
At coincidence, $\partial_y^2$ acts on
$\psi_{h,1}(y)\psi_{h,3}(y)$, with $1,3$ labeling the
external legs. The $2,4$ pair supplies the other $\psi_h^2$. Thus
\begin{equation}
 \widehat B_{2,h}^{\rm loc}
 =q_{\max}
   \int_0^{q_{\max}}\dd q_\parallel\,\phi
   \int_0^\infty\sqrt{\frac{2}{\pi}}\sigma\,\dd y\,\psi_h^2
 \left[\frac{2g_2}{\Lambda_{\rm EFT}^4}
       +\frac{g_3(q_\parallel^2-\partial_y^2)}{\Lambda_{\rm EFT}^6}
       +\cdots\right]\psi_h^2.
\label{eq:contact-mixed-contour-overlap}
\end{equation}
Substituting the Gaussian $\psi_h$ in
\eqref{eq:common-dq-mixed-gaussian} into the normalized $y$ integral
and neglecting relative $\mathcal{O}(\e^{-\sigma^2y_0^2})$ corrections,
the constant overlap gives one and the $-\partial_y^2$ overlap gives
$\sigma^2$. The finite-width $g_3$ coefficient therefore samples
$q_\parallel^2+\sigma^2$. Integrating next with
$\phi(-q_\parallel^2)=(1-q_\parallel/q_{\max})^3$ gives
\begin{equation}
 \widehat B_{2,h}^{\rm loc}=\frac{q_{\max}^2}{2\Lambda_{\rm EFT}^4}
 \left[g_2+\frac{q_{\max}^2/30+\sigma^2/2}{\Lambda_{\rm EFT}^2}g_3
       +\cdots+\mathcal{O}(\e^{-\sigma^2y_0^2})\right].
\label{eq:contact-cubic-coefficients}
\end{equation}
Each EFT contact term has its own
$\mathcal{O}(\e^{-\sigma^2y_0^2})$ relative correction.
The single $\mathcal{O}$ term schematically collects these effects,
weighted by the magnitudes of the individual contributions, and
does not represent a relative correction to their sum.
The exact Gaussian image overlaps first contribute at
$\e^{-3\sigma^2y_0^2/2}$.

The ordinary dispersion relation used here retains infinitely many
local EFT coefficients, without the finite truncation provided by
improved dispersion relations \cite{Caron-Huot:2021rmr,Caron-Huot:2022ugt}
or CSDRs \cite{Sinha:2020win}.
For $t$--$u$-symmetric amplitudes, we conjecture that the prescription
of Ref.~\cite{Dong:2025dpy} can be adapted
to half-space by using $w_\parallel^{-1}=t_\parallel^{-1}+u_\parallel^{-1}$,
$q_{\parallel,w}^2=-w_\parallel$, and smearing in $\dd q_{\parallel,w}$
as in \eqref{eq:common-dq-smeared-amplitude}.
A suitable profile may then preserve positivity while isolating
finitely many low-energy Wilson coefficients.

We next compute the gravitational tree-level contribution to the
low-energy side of the 2SDR. With all external momenta incoming,
the scalar--graviton vertex tensor and de Donder numerator are
\begin{equation}
\begin{aligned}
 V_{ij}^{\mu\nu}(k_i,k_j)
 &=p_i^\mu p_j^\nu+p_i^\nu p_j^\mu
       -\eta^{\mu\nu}(p_i\!\cdot p_j),\\
 \mathcal P_{\mu\nu,\rho\sigma}
 &=\tfrac12(\eta_{\mu\rho}\eta_{\nu\sigma}
           +\eta_{\mu\sigma}\eta_{\nu\rho}
           -\eta_{\mu\nu}\eta_{\rho\sigma}).
\end{aligned}
\label{eq:tree-scalar-gravity-vertex}
\end{equation}
In the mixed representation, with momenta along $t,z,x$ and
positions along $y$, we use the linearized de Donder Green function
obtained from the flat-space propagator by the tensor parity
extension specified in \eqref{eq:spin-tensor-boundary-components},
\begin{equation}
G_{h\,\mu\nu,\rho\sigma}(q_\parallel;y,y')
 =\frac{\mathcal P_{\mu\nu,\rho\sigma}
             \e^{-q_\parallel|y-y'|}
 -\mathcal P_{\mu\nu,\alpha\beta}
    (r_y)^\alpha{}_{\rho}(r_y)^\beta{}_{\sigma}
             \e^{-q_\parallel(y+y')}}{2q_\parallel}.
\label{eq:tree-tensor-green}
\end{equation}
Here $r_y$ acts separately on each Lorentz index, as in
\eqref{eq:tensor-reflection-ch4}.

The external states in the half-space amplitude use the boundary
modes $u_{k_i}(y)=2\ii\sin(k_i y)$. The two vertex currents are
\begin{equation}
\begin{aligned}
 V_{13}^{\mu\nu}(y)
 &=\left.V_{13}^{\mu\nu}(-\ii\partial_{y_1},-\ii\partial_{y_3})
    u_{k_1}(y_1)u_{k_3}(y_3)\right|_{y_1=y_3=y},\\
 V_{24}^{\rho\sigma}(y')
 &=\left.V_{24}^{\rho\sigma}(-\ii\partial_{y_2},-\ii\partial_{y_4})
    u_{k_2}(y_2)u_{k_4}(y_4)\right|_{y_2=y_4=y'}.
\end{aligned}
\label{eq:tree-external-currents}
\end{equation}
The arguments of $V$ replace its explicit normal components
$p_i^y=k_i$ by $-\ii\partial_{y_i}$, while the wall-parallel
components retain their external momentum values. Each derivative
acts on its own mode before the positions are identified.
These products already include the $\Pi_h$ projection and need
no further image sum. Removing the momentum-conserving delta
function along $t,z,x$ gives
\begin{equation}
 \M_h^{t\text{-tree}}
 =8\pi G_N\int_0^\infty\!\dd y\,\dd y'\,
   V_{13}^{\mu\nu}(y)
   G_{h\,\mu\nu,\rho\sigma}(q_\parallel;y,y')
   V_{24}^{\rho\sigma}(y').
\label{eq:tree-halfspace-compact}
\end{equation}
After inserting the two vertex currents and the half-space Green
function, we can perform the $s_h$ contour integral before the
external-state integrals. In momentum space, this means holding
$k_i,q_\parallel$ fixed. Contracting the two vertices with the direct
and reflected de Donder numerators gives $s_\parallel^2$ with
coefficient $1$, together with lower powers of $s_\parallel$.
Since $s_\parallel=s_h+k_s^2$ and $k_s$ is fixed, each contraction
also has unit $s_h^2$ coefficient. The 2SDR contour enclosing both
$s_h=0$ and $u_h=0$ therefore returns $1$, independently of the
lower-order terms. This is exact coefficient selection by the 2SDR,
requiring no small-$k_i^2/s_h$ expansion or other approximation.
Transforming the normal momentum to position space preserves
this conclusion. For these quadratic numerators, the contour
selection can equivalently be written in derivative form,
\begin{equation}
 \frac12\partial_{s_\parallel}^{2}[V_{13}(y)\mathcal P V_{24}(y')]
 =\frac12\partial_{s_\parallel}^{2}[V_{13}(y)\mathcal P(R_yV_{24}(y'))]
 =u_{k_1}(y)u_{k_3}(y)u_{k_2}(y')u_{k_4}(y').
\label{eq:tree-tensor-contractions}
\end{equation}
This identity gives the low-energy contour contribution when
\eqref{eq:tree-halfspace-compact} is inserted into
\eqref{eq:common-dq-smeared-amplitude} and integrated as in
\eqref{eq:general-eft-contour}.

Fourier transforming the complete
mode product with the correlated weight $W$, and setting
$y_1=y_3=y$, $y_2=y_4=y'$ in \eqref{eq:common-dq-mixed-gaussian},
gives $4\pi\sigma^2\psi_h(y)^2\psi_h(y')^2$. The gravitational
contribution to the normalized low-energy dispersion moment is therefore
\begin{equation}
\begin{aligned}
 \widehat B_{2,h}^{\rm grav}&=8\pi G_Nq_{\max}
 \int_0^{q_{\max}}\dd q_\parallel\,
 (1-q_\parallel/q_{\max})^3G_\sigma(q_\parallel;y_0),\\
 G_\sigma(q_\parallel;y_0)&\equiv\sqrt{\frac{2}{\pi}}\sigma
 \int_0^\infty\dd y\,\dd y'\,\psi_h(y)^2
 \frac{\e^{-q_\parallel|y-y'|}-\e^{-q_\parallel(y+y')}}{2q_\parallel}
 \psi_h(y')^2.
\end{aligned}
\label{eq:tree-mixed-gaussian-moment}
\end{equation}
This expression retains the complete half-space external modes,
including their original and reflected components, and both direct
propagation and propagation reflected from the wall in the Green
function.

For $\sigma y_0\geq10$, we neglect the exponentially small
external tails on $y>0$ and retain the density
$\e^{-\sigma^2(y-y_0)^2}$ centered at $+y_0$. Keeping both parts
of the long-range Green function and extending the two Gaussian
distributions to the whole real line gives the following approximation
\begin{equation}
 G_\sigma=
 \frac{\sqrt{2\pi}\,\e^{q_\parallel^2/(2\sigma^2)}}{2\sigma q_\parallel}
 \left[\operatorname{erfc}\left(\frac{q_\parallel}{\sqrt2\sigma}\right)
 -\e^{-2q_\parallel y_0}\right]
 \left(1+\mathcal O(\e^{-\sigma^2y_0^2})\right).
\label{eq:tree-gaussian-closed-convolution}
\end{equation}
Here $\operatorname{erfc}(x)=1-\operatorname{erf}(x)$, with
$\operatorname{erfc}(x)\simeq1-2x/\sqrt\pi$ as $x\to0$.
The two terms in square brackets arise from direct propagation and
propagation reflected from the wall. As $q_\parallel\to0$, their
constant terms cancel, and the difference of their linear terms
cancels the $1/q_\parallel$ singularity outside the brackets,
so $G_\sigma$ approaches $\sqrt{2\pi}\,y_0/\sigma-1/\sigma^2$.
Thus, even when the reflected component of the external wave packet
can be neglected, propagation reflected from the wall in the Green
function remains essential to IR finiteness.

Alternatively, we can evaluate the exact integral
\eqref{eq:tree-mixed-gaussian-moment} numerically, using the convolution
of the full half-space wave packets and half-space Green function.
For $\sigma/q_{\max}=10^{-2}$ and $\sigma y_0=10$, the result to three
significant figures is
\begin{equation}
 \widehat B_{2,h}^{\rm grav}\simeq1.11\times10^4\,G_N
 \left[1+\mathcal O(\e^{-\sigma^2y_0^2})\right].
\label{eq:tree-gravity-numeric-benchmark}
\end{equation}
The relative correction is the boundary EFT uncertainty discussed
above. Diverging curvature invalidates the weak-field description of
scalar--graviton minimal coupling and other local three-point
interactions, but the vertex overlaps reach that region only through
exponentially small tails. This uncertainty is distinct from the
numerical integration error. Equation~\eqref{eq:tree-halfspace-compact} includes only
$t$-channel graviton exchange because the flat-space $s$- and $u$-channel
amplitudes, $8\pi G_Ntu/s$ and $8\pi G_Nst/u$, give zero net residue
on $\Gamma_{\rm EFT}$ with the dispersive weight $1/(s_h^2u_h)$.
Their half-space amplitudes likewise grow at most linearly in $s_h$,
so their contour contributions vanish when the low-energy singularities
are enclosed.

Inserting the four-point local EFT and gravitational contributions
into \eqref{eq:channel-wise-sum-rule}, a functional with nonnegative
UV responses obeys
\begin{equation}
 \frac{q_{\max}^2}{2\Lambda_{\rm EFT}^4}
 \left[g_2+\frac{q_{\max}^2/30+\sigma^2/2}{\Lambda_{\rm EFT}^2}g_3+\cdots\right]
 +\widehat B_{2,h}^{\rm grav}+\mathcal{O}(\e^{-\sigma^2y_0^2})\geq0.
\label{eq:halfspace-computed-bound}
\end{equation}
As in \eqref{eq:contact-cubic-coefficients}, the schematic
$\mathcal O$ term combines the separate vertex tails, now including
the three-point couplings in graviton exchange. Their unequal
coefficients leave an additive uncertainty, not a common factor
that can be canceled.
Including loops would also require evaluating the relative correction
described in \eqref{eq:beam-logarithmic-scales} on the DN strip.

If the $g_2$ contribution in \eqref{eq:halfspace-computed-bound}
dominates those from $g_3,g_4,\ldots$, taking
$q_{\max}=\Lambda_{\rm EFT}/1.1$ and inserting
\eqref{eq:tree-gravity-numeric-benchmark} into
\eqref{eq:halfspace-computed-bound} gives the approximate bound
\begin{equation}
 g_2\gtrsim-2.68\times10^4\,G_N\Lambda_{\rm EFT}^2
 +\mathcal O(\e^{-\sigma^2y_0^2})+\mathcal O(g_3,\ldots).
\label{eq:halfspace-numeric-eft-bound}
\end{equation}
The two $\mathcal O$ terms indicate the exponential tails and the
higher EFT contributions at fixed dimensionless wave-packet parameters.
With these dimensionless wave-packet parameters held fixed, the
coefficient multiplying $G_N\Lambda_{\rm EFT}^2$ is independent of
the EFT scale.

Finally, we examine the origin of the gravitational contribution in
\eqref{eq:halfspace-computed-bound}. The local EFT contribution is of
order $q_{\max}^2/\Lambda_{\rm EFT}^4$, as for flat-space smearing,
while the gravitational contribution differs. For an analytic
description, we consider $y_0^{-1}\ll\sigma\ll q_{\max}$ instead
of the fixed ratios $y_0^{-1}:\sigma:q_{\max}=10^{-3}:10^{-2}:1$
used for the numerical bound. The dominant contribution to the
tree-level smearing integral \eqref{eq:tree-mixed-gaussian-moment}
then comes from $q_\parallel\in((2y_0)^{-1},\sigma)$. In the interior
of this interval, \eqref{eq:tree-gaussian-closed-convolution} reduces
to $G_\sigma\simeq\sqrt{2\pi}/(2\sigma q_\parallel)$ and the smearing
profile satisfies $\phi(-q_\parallel^2)\simeq1$. The cancellation
due to wall reflection sets the effective IR cutoff $(2y_0)^{-1}$,
while the one-dimensional convolution of the wavefunctions through
the Green function in \eqref{eq:tree-mixed-gaussian-moment}
suppresses the contribution from $q_\parallel>\sigma$. The integral
therefore has the asymptotic behavior
\begin{equation}
 \widehat B_{2,h}^{\rm grav}
 \underset{\substack{\sigma y_0\to\infty\\q_{\max}/\sigma\to\infty}}{\sim}
 4\pi\sqrt{2\pi}\,G_N\frac{q_{\max}}{\sigma}\log(2\sigma y_0).
\label{eq:tree-gravity-leading-log}
\end{equation}
The relative enhancement comes from the extra normalization factor
$q_{\max}/\sigma$ in \eqref{eq:common-dq-normalization}. It compensates
for suppression of the half-space four-point contact response,
bringing the smeared local EFT contribution to the same order as
in flat space, at the cost of enhancing the gravitational contribution
by the same factor. The contact suppression arises because these
external states include configurations without normal momentum
conservation, whereas even in half-space the four-point contact
contribution depends only on the smearing data on the effective
momentum-conserving slice selected by
\eqref{eq:contact-four-sine-overlap}. Graviton exchange instead
connects the overlaps of legs $1,3$ and $2,4$ through two three-point
couplings and the Green function, without the same four-point
constraint. Before the common normalization, the contact response
is therefore suppressed by $\sigma/q_{\max}$ relative to flat-space
smearing, while the gravitational response remains of order
$G_N\log(2\sigma y_0)$. For low-energy EFTs at electroweak or QCD
scales, $G_N\Lambda_{\rm EFT}^2\ll1$, but taking $\sigma/q_{\max}$
arbitrarily close to zero would offset this weak-coupling suppression
and make gravity compete with the EFT contribution. A smaller width
ratio makes UV positivity easier to satisfy, so the finite ratio
used for our numerical bound balances this requirement against the
low-energy gravitational contribution and can be further optimized.

\section{Geometric regulation of UV-completed gravity}
\label{sec:extended-uv-states}

Under the channel-positivity and dispersive conditions of
Sec.~\ref{sec:halfspace-positive-functionals}, the bound in
\eqref{eq:halfspace-computed-bound} includes both local EFT interactions
and IR-finite graviton exchange.
We now explain how the bound obtained with this geometric regulation
is consistent with a general UV completion of gravity, and why
deriving an analogous bound for FK dressed states is
more difficult. The key is how the IR prescription acts on the
intermediate states supplying the amplitude discontinuity, as well
as on the low-energy graviton exchange.

In flat space, the 2SDR expresses the low-energy graviton pole in
terms of high-energy absorptive data. Its forward IR singularity
must therefore also be reproduced by the weighted spectral sum
on the other side. Ref.~\cite{Haring:2024wyz} distinguishes two
mechanisms according to impact parameter. At large impact
parameter, the leading mechanism is eikonal scattering in GR.
As the Born exchange grows with energy, the ladder diagrams must
be resummed to satisfy unitarity. In the leading eikonal
approximation, cutting these ladders gives the two-particle
intermediate states of the original scattering process. This
contribution does not require new UV particles. It is the
unitarization of the long-range interaction and does not by
itself supply a microscopic UV completion.

At smaller impact parameter, scattering can probe the UV
completion. In weakly coupled string theory, higher-spin string
resonances already contribute absorptive terms at tree level.
For scalar external particles, the angular dependence of a pole
with spin $J$ and mass $M$ is $P_J(1+2t/M^2)$. In the eikonal
transverse Fourier representation, this finite polynomial gives
local interactions proportional to $\delta^{(2)}(\boldsymbol b)$
and its successive Laplacians, compactly
$P_J(1+2\Delta_{\boldsymbol b}/M^2)\delta^{(2)}(\boldsymbol b)$.
The collective sum over the infinite tower instead reproduces
the graviton pole, whose Fourier transform has the long-range
gravitational profile and its IR divergence. Thus the IR problem
is not confined to the low-energy side of the 2SDR. To obtain an
IR-finite EFT bound by bootstrap, one must consistently change
the contribution of these UV intermediate states to the
amplitude discontinuity.

Summing the string resonances yields a Regge exchange whose
transverse interaction profile diffuses as the energy increases.
Collectively it therefore behaves as an extended exchange object
rather than a point particle. When the incoming and outgoing
packets are localized near $y_0$, the local three-point vertices
for production and decay of each exchanged $s$-channel resonance
are evaluated through overlaps of its mode with the external
packets in that same region. Once transverse diffusion reaches
distances larger than the distance to the wall, the direct and
reflected contributions largely cancel there, suppressing the
contribution to the chosen matrix element. Translating the
packets to $y_0\to\infty$ restores the flat-space IR divergence
of the low-energy gravitational exchange. At the same time,
the contributions of the extended UV intermediate states
that were suppressed near the wall return.

The examples with a zero mode in \cref{sec:transverse-modes}
show that positivity of the discontinuity alone is insufficient.
In the $\mathbb M_{1,1}\times S^2$ model, the zero-mode-subtracted
Green function \eqref{eq:S2-pseudoinverse-closed} has both signs.
External wave packets can sample its negative region, giving a
negative low-energy eikonal phase and a negative relative time
delay at leading eikonal order.
Nevertheless, if particles $1$ and $3$ have the same transverse
wavefunction, and likewise particles $2$ and $4$, the densities
$\rho_{13}$ and $\rho_{24}$ in \eqref{eq:transverse-packet-densities}
remain nonnegative. Equation~\eqref{eq:compact-eikonal-cut-positivity}
then gives nonnegative discontinuities in both physical channels.
Only a qualitative conclusion is needed here. We evaluate the
low-energy contour moment and the $s$- and $u$-channel
discontinuities in the eikonal regime, with the smearing support
chosen to obey $t_{\max}\ll\Lambda_{\rm EFT}^2$.
A negative low-energy contour moment together with these
nonnegative cuts requires a nonzero $\mathcal C_\infty$ contour
integral, as in \eqref{eq:compact-zero-mode-contour-diagnosis}
and \cref{fig:compact-dispersion-contours}. For this
zero-mode-subtracted response, the real-axis UV bound does not
extend to arbitrary complex directions as it does for ordinary
two-particle scattering under the assumptions accompanying
\eqref{eq:halfspace-linear-real-bound}.

FK dressing presents a similar situation. In the leading
eikonal regime, its cancellation with virtual loops is equivalent
to removing the unresolved graviton contribution below the
specified soft-energy scale $\omega$ from the graviton ladder
loops. This is analogous to the zero-mode subtraction above.
The companion work \cite{Dong:ProjectA} retains the dressing's
reference-frame data when constructing positive,
crossing-symmetric dressed-state functionals. In the specified
eikonal model, the low-energy phase and relative time delay can
be negative at impact parameters of order $\omega^{-1}$ or
larger. When the $s$- and $u$-channel pairings remain positive,
the same contour diagnosis requires a nonzero
$\mathcal C_\infty$ contribution and obstructs an EFT bound
obtained by dropping that contour. Thus positivity and crossing
alone do not give a dressed-state EFT bound.

Zero-mode subtraction and FK dressing are both legitimate IR
prescriptions, but their effect is not confined to the low-energy
response. In the eikonal description, their action appears as
conjugate phase factors in the $S$-matrix boundary values on the
upper and lower rims of the real axis. These changes can preserve
unitarity, analyticity, crossing and positivity. They nevertheless
describe hard scattering influenced by a background, so the
process cannot be treated as ordinary $2\to2$ Fock-state scattering.
In the compact transverse example, removing the zero mode
introduces the compensating uniform source in
\eqref{eq:pseudoinverse-equation}. For dressed states, a soft cloud
accompanies each hard particle. The dressing uses a
reference-frame-dependent, isotropic energy cutoff while retaining
$ISO(3)$ and time translations. The external state therefore
contains two hard particles together with coherent soft clouds
whose mean particle number diverges in the IR limit.
Consequently, the extension of the real-axis UV bound to arbitrary
complex directions for ordinary $2\to2$ Fock-state scattering does
not automatically carry over to scattering with such
backgrounds.\footnote{The negative delay discussed here concerns
low-energy group propagation in the specified background.
Such a group advance does not imply a superluminal signal front.}
A similar situation can arise even at small impact parameter
in UV-completed gravity. The sum over UV resonances reproduces the
same IR singularity as the low-energy gravitational amplitude.
Dressing the hard particles with infinitely many soft gravitons
alone does not specify the corresponding change of this UV
spectral contribution. The detailed dressed-state analysis is
given in the companion work \cite{Dong:ProjectA}.

The geometric prescription used here is inspired by AdS$_4$,
whose negatively curved transverse geometry suppresses graviton
exchange at large geodesic separation
\cite{Cornalba:2007zb}. In the string Regge description,
curvature also affects the diffusion operator. Once diffusion
probes the curvature scale, this suppresses the extended kernel's
overlap with localized sources relative to flat propagation,
including its contribution to the Lorentzian four-point
discontinuity \cite{Brower:2006ea}.
Half-space gives an analogous suppression through the direct
and reflected profiles. The $R\to0$ limit of the joined
AdS$_4$--flat geometry in \eqref{eq:warped-halfspace-metric}
imposes a similar geometric restriction on propagation.
Both geometries act on the low-energy gravitational exchange and
its UV-completed propagation. The construction with an exactly flat interaction region,
in turn, provides asymptotic scattering states for a direct
$S$-matrix bootstrap and keeps the local EFT vertices in the flat
region, avoiding the curvature corrections present at finite
AdS radius \cite{Caron-Huot:2021enk,Penedones:2010ue}.

The string Regge limit gives a calculable example in which the
boundary changes both the massless exchange and the collective
contribution of an infinite resonance tower. For helicity-preserving
elastic scattering $++\to++$, after stripping the external $U(1)$
little-group phase as in \eqref{eq:graviton-phase-only-stripping}, the
type-II four-graviton Virasoro--Shapiro amplitude is
\cite{Bern:2021ppb}
\begin{equation}
 \M^{\bcancel{U(1)}}_{\rm VS}(s,t,u)
 =8\pi G_N\frac{s^4}{stu}
 \frac{\Gamma(1-\alpha's/4)\Gamma(1-\alpha't/4)\Gamma(1-\alpha'u/4)}
      {\Gamma(1+\alpha's/4)\Gamma(1+\alpha't/4)\Gamma(1+\alpha'u/4)}.
 \label{eq:b-string-vs}
\end{equation}
The all-incoming helicities are $++--$.
In the regime $|t|\ll\alpha'^{-1}\ll s$, set $q^2=-t$.
The leading Regge form on the upper physical edge is
\cite{DiVecchia:2023frv,Haring:2024wyz}
\begin{equation}
 \M^{\bcancel{U(1)}}_{\rm VS}(s,-q^2)\simeq\frac{8\pi G_Ns^2}{q^2}
 e^{-Y_sq^2/4},
 \qquad
 Y_s=2\alpha'\log\frac{-\ii\alpha's}{4}
     =\ell_{\rm diff}^2-\ii\pi\alpha'.
 \label{eq:b-string-regge}
\end{equation}
Here $Y_s$ is the complexified effective string length squared~\cite{DiVecchia:2023frv}.
We use it as an auxiliary diffusion parameter, with
$\ell_{\rm diff}^2\equiv\operatorname{Re}Y_s
=2\alpha'\log(\alpha's/4)$ defining the Gaussian diffusion length.
We use a heat-kernel diffusion picture similar to that of
Ref.~\cite{Brower:2006ea}, applied here
to flat-space type-II string theory without the IR confinement
of its QCD-like applications.\footnote{Early holographic calculations extracted glueball spectra
from supergravity wave equations in confining geometries \cite{Csaki:1998qr}.}
The large-log regime is $\log(\alpha's/4)\gg1$.
With $S=e^{2\ii\delta}$, the transverse Fourier transform of the
tree-level $\M/(2s)$ gives $2\delta$. A small radial IR cutoff
gives the single-exchange eikonal phase
\begin{equation}
\begin{aligned}
 \delta_{\rm str,0}(s,b)
 &=\int_{|\boldsymbol q|>\lambda_{\rm IR}}
   \frac{\dd^2q}{(2\pi)^2}\,
   e^{\ii\boldsymbol q\cdot\boldsymbol b}
   \frac{\M^{\bcancel{U(1)}}_{\rm VS}(s,-q^2)}{4s}\\
 &\simeq\frac{G_Ns}{2}\left[
   \log\frac{4}{b^2\lambda_{\rm IR}^2}-2\gamma_E
   -E_1\!\left(\frac{b^2}{Y_s}\right)\right].
\end{aligned}
 \label{eq:b-regge-flat-fourier}
\end{equation}
The expression assumes $\lambda_{\rm IR}\max(b,\ell_{\rm diff})\ll1$.
Here $E_1(z)=\int_z^\infty e^{-u}\dd u/u$.

Placing the scattering in half-space cancels the IR-cutoff
dependence between the direct and reflected interactions.
Consider two points with separation $b$ along the wall and heights
$y_1,y_2>0$. Set $r_-^2=b^2+(y_1-y_2)^2$ and
$r_+^2=b^2+(y_1+y_2)^2$. The direct-minus-image difference of the phase in
\eqref{eq:b-regge-flat-fourier} gives the normalized two-point Regge
interaction Green function
\begin{equation}
 G_{\rm str,h}(s,b,y_1,y_2)
 =\frac{\delta_{\rm str,h}(s,b,y_1,y_2)}{G_Ns}
 =\frac12\left[\log\frac{r_+^2}{r_-^2}
 -E_1\!\left(\frac{r_-^2}{Y_s}\right)
 +E_1\!\left(\frac{r_+^2}{Y_s}\right)\right].
 \label{eq:b-regge-two-point}
\end{equation}
For the comparisons below we set $y_1=y_2=y_0$.
In the large-log regime $\log(\alpha's/4)\gg1$, expanding the
complex Green function gives the leading profiles
\begin{equation}
\begin{aligned}
 \operatorname{Re}G_{\rm str,h}
 &\simeq\frac12\left[\log\!\left(1+\frac{4y_0^2}{b^2}\right)
 -E_1\!\left(\frac{b^2}{\ell_{\rm diff}^2}\right)
 +E_1\!\left(\frac{b^2+4y_0^2}{\ell_{\rm diff}^2}\right)\right],\\
 \operatorname{Im}G_{\rm str,h}
 &\simeq\frac{\pi\alpha'}{2\ell_{\rm diff}^2}
 e^{-b^2/\ell_{\rm diff}^2}
 \left(1-e^{-4y_0^2/\ell_{\rm diff}^2}\right)>0.
\end{aligned}
 \label{eq:b-regge-positive-phases}
\end{equation}
At this order, the positive imaginary part of $G_{\rm str,h}$
implies a positive imaginary part of the eikonal phase
$\delta_{\rm str,h}$ at physical $s$-channel energy $s>0$.
This gives a positive tree-level absorptive contribution, as
required by the optical theorem. The smooth Regge expression
is obtained by summing the narrow string poles and then
averaging over energy \cite{Haring:2024wyz,DiVecchia:2023frv}.

As $\ell_{\rm diff}$ becomes comparable to and then exceeds $y_0$,
the imaginary part of $G_{\rm str,h}$ is reduced relative to flat
space by the factor $1-e^{-4y_0^2/\ell_{\rm diff}^2}$.
For $\ell_{\rm diff}\gg y_0$, this becomes the relative power
suppression $4y_0^2/\ell_{\rm diff}^2$, realizing the mechanism
described above. The string exchange has an effective transverse
size of order $\ell_{\rm diff}$, and the wall suppresses its overlap
with the incoming pair localized near $y_0$ through cancellation
between the direct and reflected profiles. This is the width of
the interaction profile, distinct from free spreading of the
external packets.

The same Regge diffusion that produces this absorptive part also
makes $\operatorname{Re}G_{\rm str,h}$ decrease as
$\ell_{\rm diff}$ grows. At fixed $b,y_0$, the large-log expansion
$\log(\alpha's/4)\gg1$ gives
\begin{equation}
 \operatorname{Im}G_{\rm str,h}
 \simeq-\frac{\pi}{2}
 \frac{\partial\operatorname{Re}G_{\rm str,h}}{\partial\log s}.
 \label{eq:b-regge-real-imag-derivative}
\end{equation}
The positive imaginary part thus entails a decreasing real part.
This is the stringy softening of the pointlike gravitational
exchange by the Regge tower
\cite{Amati:1987uf,DiVecchia:2023frv}.
We examine the asymptotic regimes determined by three lengths,
the impact parameter $b$, the wall distance $y_0$ that provides the
effective IR length, and the diffusion length $\ell_{\rm diff}$.
To track the dependence on the dispersive variable, we hold
$b,y_0$ fixed and increase $s$, equivalently increasing
$\ell_{\rm diff}$, and follow the real and imaginary parts of
$G_{\rm str,h}$. For $b\ll y_0$, these leading profiles have
three asymptotic regions as $\ell_{\rm diff}$ increases
\begin{equation}
 \operatorname{Re}G_{\rm str,h}+\ii\operatorname{Im}G_{\rm str,h}
 \simeq\begin{cases}
 \displaystyle\log\frac{2y_0}{b}
    +\ii\frac{\pi\alpha'}{2\ell_{\rm diff}^2}e^{-b^2/\ell_{\rm diff}^2},
      &\ell_{\rm diff}\ll b\ll y_0,\\[7pt]
 \displaystyle\log\frac{2y_0}{\ell_{\rm diff}}
    +\ii\frac{\pi\alpha'}{2\ell_{\rm diff}^2},
      &b\ll\ell_{\rm diff}\ll y_0,\\[7pt]
 \displaystyle\frac{2y_0^2}{\ell_{\rm diff}^2}
    +\ii\frac{2\pi\alpha' y_0^2}{\ell_{\rm diff}^4},
      &b\ll y_0\ll\ell_{\rm diff}.
 \end{cases}
 \label{eq:b-regge-short-diffusion}
\end{equation}
At fixed $y_0\ll b$, there are two asymptotic regions
\begin{equation}
 \operatorname{Re}G_{\rm str,h}+\ii\operatorname{Im}G_{\rm str,h}
 \simeq\begin{cases}
 \displaystyle\frac{2y_0^2}{b^2}
    +\ii\frac{\pi\alpha'}{2\ell_{\rm diff}^2}e^{-b^2/\ell_{\rm diff}^2}
    \left(1-e^{-4y_0^2/\ell_{\rm diff}^2}\right),&\ell_{\rm diff}\ll b,\\[7pt]
 \displaystyle\frac{2y_0^2}{\ell_{\rm diff}^2}
    +\ii\frac{2\pi\alpha' y_0^2}{\ell_{\rm diff}^4},&b\ll\ell_{\rm diff}.
 \end{cases}
 \label{eq:b-regge-broad-diffusion}
\end{equation}
These expressions describe the two-point interaction Green function
$G_{\rm str,h}=\delta_{\rm str,h}/(G_Ns)$ as $s$, and hence
$\ell_{\rm diff}(s)$, increases. The full eikonal phase retains
the factor $G_Ns$. At fixed $b,y_0$ and sufficiently large $s$,
its real and imaginary parts grow schematically as $s/\log s$ and
$s/\log^2s$ within this tree-level Regge description.
Thus the string resonances with $y_0\ll\ell_{\rm diff}(s)$
still contribute to the scattering amplitude.
Their contribution is instead suppressed in the 2SDR after
including the dispersive denominator. At fixed geometry, the
high-energy absorptive tail in this model scales as
$\dd s/[s\log^2s]$ and is integrable. In flat space,
the tail scales as $\dd s/[s\log s]$ and gives an IR-divergent dispersive integral,
reproducing the low-energy graviton IR singularity
\cite{Haring:2024wyz}. The same half-space Green function therefore
implements the geometric suppression on both the low-energy
exchange and its high-energy dispersive representation.

\begin{figure}[H]
 \centering
 \includegraphics[width=.94\textwidth]{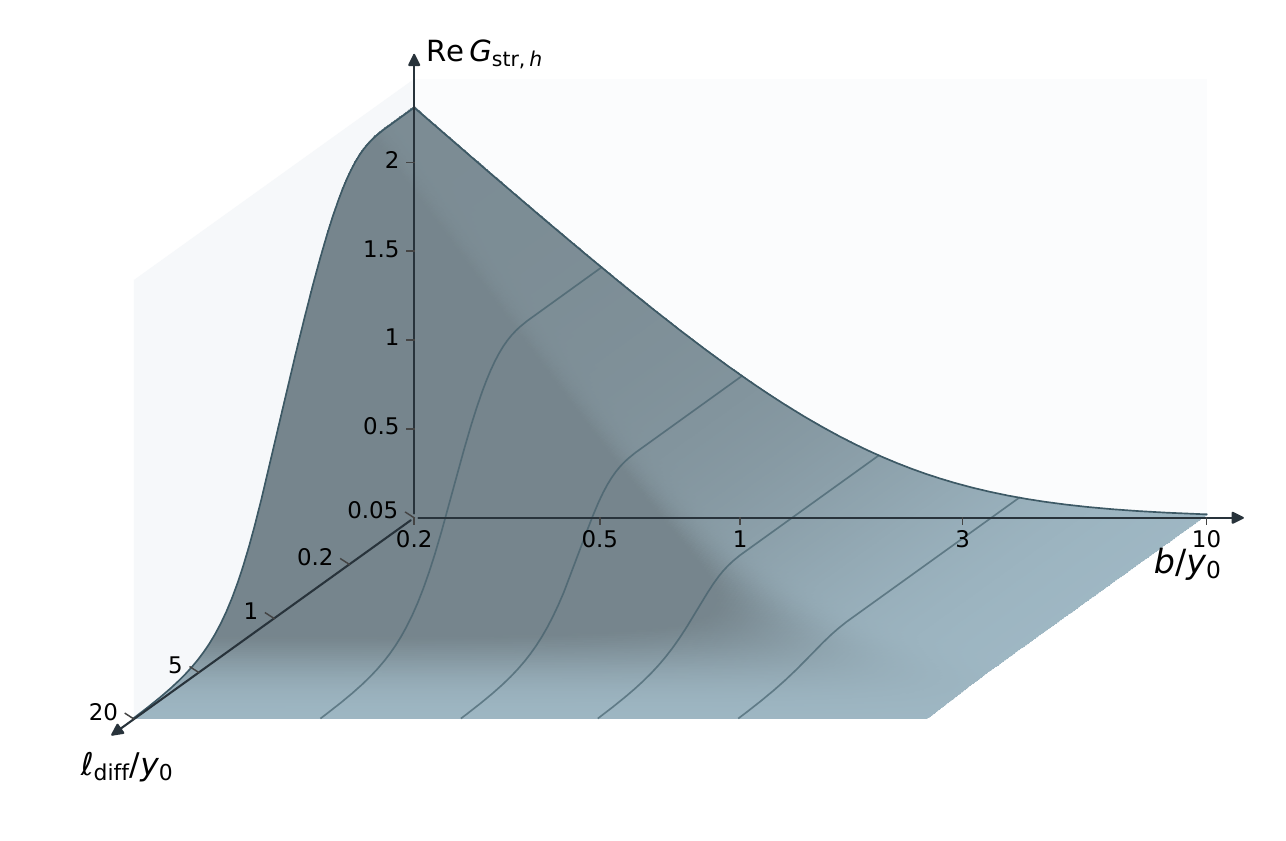}
 \caption{Real part of the interaction Green function from
 \eqref{eq:b-regge-positive-phases} at $y_1=y_2=y_0$.
 The $b/y_0$ and $\ell_{\rm diff}/y_0$ axes use logarithmic spacing.}
 \label{fig:b-regge-real-surface}
\end{figure}

\begin{figure}[H]
 \centering
 \includegraphics[width=.94\textwidth]{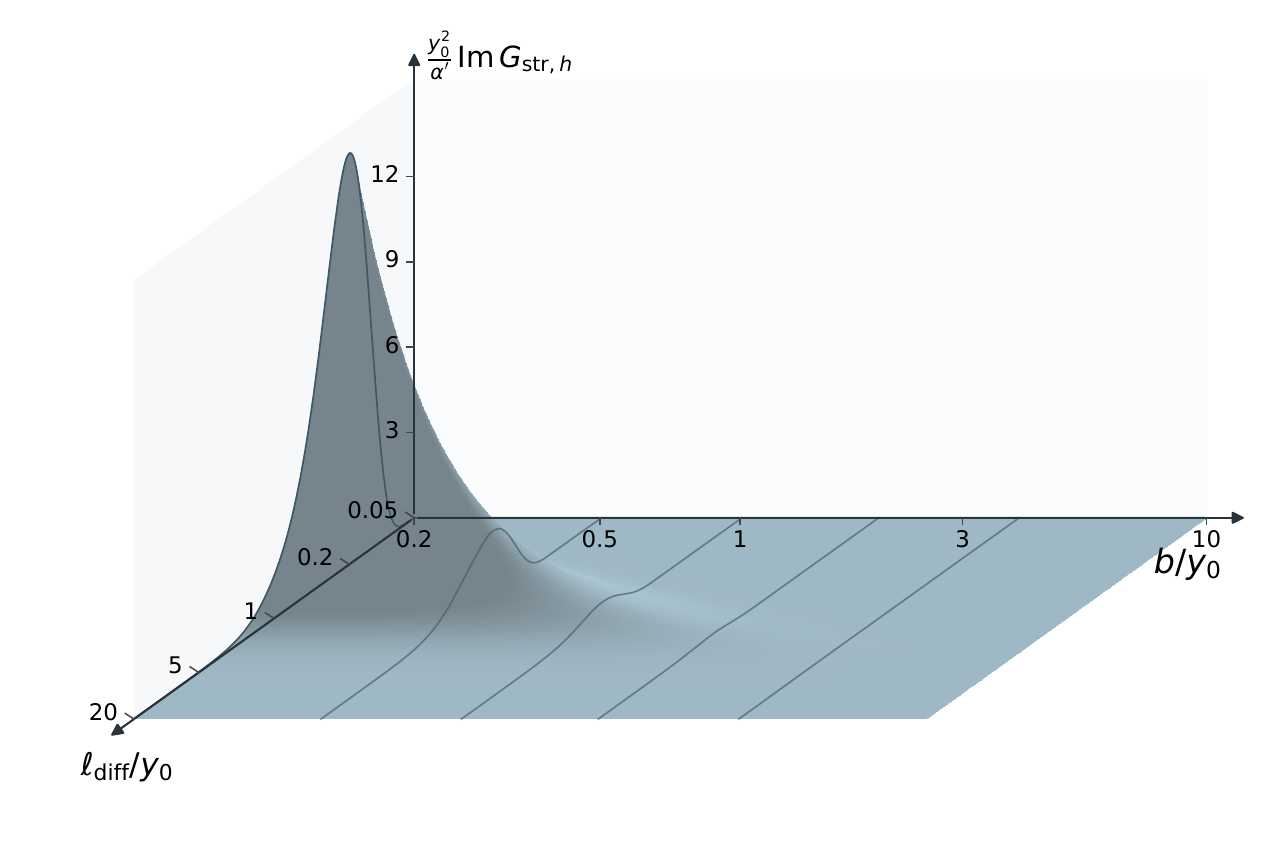}
 \caption{Leading imaginary part of the interaction Green function
 from \eqref{eq:b-regge-positive-phases}. The surface uses the same
 coordinates and projection as Figure~\ref{fig:b-regge-real-surface},
 and the vertical axis is
 $[y_0^2/\alpha']\operatorname{Im}G_{\rm str,h}$.}
 \label{fig:b-regge-abs-surface}
\end{figure}

The long-diffusion regime in \eqref{eq:b-regge-short-diffusion}
has only power-law suppression in $\ell_{\rm diff}$. Half-space
restricts one transverse direction to a half-line and leaves
the transverse spectrum gapless. By contrast, the AdS$_4$ shock
Green function decays exponentially with geodesic separation
\cite{Cornalba:2007zb}, and the gap of its transverse
operator also gives exponential suppression in
$\ell_{\rm diff}^2/R^2$, as in
\eqref{eq:b-ads-three-regimes}.
Figures~\ref{fig:b-regge-real-surface} and
\ref{fig:b-regge-abs-surface} show three-dimensional plots of
the real and imaginary parts of $G_{\rm str,h}$ against $b$ and
$\ell_{\rm diff}$ at fixed $y_0$. The real part of the Green function in
Figure~\ref{fig:b-regge-real-surface} decreases as the diffusion
length grows.

For a shock with fixed source momentum and a narrow probe
packet of energy $E$, so $s\propto E$, the group delay is the
energy derivative of the phase $2\operatorname{Re}\delta$
\cite{DiVecchia:2023frv}
\begin{equation}
 \Delta T=2\frac{\partial s}{\partial E}
              \partial_s\operatorname{Re}\delta_{\rm str,h}
 \simeq2G_N\frac{\partial s}{\partial E}
 \left[\operatorname{Re}G_{\rm str,h}
             -\frac2\pi\operatorname{Im}G_{\rm str,h}\right].
 \label{eq:b-regge-group-delay}
\end{equation}
For $\log(\alpha's/4)\gg1$, the delay is controlled by
$\operatorname{Re}G_{\rm str,h}$.
The GR limit gives the usual Shapiro delay.
When $\ell_{\rm diff}\ll b$, the transverse string interaction
profile does not appreciably overlap the other trajectory, and
the absorptive contribution is Gaussian suppressed. Once the
profile reaches $b$, string corrections become appreciable.
The positive absorptive part accompanies a decrease of the real
Green function through \eqref{eq:b-regge-real-imag-derivative}.
Further diffusion beyond $y_0$ drives this Green function
towards zero.

If this regime is reached before black-hole formation while
the weak-field eikonal description remains valid, $s\propto E$
makes the leading delay proportional to
$G_N\operatorname{Re}G_{\rm str,h}$ times the fixed source energy
scale. For a source of mass $M$, the gravitational prefactor is
of order $G_NM$. A high-energy narrow packet then has a smaller
delay than a low-energy one. Its peak approaches the arrival
time of free null propagation but remains delayed in the
stated large-log approximation. Note that the velocity discussed here is the packet peak's group
velocity, not the wavefront velocity.

The imaginary part has a different dependence on the diffusion
length. At any fixed $b>0$, $\operatorname{Im}G_{\rm str,h}
=\operatorname{Im}\delta_{\rm str,h}/(G_Ns)$ first rises from
Gaussian suppression and then falls, with a single maximum at
$b/\sqrt2<\ell_{\rm diff}<b$. For $b\ll y_0$, the peak approaches
$\ell_{\rm diff}=b$. At fixed diffusion length, it decreases
monotonically with $b$, with a Gaussian tail outside the
interaction width.
For $b\ll y_0$, the scattering separation lies below the
effective IR length. The intermediate region
$b\ll\ell_{\rm diff}\ll y_0$ in
\eqref{eq:b-regge-short-diffusion} has neither the short-diffusion
Gaussian suppression nor the additional suppression from the
wall, although the imaginary part still decreases as
$1/\ell_{\rm diff}^2$. This is the region of appreciable
absorption associated with the softening of the real part.
For $y_0\ll b$, no such intermediate window exists, as in
\eqref{eq:b-regge-broad-diffusion}. Once
$\max(b,y_0)\ll\ell_{\rm diff}$, the wall gives the common
long-diffusion falloff $2\pi\alpha'y_0^2/\ell_{\rm diff}^4$.

The $AdS_4$ shockwave provides a direct comparison
\cite{Cornalba:2006xk,Cornalba:2007zb}. Its transverse space
is $H_2$, and the transverse shockwave equation contains the operator
$-\Delta_{H_2}+2/R^2$. The gravitational shock-wave Green function
$G_{\rm AdS}(b)$ recovers the flat-space logarithm
$\log(2R/b)$ for $b\ll R$ and falls as $\exp(-2b/R)$ for
$R\ll b$, where $b$ is the geodesic separation on $H_2$.
Applying the same transverse diffusion in the Regge limit for
$\alpha'/R^2\ll1$ gives three regimes at fixed $b\ll R$, as derived in
Appendix~\ref{app:regge-kernel-checks}.
\begin{equation}
 \frac{\operatorname{Re}\delta_{\rm str,AdS}
              +\ii\operatorname{Im}\delta_{\rm str,AdS}}{G_Ns}
 \simeq\begin{cases}
 \displaystyle\log\frac{2R}{b}
    +\ii\frac{\pi\alpha'}{2\ell_{\rm diff}^2}e^{-b^2/\ell_{\rm diff}^2},
      &\ell_{\rm diff}\ll b\ll R,\\[7pt]
 \displaystyle\log\frac{2R}{\ell_{\rm diff}}
    +\ii\frac{\pi\alpha'}{2\ell_{\rm diff}^2},
      &b\ll\ell_{\rm diff}\ll R,\\[7pt]
 \displaystyle\frac{8\pi^{3/2}R^3}{9\ell_{\rm diff}^3}
    e^{-9\ell_{\rm diff}^2/(16R^2)}
    \bigl(1+\ii\frac{9\pi\alpha'}{16R^2}\bigr),
      &b\ll R\ll\ell_{\rm diff}.
 \end{cases}
 \label{eq:b-ads-three-regimes}
\end{equation}
The first two regions have the same logarithmic and Gaussian
behavior as \eqref{eq:b-regge-short-diffusion}. At long diffusion
length, the half-space real and imaginary parts fall as
$\ell_{\rm diff}^{-2}$ and $\ell_{\rm diff}^{-4}$, while both
AdS parts contain the suppression factor
$\ell_{\rm diff}^{-3}\exp[-9\ell_{\rm diff}^2/(16R^2)]$.
This exponential suppression follows from the gap $9/(4R^2)$
in the transverse spectrum entering the AdS Green function.
Geometrically, the hyperbolic metric makes the transverse area
grow exponentially with distance, diluting the long-range Green
function and giving the transverse Laplacian a spectral gap.
In contrast, half-space reflection from a wall in the single $y$
direction cancels the direct and image contributions while leaving
the transverse spectrum gapless, hence the power-law tail in
\eqref{eq:b-regge-broad-diffusion}. The DN strip, in turn, has a
gap and therefore gives an analogous exponential factor
$\exp[-\pi^2\ell_{\rm diff}^2/(16Y_0^2)]$.
Besides the $R\to0$ limit of the AdS$_4$--flat gluing, this
half-space suppression can also be understood directly through
the Dirichlet boundary condition. Analogously to the random-walk
picture of string diffusion in AdS/QCD
\cite{Brower:2006ea,Brower:2010wf},
the transverse diffusion here can be viewed as a random walk
with an absorbing Dirichlet boundary. By the reflection principle,
the image term accounts for Brownian bridges that reach the wall,
so subtracting it leaves the positive propagation weight of paths
that remain in the half-space \cite{Serafin:2022Dirichlet}.
This absorbing-boundary picture applies to the auxiliary heat kernel
and does not imply loss of physical scattering probability at the reflecting wall.

This extended response can be traced directly to the $s$-channel
resonances. The massive $s$-channel poles of \eqref{eq:b-string-vs}
lie at $s=4n/\alpha'$, with $n=1,2,\ldots$.
The residue $R_n(t)$, defined as the coefficient of
$(4n/\alpha'-s)^{-1}$, sums finitely many spin-$J$ contributions
at each mass level in the chosen helicity channel. Each such
contribution is proportional to $P_J(1+\alpha't/(2n))$, whose
$t^r$ term formally Fourier transforms into
$(\nabla_b^2)^r\delta^{(2)}(\boldsymbol b)$.
Normalizing the exact residue by its value at $t=0$, and then taking
the leading-logarithmic limit at $|\alpha't|\ll1$ and $\log n\gg1$,
gives
\begin{equation}
\begin{aligned}
 \frac{R_n(t)}{R_n(0)}
 &=\prod_{k=1}^{n-1}\biggl(1+\frac{\alpha't}{4k}\biggr)^2\\
 &\simeq\sum_{r=0}^{\infty}
 \frac{\bigl[(\alpha't/2)\log n\bigr]^r}{r!}
 =e^{\ell_{\rm diff}^2t/4}.
\end{aligned}
 \label{eq:b-resonance-regge-profile}
\end{equation}
The smooth approximation above includes energy averaging over nearby
poles. When $|\alpha't\log n|/2\ll1$, successive powers decrease,
so a few local contact terms suffice.
When $|\alpha't\log n|/2$ is a finite number much larger than one,
the series still converges because $r!$ eventually outgrows the
numerator's powers, but a low-order contact expansion is no longer
controlled. The full Fourier transform is a Gaussian in impact parameter
with width $\ell_{\rm diff}$ rather than a local contact term.
This width grows with $s$ in the weak-gravity Regge regime.
The averaged $s$-channel poles reproduce the Regge absorptive part,
realizing resonance--Regge duality \cite{DiVecchia:2023frv,Haring:2024wyz}.

The wall correction to local EFT data from a gapped sector is
instead exponentially small for packets localized far from the wall.
For a representative exchange of mass $m\geq\Lambda_{\rm gap}$,
the transverse Green function at equal heights $y_0$ is
\begin{equation}
 2\pi G_{h,m}^{0}(b,y_0,y_0)
 =K_0(mb)-K_0\!\left(m\sqrt{b^2+4y_0^2}\right).
 \label{eq:b-gap-image-tail}
\end{equation}
For $my_0\gg1$, the second term is exponentially suppressed by
$e^{-2my_0}$ uniformly in $b$, since its distance is at least $2y_0$
and $K_0$ has an exponential tail.
If the external-wavefunction overlap is concentrated near height $y_0$
with $\Lambda_{\rm gap}y_0\gg1$, convolution with this Green function
retains the exponential suppression of the image correction.
The geometry can therefore control the gapless gravitational response
while probing massive-sector EFT data with exponentially small
boundary corrections.

Dressed-state scattering that modifies only the soft-graviton sector
leaves the tree-level UV resonance spectrum unchanged.
At leading conservative eikonal order, only the FK phase factor
contributes to the leading dressing correction, as discussed around
\eqref{eq:transverse-resolution-validity}.
For purely gravitational scattering at $b\ll\omega^{-1}$, its effect
on the eikonal phase is, to logarithmic accuracy,
\begin{equation}
 \frac{\delta(s,b)}{G_Ns}\simeq\log\frac{1}{b\lambda_{\rm IR}}
 \quad\longmapsto\quad
 \frac{\delta_{\rm dr}(s,b)}{G_Ns}\simeq\log\frac{1}{b\omega}.
 \label{eq:b-soft-phase-shift}
\end{equation}
This shifts the low-energy moment in the 2SDR while leaving the
tree-level $s$-channel discontinuity unchanged. The same Regge
diffusion therefore continues to decrease the real profile, now
with the divergent IR constant removed. It becomes negative
when $\ell_{\rm diff}$ reaches order $\omega^{-1}$, whereas the
half-space and AdS profiles in \eqref{eq:b-regge-short-diffusion}
and \eqref{eq:b-ads-three-regimes} approach zero from above.
If eikonal exponentiation remains valid, this IR-finite result is
obtained by dividing the upper and lower real-axis boundary values
of the exclusive $S$ matrix by mutually conjugate soft phase factors.
This is not a common overall factor of the connected amplitude
on both edges.
At loop level, the modified tree input also changes the eikonal
ladder diagrams and hence the contribution of their two-particle
cuts to the discontinuity.

String theory illustrates how summing a UV spectrum produces a
long-range absorptive profile and how the geometry modifies its
contribution. For other UV completions satisfying the boundary conditions
and $S$-matrix assumptions of Sec.~\ref{sec:halfspace-s-matrix-properties},
the same optical theorem and 2SDR apply to the chosen external states
without specifying the intermediate UV spectrum. The geometry
consistently constrains both low-energy graviton exchange and the
long-range UV spectral contributions.

\section{Conclusion}
\label{sec:conclusion}

We have obtained an infrared-finite dispersive bound on a local four-scalar EFT interaction in a
scattering problem with a flat bulk and specified long-distance geometry. The construction uses the
S matrix between asymptotic states satisfying the boundary conditions, with correlated external
wavefunctions that localize the scattering region. Its positivity is supported by the construction
of positive state densities in physical channels, numerical tests of the common smearing functional,
and the analytic and asymptotic checks in Section~\ref{sec:halfspace-positive-functionals}. With the
dimensionless parameters specifying the wave-packet profiles held fixed, the gravitational allowance
in the EFT bound is finite and of order $G_N\Lambda_{\rm EFT}^2$. This gives a constraint on local
bulk EFT coefficients in four dimensions while the long-range gravitational contribution is modified
by the specified spacetime geometry.

The distinction between local interactions and long-range propagation is central to this result. A
local EFT vertex is probed through the overlap of its external wavefunctions. When this overlap is
concentrated in the flat interior, the original bulk Wilson coefficients enter calculable matrix
elements, while the exponentially suppressed overlap tails limit sensitivity to the interface with
the curved exterior. Changing the external states then changes the local contribution to the
dispersion relation through these calculable overlap integrals. Graviton exchange connects separated
vertices and remains sensitive to the exterior geometry. Its contribution must therefore be
evaluated with the complete graviton propagator satisfying the boundary conditions, even when both
vertices lie well inside the flat region.

The ultraviolet analysis shows how this distinction extends to the spectral side of the dispersion
relation. Even when the local bulk vertices remain unchanged, the intermediate states must satisfy
the same geometric boundary conditions as the low-energy calculation, which changes their spectral
contributions and their scattering with the external wave packets. Since high-energy absorptive
contributions can extend over long transverse distances, consistency concerns the complete spectral
integral. Removing an infrared-divergent contribution from the low-energy amplitude alone does not
establish an infrared-finite sum rule. The corresponding ultraviolet spectral integral must be
controlled as well. This is the ultraviolet consistency requirement accompanying infrared-finite
bounds from AdS correlators or from scattering geometries with asymptotic states.

The string calculation gives an explicit realization of this requirement. In the Regge description
studied here, half-space reflection produces power-law suppression at large diffusion length, making
the ultraviolet tail of the twice-subtracted spectral integral convergent. The spectral gaps of
AdS$_4$ and the DN strip instead give exponential suppression at sufficiently large diffusion
length. Moving the scattering region far from the half-space wall recovers both the low-energy
infrared sensitivity and the long-range spectral contribution without changing the spacetime
background. The microscopic resonance degrees of freedom present in flat space remain in the
half-space spectrum, whereas their contributions to half-space scattering observables change. For
other candidate UV completions, such as asymptotic safety \cite{Reuter:1996cp,Bonanno:2020bil} and
loop quantum gravity \cite{Ashtekar:2004eh,Engle:2007wy}, a corresponding application would require
calculating the ultraviolet absorptive response in the chosen geometry and relating it to the local
EFT through a dispersion relation defined by the same external states.

The explicit EFT bound in Section~\ref{sec:sum-rules-examples} provides a starting point for further
optimization. The correlated wave packets used there localize the scattering region near the wall
while keeping the local vertex overlaps inside the flat region. Once the wall distance exceeds the
packet's normal position-space width, increasing that distance at fixed packet widths changes the
leading gravitational contribution only logarithmically in the regime of
\eqref{eq:tree-gravity-leading-log}. The spatial profiles and their distance from the wall therefore
provide adjustable degrees of freedom for improving the relative contributions of the local EFT
interaction and graviton exchange. A natural next step is to optimize these correlated wave-packet
profiles and positions while preserving positivity in both the $s$- and $u$-channels, seeking the
strongest bound on the local EFT coefficients.

Photon scattering offers a direct physical application. The state and reflection conventions
developed here provide a starting point for polarization-dependent smearing functionals that probe
$F^4$ contact interactions and effective graviton--photon--photon couplings. These three-point
couplings modify the vertices connecting the external photons to the graviton exchanged in the
$t$-channel. In a Weyl-tensor basis, they arise by expanding
$W_{\mu\nu\rho\sigma}F^{\mu\nu}F^{\rho\sigma}$, schematically written as $RFF$, in metric
perturbations about flat spacetime. Their coefficients enter the combinations governing corrections
to charged-black-hole extremality and the associated inequalities motivated by the weak gravity
conjecture \cite{Henriksson:2022oeu}. The corresponding positive smearing functional must first be
constructed and tested. Its finite geometric gravitational correction could then be computed and
optimized in Wilson-coefficient bounds on departures from those inequalities, replacing the
arbitrary infrared logarithm.

Species bounds provide another application, following the dispersive approach of
\cite{Caron-Huot:2024lbf}, where positivity of calculable matter-loop contributions to graviton
scattering constrains the number of light species relative to the gravitational cutoff. The one-loop
amplitudes and their physical cuts encode the masses, spins and couplings of the matter running in
the loop. Recomputing these bulk one-loop amplitudes with half-space external states and propagators
satisfying the same boundary conditions would connect this information to infrared-finite EFT
bounds. Comparing graviton scattering with electromagnetic matter-loop contributions to photon
scattering could distinguish all gravitationally coupled species from their electrically charged
subset, while axion channels could probe matter responsible for anomalous couplings. These
comparisons would yield species bounds weighted by the relevant couplings at one-loop order. Each
application requires a positive ultraviolet absorptive contribution for the chosen smearing
functional and a calculation of the corresponding low-energy bulk one-loop amplitude, which the
dispersion relation then relates.

The joined AdS--flat geometry also suggests a way to study the relation between EFT bounds from bulk
S-matrix scattering and from AdS correlators \cite{Caron-Huot:2021enk,Fitzpatrick:2011jn}, together
with a different route to the flat-space limit. Instead of sending the AdS radius to infinity, one
can keep the exterior radius fixed in Fig.~\ref{fig:halfspace-beam-completion}, increase the length
of the inserted flat interval, and place the wave-packet centers in this exactly flat region. At
each finite interval length, the exterior AdS geometry regulates long-distance propagation while the
size of the exactly flat interaction region can be increased independently of the exterior curvature
scale. Following the scattering states, soft sector and dispersive data along this family of
geometries may clarify how flat-space Fock states are recovered in this limit and how the
information extracted from the flat-space S matrix is related to that from AdS correlators, without
requiring the exterior curvature to vanish.

The construction could be extended to graviton and general matter scattering, including higher-spin
interactions and new-particle spectra. Relating the resulting Wilson-coefficient bounds to
observables may then constrain particle phenomenology, black-hole physics, and electromagnetic and
gravitational-wave propagation in astrophysical and cosmological settings.

\acknowledgments
I thank Ashoke Sen, Ahmadullah Zahed, and Zimo Sun for helpful discussions.
Z.D. is supported by the European Research Council, grant agreement n. 101039756.
The author developed the research, used ChatGPT as an auxiliary tool for discussion and manuscript
polishing, and takes full responsibility for the final content.

\appendix
\crefname{appendix}{App.}{Apps.}
\Crefname{appendix}{App.}{Apps.}
\section{Conventions and notation}
\label{app:conventions}

We use the mostly-minus metric
$\eta_{\mu\nu}=\operatorname{diag}(1,-1,-1,-1)$ and take all four hard
momenta to be incoming.
\begin{equation}
 p_i^2=0,\qquad
 \sum_{i=1}^4p_i=0,\qquad
 s=(p_1+p_2)^2,\quad
 t=(p_1+p_3)^2,\quad
 u=(p_1+p_4)^2,\quad
 s+t+u=0 .
 \label{eq:conventions-mandelstam}
\end{equation}
\noindent
The physical incoming momenta are $p_1,p_2$, while the physical outgoing
momenta are $-p_3,-p_4$.

For the frame dependence of Faddeev--Kulish
dressing, a unit future-timelike vector $n$ specifies the detector frame
whose scattering resolution the dressing describes. We write
$|\boldsymbol k|_n^2=(n\cdot k)^2-k^2$ for the spatial momentum squared
in this frame. In particular, $|\boldsymbol k|_n=n\cdot k$ for a
future-directed null momentum.

We list initial and final helicity labels in that order,
$d^J_{h_ih_f}(\theta)
=\langle Jh_f|\e^{-\ii\theta J_y}|Jh_i\rangle$.
This fixes the rotation sign in \eqref{eq:eikonal-common-weight}.

\section{Wave-packet spreading and infrared logarithms}
\label{app:beam-log-estimates}

The estimate of $\log W$ in \eqref{eq:beam-logarithmic-scales}
describes gravitational corrections at arbitrary loop order in the
leading-soft limit. Here we retain only its infrared-divergent terms.
These are absent for frozen wave packets at every loop order, but can
arise when the packets spread. Tree-level exchange is strictly
infrared finite in both cases, as discussed in
Sec.~\ref{sec:sum-rules-examples}.
The position-space width in \eqref{eq:halfspace-diffracting-packet} grows towards both
future and past infinity, so the fixed-path image cancellation cannot
be used throughout the Wilson-line integral. We explain below how
the finite DN strip cuts off the soft and collinear regions.
The reflected pair and self-image coefficients are given explicitly
in \eqref{eq:beam-soft-collinear-coefficient}.

For the common DN spectrum in \eqref{eq:far-dn-spectrum}, the exchanged
mode has wall-parallel mass at least $\pi/(2Y_0)$. Its propagator is
therefore bounded in the zero-energy region. Above this gap, the
logarithmic part of the soft integral is represented by
\begin{equation}
 \int_{\pi/(2Y_0)}^{\Lambda_{\rm loop}}\frac{\dd\ell}{\ell}
 =\log(Y_0\Lambda_{\rm loop})+\log\frac{2}{\pi}.
 \label{eq:beam-soft-log-app}
\end{equation}
\noindent
The discrete mode sum determines the finite terms at the lower endpoint.
The upper endpoint is the soft matching scale, for which the main text
uses the characteristic transfer $q_{\max}$. Thus the soft logarithm
has size $\log(Y_0q_{\max})$.

The collinear cutoff follows from the same spectrum, but is not
numerically identical to the soft cutoff. In wall-parallel kinematics,
the singular configuration requires the external, emitted and
intermediate on-shell modes to have the same wall-parallel velocity.
Their effective $(2+1)$-dimensional masses would then obey
$k_a=k_b+k_c$, whereas
\begin{equation}
 k_a-k_b-k_c=\frac{\pi}{Y_0}
 \left(a-b-c-\frac12\right)\neq0,
 \qquad a,b,c=0,1,\ldots .
 \label{eq:beam-dn-collinear-mismatch}
\end{equation}
\noindent
The virtualities that cut off the collinear angular integral are
therefore set by $Y_0^{-2}$, with dimensionless factors depending on
the quantized normal momenta and the wall-parallel kinematics,
including the energy sharing. The collinear angular measure
$\dd\theta/\theta$ supplies the second logarithm. At the reference
matching scale $q_{\max}$, its enhanced part is
$\log(q_{\max}/\mu_{\rm col})\sim\log(Y_0q_{\max})$.
The common DN spectrum thus removes the soft and collinear infrared
divergences of the massless fields considered here.

Multiplying these logarithmic enhancements by the normal-momentum
coefficient in \eqref{eq:beam-soft-collinear-coefficient}, and then
contracting the external normal-momentum profiles of width $\sigma$, gives the
$G_N\sigma^2\log^2(Y_0q_{\max})$ size in
\eqref{eq:beam-logarithmic-scales}. Both infrared cutoffs are supplied
by the strip spectrum, while their dimensionless factors and the
overall coefficient remain in the mode sums and packet overlaps.
This gives the double-logarithmic scale estimate for the common
DN spectrum used here.

\section{Convolution and multiplication in smearing}
\label{sec:archive531-relative-state-envelopes}
\label{sec:relative-state-envelopes}

To implement the multiplicative construction described in
Sec.~\ref{sec:congruence-construction}, we place the restrictions
on the $y$ components of $\mathbf q,\mathbf q'$ inside the
momentum-conserving seed, while applying the linear pair-center shaking
in the initial and final congruence maps. In the pair variables of
\eqref{eq:scheme-II-s-pair-variables}, the Gaussian weight
\eqref{eq:common-su-rank-one-weight} factorizes as
\begin{equation}
 W(k_i)=\bigl[C_K(K)C_q(q_y)\bigr]^*
        \bigl[C_K(K')C_q(q_y')\bigr].
 \label{eq:relative-center-factorization}
\end{equation}
\noindent
The left conjugation specifies the phase convention while preserving
the original $W$. The center and relative profiles are
\begin{equation}
 C_K(K)=\frac{\e^{-K^2/(4\sigma^2)+\ii(y_1+y_2)K/2}}
                   {(\pi\sigma^2)^{1/4}},\qquad
 C_q(q_y)=\frac{\e^{-q_y^2/(4\sigma^2)+\ii(y_1-y_2)q_y/2}}
                    {(\pi\sigma^2)^{1/4}}.
 \label{eq:relative-center-profiles}
\end{equation}
\noindent
The pair-variable Jacobian is
$\dd k_1\dd k_2=\dd K\dd q_y/2$, and similarly on the outgoing side.
The two profiles are therefore normalized symmetrically with measures
$\dd K/\sqrt2$ and $\dd q_y/\sqrt2$, whose product is the original pair
measure. Their Fourier phases localize the pair center at
$(y_1+y_2)/2$ and the particle separation at $y_1-y_2$.

For given seed Mandelstam invariants, $K_{\text{pw-2}}$ gives
equal weight to common $SO(3)$ rotations of the two directions that
preserve their relative angle.
The envelopes reweight these orientations
through $q_y=\sqrt{s}\,\Omega_y$ and $q_y'=\sqrt{s}\,\Omega_y'$.
As $\sigma/\sqrt{s}\to0$, the directions concentrate near the $zx$ great
circle, leaving their common $SO(2)$ rotation at fixed angle.
At nonzero normal-momentum width $\sigma$, corresponding to a position-space extent of order
$1/\sigma$ along $y$, the envelopes act on the full spherical states, giving
\begin{equation}
 K_{\text{pw-2}}'(s,\Omega,\Omega')
 =\sum_{J,m}a_J(s)
 [\e^{\ii P_s\cdot X}C_q(q_y)Y_{Jm}(\Omega)]^*
 [\e^{\ii P_s\cdot X}C_q(q_y')Y_{Jm}(\Omega')].
 \label{eq:scheme-II-middle-kernel}
\end{equation}
\noindent
This is a wavefunction expansion with the original nonnegative
coefficients $a_J$. The enveloped states need not be orthogonal or
normalized, so no new partial-wave diagonalization is required, and
$a_J$ need not be angular-momentum eigenvalues of the enveloped kernel.

The seed $K_{\text{pw-2}}'$ has $K=K'=0$ but retains $q_y,q_y'$ and hence
nonzero $k_t,k_u$. Multiplication alone leaves it on the conserving
slice $k_0=0$. The linear map $V_s'$ shakes the full two-particle states,
keeping $P_{s\parallel}$, $q_y$ and the oriented relative-motion plane
within the wall-parallel $(t,z,x)$ space fixed on the incoming side,
with analogous outgoing constraints. It generates independent $K,K'$
components and adjusts the parallel momenta to keep each leg on shell.
This paired action preserves positivity while producing nonconserving
total normal momentum. Applying the corresponding boundary projection
then supplies a positive kernel for the half-space or DN strip.
Nonzero $K$ singles out the normal direction and couples the center
motion to the relative normal component $q_y$. The shaking is therefore
convolution-like rather than a rotationally invariant convolution.
The exact reconstruction and invariant map of
$V_s'K_{\text{pw-2}}'(V_s')^\dagger$ are
\begin{equation}
 \begin{aligned}
 s^{\mathrm{sd}'}&=s_\parallel,\\
 t^{\mathrm{sd}'}&=-\frac{s_\parallel}{2}+\frac{(k_1-k_2)(k_4-k_3)}{2}
 +\frac{s_\parallel}
 {\sqrt{[s_\parallel-(k_1+k_2)^2][s_\parallel-(k_3+k_4)^2]}}
 \\
 &\qquad\times\left[t_\parallel+\frac{s_\parallel-\sum k_i^2}{2}
 +\frac{(k_1^2-k_2^2)(k_3^2-k_4^2)}{2s_\parallel}\right],\\
 u^{\mathrm{sd}'}&=-s^{\mathrm{sd}'}-t^{\mathrm{sd}'}.
 \end{aligned}
 \label{eq:relative-seed-exact-map}
\end{equation}
\noindent
The square root is continued from infinity with asymptotic value
$s_\parallel$, as in \eqref{eq:forward-sheet-infinity}.

To display its limiting behavior, consider
$k_i^2\sim\sigma^2\ll|s_h|$.
Retaining the leading and next-to-leading shaking terms and extracting
their common rescaling gives
\begin{equation}
 (s^{\mathrm{sd}'},t^{\mathrm{sd}'},u^{\mathrm{sd}'})
 =\frac{s_h+k_s^2}{s_h-k_0^2}
 \Bigl(s_h-k_0^2,\,
 t_h+\mathcal O(\tfrac{k^4}{s_h}),\,
 u_h-\mathcal O(\tfrac{k^4}{s_h})\Bigr).
 \label{eq:relative-seed-factorized-map}
\end{equation}
\noindent
Here $s_h=s_\parallel-k_s^2$, $t_h=t_\parallel-k_t^2$ and
$u_h=u_\parallel-k_u^2$, with $s_h+t_h+u_h=k_0^2$.

Using the seed energy and transfer as the outer integration variables,
and moving the boundary projectors to the half-space amplitude as in
\eqref{eq:spin-projector}, gives the smeared dispersion relation in
the physical $s$-channel regime,
\begin{equation}
 \begin{gathered}
 (V_s'K_{\text{pw-2}}'(V_s')^\dagger)[\M_h]
 =-\frac{1}{4\pi^2}\int\frac{\dd s^{\mathrm{sd}'}}{2\pi\ii}
 \int_{-t_{\max}}^0\dd t^{\mathrm{sd}'}\\
 \int_{\mathcal D_s}\prod\limits_{i=1}^4\dd k_i\,
 \frac{W(k_i)\phi(t^{\mathrm{sd}'})\M_h(s_\parallel,t_\parallel;k_i)}
 {s^{\mathrm{sd}'}u^{\mathrm{sd}'}
 \sqrt{(t^{\mathrm{sd}'}+k_t^2)u^{\mathrm{sd}'}
       +t^{\mathrm{sd}'}k_u^2}}.
 \end{gathered}
 \label{eq:relative-seed-sixfold-nlo}
\end{equation}
\noindent
The Gaussian integrations use the $-10\sigma$ to $10\sigma$ interval
as in \eqref{eq:planar-seed-sixfold-nlo}, restricted to the physical
spherical domain $\mathcal D_s$, where the radicand in
\eqref{eq:relative-seed-sixfold-nlo} is positive and real on the physical
cut. At leading order in \eqref{eq:relative-seed-factorized-map}, this
transfer support becomes $t_\parallel=t^{\mathrm{sd}'}+k_t^2\leq0$.
Inverting \eqref{eq:relative-seed-exact-map} on the physical sheet and
substituting the resulting half-space amplitude arguments gives this
smearing action.
It preserves strict $s$-channel positivity. As for the
previous construction, its crossed discontinuity need not be positive
or arise solely from physical intermediate states.

The measure in \eqref{eq:relative-seed-sixfold-nlo} shows why this
construction is less convenient for a crossing-symmetric prescription.
Using the leading terms of \eqref{eq:relative-seed-factorized-map},
the measure and weight of the transfer integral are proportional to
$\dd t_h\,\phi(t_h)/\sqrt{-t_\parallel}$ times other smooth factors.
This couples parallel transfer with normal momentum, rather than
giving a simple $\dd q_\parallel$ weight as in
\eqref{eq:planar-seed-sixfold-nlo}.

This difference favors the common
$\dd q_\parallel\,\phi(-q_\parallel^2)$ prescription used in
Sec.~\ref{sec:general-halfspace-dispersion}.
As the mixed-representation calculation there shows, the parallel-transfer
integral stays in momentum space while the normal mode overlaps are evaluated
in position space, for both local contact interactions and tree-level
graviton exchange.
The $\phi(t_h)$ weight and $\dd q_h$ measure instead couple the
parallel and normal components, losing that separation.

\section{Shock waves and Regge diffusion in AdS$_4$}
\label{app:regge-kernel-checks}

For the AdS$_4$ comparison in Sec.~\ref{sec:extended-uv-states},
$b$ is the geodesic distance on
$H_2$ of radius $R$. General relativity gives the shock-wave equation
$(-\Delta_{H_2}+2/R^2)G_{\rm AdS}=\delta_{H_2}$, with the
delta function normalized using the hyperbolic area measure
\cite{Cornalba:2006xk,Cornalba:2007zb}. Its exact form is
\begin{equation}
 G_{\rm AdS}(b)
 =\frac{Q_1(\cosh(b/R))}{2\pi}
 =\frac{1}{2\pi}\left[
 \cosh\frac bR\log\coth\frac{b}{2R}-1\right].
 \label{eq:b-ads-einstein-green}
\end{equation}
\noindent
Here $Q_1(z)$ is the Legendre function of the second kind.
Thus $2\pi G_{\rm AdS}$ approaches $\log(2R/b)-1$ at $b\ll R$
and $(4/3)e^{-2b/R}$ at $R\ll b$.

Writing the reciprocal eigenvalues of the shock-wave operator as Laplace integrals
gives the Schwinger proper-time representation of $G_{\rm AdS}$.
The integration variable $\tau$ is an auxiliary proper time,
while $b$ remains the transverse separation.
Here $H_{H_2}^{\tau}$ is the heat kernel on the transverse $H_2$ of AdS,
generated by its Laplacian. We define
$\lambda_\nu^{\rm AdS}\equiv(\nu^2+9/4)/R^2$ as the eigenvalue
of the full shock-wave operator $-\Delta_{H_2}+2/R^2$.
Regge diffusion multiplies each propagator spectral component by
$\exp[-\lambda_\nu^{\rm AdS}\ell_{\rm diff}^2/4]$,
where $\ell_{\rm diff}$ follows \eqref{eq:b-string-regge}.
This shifts the lower endpoint of the proper-time integral to
$\ell_{\rm diff}^2/4$, with the curvature term supplying the factor
$e^{-2\tau/R^2}$ below,
\begin{equation}
\begin{aligned}
 G_{\rm AdS}^{\ell_{\rm diff}^2/4}(b)
 &=\int_{\ell_{\rm diff}^2/4}^{\infty}\!\dd\tau\,
 e^{-2\tau/R^2}H_{H_2}^{\tau}(b)>0,\\
 \partial_{\ell_{\rm diff}^2}G_{\rm AdS}^{\ell_{\rm diff}^2/4}(b)
 &=\frac{s}{2\alpha'}\,\partial_s
 G_{\rm AdS}^{\ell_{\rm diff}^2/4}(b)<0.
\end{aligned}
 \label{eq:b-ads-regge-diffusion}
\end{equation}
\noindent
The $s$ derivative is taken at fixed $b$, $R$ and $\alpha'$.
Continuing to the complex diffusion parameter $Y_s$ in
\eqref{eq:b-string-regge}, the Green function obeys the same leading
relation between its imaginary part and the logarithmic energy derivative
of its real part as \eqref{eq:b-regge-real-imag-derivative},
for $\log(\alpha's/4)\gg1$ and $\alpha'/R^2\ll1$.
At this order, its real part decreases towards zero from above,
as in the half-space profile of Figure~\ref{fig:b-regge-real-surface}.
For $b\ll R$, the imaginary part is appreciable in the window
$b\lesssim\ell_{\rm diff}\ll R$, between the short-diffusion
Gaussian suppression and the long-diffusion curvature suppression.
This is analogous to the absorption window illustrated in
Figure~\ref{fig:b-regge-abs-surface}.

The heat kernel is normalized with the hyperbolic area measure,
and the eikonal phase is
$\delta_{\rm str,AdS}/(G_Ns)=2\pi G_{\rm AdS}^{Y_s/4}$.
Expanding the heat kernel with spectral measure $\nu\tanh(\pi\nu)$
and performing the proper-time integral gives
\begin{equation}
 2\pi G_{\rm AdS}^{\ell_{\rm diff}^2/4}(b)
 =\int_0^\infty\!\dd\nu\,
 \frac{\nu\tanh(\pi\nu)}{\nu^2+9/4}
 P_{-1/2+\ii\nu}\!\left(\cosh\frac bR\right)
 \exp\!\left[-\frac{\ell_{\rm diff}^2}{4R^2}
                   \left(\nu^2+\frac94\right)\right].
 \label{eq:b-ads-spectral-integral}
\end{equation}
\noindent
The $2/R^2$ shift in the shock-wave equation places the lower end
of this spectrum at
$9/(4R^2)$. At $b\ll R\ll\ell_{\rm diff}$, the integral is
dominated by $\nu\ll1$, where $\tanh(\pi\nu)\simeq\pi\nu$
and $P_{-1/2+\ii\nu}(\cosh(b/R))\simeq1$. Hence
\begin{equation}
\begin{aligned}
 2\pi G_{\rm AdS}^{\ell_{\rm diff}^2/4}(b)
 &\simeq\frac{4\pi}{9}
 e^{-9\ell_{\rm diff}^2/(16R^2)}
 \int_0^\infty\!\dd\nu\,\nu^2
 e^{-\nu^2\ell_{\rm diff}^2/(4R^2)}\\
 &=\frac{8\pi^{3/2}R^3}{9\ell_{\rm diff}^3}
 e^{-9\ell_{\rm diff}^2/(16R^2)},
 \qquad b\ll R\ll\ell_{\rm diff}.
\end{aligned}
 \label{eq:b-ads-long-diffusion-coefficient}
\end{equation}
\noindent
The relative corrections are of order $b^2/R^2$ and
$R^2/\ell_{\rm diff}^2$. For $R\ll\ell_{\rm diff}$, the corresponding
leading absorptive coefficient is the one in \eqref{eq:b-ads-three-regimes}.
These coefficients follow from this heat-kernel prescription
at $\alpha'/R^2\ll1$, rather than an exact complete AdS string
amplitude.

In the first two regimes of \eqref{eq:b-ads-three-regimes},
the real parts before dropping constants are
$\log(2R/b)-1$ and
$\log(2R/\ell_{\rm diff})-1+\gamma_E/2$, respectively.
The corresponding half-space expression in
\eqref{eq:b-regge-short-diffusion} is
$\log(2y_0/\ell_{\rm diff})+\gamma_E/2$ in its intermediate
regime. The main text omits these constants relative to the
large logarithms.

\begingroup
\footnotesize
\bibliographystyle{JHEP}
\bibliography{refs_merged}

@article{Henning:2015daa,
    author = "Henning, Brian and Lu, Xiaochuan and Melia, Tom and Murayama, Hitoshi",
    title = "{Hilbert series and operator bases with derivatives in effective field theories}",
    eprint = "1507.07240",
    archivePrefix = "arXiv",
    primaryClass = "hep-th",
    doi = "10.1007/s00220-015-2518-2",
    journal = "Commun. Math. Phys.",
    volume = "347",
    number = "2",
    pages = "363--388",
    year = "2016"
}

@article{Henning:2019enq,
    author = "Henning, Brian and Melia, Tom",
    title = "{Constructing effective field theories via their harmonics}",
    eprint = "1902.06754",
    archivePrefix = "arXiv",
    primaryClass = "hep-ph",
    doi = "10.1103/PhysRevD.100.016015",
    journal = "Phys. Rev. D",
    volume = "100",
    number = "1",
    pages = "016015",
    year = "2019"
}

@article{Dong:2021vxo,
    author = "Dong, Zi-Yu and Ma, Teng and Shu, Jing",
    title = "{Constructing on-shell operator basis for all masses and spins}",
    eprint = "2103.15837",
    archivePrefix = "arXiv",
    primaryClass = "hep-ph",
    doi = "10.1103/PhysRevD.107.L111901",
    journal = "Phys. Rev. D",
    volume = "107",
    number = "11",
    pages = "L111901",
    year = "2023"
}

@article{Dong:2022mcv,
    author = "Dong, Zi-Yu and Ma, Teng and Shu, Jing and Zheng, Yu-Hui",
    title = "{Constructing generic effective field theory for all masses and spins}",
    eprint = "2202.08350",
    archivePrefix = "arXiv",
    primaryClass = "hep-ph",
    doi = "10.1103/PhysRevD.106.116010",
    journal = "Phys. Rev. D",
    volume = "106",
    number = "11",
    pages = "116010",
    year = "2022"
}

@article{Harlander:2023ozs,
    author = "Harlander, Robert V. and Schaaf, Magnus C.",
    title = "{AutoEFT: Automated operator construction for effective field theories}",
    eprint = "2309.15783",
    archivePrefix = "arXiv",
    primaryClass = "hep-ph",
    reportNumber = "TTK-23-25; P3H-23-066",
    doi = "10.1016/j.cpc.2024.109198",
    journal = "Comput. Phys. Commun.",
    volume = "300",
    pages = "109198",
    year = "2024"
}

@article{Dong:2026swi,
    author = "Dong, Ziyu and Ma, Teng and Yang, Chengjie and Zhou, Zizheng",
    title = "{Static-Recoil Factorization in Heavy-Baryon Chiral EFTs}",
    eprint = "2607.12037",
    archivePrefix = "arXiv",
    primaryClass = "hep-ph",
    month = "7",
    year = "2026"
}

@article{Adams:2006sv,
    author = "Adams, Allan and Arkani-Hamed, Nima and Dubovsky, Sergei and Nicolis, Alberto and Rattazzi, Riccardo",
    title = "{Causality, analyticity and an IR obstruction to UV completion}",
    eprint = "hep-th/0602178",
    archivePrefix = "arXiv",
    reportNumber = "CERN-PH-TH-2006-033, HUTP-06-A0005",
    doi = "10.1088/1126-6708/2006/10/014",
    journal = "JHEP",
    volume = "10",
    pages = "014",
    year = "2006"
}

@article{Arkani-Hamed:2020blm,
    author = "Arkani-Hamed, Nima and Huang, Tzu-Chen and Huang, Yu-tin",
    title = "{The EFT-Hedron}",
    eprint = "2012.15849",
    archivePrefix = "arXiv",
    primaryClass = "hep-th",
    reportNumber = "NCTS-TH/2014, CALT-TH 2020-061",
    doi = "10.1007/JHEP05(2021)259",
    journal = "JHEP",
    volume = "05",
    pages = "259",
    year = "2021"
}

@article{Caron-Huot:2021rmr,
    author = "Caron-Huot, Simon and Mazac, Dalimil and Rastelli, Leonardo and Simmons-Duffin, David",
    title = "{Sharp boundaries for the swampland}",
    eprint = "2102.08951",
    archivePrefix = "arXiv",
    primaryClass = "hep-th",
    doi = "10.1007/JHEP07(2021)110",
    journal = "JHEP",
    volume = "07",
    pages = "110",
    year = "2021"
}

@article{Camanho:2014apa,
    author = "Camanho, Xian O. and Edelstein, Jose D. and Maldacena, Juan and Zhiboedov, Alexander",
    title = "{Causality Constraints on Corrections to the Graviton Three-Point Coupling}",
    eprint = "1407.5597",
    archivePrefix = "arXiv",
    primaryClass = "hep-th",
    doi = "10.1007/JHEP02(2016)020",
    journal = "JHEP",
    volume = "02",
    pages = "020",
    year = "2016"
}

@article{Caron-Huot:2022ugt,
    author = "Caron-Huot, Simon and Li, Yue-Zhou and Parra-Martinez, Julio and Simmons-Duffin, David",
    title = "{Causality constraints on corrections to Einstein gravity}",
    eprint = "2201.06602",
    archivePrefix = "arXiv",
    primaryClass = "hep-th",
    reportNumber = "CALT-TH 2021-003",
    doi = "10.1007/JHEP05(2023)122",
    journal = "JHEP",
    volume = "05",
    pages = "122",
    year = "2023"
}

@article{Henriksson:2022oeu,
    author = "Henriksson, Johan and McPeak, Brian and Russo, Francesco and Vichi, Alessandro",
    title = "{Bounding violations of the weak gravity conjecture}",
    eprint = "2203.08164",
    archivePrefix = "arXiv",
    primaryClass = "hep-th",
    doi = "10.1007/JHEP08(2022)184",
    journal = "JHEP",
    volume = "08",
    pages = "184",
    year = "2022"
}

@article{Remmen:2019cyz,
    author = "Remmen, Grant N. and Rodd, Nicholas L.",
    title = "{Consistency of the Standard Model Effective Field Theory}",
    eprint = "1908.09845",
    archivePrefix = "arXiv",
    primaryClass = "hep-ph",
    doi = "10.1007/JHEP12(2019)032",
    journal = "JHEP",
    volume = "12",
    pages = "032",
    year = "2019"
}

@article{Albert:2023jtd,
    author = "Albert, Jan and Rastelli, Leonardo",
    title = "{Bootstrapping pions at large N. Part II. Background gauge fields and the chiral anomaly}",
    eprint = "2307.01246",
    archivePrefix = "arXiv",
    primaryClass = "hep-th",
    reportNumber = "YITP-SB-2023-15",
    doi = "10.1007/JHEP09(2024)039",
    journal = "JHEP",
    volume = "09",
    pages = "039",
    year = "2024"
}

@article{Dong:2024omo,
    author = "Dong, Zi-Yu and Ma, Teng and Pomarol, Alex and Sciotti, Francesco",
    title = "{Bootstrapping the chiral-gravitational anomaly}",
    eprint = "2411.14422",
    archivePrefix = "arXiv",
    primaryClass = "hep-th",
    doi = "10.1007/JHEP05(2025)114",
    journal = "JHEP",
    volume = "05",
    pages = "114",
    year = "2025"
}

@article{Dong:2025dpy,
    author = "Dong, Ziyu and Jeong, Jaehoon and Pomarol, Alex",
    title = "{Causal bounds on EFTs with anomalies with a pseudoscalar, photons, and gravitons}",
    eprint = "2510.12138",
    archivePrefix = "arXiv",
    primaryClass = "hep-th",
    reportNumber = "KIAS-Q25016",
    doi = "10.1007/JHEP02(2026)102",
    journal = "JHEP",
    volume = "02",
    pages = "102",
    year = "2026"
}

@article{Jacob:1959at,
    author = "Jacob, M. and Wick, G. C.",
    title = "{On the General Theory of Collisions for Particles with Spin}",
    doi = "10.1006/aphy.2000.6022",
    journal = "Annals Phys.",
    volume = "7",
    pages = "404--428",
    year = "1959"
}

@article{deRham:2017avq,
    author = "de Rham, Claudia and Melville, Scott and Tolley, Andrew J. and Zhou, Shuang-Yong",
    title = "{Positivity bounds for scalar field theories}",
    eprint = "1702.06134",
    archivePrefix = "arXiv",
    primaryClass = "hep-th",
    doi = "10.1103/PhysRevD.96.081702",
    journal = "Phys. Rev. D",
    volume = "96",
    number = "8",
    pages = "081702",
    year = "2017"
}

@article{Tolley:2020gtv,
    author = "Tolley, Andrew J. and Wang, Zi-Yue and Zhou, Shuang-Yong",
    title = "{New positivity bounds from full crossing symmetry}",
    eprint = "2011.02400",
    archivePrefix = "arXiv",
    primaryClass = "hep-th",
    doi = "10.1007/JHEP05(2021)255",
    journal = "JHEP",
    volume = "05",
    pages = "255",
    year = "2021"
}

@article{Sinha:2020win,
    author = "Sinha, Aninda and Zahed, Ahmadullah",
    title = "{Crossing Symmetric Dispersion Relations in Quantum Field Theories}",
    eprint = "2012.04877",
    archivePrefix = "arXiv",
    primaryClass = "hep-th",
    doi = "10.1103/PhysRevLett.126.181601",
    journal = "Phys. Rev. Lett.",
    volume = "126",
    number = "18",
    pages = "181601",
    year = "2021"
}

@article{EliasMiro:2022xaa,
    author = "Elias Miro, Joan and Guerrieri, Andrea and Gumus, Mehmet Asim",
    title = "{Bridging positivity and S-matrix bootstrap bounds}",
    eprint = "2210.01502",
    archivePrefix = "arXiv",
    primaryClass = "hep-th",
    doi = "10.1007/JHEP05(2023)001",
    journal = "JHEP",
    volume = "05",
    pages = "001",
    year = "2023"
}

@article{Bonnefoy:2025uzf,
    author = "Bonnefoy, Quentin and Cort{\'e}s, Vicente and Gendy, Emanuele and Grojean, Christophe and von Merkl, Karim Ritter and Pilatus, Paula Naomi",
    title = "{Geometry of effective field theory positivity cones}",
    eprint = "2508.18165",
    archivePrefix = "arXiv",
    primaryClass = "math-ph",
    doi = "10.1007/s00023-026-01687-y",
    month = "8",
    year = "2025"
}

@article{Bellazzini:2021oaj,
    author = "Bellazzini, Brando and Riembau, Marc and Riva, Francesco",
    title = "{IR side of positivity bounds}",
    eprint = "2112.12561",
    archivePrefix = "arXiv",
    primaryClass = "hep-th",
    doi = "10.1103/PhysRevD.106.105008",
    journal = "Phys. Rev. D",
    volume = "106",
    number = "10",
    pages = "105008",
    year = "2022"
}

@article{Chang:2025cxc,
    author = "Chang, Cyuan-Han and Parra-Martinez, Julio",
    title = "{Graviton loops and negativity}",
    eprint = "2501.17949",
    archivePrefix = "arXiv",
    primaryClass = "hep-th",
    doi = "10.1007/JHEP08(2025)175",
    journal = "JHEP",
    volume = "08",
    pages = "175",
    year = "2025"
}

@article{Weinberg:1965nx,
    author = "Weinberg, Steven",
    title = "{Infrared photons and gravitons}",
    doi = "10.1103/PhysRev.140.B516",
    journal = "Phys. Rev.",
    volume = "140",
    pages = "B516--B524",
    year = "1965"
}

@article{Akhoury:2011kq,
    author = "Akhoury, Ratindranath and Saotome, Ryo and Sterman, George",
    title = "{Collinear and Soft Divergences in Perturbative Quantum Gravity}",
    eprint = "1109.0270",
    archivePrefix = "arXiv",
    primaryClass = "hep-th",
    reportNumber = "YITP-SB-11-31",
    doi = "10.1103/PhysRevD.84.104040",
    journal = "Phys. Rev. D",
    volume = "84",
    pages = "104040",
    year = "2011"
}

@article{Naculich:2011ry,
    author = "Naculich, Stephen G. and Schnitzer, Howard J.",
    title = "{Eikonal methods applied to gravitational scattering amplitudes}",
    eprint = "1101.1524",
    archivePrefix = "arXiv",
    primaryClass = "hep-th",
    reportNumber = "BOW-PH-148, BRX-TH-628",
    doi = "10.1007/JHEP05(2011)087",
    journal = "JHEP",
    volume = "05",
    pages = "087",
    year = "2011"
}

@article{Donoghue:1994dn,
    author = "Donoghue, John F.",
    title = "{General relativity as an effective field theory: The leading quantum corrections}",
    eprint = "gr-qc/9405057",
    archivePrefix = "arXiv",
    reportNumber = "UMHEP-408",
    doi = "10.1103/PhysRevD.50.3874",
    journal = "Phys. Rev. D",
    volume = "50",
    pages = "3874--3888",
    year = "1994"
}

@article{Shapiro:1964uw,
    author = "Shapiro, Irwin I.",
    title = "{Fourth Test of General Relativity}",
    doi = "10.1103/PhysRevLett.13.789",
    journal = "Phys. Rev. Lett.",
    volume = "13",
    pages = "789--791",
    year = "1964"
}

@article{Bertotti:2003rm,
    author = "Bertotti, B. and Iess, L. and Tortora, P.",
    title = "{A test of general relativity using radio links with the Cassini spacecraft}",
    doi = "10.1038/nature01997",
    journal = "Nature",
    volume = "425",
    pages = "374--376",
    year = "2003"
}

@article{Kulish:1970ut,
    author = "Kulish, P. P. and Faddeev, L. D.",
    title = "{Asymptotic conditions and infrared divergences in quantum electrodynamics}",
    reportNumber = "D70-07927",
    doi = "10.1007/BF01066485",
    journal = "Theor. Math. Phys.",
    volume = "4",
    pages = "745",
    year = "1970"
}

@article{Ware:2013zja,
    author = "Ware, John and Saotome, Ryo and Akhoury, Ratindranath",
    title = "{Construction of an asymptotic S matrix for perturbative quantum gravity}",
    eprint = "1308.6285",
    archivePrefix = "arXiv",
    primaryClass = "hep-th",
    doi = "10.1007/JHEP10(2013)159",
    journal = "JHEP",
    volume = "10",
    pages = "159",
    year = "2013"
}

@article{Sahoo:2018lxl,
    author = "Sahoo, Biswajit and Sen, Ashoke",
    title = "{Classical and Quantum Results on Logarithmic Terms in the Soft Theorem in Four Dimensions}",
    eprint = "1808.03288",
    archivePrefix = "arXiv",
    primaryClass = "hep-th",
    doi = "10.1007/JHEP02(2019)086",
    journal = "JHEP",
    volume = "02",
    pages = "086",
    year = "2019"
}

@article{Haring:2022cyf,
    author = {H{\"a}ring, Kelian and Zhiboedov, Alexander},
    title = "{Gravitational Regge bounds}",
    eprint = "2202.08280",
    archivePrefix = "arXiv",
    primaryClass = "hep-th",
    reportNumber = "CERN-TH-2022-016",
    doi = "10.21468/SciPostPhys.16.1.034",
    journal = "SciPost Phys.",
    volume = "16",
    number = "1",
    pages = "034",
    year = "2024"
}

@article{Creminelli:2022onn,
    author = "Creminelli, Paolo and Janssen, Oliver and Senatore, Leonardo",
    title = "{Positivity bounds on effective field theories with spontaneously broken Lorentz invariance}",
    eprint = "2207.14224",
    archivePrefix = "arXiv",
    primaryClass = "hep-th",
    doi = "10.1007/JHEP09(2022)201",
    journal = "JHEP",
    volume = "09",
    pages = "201",
    year = "2022"
}

@article{Creminelli:2023kze,
    author = "Creminelli, Paolo and Delladio, Matteo and Janssen, Oliver and Longo, Alessandro and Senatore, Leonardo",
    title = "{Non-analyticity of the S-matrix with spontaneously broken Lorentz invariance}",
    eprint = "2312.08441",
    archivePrefix = "arXiv",
    primaryClass = "hep-th",
    doi = "10.1007/JHEP06(2024)201",
    journal = "JHEP",
    volume = "06",
    pages = "201",
    year = "2024"
}

@article{Witten:1998qj,
    author = "Witten, Edward",
    title = "{Anti de Sitter space and holography}",
    eprint = "hep-th/9802150",
    archivePrefix = "arXiv",
    reportNumber = "IASSNS-HEP-98-15",
    doi = "10.4310/ATMP.1998.v2.n2.a2",
    journal = "Adv. Theor. Math. Phys.",
    volume = "2",
    pages = "253--291",
    year = "1998"
}

@article{Heemskerk:2009pn,
    author = "Heemskerk, Idse and Penedones, Joao and Polchinski, Joseph and Sully, James",
    title = "{Holography from Conformal Field Theory}",
    eprint = "0907.0151",
    archivePrefix = "arXiv",
    primaryClass = "hep-th",
    reportNumber = "NSF-KITP-09-110",
    doi = "10.1088/1126-6708/2009/10/079",
    journal = "JHEP",
    volume = "10",
    pages = "079",
    year = "2009"
}

@article{Penedones:2010ue,
    author = "Penedones, Joao",
    title = "{Writing CFT correlation functions as AdS scattering amplitudes}",
    eprint = "1011.1485",
    archivePrefix = "arXiv",
    primaryClass = "hep-th",
    doi = "10.1007/JHEP03(2011)025",
    journal = "JHEP",
    volume = "03",
    pages = "025",
    year = "2011"
}

@article{Caron-Huot:2021enk,
    author = "Caron-Huot, Simon and Mazac, Dalimil and Rastelli, Leonardo and Simmons-Duffin, David",
    title = "{AdS bulk locality from sharp CFT bounds}",
    eprint = "2106.10274",
    archivePrefix = "arXiv",
    primaryClass = "hep-th",
    doi = "10.1007/JHEP11(2021)164",
    journal = "JHEP",
    volume = "11",
    pages = "164",
    year = "2021"
}

@article{Allen:1987tz,
    author = "Allen, Bruce and Folacci, Antoine",
    title = "{The Massless Minimally Coupled Scalar Field in De Sitter Space}",
    reportNumber = "Print-87-0163 (MEUDON)",
    doi = "10.1103/PhysRevD.35.3771",
    journal = "Phys. Rev. D",
    volume = "35",
    pages = "3771",
    year = "1987"
}

@article{Folacci:1992xc,
    author = "Folacci, Antoine",
    title = "{BRST quantization of the massless minimally coupled scalar field in de Sitter space: Zero modes, euclideanization and quantization}",
    eprint = "0911.2064",
    archivePrefix = "arXiv",
    primaryClass = "gr-qc",
    doi = "10.1103/PhysRevD.46.2553",
    journal = "Phys. Rev. D",
    volume = "46",
    pages = "2553--2559",
    year = "1992"
}

@article{Podolsky:2001vu,
    author = "Podolsky, Jiri and Ortaggio, M.",
    title = "{Symmetries and geodesics in (anti-)de Sitter space-times with nonexpanding impulsive waves}",
    eprint = "gr-qc/0105065",
    archivePrefix = "arXiv",
    doi = "10.1088/0264-9381/18/14/307",
    journal = "Class. Quant. Grav.",
    volume = "18",
    pages = "2689--2706",
    year = "2001"
}

@article{Noumi:2022wwf,
    author = "Noumi, Toshifumi and Tokuda, Junsei",
    title = "{Finite energy sum rules for gravitational Regge amplitudes}",
    eprint = "2212.08001",
    archivePrefix = "arXiv",
    primaryClass = "hep-th",
    reportNumber = "KOBE-COSMO-22-20, CTPU-PTC-22-29",
    doi = "10.1007/JHEP06(2023)032",
    journal = "JHEP",
    volume = "06",
    pages = "032",
    year = "2023"
}

@article{Bellazzini:2025bay,
    author = "Bellazzini, B. and Berman, J. and Isabella, G. and Riva, F. and Romano, M. and Sciotti, F.",
    title = "{Positivity with long-range interactions}",
    eprint = "2512.13780",
    archivePrefix = "arXiv",
    primaryClass = "hep-th",
    doi = "10.1007/JHEP07(2026)235",
    journal = "JHEP",
    volume = "07",
    pages = "235",
    year = "2026"
}

@article{Mazzeo:1987resolvent,
    author = {Mazzeo, Rafe R. and Melrose, Richard B.},
    title = {Meromorphic extension of the resolvent on complete spaces with asymptotically constant negative curvature},
    doi = {10.1016/0022-1236(87)90097-8},
    journal = {J. Funct. Anal.},
    volume = {75},
    number = {2},
    pages = {260--310},
    year = {1987}
}

@article{Guillarmou:2005resolvent,
    author = {Guillarmou, Colin},
    title = {Meromorphic properties of the resolvent on asymptotically hyperbolic manifolds},
    eprint = {math/0311424},
    archivePrefix = {arXiv},
    primaryClass = {math.SP},
    doi = {10.1215/S0012-7094-04-12911-2},
    journal = {Duke Math. J.},
    volume = {129},
    number = {1},
    pages = {1--37},
    year = {2005}
}

@article{Oda:2021tiv,
    author = "Oda, Kin-ya and Wada, Juntaro",
    title = "{A complete set of Lorentz-invariant wave packets and modified uncertainty relation}",
    eprint = "2104.01798",
    archivePrefix = "arXiv",
    primaryClass = "hep-th",
    doi = "10.1140/epjc/s10052-021-09482-1",
    journal = "Eur. Phys. J. C",
    volume = "81",
    number = "8",
    pages = "751",
    year = "2021",
    note = "[Erratum: Eur.Phys.J.C 83, 135 (2023)]"
}

@article{Oda:2023qek,
    author = "Oda, Kin-ya and Wada, Juntaro",
    title = "{Lorentz-covariant spinor wave packet}",
    eprint = "2307.05932",
    archivePrefix = "arXiv",
    primaryClass = "hep-th",
    doi = "10.1103/PhysRevD.110.076001",
    journal = "Phys. Rev. D",
    volume = "110",
    number = "7",
    pages = "076001",
    year = "2024"
}

@article{Haring:2024wyz,
    author = {H{\"a}ring, Kelian and Zhiboedov, Alexander},
    title = "{What is the graviton pole made of?}",
    eprint = "2410.21499",
    archivePrefix = "arXiv",
    primaryClass = "hep-th",
    reportNumber = "CERN-TH-2024-178",
    month = "10",
    year = "2024"
}

@article{Amati:1987uf,
    author = "Amati, D. and Ciafaloni, M. and Veneziano, G.",
    title = "{Classical and Quantum Gravity Effects from Planckian Energy Superstring Collisions}",
    reportNumber = "CERN-TH-4886/87",
    doi = "10.1142/S0217751X88000710",
    journal = "Int. J. Mod. Phys. A",
    volume = "3",
    pages = "1615--1661",
    year = "1988"
}

@article{Brower:2006ea,
    author = "Brower, Richard C. and Polchinski, Joseph and Strassler, Matthew J. and Tan, Chung-I",
    title = "{The Pomeron and gauge/string duality}",
    eprint = "hep-th/0603115",
    archivePrefix = "arXiv",
    reportNumber = "BROWN-HET-1462",
    doi = "10.1088/1126-6708/2007/12/005",
    journal = "JHEP",
    volume = "12",
    pages = "005",
    year = "2007"
}

@article{DiVecchia:2023frv,
    author = "Di Vecchia, Paolo and Heissenberg, Carlo and Russo, Rodolfo and Veneziano, Gabriele",
    title = "{The gravitational eikonal: From particle, string and brane collisions to black-hole encounters}",
    eprint = "2306.16488",
    archivePrefix = "arXiv",
    primaryClass = "hep-th",
    reportNumber = "CERN-TH-2023-108, NORDITA 2023-026, QMUL-PH-23-09, UUITP-14/23",
    doi = "10.1016/j.physrep.2024.06.002",
    journal = "Phys. Rept.",
    volume = "1083",
    pages = "1--169",
    year = "2024"
}

@article{Beadle:2025cdx,
    author = "Beadle, Carl and Isabella, Giulia and Perrone, Davide and Ricossa, Sara and Riva, Francesco and Serra, Francesco",
    title = "{The EFT bootstrap at finite M$_{PL}$}",
    eprint = "2501.18465",
    archivePrefix = "arXiv",
    primaryClass = "hep-th",
    doi = "10.1007/JHEP06(2025)209",
    journal = "JHEP",
    volume = "06",
    pages = "209",
    year = "2025"
}

@article{White:2011yy,
    author = "White, Chris D.",
    title = "{Factorization Properties of Soft Graviton Amplitudes}",
    eprint = "1103.2981",
    archivePrefix = "arXiv",
    primaryClass = "hep-th",
    doi = "10.1007/JHEP05(2011)060",
    journal = "JHEP",
    volume = "05",
    pages = "060",
    year = "2011"
}

@article{Choi:2017ylo,
    author = "Choi, Sangmin and Akhoury, Ratindranath",
    title = "{BMS Supertranslation Symmetry Implies Faddeev-Kulish Amplitudes}",
    eprint = "1712.04551",
    archivePrefix = "arXiv",
    primaryClass = "hep-th",
    doi = "10.1007/JHEP02(2018)171",
    journal = "JHEP",
    volume = "02",
    pages = "171",
    year = "2018"
}

@article{Choi:2019rlz,
    author = "Choi, Sangmin and Akhoury, Ratindranath",
    title = "{Subleading soft dressings of asymptotic states in QED and perturbative quantum gravity}",
    eprint = "1907.05438",
    archivePrefix = "arXiv",
    primaryClass = "hep-th",
    doi = "10.1007/JHEP09(2019)031",
    journal = "JHEP",
    volume = "09",
    pages = "031",
    year = "2019"
}

@article{DiVecchia:2021bdo,
    author = "Di Vecchia, Paolo and Heissenberg, Carlo and Russo, Rodolfo and Veneziano, Gabriele",
    title = "{The eikonal approach to gravitational scattering and radiation at $ \mathcal{O} $(G$^{3}$)}",
    eprint = "2104.03256",
    archivePrefix = "arXiv",
    primaryClass = "hep-th",
    reportNumber = "CERN-TH-2021-046, NORDITA 2021-028, QMUL-PH-21-17",
    doi = "10.1007/JHEP07(2021)169",
    journal = "JHEP",
    volume = "07",
    pages = "169",
    year = "2021"
}

@article{Csaki:2022tvb,
    author = "Cs{\'a}ki, Csaba and Dong, Zi-Yu and Telem, Ofri and Terning, John and Yankielowicz, Shimon",
    title = "{Dressed vs. pairwise states, and the geometric phase of monopoles and charges}",
    eprint = "2209.03369",
    archivePrefix = "arXiv",
    primaryClass = "hep-th",
    doi = "10.1007/JHEP02(2023)211",
    journal = "JHEP",
    volume = "02",
    pages = "211",
    year = "2023"
}

@article{Donnay:2026urd,
    author = "Donnay, Laura and Herfray, Yannick",
    title = "{Infrared physics of QED and gravity from representation theory}",
    eprint = "2603.06297",
    archivePrefix = "arXiv",
    primaryClass = "hep-th",
    month = "3",
    year = "2026"
}

@article{Duary:2022afn,
    author = "Duary, Sarthak",
    title = "{AdS correction to the Faddeev-Kulish state: migrating from the flat peninsula}",
    eprint = "2212.09509",
    archivePrefix = "arXiv",
    primaryClass = "hep-th",
    doi = "10.1007/JHEP05(2023)079",
    journal = "JHEP",
    volume = "05",
    pages = "079",
    year = "2023"
}

@article{Wald:1983ky,
    author = "Wald, Robert M.",
    title = "{Asymptotic behavior of homogeneous cosmological models in the presence of a positive cosmological constant}",
    doi = "10.1103/PhysRevD.28.2118",
    journal = "Phys. Rev. D",
    volume = "28",
    pages = "2118--2120",
    year = "1983"
}

@article{Cornalba:2006xk,
    author = "Cornalba, Lorenzo and Costa, Miguel S. and Penedones, Joao and Schiappa, Ricardo",
    title = "{Eikonal Approximation in AdS/CFT: From Shock Waves to Four-Point Functions}",
    eprint = "hep-th/0611122",
    archivePrefix = "arXiv",
    reportNumber = "ROM2F-2006-25, LPTENS-06-49, CERN-PH-TH-2006-233",
    doi = "10.1088/1126-6708/2007/08/019",
    journal = "JHEP",
    volume = "08",
    pages = "019",
    year = "2007"
}

@article{Bellazzini:2022wzv,
    author = "Bellazzini, Brando and Isabella, Giulia and Riva, Massimiliano Maria",
    title = "{Classical vs quantum eikonal scattering and its causal structure}",
    eprint = "2211.00085",
    archivePrefix = "arXiv",
    primaryClass = "hep-th",
    doi = "10.1007/JHEP04(2023)023",
    journal = "JHEP",
    volume = "04",
    pages = "023",
    year = "2023"
}

@article{Cornalba:2007zb,
    author = "Cornalba, Lorenzo and Costa, Miguel S. and Penedones, Joao",
    title = "{Eikonal approximation in AdS/CFT: Resumming the gravitational loop expansion}",
    eprint = "0707.0120",
    archivePrefix = "arXiv",
    primaryClass = "hep-th",
    reportNumber = "ROM2F-2007-11, LPTENS-07-27",
    doi = "10.1088/1126-6708/2007/09/037",
    journal = "JHEP",
    volume = "09",
    pages = "037",
    year = "2007"
}

@article{Johnson:2015bump,
    author = {Johnson, Steven G.},
    title = {Saddle-point integration of {$C_\infty$} ``bump'' functions},
    eprint = {1508.04376},
    archivePrefix = {arXiv},
    primaryClass = {math.CV},
    year = {2015}
}

@article{Shevtsova:2014exponent,
    author = {Shevtsova, Irina},
    title = {On the accuracy of the approximation of the complex exponent by the first terms of its Taylor expansion with applications},
    eprint = {1301.2783},
    archivePrefix = {arXiv},
    primaryClass = {math.PR},
    doi = {10.1016/j.jmaa.2014.03.075},
    journal = {J. Math. Anal. Appl.},
    volume = {418},
    number = {1},
    pages = {185--210},
    year = {2014}
}

@article{Bressloff:2024diffusion,
    author = {Bressloff, Paul C.},
    title = {Cellular diffusion processes in singularly perturbed domains},
    eprint = {2407.05747},
    archivePrefix = {arXiv},
    primaryClass = {math.AP},
    doi = {10.1007/s00285-024-02160-2},
    journal = {Journal of Mathematical Biology},
    volume = {89},
    number = {6},
    pages = {58},
    year = {2024}
}

@misc{Dong:ProjectA,
    author = "Dong, Ziyu",
    title = "{Crossing and Positivity for Finite-Resolution Dressed Amplitudes}",
    howpublished = "In preparation"
}

@article{Cohl:2012hyperbolic,
    author = {Cohl, Howard S. and Kalnins, Ernie G.},
    title = {Fundamental solution of the Laplacian in the hyperboloid model of hyperbolic geometry},
    eprint = {1201.4406},
    archivePrefix = {arXiv},
    primaryClass = {math-ph},
    year = {2012}
}

@article{DeWolfe:1999cp,
    author = "DeWolfe, O. and Freedman, D. Z. and Gubser, S. S. and Karch, A.",
    title = "{Modeling the fifth-dimension with scalars and gravity}",
    eprint = "hep-th/9909134",
    archivePrefix = "arXiv",
    reportNumber = "HUTP-99-A048, MIT-CTP-2903",
    doi = "10.1103/PhysRevD.62.046008",
    journal = "Phys. Rev. D",
    volume = "62",
    pages = "046008",
    year = "2000"
}

@article{Pasiecznik:2025eqc,
    author = "Pasiecznik, Celina",
    title = "{Bootstrapping Gravity with Crossing Symmetric Dispersion Relations}",
    eprint = "2506.09884",
    archivePrefix = "arXiv",
    primaryClass = "hep-th",
    month = "6",
    year = "2025"
}

@article{Melville:2013qca,
    author = "Melville, S. and Naculich, S. G. and Schnitzer, H. J. and White, C. D.",
    title = "{Wilson line approach to gravity in the high energy limit}",
    eprint = "1306.6019",
    archivePrefix = "arXiv",
    primaryClass = "hep-th",
    reportNumber = "BRX-TH-667, BOW-PH-157",
    doi = "10.1103/PhysRevD.89.025009",
    journal = "Phys. Rev. D",
    volume = "89",
    number = "2",
    pages = "025009",
    year = "2014"
}

@article{Fitzpatrick:2011jn,
    author = "Fitzpatrick, A. Liam and Kaplan, Jared",
    title = "{Scattering States in AdS/CFT}",
    eprint = "1104.2597",
    archivePrefix = "arXiv",
    primaryClass = "hep-th",
    reportNumber = "SLAC-PUB-14507",
    month = "4",
    year = "2011"
}

@inproceedings{Asorey:2006ij,
    author = "Asorey, M. and Ibort, A. and Marmo, G.",
    title = "{Path integrals and boundary conditions}",
    booktitle = "{32nd International Meeting on Fundamental Physics}",
    eprint = "quant-ph/0609023",
    archivePrefix = "arXiv",
    month = "9",
    year = "2006"
}

@article{Bern:2021ppb,
    author = "Bern, Zvi and Kosmopoulos, Dimitrios and Zhiboedov, Alexander",
    title = "{Gravitational effective field theory islands, low-spin dominance, and the four-graviton amplitude}",
    eprint = "2103.12728",
    archivePrefix = "arXiv",
    primaryClass = "hep-th",
    reportNumber = "CERN-TH-2021-035",
    doi = "10.1088/1751-8121/ac0e51",
    journal = "J. Phys. A",
    volume = "54",
    number = "34",
    pages = "344002",
    year = "2021"
}

@article{Csaki:1998qr,
    author = "Csaki, Csaba and Ooguri, Hirosi and Oz, Yaron and Terning, John",
    title = "{Glueball mass spectrum from supergravity}",
    eprint = "hep-th/9806021",
    archivePrefix = "arXiv",
    reportNumber = "UCB-PTH-98-30, LBL-41875, LBNL-41875, NSF-ITP-98-068",
    doi = "10.1088/1126-6708/1999/01/017",
    journal = "JHEP",
    volume = "01",
    pages = "017",
    year = "1999"
}

@article{Brower:2010wf,
    author = "Brower, Richard C. and Djuric, Marko and Sarcevic, Ina and Tan, Chung-I",
    title = "{String-Gauge Dual Description of Deep Inelastic Scattering at Small-$x$}",
    eprint = "1007.2259",
    archivePrefix = "arXiv",
    primaryClass = "hep-ph",
    doi = "10.1007/JHEP11(2010)051",
    journal = "JHEP",
    volume = "11",
    pages = "051",
    year = "2010"
}

@article{Serafin:2022Dirichlet,
    author = {Serafin, Grzegorz},
    title = {Laplace Dirichlet heat kernels in convex domains},
    eprint = {2110.06147},
    archivePrefix = {arXiv},
    primaryClass = {math.AP},
    doi = {10.1016/j.jde.2022.01.026},
    journal = {Journal of Differential Equations},
    volume = {314},
    pages = {700--732},
    year = {2022}
}

@article{Reuter:1996cp,
    author = "Reuter, M.",
    title = "{Nonperturbative evolution equation for quantum gravity}",
    eprint = "hep-th/9605030",
    archivePrefix = "arXiv",
    reportNumber = "DESY-96-080",
    doi = "10.1103/PhysRevD.57.971",
    journal = "Phys. Rev. D",
    volume = "57",
    pages = "971--985",
    year = "1998"
}

@article{Bonanno:2020bil,
    author = "Bonanno, Alfio and Eichhorn, Astrid and Gies, Holger and Pawlowski, Jan M. and Percacci, Roberto and Reuter, Martin and Saueressig, Frank and Vacca, Gian Paolo",
    title = "{Critical reflections on asymptotically safe gravity}",
    eprint = "2004.06810",
    archivePrefix = "arXiv",
    primaryClass = "gr-qc",
    doi = "10.3389/fphy.2020.00269",
    journal = "Front. in Phys.",
    volume = "8",
    pages = "269",
    year = "2020"
}

@article{Ashtekar:2004eh,
    author = "Ashtekar, Abhay and Lewandowski, Jerzy",
    title = "{Background independent quantum gravity: A Status report}",
    eprint = "gr-qc/0404018",
    archivePrefix = "arXiv",
    doi = "10.1088/0264-9381/21/15/R01",
    journal = "Class. Quant. Grav.",
    volume = "21",
    pages = "R53",
    year = "2004"
}

@article{Engle:2007wy,
    author = "Engle, Jonathan and Livine, Etera and Pereira, Roberto and Rovelli, Carlo",
    title = "{LQG vertex with finite Immirzi parameter}",
    eprint = "0711.0146",
    archivePrefix = "arXiv",
    primaryClass = "gr-qc",
    doi = "10.1016/j.nuclphysb.2008.02.018",
    journal = "Nucl. Phys. B",
    volume = "799",
    pages = "136--149",
    year = "2008"
}

@article{Caron-Huot:2024lbf,
    author = "Caron-Huot, Simon and Li, Yue-Zhou",
    title = "{Gravity and a universal cutoff for field theory}",
    eprint = "2408.06440",
    archivePrefix = "arXiv",
    primaryClass = "hep-th",
    doi = "10.1007/JHEP02(2025)115",
    journal = "JHEP",
    volume = "02",
    pages = "115",
    year = "2025"
}
\endgroup

\end{document}